\documentclass[aps,prd,groupedaddress,preprintnumbers]{revtex4}
\usepackage{axodraw,color}
\usepackage{pstricks}
\usepackage{epsfig}

\newcommand{\vs}[1]{\rule[- #1 mm]{0mm}{#1 mm}}
\newcommand{\be}{\vs{2}\begin{equation}}
\newcommand{\eq}{\vs{2}\begin{equation}}
\newcommand{\en}{\\[2mm]\end{equation}}
\newcommand{\ee}{\\[2mm]\end{equation}}
\newcommand{\bea}{\begin{eqnarray}}
\newcommand{\ena}{\end{eqnarray}}
\newcommand{\eea}{\end{eqnarray}}

\begin{document}

\renewcommand{\theequation}{\Roman{section}.\arabic{equation}}

\preprint{LPT-ORSAY/12-18}

\preprint{CPT-P002-2012}

\title{Two-loop representations of low-energy pion form factors and $\pi\pi$ scattering phases
in the presence of isospin breaking}

\author{S. Descotes-Genon}
\email{descotes@th.u-psud.fr}
\affiliation{Laboratoire de Physique Th\'eorique, CNRS/Univ. Paris-Sud 11 
(UMR 8627), 91405 Orsay Cedex, France}

\author{M. Knecht}
\email{knecht@cpt.univ-mrs.fr} 
\affiliation{Centre de Physique Th\'eorique~-~UMR 7332\\
 Aix-Marseille Univ.~-~CNRS~-~Univ. du Sud Toulon-Var
CNRS-Luminy Case 907~-~13288 Marseille Cedex 9, France}


\date{\today}

\begin{abstract}
\noindent
Dispersive representations of the $\pi\pi$ scattering amplitudes and pion form factors,
valid at two-loop accuracy in the low-energy expansion, are constructed in the presence of isospin-breaking
effects induced by the difference between the charged and neutral pion masses.
Analytical expressions for the corresponding phases of the scalar and vector pion form factors 
are computed. It is shown that each of
these phases consists of the sum of a ``universal'' part and a form-factor dependent contribution.
The first one is entirely determined in terms of the $\pi\pi$ scattering amplitudes alone, and reduces 
to the phase satisfying Watson's theorem in the isospin limit. The second one can be sizeable,
although it vanishes in the same limit. 
The dependence of these isospin corrections with respect to 
the parameters of the subthreshold expansion of the $\pi\pi$ amplitude is studied, 
and an equivalent representation in terms
of the $S$-wave scattering lengths is also briefly presented and discussed. 
In addition, partially analytical expressions for the two-loop form factors and $\pi\pi$ scattering amplitudes
in the presence of isospin breaking are provided.

\end{abstract}

\maketitle

\section{INTRODUCTION} \label{intro}
\setcounter{equation}{0}

In recent years, our knowledge of low-energy pion-pion scattering 
has improved in a very significant way and in several respects. Firstly,
the high precision $K^+_{e4}$ experiments performed at the BNL AGS by the E865 
experiment~\cite{Pislak:2001bf,Pislak:2003sv} 
and, more recently, by the NA48/2 collaboration~\cite{Batley:2007zz,Batley:2010zza} at the CERN SPS,
have provided very accurate determinations of the
difference $\delta_0^0 - \delta_1^1$ of the  pion-pion phase 
shifts in the $S$ and $P$ waves in the energy range between threshold and the kaon mass.
Next, one should mention the measurement of the invariant mass distribution
in $K^{\pm} \to \pi^\pm \pi^0 \pi^0$ decays~\cite{Batley:2005ax,Batley:2000zz}, that
gives information on the $S$-wave $\pi\pi$ scattering lengths \cite{Cabibbo:2004gq} 
(see also \cite{Cabibbo:2005ez,Gamiz:2006km,Gasser:2011ju}). Finally, forthcoming
analyses of the data collected by the NA48/2 experiment on the $K^+_{e4}$ decay channel 
into a pair of neutral pions (for preliminary reports, see \cite{Masetti:2007de,NA48-2_Ke4_cusp}), or on the 
$K^0_L \to \pi^0 \pi^0 \pi^0$ decay mode~\cite{Madigozhin:2009zz}, 
together with the measurement of the pionium lifetime by the DIRAC collaboration~\cite{Yazkov:2009zz},
should provide additional information, and might sharpen the picture even more. 
In the meantime, the accuracy obtained on $K_{\ell 4}$ decays from NA48/2 implies 
that these data clearly drive the current determination of the difference between 
the $S$ and $P$ phase shifts at low energies, and in particular of the two scattering 
lengths $a_0^0$ and $a_0^2$, for which very accurate predictions are available~\cite{Colangelo:2001df}. This provides a 
particularly stringent test of two-flavour chiral perturbation theory~\cite{Gasser:1983yg}, and its underlying 
assumptions~\cite{DescotesGenon:2001tn}.

In order to extract relevant information on low-energy pion-pion scattering
from the above processes, it has become mandatory to take isospin violations into
account. This is certainly quite easy to understand in the case of
the $K^{\pm} \to \pi^\pm \pi^0 \pi^0$ decay, where one exploits the presence
of a unitarity cusp in the invariant $\pi^0\pi^0$ mass distribution, which occurs only
if the masses of the charged and neutral pions differ~\cite{cusp61,Cabibbo:2004gq}. Perhaps somewhat
more unexpectedly, isospin-violating corrections proved also of importance \cite{Gasser:2007de}
in the analysis of the $K^+_{e4}$ data, in order to account for the high
precision reached by the recent NA48/2 experiment, and to make comparison with theory meaningful
\cite{Colangelo:2007df,Colangelo:2008sm}.
Actually, once isospin corrections are applied also to the E865 data, there
remains a disagreement with NA48/2~\cite{Colangelo:2007df}, whose origin seems to lie in the original analysis
performed by the E865 collaboration (for details, see the errata quoted under refs.~\cite{Pislak:2001bf,Pislak:2003sv}). 
Anyway,  the analysis of the full data set collected 
by NA48/2 has by now completely superseded the E865 results, and one should focus on the former to study
pion-pion scattering from $K^+_{e4}$ decays.

In the present paper, we propose to address the issue of isospin-violating effects in
low-energy pion-pion interactions using an approach based on a dispersive
construction of the various $\pi\pi$ scattering amplitudes and pion form factors 
in the presence of isospin breaking. 
Ultimately, we wish to extend this program~\cite{wip} to the $K_{\ell 4}$ form factors
analysed in the NA48/2 experiment. Before undertaking this enterprise and hitting
the full complexity of this four-body decay, we want to demonstrate its feasibility 
and exhibit the general features of such a method by considering the somewhat simpler setting 
provided by the scalar and vector form factors of the pion.

As far as the amplitude for elastic $\pi\pi$ scattering in the isospin limit is concerned, the 
general framework has been laid down in ref.~\cite{FSS93}, and the explicit construction of the two-loop 
amplitude has subsequently been performed along these lines in detail in ref.~\cite{KMSF95}. Concerning the pion form factors,
the corresponding dispersive representations in the framework of the chiral expansion
have been studied in ref.~\cite{GasserMeissner91} in the isospin limit, but only one-loop expressions were given
in analytical form. Full two-loop expressions of the vector form factors have been obtained by 
integrating the corresponding dispersive integrals in ref.~\cite{Colangelo96}.
In ref.~\cite{Bijnens98} the two-loop expressions of the vector and scalar form-factors have also been
obtained in the absence of isospin violation by the direct evaluation of Feynman graphs generated from the effective chiral Lagrangian
at next-to-next-to-leading order. A similar calculation for the pion-pion scattering amplitude
in the isospin limit has been achieved in ref.~\cite{pi-pi2loops}. Finally, let us also mention
that the reconstruction theorem for elastic $\pi\pi$ scattering in the isospin limit of ref.~\cite{FSS93}
was extended by the authors of ref. \cite{NovZdra08} to the whole set of scattering amplitudes involving the mesons 
of the lightest pseudoscalar octet. Applications of this framework to the decay modes $P \rightarrow \pi\pi\pi$,
with $P = K , \eta$, have also been considered \cite{P_to_3pi,eta_to_3pi}.

These dispersive constructions generate subtraction polynomials with unspecified coefficients. 
The latter are in one-to-one correspondence with the appropriate combinations of low-energy
constants and chiral logarithms that would be encountered in a calculation of the corresponding Feynman
diagrams generated by the chiral lagrangian.
In the case of the form factors, these coefficients may be identified with their slopes and curvatures. 
In the case of the $\pi\pi$ scattering amplitudes, they can be expressed in terms of the subthreshold parameters 
occurring in the  expansions of these amplitudes as Taylor series around the center of the Mandelstam triangle. 
This was the option considered in the isospin-symmetric case in ref.~\cite{KMSF95}. By
no means, however, is this choice a necessity. It has, for instance, become customary to rather let the scattering lengths play a prominent role.
They have a more direct physical interpretation than the subthreshold parameters, and are thus considered 
as more ``experimentalist friendly''. We will therefore also provide expressions where the subtraction polynomials
are given in terms of the two $S$-wave $I=0$ and $I=2$ scattering lengths, $a_0^0$ and $a_0^2$, in the isospin limit.
In the isospin-symmetric situation, this provides an alternative to the choice made in ref.~\cite{KMSF95}.
Of course, taking the expressions at two-loop order provided in the latter reference, one could
convert the expressions for the $\pi\pi$ scattering amplitude given there to the one presented here in terms
of the scattering lengths. The two formulations are equivalent, up to corrections that are of higher order.
In the situation where isospin is broken, this allows us to discuss
the size of the corresponding corrections to the phases of the form factors in terms of $a_0^0$ and $a_0^2$.
This second option is of course the most interesting in the present context, where these scattering
lengths are the quantities one would eventually like to determine from the data. It is thus important
that the corrections due to isospin breaking are not studied for a fixed {\it a priori} value for them. Indeed, 
given the precision reached by the latest experiments, one ought to perform a quantitative evaluation of the 
possible bias introduced if isospin corrections are evaluated for fixed values of these scattering lengths. 
This provides another motivation for the present work.

Here we will mainly concentrate on the phases of the pion form factors. Full two-loop expressions for the scattering amplitudes 
and form factors themselves require the evaluation of dispersion integrals corresponding to specific topologies of two-loop
three-point Feynman diagrams of the non-factorizing type (``acnode'' of ``fish'' diagrams, cf. fig. \ref{fig3}). 
Explicit analytical expressions for them do not seem to be available in the literature
in the cases where several distinct masses are present. We therefore present only partially
analytical expressions for the scattering amplitudes and form factors.
Note that a similar situation arises in
the evaluation of the $SU(3)$ vector \cite{Bijnens:2002hp} and scalar \cite{Bijnens:2003xg} form factors 
at two loops without isospin breaking, but where the difference between the pion and kaon masses
has to be dealt with. These difficulties do not show up in the computation of the phases of the two-loop 
form factors, where the technically most demanding step is the computation of the projections of the one-loop
amplitudes on the $S$ and $P$ partial waves, which can be done analytically.

Coming now to the outline of this paper,
our first purpose will be to extend the frameworks of refs. \cite{FSS93,GasserMeissner91} to the 
situation where the difference between the masses of charged and neutral pions is
taken into account. The general framework leading to two-loop representations for form factors
and scattering amplitudes when isospin symmetry no longer holds is thus described in section \ref{general}. 
In section \ref{1stIteration}, we implement the program of constructing the corresponding form factors
and scattering amplitudes at the one-loop level and provide explicit expressions for them.
The second iteration, leading to two-loop representations of the form factors and the scattering amplitudes,
is discussed in section \ref{2ndIteration}. 
The issue of isospin breaking in the phases of the two-loop form factors is addressed in section \ref{IB_in_phases}. 
Section \ref{numerics} is devoted to the numerical analysis of the isospin-breaking contributions  
in the phases of the form factors.
Finally, a summary of this study and our conclusions are to be found in section \ref{conclusion}.
This main body of the text is supplemented with six appendices, where details concerning more computational or
technical aspects have been gathered.

\section{GENERAL FRAMEWORK AND PRELIMINARY REMARKS} \label{general}
\setcounter{equation}{0}

The objects of our study are the scalar and vector form factors of the pion,
defined through the following matrix elements
\begin{eqnarray}
\!\!\!
\langle \pi^0(p_1) \pi^0(p_2) \vert {\widehat m}({\overline u}u + {\overline d}d)(0)\vert\Omega\rangle
 &=&
 + F_S^{\pi^0}(s)
\quad { }
\nonumber
\\ 
\!\!\!\langle \pi^+(p_{\mbox{\tiny{$ +$}}}) \pi^-(p_{\mbox{\tiny{$ -$}}}) \vert {\widehat m}({\overline u}u + {\overline d}d) (0)\vert\Omega\rangle
 &=&
- F_S^{\pi}(s) ,
\quad { }
\label{eq:F_S}
\end{eqnarray}
with ${\widehat m}\equiv (m_u + m_d)/2$, and
\begin{eqnarray}
\!\!\!\!\!\!\!\!\!\!
\frac{1}{2}\langle \pi^+ \pi^- \vert 
({\overline u}\gamma_\mu u - 
{\overline d}\gamma_\mu d)(0)\vert\Omega\rangle 
\!
 &=&
 \!
(p_{\mbox{\tiny{$ -$}}}  - p_{\mbox{\tiny{$ +$}}})_\mu F_V^{\pi}(s) ,
\nonumber\\
&&
\label{eq:F_V}
\end{eqnarray}
respectively, in the presence of isospin-breaking corrections induced by the
difference in the masses of the charged and neutral pions. In each case, $s$ denotes
the squared invariant mass of the dipion system, $s=(p_1 + p_2)^2$ or $s=(p_{\mbox{\tiny{$ -$}}}  + p_{\mbox{\tiny{$ +$}}})^2$,
with $p_{1,2}^2= M_{\pi^0}^2$, $p_{\mbox{\tiny{$\pm$}}}^2 = M_{\pi}^2$, and $\vert\Omega\rangle$ stands
for the QCD vacuum state. The mass of the neutral pion is denoted by $M_{\pi^0}$, while
$M_{\pi}$ stands for the mass of the charged pion. We will define the isospin limit as
the case when the neutral pion mass tends to the charged pion mass, $M_{\pi^0} \to M_{\pi}$, 
while keeping the latter fixed.  This explains
the convention that we follow in this paper, namely that all quantities without superscript refer to the 
charged pion case (default case), and that we refer to quantities involving neutral pions by an explicit $0$ superscript.
The minus sign in the definition of $F_S^{\pi}(s)$ reflects a choice of phase for the charged-pion states. 
In addition, we choose the crossing phases to be $-1$ for charged pions and $+1$ for neutral pions. These
choices are compatible with the Condon and Shortley phase convention in the isospin-symmetric
situation. We further assume throughout that symmetry under charge conjugation holds.

These form factors, while
being perfectly well-defined observable quantities in QCD, are however not
observables from a strictly experimental point of view: they can only be measured 
indirectly, and should thus at best be considered as pseudo-observables. For instance, in the Standard Model, 
the vector form factor $F_V^{\pi}(s)$  appears in the physical process $e^+ e^- \to \pi^+ \pi^-$ 
through the exchange of a single neutral spin-one gauge boson, which in practice reduces 
to only photon exchange at low energies. As far as the scalar form-factors $F_S^{\pi^0}(s)$ and $F_S^{\pi}(s)$ are concerned, 
a similar statement can in principle also be made, but is of little use
in practice, since the Standard Model contributions 
to the processes $e^+ e^- \to \pi^+ \pi^-\,,\ \pi^0 \pi^0$,
arising from the exchange of a Higgs particle, are well below the level of sensitivity that one could expect for any experiment of this type
in the foreseeable future. Despite these limitations on the experimental side, these form factors
prove useful as a theoretical laboratory. They allow us to discuss and to illustrate several issues related
to the structure of isospin-breaking contributions within a rather simple context. The full complexity
of experimentally more interesting situations, like the $K_{\ell 4}$ form factors or the 
decay amplitudes of light pseudoscalar mesons (eta or kaons) into three pions, can then be addressed on the basis of these 
considerations and the general framework developed here, see  \cite{wip} and the forthcoming publication \cite{P_to_3pi} in the former or 
the latter case, respectively.

In the present section, we aim at tackling two issues, namely discussing precisely the isospin contributions that we intend to deal with, 
and describing our general theoretical framework. Then we can focus on the specific pion form factors that we use as an illustration.

\subsection{Electromagnetic corrections}\label{sec:em_corr}

At the fundamental level, isospin violations have two origins within the Standard Model: 
the electroweak interactions mediated by the gauge bosons, and the quark mass difference 
$m_u - m_d$, arising through the coupling of the light $u$ and $d$ quark flavours to the Higgs boson. Both effects
contribute to the mass difference between charged and neutral pions, although the second one
turns out to be marginal: the pion mass difference is mainly an electromagnetic effect~\cite{Weinberg77,GasserLeut82}.

Chiral perturbation theory~\cite{Wein79,Gasser:1983yg} including electromagnetism~\cite{Urech:1994hd,Neufeld:1994eg,Neufeld:1995mu,Knecht:1997jw,Meissner:1997fa,Schweizer:2002ft}
provides in principle a suitable framework
to deal with these isospin-breaking contributions in the low-energy domain. It
has been applied to the computation of several quantities at the one-loop level,
including the $\pi\pi$ scattering amplitudes \cite{Knecht:1997jw,Meissner:1997fa,KnechtNehme02} and pion form factors 
\cite{Kubis:1999db} in the two-flavour case. Unfortunately, from a practical point of view, this is
not a level of accuracy able to match the experimental one in several cases of
interest (low-energy pion-pion scattering or $K_{l4}$ decay, for instance). One might
of course contemplate the extension of this effective Lagrangian framework to next-to-next-to-leading
order, but this is a more ambitious program, the interest of which might 
be limited eventually by the proliferation of low-energy constants. 
We will therefore not pursue this issue here.

\begin{figure}[t]
\center\epsfig{figure=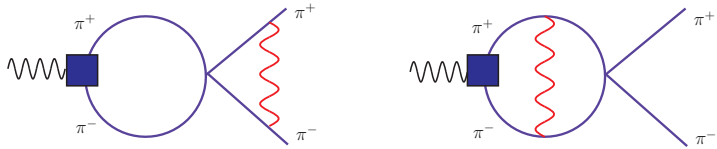,height=2.8cm}
\caption{Examples of radiative corrections usually included (left) and not included (right) in the data analyses.\label{fig1}}
\end{figure}

Instead, we will rather consider the point of view described in refs. \cite{Gasser:2007de,Colangelo:2008sm}:
we thus assume a situation, as is, actually, often the case in the
analyses of experimental data (such as for E865 and NA48/2), where part of the radiative corrections 
due to real and virtual corrections have already been dealt with in some manner,
while those that may remain are supposed to be negligible. 
In this kind of procedure, radiative corrections
of the type shown on the left-hand side of fig.~\ref{fig1}, for instance, together with emission of soft photons, 
are subtracted away, but other photonic effects, like the one shown on the right-hand side of the same figure,
are not taken into account, but are considered to be negligible. 
Notice that the latter would be included in a genuine two-loop calculation in the framework of the QCD+QED effective theory. 
One might also think of considering the possibility of treating them in the dispersive framework that we are using here,
for instance upon including also photons among the possible intermediate states in the unitarity 
conditions for the relevant partial waves. While this remains an interesting issue, it would however
lead us beyond the purposes of the present work. Within the framework assumed here, this leaves the difference
in the pion masses as the only remaining source of isospin breaking that we have to consider. 
In practice, it means that the charged and neutral pion
masses will be kept at their experimental values, but the origin of the
difference in their masses will not be addressed. In other words, we assume that general
properties like analyticity, unitarity, and crossing, together with chiral counting, can be
used to describe a world where the charged and neutral pion masses differ, even though the
interaction at the origin of this difference is not explicitly accounted for.

\subsection{Dispersive construction of the form factors}

\begin{figure}[t]
\center\epsfig{figure=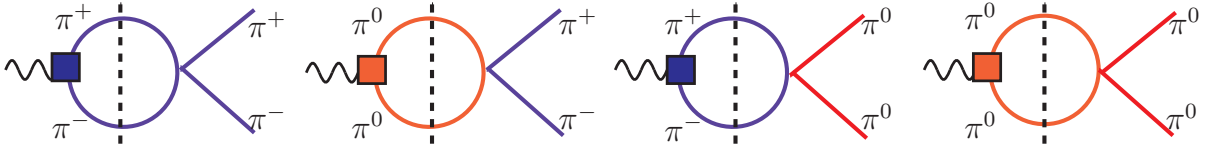,height=2.4cm}
\caption{Diagrammatic representation of the unitarity relations for the form factors.
In the case of the vector form factor, only the first diagram contributes.\label{unitaritydiag}} 
\end{figure}

The starting point of our study is provided by the dispersion relations satisfied by
the pion form factors and scattering amplitudes. Here we will only be interested in
the structure of these quantities in the low-energy region. In order to obtain dispersive 
representations that agree with their analytic structures up to two loops in the low-energy
expansion, it is convenient to consider thrice subtracted dispersion relations
(for a discussion of this issue in the isospin limit, see e.g. ref.~\cite{GasserMeissner91} for the pion form factors, 
and ref.~\cite{FSS93} for the $\pi\pi$ scattering amplitude). Let us start with the form factors,
for which these dispersive representations read
\begin{eqnarray}
F_S^{\pi^0}\!(s) &=& F_S^{\pi^0}\!(0) \! \left[
1 + \frac{1}{6}\langle r^2\rangle_S^{\pi^0}s + c_S^{\pi^0} \! s^2
+ U_S^{\pi^0}\!(s) 
\right]
\nonumber\\
F_S^{\pi}(s) &=& F_S^{\pi}(0) \! \left[
1 + \frac{1}{6}\langle r^2\rangle_S^{\pi}\,s + c_S^{\pi} \, s^2
+ U_S^{\pi}(s)
\right]
\nonumber\\
F_V^{\pi}(s) &=& 
1 + \frac{1}{6}\langle r^2\rangle_V^{\pi}\,s + c_V^{\pi} s^2
+ U_V^{\pi}(s)
.
\label{eq:dispersiveff}
\end{eqnarray}
In the last of these relations, the condition $F_V^{\pi}(0) = 1$, due to the conservation of
the electromagnetic current and thus valid to all orders,
has been used. Through crossing, $F_S^{\pi^0}(0)$ and $F_S^{\pi}(0)$ are equal to the corresponding 
sigma-term type form-factors, 
$\langle \pi^0(p)  \vert {\widehat m}({\overline u}u + {\overline d}d)(0)\vert \pi^0(p)\rangle$ and
$\langle \pi^\pm(p)  \vert {\widehat m}({\overline u}u + {\overline d}d)(0)\vert \pi^\pm(p)\rangle$,
respectively, for which there also exist relations~\cite{GasserZepeda} valid to all orders, that follow 
from the Feynman-Hellmann theorem~\cite{FeynHell},
\begin{equation}
F_S^{\pi^0}\!(0)\,=\,{\widehat m}\,\frac{\partial M_{\pi^0}^2}{\partial {\widehat m}}  ,
\ F_S^{\pi}(0)\,=\,{\widehat m}\,\frac{\partial M_{\pi}^2}{\partial {\widehat m}} .
\end{equation}
Since the dominant contribution to the pion mass difference is purely of electromagnetic origin
and independent of the quark masses \cite{Eckeretal89}, one has
\begin{equation}
\frac{F_S^{\pi}(0)}{F_S^{\pi^0}\!(0)} \,=\, 1 \,+\,\dots ,
\label{ratioF_S}
\end{equation}
where the ellipsis denotes higher order terms.
The unitarity parts are given in terms of dispersion integrals,
\begin{eqnarray}
U_S^{\pi^0}(s) &=&\frac{s^3}{\pi}\,\int\frac{dx}{x^{ 3}}\,
\frac{{\mbox{Im}}F_S^{\pi^0}\!(x)/F_S^{\pi^0}\!(0)}{x - s -i0}
\nonumber\\
U_S^{\pi}(s) &=&  \frac{s^3}{\pi}\,\int\frac{dx}{x^{ 3}}\,
\frac{{\mbox{Im}}F_S^{\pi}(x)/F_S^{\pi}(0)}{x - s -i0}
\nonumber\\
U_V^{\pi}(s) &=& \frac{s^3}{\pi}\,\int\frac{dx}{x^{  3}}\,
\frac{{\mbox{Im}}F_V^{\pi}(x)}{x - s -i0}\,.
\end{eqnarray}
In the low-energy region, the form factors are analytical functions in the complex $s$-plane, 
except for cut singularities on the positive real axis, starting at $s=4 M_{\pi}^2$ in the case of 
$F_V^{\pi}(s)$, and at $s=4 M_{\pi^0}^2$ in the cases of  $F_S^{\pi}(s)$
and $F_S^{\pi^0}(s)$. For $s$ real and below these cuts, the form factors are real.
In the chiral expansion, the form factors behave dominantly as
\begin{eqnarray}
&{\mbox{Re}}F_S^{\pi(\pi^0)}(s) \sim {\cal O}(E^2),\quad  
&{\mbox{Im}}F_S^{\pi(\pi^0)}(s) \sim {\cal O}(E^4),
\nonumber\\
&{\mbox{Re}}F_V^{\pi}(s) \sim {\cal O}(E^0),\quad  
&{\mbox{Im}}F_V^{\pi}(s) \sim {\cal O}(E^2) ,
\label{countingFF}
\end{eqnarray}
where $E$ denotes either a pion momentum or a pion mass.
Furthermore, intermediate states with more than two pions
contribute only from the three-loop level onwards.
Therefore, below the thresholds involving other states than the pions,
and up to and including two loops in the  two-flavour chiral expansion, only discontinuities 
arising from two-pion intermediate states need to be 
retained~\cite{GasserMeissner91}, as illustrated in fig.~\ref{unitaritydiag}. 
In order to distinguish among the different $\pi \pi$ scattering channels, we use the following
superscripts: $00$ for $\pi^0 \pi^0 \to \pi^0 \pi^0$, $++$ for $\pi^+ \pi^+ \to \pi^+ \pi^+$, $+-$ for
$\pi^+ \pi^- \to \pi^+ \pi^-$, $+0$ for $\pi^+ \pi^0 \to \pi^+ \pi^0$,
and $x$ for the inelastic channels $\pi^0 \pi^0 \to \pi^+ \pi^-$
and $\pi^+ \pi^- \to \pi^0 \pi^0$. We have 
\begin{eqnarray}
{\mbox{Im}}F_S^{\pi^0}(s) &=& {\mbox{Re}}\bigg\{
\frac{1}{2}\,\sigma_{0}(s)f_0^{00}(s) F_S^{\pi^0\! *}(s) \theta(s-4M_{\pi^0}^2)
-
\sigma (s)f_0^{x}(s)F_S^{\pi *}(s) \theta(s-4M_{\pi}^2)
\bigg\} + {\cal O}(E^8)
\nonumber\\
{\mbox{Im}}F_S^{\pi}(s) &=& {\mbox{Re}}\bigg\{
\sigma (s)f_0^{\mbox{\tiny{$ +-$}}}(s)F_S^{\pi *}(s)\theta(s-4M_{\pi}^2)
- 
\frac{1}{2}\,\sigma_{0}(s)f_0^{x}(s)F_S^{\pi^0\! *}(s)\theta(s-4M_{\pi^0}^2)
\bigg\} + {\cal O}(E^8)
\nonumber\\
{\mbox{Im}}F_V^{\pi}(s) &=& {\mbox{Re}}\bigg\{
\sigma (s)f_1^{\mbox{\tiny{$ +-$}}}(s)F_V^{\pi *}(s)\theta(s-4M_{\pi}^2) \bigg\}
+ {\cal O}(E^6) ,
\label{discFF}
\end{eqnarray}
where we define the phase-space functions
\begin{equation}
\sigma_{0}(s)\,=\,\sqrt{1 - \frac{4M_{\pi^0}^2}{s}}\,,\ \sigma(s)\,=\,\sqrt{1 - \frac{4M_{\pi}^2}{s}} .
\label{def_sigma}
\end{equation}
%

\noindent
In these expressions, $f_0(s)$ and $f_1(s)$ denote the $S$ and $P$ partial waves, respectively, 
of the $\pi\pi$ scattering amplitudes $A(s,t)$ in the corresponding channels. These partial waves have 
been defined as usual by projections of the corresponding amplitudes,
\begin{equation}
f_\ell (s) \,=\, \frac{1}{32\pi} \int_{-1}^{+1} dz A(s,t) P_\ell (z) ,
\label{PWproj}
\end{equation}
with $P_0(z)=1$ and $P_1(z)=z$ the appropriate Legendre polynomials, and $z=\cos \theta$, where $\theta$ denotes the scattering angle
in the center-of-mass frame. The relation with the Mandelstam
variables $s,t,u$ (summing up to the squared masses of the incoming and outgoing particles) depends on the process under consideration. For processes involving four particles with the same mass, it is simply given by
\begin{equation}
t \,=\, - \frac{s - 4M^2}{2}\,(1-z) .
\end{equation}
This is, in particular, the case for $A^{00}(s,t)$ (with $M=M_{\pi^0}$), as well as for $A^{\mbox{\tiny{$++$}}}(s,t)$ and
for $A^{\mbox{\tiny{$+-$}}}(s,t)$ (with now $M=M_\pi$). In the reactions involving both charged
and neutral pions, it becomes 
\begin{equation}
t \,=\, - \frac{1}{2}\,(s - 2 M_{\pi^0}^2 - 2 M_\pi^2) + \frac{z}{2}
\sqrt{(s - 4 M_{\pi^0}^2)(s - 4 M_\pi^2)}
\end{equation}
for $A^x(s,t)$, and 
\begin{equation}
t \,=\, - \frac{\lambda(s)}{2s}\,(1-z) \,,\quad \lambda(s) = [s - (M_{\pi} + M_{\pi^0})^2][s - (M_{\pi} - M_{\pi^0})^2] .
\end{equation}
for $A^{\mbox{\tiny{$+$}} 0}(s,t)$.
These expressions hold above the kinematical threshold, $s\ge 4 M_{\pi}^2$
for $A^x(s,t)$, and $s\ge (M_{\pi} + M_{\pi^0})^2$ for $A^{\mbox{\tiny{$+$}} 0}(s,t)$,
for instance.
Let us point out that the normalization
in (\ref{PWproj}) differs from the usual definition of the $\pi\pi$ partial-wave amplitudes
by a factor of $2$: it would correspond to a decomposition of the scattering amplitudes given by
\begin{equation}
A(s,t) \,=\, 16\pi \sum_{\ell\ge 0} (2\ell +1) P_{\ell}(\cos\theta) f_{\ell}(s) ,
\label{PWdecomp}
\end{equation}
i.e. with the normalization factor $16\pi$ instead of the usual $32\pi$. 
This modification is motivated by the fact that, in the presence of isospin breaking, the two-pion states do no 
longer obey (generalized) Bose symmetry, except in the cases of two neutral pions, or of two 
identically charged pions. In these cases, the appropriate symmetry factor has been included 
instead in the expressions of the corresponding phase spaces, cf. eq. (\ref{discFF}). 
In the chiral expansion, the $\pi\pi$ partial waves behave as
\begin{eqnarray}
&&
{\mbox{Re}}f_{\ell}(s)  \sim {\cal O}(E^2),
\, {\mbox{Im}}f_{\ell}(s)  \sim {\cal O}(E^4),
\, \ell = 0,1 , 
\nonumber\\
&&
{\mbox{Re}}f_{\ell}(s)  \sim {\cal O}(E^4),
\, {\mbox{Im}}f_{\ell}(s)  \sim {\cal O}(E^8),
\, \ell \ge 2 .
\label{countingPW}
\end{eqnarray}
This is in perfect agreement with the chiral counting (\ref{countingFF}) of the form factors,
together with the expressions (\ref{discFF}) of their low-energy discontinuities. 

\subsection{Dispersive construction of $\pi\pi$ scattering amplitudes}\label{sec:dispconstpipi}

It is actually possible to obtain dispersive representations for the two-loop $\pi\pi$ scattering amplitudes
themselves, following the same procedure as for the form factors.
As in the isospin-symmetric case~\cite{FSS93}, they follow from fixed-$t$ dispersion relations, combined
with very general properties like relativistic invariance, unitarity, analyticity, crossing, and from the chiral 
counting properties that have just been recalled. 
Isospin breaking is not expected to modify the asymptotic high-energy behaviour
of the amplitudes, so that two subtractions should be enough in order to obtain convergent
dispersion relations. We start with three subtractions in order to construct  low-energy expressions
of the scattering amplitudes that are valid up to and including two loops in the chiral expansion.

Whereas the various channels can all be described in terms of a single amplitude $A(s\vert t,u)$
as long 
as isospin symmetry holds, several independent amplitudes, not related by crossing, are necessary in order 
to deal with all the different available channels once isospin is broken. Otherwise, the derivation proceeds 
as in the isospin-symmetric case \cite{Knecht97}, up to the kinematical peculiarities 
due to the presence of particles with different masses. The relevant features can be inferred
from the discussion of ref. \cite{NovZdra08}, devoted to the extension of
the results of ref. \cite{FSS93} to the scattering amplitudes of the mesons belonging
to the octet of lightest pseudoscalar states, pions, kaons, and eta. 
We will therefore directly write down the resulting expressions,
and then provide a few additional comments on their structure.

\emph{i) Elastic scattering involving only
neutral pions} remains the simplest case, with a single, fully crossing invariant amplitude, 
which has the following two-loop structure (for convenience, we display, from now on, the dependence on the 
three Mandelstam variables $s,t,u$, although they are not independent)
\begin{eqnarray}
A^{00}(s,t,u) &=& 
P^{00}(s,t,u) \,+\, W^{00}_0(s) \,+\, W^{00}_0(t) \,+\, W^{00}_0(u) \,+\, {\cal O}(E^8) .
\label{A00}
\end{eqnarray}
It involves a polynomial $P^{00}(s,t,u)$ of third order in $s,t,u$, symmetric under any
permutation of its variables, together with a dispersive
integral $W^{00}(s)$. This function has a discontinuity on the positive real $s$-axis 
starting at $s=4M_{\pi^0}^2$, and specified by the $\ell = 0$ partial-wave amplitude $f^{00}_0(s)$,
\begin{equation}
{\mbox{Im}} W^{00}_0(s) \,=\, 16\pi\, {\mbox{Im}}f^{00}_0 (s) \,\theta (s-4M_{\pi^0}^2) .
\end{equation}
Again, at the order under consideration, this discontinuity is provided by the
unitarity condition, in terms of the $\ell = 0$ partial waves in the relevant channels,
$f^{00}_0(s)$ and $f^{x}_0(s)$,
\begin{equation}
\frac{1}{16\pi}\,{\mbox{Im}}W^{00}_0(s) \,=\,
\frac{1}{2}\,\sigma_0 (s)\,
\left\vert f^{00}_0(s) \right\vert ^2
\theta (s-4M_{\pi^0}^2) 
\,+\,
\sigma (s)\,
\left\vert f^{x}_0(s) \right\vert ^2
\theta (s-4M_{\pi}^2) + {\cal O}(E^8).\ \ \quad{ }
\label{Im_W00}
\end{equation}

\emph{ii) The processes involving exactly two neutral pions,} i.e. $\pi^\pm \pi^0 \to \pi^\pm \pi^0$ and $\pi^+ \pi^- \to \pi^0 \pi^0$, 
provide the next family of amplitudes that are related under crossing. They display the following structure at two loops
in the chiral expansion:
\begin{eqnarray} 
A^{x}(s,t,u) &=&
- P^{x}(s,t,u) - W^{x}_{0}(s) 
- \left[W^{\mbox{\tiny{$ +$}}0}_{0}(t) +3(s-u)W^{\mbox{\tiny{$ +$}}0}_{1}(t)\right] 
- \left[W^{\mbox{\tiny{$ +$}}0}_{0}(u) +3(s-t)W^{\mbox{\tiny{$ +$}}0}_{1}(u)\right]
+ {\cal O}(E^8) ,
\nonumber\\
&& \qquad{ }
\label{Ax}
\end{eqnarray}
whereas $A^{\mbox{\tiny{$ +$}}0}(s,t,u) = - A^x(t,s,u)$ through crossing. In the above expression,
$P^{x}(s,t,u)$ represents a polynomial of third order in the Mandelstam variables,
symmetric under exchange of $t$ and $u$ (Bose symmetry).
The functions $W^{\mbox{\tiny{$ +$}}0}_{0,1}(s)$ and $W^{x}_{0}(s)$ have discontinuities on the 
positive real $s$-axis, starting at $s=(M_{\pi} + M_{\pi^0})^2$ and at $s=4M_{\pi^0}^2$,
respectively. These discontinuities are again given in terms of the appropriate lowest ($S$ and $P$)
$\pi\pi$ partial waves
\begin{eqnarray} 
{\mbox{Im}}W^{\mbox{\tiny{$ +$}}0}_{0}(s)
&=&
16\pi
\bigg[{\mbox{Im}}f^{\mbox{\tiny{$ +$}}0}_0(s) + \frac{3\left(M_{\pi}^2 - M_{\pi^0}^2\right)^2}{\lambda (s)}
\,{\mbox{Im}}f^{\mbox{\tiny{$ +$}}0}_1(s)
\bigg]
\theta \!\left(s-(M_{\pi} + M_{\pi^0})^2\right)
\nonumber\\
{\mbox{Im}}W^{\mbox{\tiny{$ +$}}0}_{1}(s)
&=&
16\pi 
\frac{s}{\lambda (s)} \, {\mbox{Im}}f^{\mbox{\tiny{$ +$}}0}_1(s)
\theta \!\left(s-(M_{\pi} + M_{\pi^0})^2\right)
\nonumber\\
{\mbox{Im}}W^{x}_{0}(s) &=& 
- 16\pi \,{\mbox{Im}}f^{x}_0(s) \,\theta (s-4M_{\pi^0}^2)
.
\label{Im_Wx}
\end{eqnarray}
Up to higher-order contributions, the unitarity condition allows
to rewrite these expressions in terms of the same lowest partial waves.
In the elastic channel, there is only one contribution, arising from the $\pi^+ \pi^0$ intermediate state, 
whereas the inelastic channel involves two contributions:
\begin{eqnarray} 
\frac{1}{16\pi}\,{\mbox{Im}}W^{\mbox{\tiny{$ +$}}0}_{0}(s)
&=&
\bigg[
\frac{\lambda^{1/2}(s)}{s}
\left\vert f^{\mbox{\tiny{$ +$}}0}_0(s) \right\vert ^2 + 
\frac{3\left(M_{\pi}^2 - M_{\pi^0}^2\right)^2}{s \lambda^{1/2} (s)}
\, \left\vert f^{\mbox{\tiny{$ +$}}0}_1(s) \right\vert ^2
\bigg]
\theta \!\left(s-(M_{\pi} + M_{\pi^0})^2\right)
\,+\, {\cal O}(E^8)
\nonumber\\
\frac{1}{16\pi}\,{\mbox{Im}}W^{\mbox{\tiny{$ +$}}0}_{1}(s)
&=&
\frac{1}{\lambda^{1/2} (s)} \, \left\vert f^{\mbox{\tiny{$ +$}}0}_1(s) \right\vert ^2
\,\theta \!\left(s-(M_{\pi} + M_{\pi^0})^2\right)
\,+\, {\cal O}(E^8)
\nonumber\\
\frac{1}{16\pi}\,{\mbox{Im}}W^{x}_{0}(s) &=& 
- \frac{1}{2}\,\sigma_0 (s)\,
f^{x}_0(s) \left[ f^{00}_0(s) \right] ^\star
\theta (s-4M_{\pi^0}^2) -
\sigma (s)\,
f^{\mbox{\tiny{$ +-$}}}_0(s) 
\left[ f^{x}_0(s) \right] ^\star
\theta (s-4M_{\pi}^2)
\,+\, {\cal O}(E^8) . \qquad{ }
\end{eqnarray}

\emph{iii) Finally, the subset of elastic scattering processes
involving only charged pions} remains to be considered. Their amplitudes being all related by crossing,
it is enough to display explicitly one of them, for instance,
\begin{eqnarray}
A^{\mbox{\tiny{$ +-$}}}(s,t,u) &=&
P^{\mbox{\tiny{$ +-$}}}(s,t,u) +
\left[ W^{\mbox{\tiny{$ +-$}}}_{0}(s) + 3 (t-u)W^{\mbox{\tiny{$ +-$}}}_{1}(s) \right] +
\left[ W^{\mbox{\tiny{$ +-$}}}_{0}(t) + 3 (s-u)W^{\mbox{\tiny{$+-$}}}_{1}(t) \right] +
W^{\mbox{\tiny{$ ++$}}}_{0}(u) +  {\cal O}(E^8) ,
\nonumber\\
&& \ \quad{ }
\label{A+-}
\end{eqnarray}
where the third order polynomial $P^{\mbox{\tiny{$ +-$}}}(s,t,u)$ is symmetric under
exchange of $s$ and $t$ (Bose symmetry in the crossed $u$-channel). 
The three functions $W^{\mbox{\tiny{$ +-$}}}_{0,1}(s)$ and $W^{\mbox{\tiny{$ ++$}}}_{0}(s)$ have cut
singularities along the real $s$ axis, starting at $s=4M_{\pi^0}^2$
or at $s=4 M_{\pi}^2$. The corresponding discontinuities
read
\begin{eqnarray}
{\mbox{Im}} W^{\mbox{\tiny{$ +-$}}}_{0}(s) &=&  16\pi {\mbox{Im}}f^{\mbox{\tiny{$ +-$}}}_0(s) \theta (s-4M_{\pi^0}^2)
\nonumber\\
{\mbox{Im}} W^{\mbox{\tiny{$ +-$}}}_{1}(s) &=&  16\pi {\mbox{Im}}f^{\mbox{\tiny{$ +-$}}}_1(s) \theta (s-4M_{\pi}^2)
\nonumber\\
{\mbox{Im}} W^{\mbox{\tiny{$ ++$}}}_{0}(s) &=&  16\pi {\mbox{Im}}f^{\mbox{\tiny{$ ++$}}}_0(s) \theta (s-4M_{\pi}^2) .
\end{eqnarray}
The unitarity condition for the three $\pi\pi$ partial waves involved then leads to
\begin{eqnarray}
\frac{1}{16\pi}\,{\mbox{Im}}
W^{\mbox{\tiny{$ +-$}}}_{0}(s) &=& \sigma (s)\,
\left\vert f^{\mbox{\tiny{$ +-$}}}_0(s) \right\vert ^2
\theta (s-4M_{\pi}^2)
\,+\,
\frac{1}{2}\,
\sigma_0 (s)\,
\left\vert f^{x}_0(s) \right\vert ^2
\theta (s-4M_{\pi^0}^2)
\,+\, {\cal O}(E^8) 
\nonumber\\
\frac{1}{16\pi}\,{\mbox{Im}}
W^{\mbox{\tiny{$ +-$}}}_{1}(s) &=& \sigma (s)\,
\left\vert f^{\mbox{\tiny{$ +-$}}}_1(s) \right\vert ^2
\theta (s-4M_{\pi}^2)
\,+\, {\cal O}(E^8) 
\nonumber\\
\frac{1}{16\pi}\,{\mbox{Im}}
W^{\mbox{\tiny{$ ++$}}}_{0}(s) &=& 
\frac{1}{2}\,\sigma (s)\,
\left\vert f^{\mbox{\tiny{$ ++$}}}_0(s) \right\vert ^2
\theta (s-4M_{\pi}^2) 
\,+\, {\cal O}(E^8) .
\label{Im_Wpm}
\end{eqnarray}

The reason why only the lowest $S$ and $P$ partial waves play a role in these expressions
follows again from the chiral counting (\ref{countingPW}) for the partial waves.
In the following, we will make use of the chiral expansion for the real parts of the $\ell=0,1$ partial waves,
for values of $s$ corresponding to the cut along the positive real axis, 
\begin{equation} \label{partialwavesamplitude}
{\mbox{Re}} f_\ell (s) \,=\, \varphi_\ell (s) + \psi_\ell (s) + {\cal O}(E^6) ,
\end{equation}
with $\varphi_\ell (s) \sim {\cal O}(E^2)$ and $\psi_\ell (s) \sim {\cal O}(E^4)$, so that
\begin{equation}
\left\vert f_\ell (s) \right\vert ^2 \,=\, \left[ {\mbox{Re}} f_\ell (s) \right]^2 
\,+\, {\cal O}(E^8)\,=\, \left[ \varphi_\ell (s) \right]^2 + 2 \varphi_\ell (s) \psi_\ell(s)
\,+\, {\cal O}(E^8)  , \ \ell=0,1 .
\end{equation} 
Let us also emphasize that the functions $W(s)$ only have
a right-hand cut, that coincides with the right-hand cut of the corresponding $S$ and $P$
$\pi\pi$ partial-wave projections~\cite{KMSF95}. This structure is in agreement with the analyticity properties
of the $\pi\pi$ scattering amplitudes $A(s,t,u)$ required by unitarity and crossing.
The decompositions (\ref{A00}), (\ref{Ax}), and (\ref{A+-}) satisfy these 
constraints, to the given order in the low-energy expansion.
Of course, the partial-wave amplitudes
have a more complicated analytical structure, coming from the projection 
in eq.~(\ref{PWproj}), 
 with also a left-hand cut, 
and even a circular cut in the case of the $\pi^\pm \pi^0 \to \pi^\pm \pi^0$
channel (for a description of the analytic structure of partial-wave amplitudes
in a general context, see e.g.~\cite{PWanalyticity}). At this stage
we should also stress that a full partial-wave decomposition (\ref{PWdecomp}) of the $\pi\pi$ amplitudes
is actually not required. For our purposes, it is sufficient to know that the discontinuity of the latter 
in the complex $s$-plane can be written, in the low-energy region of interest here, as
\begin{equation}
{\mbox{Im}} A(s,t) \,=\, 16\pi \left[ {\mbox{Im}} f_0(s) + 3 z {\mbox{Im}} f_{1}(s) \right] + \Phi_{\ell\ge 2}(s,t) ,
\label{PWdecomp2}
\end{equation}
with $\Phi_{\ell\ge 2}(s,t) \sim {\cal O}(E^8)$  as its dominant chiral behaviour.

\begin{figure}
\begin{center}
\begin{picture}(375,130)(0,130)

\SetWidth{1.5}

\Boxc(0,200)(70,20)
\Text(0,200)[]{$A$ at order {$E^{2k}$}}

\ArrowLine(35,200)(90,200)

\Text(62.5,240)[]{projection}
\Text(62.5,230)[]{over partial waves}

\Boxc(125,200)(70,20)
\Text(125,200)[]{$f$ at order {$E^{2k}$}}

\ArrowLine(160,200)(202.5,200)

\Text(180,230)[]{unitarity}

\Boxc(250,200)(95,20)
\Text(250,200)[]{Im $f$ at order {$E^{2k+2}$}}

\ArrowLine(297.5,200)(335,200)

\Text(317.5,230)[]{dispersion relation}

\Boxc(375,200)(80,20)
\Text(375,200)[]{$A$ at order {$E^{2k+2}$}}

\ArrowLine(375,190)(375,150)
\ArrowLine(375,150)(0,150)
\ArrowLine(0,150)(0,190)

\end{picture}
\end{center}
\caption{Recursive construction of two-loop representations for the pion form factors and scattering amplitudes in the low-energy
regime. One starts with polynomial expressions at order ${\cal O}(E^2)$, $k=1$, (resp. ${\cal O}(E^0)$, $k=0$) for the $\pi\pi$ amplitudes
and scalar form factors (resp. for the vector form factor), and obtains the two-loop representations after two iterations.\label{iterconst}}
\end{figure}
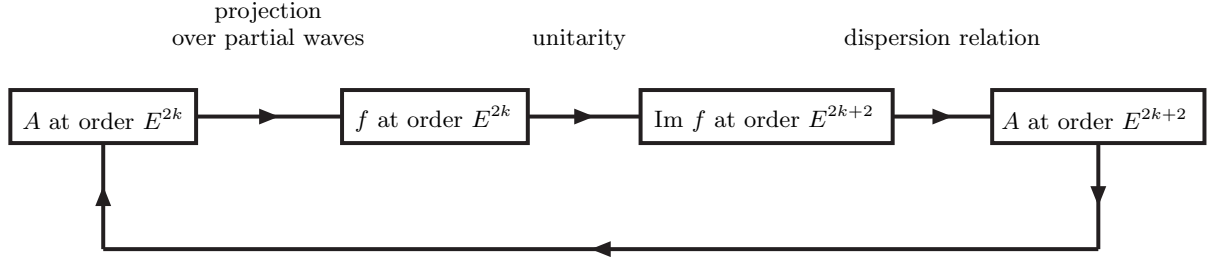

The very general features and the results that have just been presented are at the basis of the construction
of two-loop representations of the pion form factors and scattering amplitudes in the low-energy
regime. This construction is achieved trough a two-step recursive process of which we now give a 
short outline, summarised also in fig.~\ref{iterconst}. 
Chiral counting provides the initial information, namely that at lowest order the form factors 
reduce to real constants, to be identified with their values at $s=0$, while the $\pi\pi$ scattering
amplitudes consist of ${\cal O}(E^2)$ polynomials of at most first order in the Mandelstam variables.
This initial input, together with unitarity, fixes the discontinuities of the form factors and of the 
amplitudes, through the expressions of the functions $\varphi_\ell (s)$ at next-to-leading
order. The complete one-loop expressions are then recovered up to a subtraction polynomial
of at most first order (second order) in $s$ (in the Mandelstam variables) in the case of the form factors  
(of the scattering amplitudes). In turn, these one-loop expressions then provide the discontinuities at 
next-to-next-to-leading order, and thus the form factors (and amplitudes) themselves at order ${\cal O}(E^6)$,
up to a polynomial ambiguity of second order in $s$ (third order in the Mandelstam variables). In the case
of the $\pi\pi$ scattering amplitudes, crossing imposes further restrictions on the possible terms
that may appear in these polynomials. Notice that the presence of these polynomials reflect 
the fact that the functions $W^{00}_{0}(s)$, $W^{+0}_{0,1}(s)$, etc. are only specified by their analytical 
properties, in particular as far as their discontinuities are concerned. This leaves room for polynomial
ambiguities in the expressions of these functions, and the maximal degree of the polynomials is then limited by 
chiral power counting. 

This second iteration relies on the possibility to obtain analytical expressions for the
${\cal O}(E^4)$ pieces $\psi_\ell(s)$ of the real parts of the lowest partial waves from the one-loop $\pi\pi$ amplitudes, 
whose structures are no longer polynomial. This represents the technically most demanding step.
The following sections are therefore devoted to the detailed implementation of this program. Since
our main interest lies in discussing the effects of isospin breaking on the phases of the pion form factors,
we will provide explicit two-loop expressions for the latter only. This means that we will stop 
in the middle of the second iteration of the recursive procedure. Completing this second iteration
would provide full two-loop expressions for the form factors, and not only that of their imaginary part as 
needed for the phase shifts. This in turn would require one to obtain analytical expressions for the
corresponding dispersion integrals, a daunting task as already explained in the introduction.
We therefore defer this remaining step to future work.

\subsection{The subtraction polynomials: subthreshold parameters vs. scattering lengths}\label{sub_poly}

The two-loop dispersive construction provides representations of the $\pi\pi$ scattering amplitudes
that involve subtraction polynomials of at most third order in the Mandelstam variables. 
These polynomials depend on a certain number of parameters that
are not fixed by the general properties, listed at the beginning of subsection \ref{sec:dispconstpipi},
on which the representations for the $\pi\pi$ scattering amplitudes rest. Furthermore, beyond
general constraints coming, for instance, from the crossing property, there is, of course, a certain 
degree of arbitrariness in the form of these polynomials, and thus on the physical meaning of the
corresponding coefficients. 

In the isospin-symmetric case treated in refs.~\cite{FSS93, KMSF95}, the form of the subtraction polynomial 
$P(s \vert t,u)$ was chosen so that some
of its coefficients were identified with the subthreshold parameters of the amplitudes (the coefficients of 
its Taylor expansion around the center of the Mandelstam triangle). Among other reasons, this was motivated by 
the fact that the chiral expansions of these quantities show  better convergence properties than,
for instance, the scattering lengths. The latter can then be obtained from their expressions in
terms of these subthreshold parameters (see, e.g., the corresponding two-loop expressions in
Appendix B of ref.~\cite{KMSF95} and the discussion in ref.~\cite{DescotesGenon:2001tn}).
Along the same line of thought, in the presence of isospin breaking 
the most general subtraction polynomials of third order in the Mandelstam variables, and
compatible with the symmetries of the amplitudes under crossing, can then be written as
\begin{eqnarray}
P^{00}(s,t,u) &=& \frac{\alpha_{00} M_{\pi^0}^2}{F_{\pi}^2}
\nonumber\\
&& +\,
\frac{3\lambda_{00}^{(1)}}{F_{\pi}^4}\left[
(s-2M_{\pi^0}^2)^2 + (t-2M_{\pi^0}^2)^2 + (u-2M_{\pi^0}^2)^2   
\right]
\nonumber\\
&& +\,
\frac{3\lambda_{00}^{(2)}}{F_{\pi}^6}\left[
(s-2M_{\pi^0}^2)^3 + (t-2M_{\pi^0}^2)^3 + (u-2M_{\pi^0}^2)^3
\right]
\nonumber\\
P^x(s,t,u) &=& \frac{\beta_{x}}{F_{\pi}^2}\,
\left( s - \frac{2}{3}M_{\pi}^2 - \frac{2}{3}M_{\pi^0}^2\right)\,+\,
\frac{\alpha_{x} M_{\pi^0}^2}{3F_{\pi}^2}    
\nonumber\\
&& +\,
\frac{\lambda_{x}^{(1)}}{F_{\pi}^4}(s-2M_{\pi^0}^2)(s-2M_{\pi}^2)  
\,+\,
\frac{\lambda_{x}^{(2)}}{F_{\pi}^4}\left[
(t - M_{\pi}^2 - M_{\pi^0}^2)^2 + (u - M_{\pi}^2 - M_{\pi^0}^2)^2
\right]   
\nonumber\\
&& +\,
\frac{\lambda_{x}^{(3)}}{F_{\pi}^6}\left[
(s-2M_{\pi^0}^2)(s-2M_{\pi}^2)^2 + (s-2M_{\pi^0}^2)^2(s-2M_{\pi}^2)  
\right]
\nonumber\\
&& +\,
\frac{\lambda_{x}^{(4)}}{F_{\pi}^6}\left[
(t - M_{\pi}^2 - M_{\pi^0}^2)^3 + (u - M_{\pi}^2 - M_{\pi^0}^2)^3
\right]   
\nonumber\\
P^{\mbox{\tiny{$ +-$}}}(s,t,u) &=&
\frac{\beta_{\mbox{\tiny{$ +-$}}}}{F_{\pi}^2}\,
\left( s + t - \frac{8}{3}M_{\pi}^2 \right)\,+\,
\frac{2\alpha_{\mbox{\tiny{$ +-$}}} M_{\pi^0}^2}{3F_{\pi}^2}
\nonumber\\
&& +\,
\frac{\lambda_{\mbox{\tiny{$ +-$}}}^{(1)} + \lambda_{\mbox{\tiny{$ +-$}}}^{(2)}}{F_{\pi}^4}\left[
(s- 2M_{\pi}^2 )^2 + (t - 2M_{\pi}^2)^2
\right]
\,+\,
\frac{2 \lambda_{\mbox{\tiny{$ +-$}}}^{(2)}}{F_{\pi}^4}
(u- 2M_{\pi}^2 )^2   
\nonumber\\
&& +\,
\frac{\lambda_{\mbox{\tiny{$ +-$}}}^{(3)} + \lambda_{\mbox{\tiny{$ +-$}}}^{(4)}}{F_{\pi}^6}\left[
(s- 2M_{\pi}^2 )^3 + (t - 2M_{\pi}^2)^3
\right]
\,+\,
\frac{2 \lambda_{\mbox{\tiny{$ +-$}}}^{(4)}}{F_{\pi}^6}
(u- 2M_{\pi}^2 )^3   .
\label{polynomials_P}
\end{eqnarray}
In each of these polynomials, the first line is of at most first order in $s$, $t$, and $u$, and
corresponds to the tree-level amplitudes. The second line corresponds
to the subtraction terms at one-loop level: to construct the scattering amplitudes at
one-loop precision, it is enough to consider only twice-subtracted dispersion relations \cite{FSS93, KMSF95}.

On the other hand, it has become customary to rather let the scattering lengths play a prominent role,
since they are usually considered to have a more direct physical interpretation than subthreshold parameters. 
The point we would like to stress here is that the framework developed in refs.~\cite{FSS93, KMSF95},
and that we extend in the present work to the isospin-violating situation, is rather flexible
from this point of view, and can accommodate several choices of parameters, according to one's needs
or purposes. It is simply a matter of appropriately choosing the forms of the lowest-order amplitudes
and of the subtraction polynomials introduced at each of the two iterations.
Different sets of parameters can be related, order by order in the chiral expansion, and the corresponding
two-loop amplitudes differ only by higher-order terms. In the present article, we will use the representation
in terms of subthreshold parameters, with the subtraction polynomials given in eq. (\ref{polynomials_P}) above.
In appendix \ref{app:scatt_lengths}, we provide the corresponding expressions in terms of the scattering lengths,
and give an outline of how the present analysis can be reformulated in terms of the latter quantities. For more
details, we refer the reader to our forthcoming work \cite{wip}.

Whatever choice is eventually considered, these polynomials altogether depend on fifteen independent subtraction constants. In the isospin
limit, only six independent subtraction constants are required: isospin symmetry
induces nine linear relations among these constants. These relations can be summarised by the
statements that, as $M_{\pi^0}\rightarrow M_\pi$, one has
\begin{eqnarray}
P^{00}(s,t,u) & \rightarrow & P(s \vert t,u) + P(t \vert s,u) + P(u \vert s,t)
\nonumber\\
P^x(s,t,u) & \rightarrow & P(s \vert t,u)
\nonumber\\
P^{\mbox{\tiny{$ +-$}}}(s,t,u)  & \rightarrow & P(s \vert t,u) + P(t \vert s,u)
,
\end{eqnarray}
where
\begin{eqnarray}
P(s \vert t,u) &=& \frac{\beta}{F_{\pi}^2}\,
\left( s - \frac{4}{3}M_{\pi}^2 \right)\,+\,
\frac{\alpha M_{\pi}^2}{3F_{\pi}^2}    
\nonumber\\
&& +\,
\frac{\lambda_{1}}{F_{\pi}^4} (s-2M_{\pi}^2)^2
\,+\,
\frac{\lambda_{2}}{F_{\pi}^4}\left[
(t - 2 M_{\pi}^2 )^2 + (u - 2 M_{\pi}^2 )^2
\right]   
\nonumber\\
&& +\,
\frac{\lambda_{3}}{F_{\pi}^6} (s-2 M_{\pi}^2)^3  
\,+\,
\frac{\lambda_{4}}{F_{\pi}^6} \left[
(t - 2 M_{\pi}^2 )^3 + (u - 2 M_{\pi}^2 )^3
\right] 
\end{eqnarray}
is the subtraction polynomial for the isospin-symmetric scattering amplitude $A(s \vert t,u)$, 
cf. reference \cite{KMSF95}. Some relations between these subtraction constants are 
given explicitly below [see, for instance, section \ref{FF_and_Amp_tree} and the end of section \ref{Amp_1loop}].

\section{FIRST ITERATION: ONE-LOOP EXPRESSIONS} \label{1stIteration}
\setcounter{equation}{0}

In this section, we discuss the pion form factors and the $\pi\pi$ scattering amplitudes
at leading order, and then proceed with the construction of the corresponding one-loop 
expressions along the lines described above.

\subsection{Leading-order form factors and $\pi\pi$ amplitudes}\label{FF_and_Amp_tree}

At lowest order in the chiral expansion, the form factors are constants, that
may be identified with their values at $s=0$, $F_S^{\pi}(0)$, $F_S^{\pi^0}(0)$,
and $F_V^{\pi}(0) = 1$. 

At the same order, the $\pi\pi$ scattering amplitudes in the 
relevant channels read [cf. eq. (\ref{polynomials_P})]
\begin{eqnarray}
A^{x}(s,t) &=& -\frac{\beta_{x}}{F_{\pi}^2}\!
\left( s - \frac{2}{3}M_{\pi}^2 - \frac{2}{3}M_{\pi^0}^2\right)\! -
\frac{\alpha_{x} M_{\pi^0}^2}{3F_{\pi}^2}    
\nonumber\\
A^{\mbox{\tiny{$ +-$}}}(s,t) &=& \frac{\beta_{\mbox{\tiny{$\! +-$}}}}{F_{\pi}^2}\!
\left( s + t - \frac{8}{3}M_{\pi}^2 \right)\! +
\frac{2\alpha_{\mbox{\tiny{$\! +-$}}} M_{\pi^0}^2}{3F_{\pi}^2}
\nonumber\\
A^{00}(s,t) &=& \frac{\alpha_{00} M_{\pi^0}^2}{F_{\pi}^2} .
\label{AmpTree1}
\end{eqnarray}
From these amplitudes, the partial wave projections are obtained as
${\mbox{Re}}f_{\ell}(s) = \varphi_{\ell}(s) + {\cal O}(E^4)$, $\ell = 0,1$,
with
\begin{eqnarray}
\varphi_0^{x}(s) &=& -\frac{\beta_{x}}{16\pi F_{\pi}^2}
\left( s - \frac{2}{3}M_{\pi}^2 - \frac{2}{3}M_{\pi^0}^2\right) -
\frac{\alpha_{x} M_{\pi^0}^2}{48\pi F_{\pi}^2}  
\nonumber\\
\varphi_0^{\mbox{\tiny{$ +-$}}}(s) &=& \frac{\beta_{\mbox{\tiny{$\! +-$}}}}{32\pi F_{\pi}^2}
\left( s  - \frac{4}{3}M_{\pi}^2 \right) +
\frac{\alpha_{\mbox{\tiny{$\! +-$}}} M_{\pi^0}^2}{24\pi F_{\pi}^2}
\nonumber\\
\varphi_1^{\mbox{\tiny{$ +-$}}}(s) &=& \frac{\beta_{\mbox{\tiny{$\! +-$}}}}{96\pi F_{\pi}^2}
\left( s  - {4}M_{\pi}^2 \right)
\nonumber\\
\varphi_0^{00}(s) &=& \frac{\alpha_{00} M_{\pi^0}^2}{16\pi F_{\pi}^2} .
\label{PWTree1}
\end{eqnarray}
The various parameters, like $\beta_{x}$ or $\alpha_{00}$, that occur in these
expressions are free, in the sense that they are not fixed by the general
principles (analyticity, unitarity, crossing, and chiral symmetry).
The occurrence, in these expressions, of $M_{\pi^0}^2$ rather than $M_{\pi}^2$
in the terms proportional to $\alpha_{x}$, $\alpha_{\mbox{\tiny{$\! +-$}}}$, or $\alpha_{00}$
is a pure matter of choice, and can be considered as part of the definition of these parameters.
The presence, in the denominator, of $F_\pi$, the pion decay constant in the isospin limit,
is likewise a matter of convention.

As discussed in subsection \ref{sub_poly}, in the isospin limit $\alpha_{x}$, $\alpha_{\mbox{\tiny{$\! +-$}}}$, 
and $\alpha_{00}$ take a common value $\alpha$. Similarly, the parameters
$\beta_{x}$ and $\beta_{\mbox{\tiny{$\! +-$}}}$ become equal to the same quantity $\beta$
in this limit. Let us notice that there 
is no analogous quantity $\beta_{00}$ in the case of elastic $\pi^0\pi^0$
scattering, due to the Bose symmetry and the identity $s+t+u = 4 M_{\pi^0}^2$. 
Both $\alpha$ and $\beta$ were introduced in ref.~\cite{FSS93}. They remain finite
in the chiral limit, and describe the $I=0$ and $I=2$ $S$-wave scattering
lengths $a_0^I$ in the isospin limit \cite{FSS93,KMSF95} at lowest order,
\begin{equation}
a_0^0 \,=\, \frac{M_{\pi}^2}{96 \pi F_\pi^2}\,(5\alpha + 16 \beta)\ ,\quad
a_0^2 \,=\, \frac{M_{\pi}^2}{48 \pi F_\pi^2}\,(\alpha - 4 \beta) .
\label{alpha_beta_a00_a02}
\end{equation}
On the other hand, the lowest-order $S$-wave scattering lengths corresponding to the amplitudes 
$A^{x}(s,t)$, $A^{\mbox{\tiny{$ +-$}}}(s,t)$, and $A^{00}(s,t)$ were computed in ref. \cite{Knecht:1997jw}
and read
\begin{eqnarray} 
a_0^{x} &=& \frac{2}{3} \left(- a_0^0 +  a_0^2 \right)
- (4 \beta - \alpha) \,\frac{M_{\pi}^2 - M_{\pi^0}^2}{48 \pi F_\pi^2}
\nonumber\\
a_0^{\mbox{\tiny{$ +-$}}} &=& \frac{1}{3} \left(2 a_0^0 +  a_0^2 \right)
+ (4 \beta - \alpha) \,\frac{M_{\pi}^2 - M_{\pi^0}^2}{24 \pi F_\pi^2}
\nonumber\\
a_0^{00} &=& \frac{2}{3} \left( a_0^0 + 2 a_0^2 \right) - \alpha \,\frac{M_{\pi}^2 - M_{\pi^0}^2}{16 \pi F_\pi^2} ,
\end{eqnarray}
As compared to ref. \cite{Knecht:1997jw}, where $\beta = 1$ was taken at lowest order, we have kept the dependence
with respect to $\beta$ in the isospin-violating correction terms. 
These expressions also account for the difference in the normalization of the partial-wave amplitudes as compared to \cite{Knecht:1997jw}, 
see eq. (\ref{PWdecomp}). Upon comparing these formulae with
the expressions of the scattering lengths computed directly from the amplitudes displayed in (\ref{AmpTree1}), we obtain
the following identifications:
\begin{eqnarray}
\beta_{x} &=& \beta_{\mbox{\tiny{$\! +-$}}} \ =\ \beta
\nonumber
\end{eqnarray}
\begin{eqnarray}
\alpha_{x} &=&  \alpha \,+\, 2\beta\,\frac{M_{\pi}^2 - M_{\pi^0}^2}{M_{\pi^0}^2}\,, \quad
\alpha_{\mbox{\tiny{$\! +-$}}} \ =\ \alpha \,+\, 4\beta\,\frac{M_{\pi}^2 - M_{\pi^0}^2}{M_{\pi^0}^2}
\,,\quad \alpha_{00} \ =\ \alpha \,.
\label{alphabetaLO}
\end{eqnarray}
At higher orders, and in the absence of isospin symmetry, all these coefficients become independent,
and these simple expressions receive additional contributions. The computation of the corresponding 
isospin-breaking corrections at next-to-leading order will be addressed below, see subsection \ref{sub_csts_at_NLO} 
and appendix \ref{app:subtraction}.

Finally, let us recall, from ref. \cite{Knecht:1997jw}, that the lowest-order $\pi \pi$ amplitudes
(\ref{AmpTree1}) take the form
\begin{eqnarray}
 A^{x}(s,t) &=& - A(s\vert t,u)\qquad\qquad\qquad\ \qquad [s+t+u = 2 M_{\pi^0}^2 + 2 M_{\pi}^2 ]
\nonumber\\
 A^{\mbox{\tiny{$ +-$}}}(s,t) &=& A(s\vert t,u) \,+\, A(t\vert s,u)\qquad\qquad\qquad\,\ \quad [s+t+u = 4 M_{\pi}^2 ]
\nonumber\\
 A^{00}(s,t) &=& A(s\vert t,u) \,+\, A(t\vert u,s) \,+\, A(u\vert s,t)\qquad [s+t+u = 4 M_{\pi^0}^2 ] ,
\end{eqnarray}
with
\begin{equation}
 A(s\vert t,u) \,=\, \frac{s - 2{\widehat m} B}{F^2} 
,
\label{pipiAmp_LO}
\end{equation}
and one has to be aware that
the variable $u$ that appears in $A(s\vert t,u)$ takes a different meaning in each case, as indicated between brackets.
Identifying these expressions with the ones in eq. (\ref{AmpTree1}) then gives
\begin{equation}
\alpha \,=\, \frac{F_\pi^2 }{F^2} \left(4  - 3 \, \frac{2{\widehat m} B}{ M_{\pi^0}^2} \right)
 , \ \beta \,=\, \frac{F_\pi^2 }{F^2}
\label{alpha_beta_LO}  
\end{equation}
at this order. Beyond leading order, the expressions (\ref{alpha_beta_LO}) involve the low-energy
constants ${\bar \ell}_3$ and ${\bar \ell}_4$ of ref. \cite{Gasser:1983yg}, the appropriate formulae can be
found in ref. \cite{KMSF95}. At this point, it may be useful to make briefly contact with the discussion towards the end of 
subsection \ref{sub_poly}, after eq. (\ref{polynomials_P}). Indeed, one might actually consider
three sets of independent quantities, $(\alpha, \beta )$, $(a_0^0, a_0^2)$, $({\bar \ell}_3 , {\bar \ell}_4)$
as the unknowns of the problem. From a theoretical point of view, they are, to some extent, interchangeable.
The last set naturally arises in the quark mass expansion that is implemented through the calculation
of Feynman graphs generated by the effective chiral lagrangian. The two first sets are better suited for 
addressing phenomenological issues related to the analysis of experimental data. Here, we choose to organize
the discussion in terms of the set $(\alpha, \beta )$. The transcription in terms of
the two $S$-wave scattering lengths can be found, as already mentioned, in appendix \ref{app:scatt_lengths}
and in a forthcoming article \cite{wip}.

\subsection{Pion form factors at one loop}\label{FF_1loop}

We can now start the procedure described in figure~\ref{iterconst}.
At this stage, the unitarity conditions take then the following form:
\begin{eqnarray}\label{unitaroneloop1}
{\mbox{Im}}F_S^{\pi^0}(s) &=&  
\frac{1}{2}\,\sigma_{0}(s)\varphi_0^{00}(s)F_S^{\pi^0}(0)\theta(s-4M_{\pi^0}^2)
-
\sigma (s)\varphi_0^{x}(s)F_S^{\pi}(0)\theta(s-4M_{\pi}^2)
\,+\,{\cal O}(E^6)
\nonumber\\
{\mbox{Im}}F_S^{\pi}(s) &=&  
\sigma (s)\varphi_0^{\mbox{\tiny{$ +-$}}}(s)F_S^{\pi}(0)\theta(s-4M_{\pi}^2)
-
\frac{1}{2}\,\sigma_{0}(s)\varphi_0^{x}(s)F_S^{\pi^0}(0)\theta(s-4M_{\pi^0}^2)
\,+\,{\cal O}(E^6)
\nonumber\\
{\mbox{Im}}F_V^{\pi}(s) &=&  
\sigma (s)\varphi_1^{\mbox{\tiny{$ +-$}}}(s) \theta(s-4M_{\pi}^2) \,+\,{\cal O}(E^4) .
\label{unitaroneloop2}
\end{eqnarray}
We have now to determine the full form factors, exploiting the fact that we know their analytic structure, 
i.e., a cut along the positive real axis, and the value of the discontinuity along this cut.
We introduce the well-known functions ${\bar J}_0(s)$ and ${\bar J}(s)$ defined by the following dispersive
integrals
\begin{eqnarray}
{\bar J}_0 (s) &=& \frac{s}{16\pi^2}\,\int_{4M_{\pi^0}^2}^{\infty}\,\frac{dx}{x}\,\frac{1}{x-s-i0}\,\sigma_0 (x)
\nonumber\\
 {\bar J} (s) &=& \frac{s}{16\pi^2}\,\int_{4M_{\pi}^2}^{\infty}\,\frac{dx}{x}\,\frac{1}{x-s-i0}\,\sigma (x)
.
\label{JbarDisp}
\end{eqnarray}
These functions correspond to the standard one-loop integrals subtracted at $s=0$, and through a change of variable 
the integrals can be brought into the more familiar form
\begin{equation}
{\bar J} (s) \,=\, \frac{-1}{16\pi^2}\int_0^1 dx \ln\left[1 - x(1-x)\frac{s}{M_\pi^2}\right] ,
\label{Jbar1loop}
\end{equation}
and a similar expression for ${\bar J}_0 (s)$, with $M_\pi$ replaced by $M_{\pi^0}$.
In the latter form, the integration is easy to perform for, say, $s<0$. The expression
of ${\bar J}(s)$ for the remaining values of $s$ is obtained through analytic continuation,
with the $s+i0$ prescription on the cut, as made explicit in the representation (\ref{JbarDisp}).
The result is well known and reads
\begin{equation}
{\bar J} (s) \,=\, \displaystyle{\frac{1}{16\pi^2}}\,\left[2\,+\,\sigma(s)\,L (s) + i\pi\sigma(s) \theta(s - 4M_{\pi}^2)\right]
\end{equation}
Here, the function $L(s)$ is defined by
\begin{eqnarray}
L (s) \ =\  
\left\{
\begin{array}{l}
\ln\left(
{\displaystyle \frac{1 - \sigma(s) }{1 + \sigma(s) } }\right) \quad [s\ge 4 M_{\pi}^2]\\
\\
\ln\left(
{\displaystyle \frac{\sigma(s) - 1}{\sigma(s) + 1} }\right) \quad [s\le 4 M_{\pi}^2] 
\end{array}
\right.
\quad {\mbox{with}}
\quad
\sigma (s) &=& 
\left\{
\begin{array}{l}
\sqrt{1 - {\displaystyle\frac{4 M_{\pi}^2}{s}}} \quad [s\le 0\ {\mbox{or}}\ s \ge 4 M_{\pi}^2 ]\\
\\
i \sqrt{\displaystyle{\frac{4 M_{\pi}^2}{s}} - 1}  \quad [0 \le s\le 4 M_{\pi}^2] \, ,\\
\end{array}
\right.
\end{eqnarray} 
where we have performed a similar analytical continuation of the phase-space function
$\sigma(s)$, whose cut extends over the interval $0 \le s\le 4 M_{\pi}^2$. The
function $\ln (s)$ is defined as usual with its cut on the negative real axis.
Analogous functions $L_0(s)$ and $\sigma_0(s)$ are defined upon replacing $M_{\pi}$
by $M_{\pi^0}$ in the above expressions.
Then one can easily find functions with the appropriate discontinuities (\ref{unitaroneloop2}) to represent the form factors 
following the dispersive representation eq.~(\ref{eq:dispersiveff}):
\begin{eqnarray}
U_S^{\pi^0}(s) &=& P_S^{\pi^0}(s) + 16\pi\,\frac{1}{2}\,\varphi_0^{00}(s)  {\bar J}_0 (s)
- 16\pi \varphi_0^{x}(s)\,\frac{F_S^{\pi}(0)}{ F_S^{\pi^0}(0)}\, {\bar J} (s)
\,+\, {\cal O}(E^6)
\nonumber\\
U_S^{\pi}(s) &=& P_S^{\pi}(s) + 16\pi \varphi_0^{\mbox{\tiny{$ +-$}}}(s)   {\bar J} (s)
 - 16\pi\,\frac{1}{2}\,\varphi_0^{x}(s)\,
\frac{F_S^{\pi^0}(0)}{ F_S^{\pi}(0)}\,{\bar J}_0 (s)
\,+\, {\cal O}(E^6)
\nonumber\\
U_V^{\pi}(s) &=& P_V^{\pi}(s) + 16\pi \varphi_1^{\mbox{\tiny{$ +-$}}}(s)   {\bar J} (s)
\,+\, {\cal O}(E^6) . \quad{ }
\end{eqnarray}
$P_S^{\pi^0}(s)$, $P_S^{\pi}(s)$, and $P_V^{\pi}(s)$ represent calculable
polynomials at most quadratic in $s$, which are determined by the property
that the functions $U_{S,V}^{\pi}(s)$ and $U_S^{\pi^0}(s)$ have vanishing first and
second derivatives at $s=0$. These polynomials can be reabsorbed into the subtraction constants
such as to build up the (one-loop) radii and curvatures. At one loop, only one subtraction
constant is required for each form factor. The corresponding expressions can therefore be rewritten as 
\begin{eqnarray}
F_S^{\pi^0}(s) &=& F_S^{\pi^0}\! (0)\! \Bigg[
1 + a_S^{\pi^0} \! s +  
16\pi \frac{\varphi_0^{00}(s)}{2}  {\bar J}_0 (s) \! \Bigg]
  \,-\,16\pi {F_S^{\pi}(0)}\, \varphi_0^{x}(s)\,{\bar J} (s)
\nonumber\\
F_S^{\pi}(s) &=& F_S^{\pi}(0) \!\Bigg[
1 +  a_S^{\pi}\,s +  
16\pi \varphi_0^{\mbox{\tiny{$ +-$}}}(s)   {\bar J} (s)
\Bigg]
 \,-\,16\pi {F_S^{\pi^0}(0)} \,\frac{1}{2}\,\varphi_0^{x}(s)\,
{\bar J}_0 (s)
\nonumber\\
F_V^{\pi}(s) &=& 
1 + a_V^{\pi}\,s  
\,+\,16\pi \varphi_1^{\mbox{\tiny{$ +-$}}}(s)   {\bar J} (s)
 .\label{1loopFF}
\end{eqnarray}
At this order, the subtraction constants $a_S^{\pi^0}$, $a_S^{\pi}$, and $a_V^{\pi}$
are then related to the radii through
 \begin{eqnarray}
\langle r^2\rangle_S^{\pi^0} &=& 6\, a_S^{\pi^0} \,-\,
\frac{1}{48\pi^2 F_{\pi}^2}\,\left[\,
\frac{F_S^{\pi}(0)}{ F_S^{\pi^0}(0)}\,\left(
2\beta_{x}\,\frac{M_{\pi}^2 + M_{\pi^0}^2}{M_{\pi}^2}\,
-\,\alpha_{x}\,\frac{M_{\pi^0}^2}{M_{\pi}^2} \right) \,-\,
\frac{3}{2}\,\alpha_{00} \right]
\nonumber\\
\langle r^2\rangle_S^{\pi} &=& 6\, a_S^{\pi} \,-\,
\frac{1}{96\pi^2 F_{\pi}^2}\,\left[\,
\frac{F_S^{\pi^0}(0)}{ F_S^{\pi}(0)}\,\left(
2\beta_{x}\,\frac{M_{\pi}^2 + M_{\pi^0}^2}{M_{\pi^0}^2}\,
-\,\alpha_{x}\right) \,+\,
4\beta_{\mbox{\tiny{$\! +-$}}} \,-\,4\alpha_{\mbox{\tiny{$\! +-$}}}
\,\frac{M_{\pi^0}^2}{M_{\pi}^2}  \right]
\nonumber\\
\langle r^2\rangle_V^{\pi} &=& 6\, a_V^{\pi} \,-\, \frac{1}{24\pi^2 F_{\pi}^2}\,\beta_{\mbox{\tiny{$\! +-$}}} ,
\label{1loop_radii}
\end{eqnarray}
while the curvatures are given by
\begin{eqnarray}
c_S^{\pi^0} &=& 
\frac{1}{2880\pi^2 F_{\pi}^2 M_{\pi}^2}\,\left[\,
\frac{F_S^{\pi}(0)}{ F_S^{\pi^0}(0)}\,\left(
28\beta_{x}\,-\,
2\beta_{x}\,\frac{M_{\pi^0}^2}{M_{\pi}^2}\,
+ \alpha_{x}\,\frac{M_{\pi^0}^2}{M_{\pi}^2} \right) \,+\,
\frac{3}{2}\,\alpha_{00}\,\frac{M_{\pi}^2}{M_{\pi^0}^2} \right]
\nonumber\\
c_S^{\pi} &=&
\frac{1}{5760\pi^2 F_{\pi}^2 M_{\pi}^2}\,\Bigg[\,
\frac{F_S^{\pi^0}(0)}{ F_S^{\pi}(0)}\,\frac{M_{\pi}^2}{M_{\pi^0}^2}\,\left(
28\beta_{x} \,-\,
2\beta_{x}\,\frac{M_{\pi}^2}{M_{\pi^0}^2}\,
+\,\alpha_{x}\right)
\,+\,
26\beta_{\mbox{\tiny{$\! +-$}}} \,+\,4\alpha_{\mbox{\tiny{$\! +-$}}}
\,\frac{M_{\pi^0}^2}{M_{\pi}^2}  \Bigg] 
\nonumber\\
c_V^{\pi} &=&
\frac{1}{960\pi^2 F_{\pi}^2 M_{\pi}^2}\,\beta_{\mbox{\tiny{$\! +-$}}} \, .
\label{1loop_curvatures}  
\end{eqnarray}

\subsection{One-loop representation of $\pi\pi$ scattering amplitudes}\label{Amp_1loop}

The form factors at two loops are obtained once we know the real parts of the
one-loop $S$ and $P$ $\pi\pi$ partial wave projections, as shown in eq.~(\ref{unitaroneloop1}). With this aim in mind, we now
undertake the construction of the $\pi\pi$ scattering amplitudes to one-loop in the
presence of isospin breaking. The starting point is provided by the lowest-order
expressions (\ref{AmpTree1}) of these amplitudes, supplemented with two amplitudes 
that are deduced from the former ones by crossing, and that are needed to express unitarity in the crossed channels,
\begin{eqnarray}
A^{\mbox{\tiny{$ +$}}0}(s,t) &=& \frac{\beta_{x} }{F_{\pi}^2}\,
\left( t - \frac{2}{3}M_{\pi}^2 - \frac{2}{3}M_{\pi^0}^2\right)\,+\,
\frac{\alpha_{x} M_{\pi^0}^2}{3F_{\pi}^2}    
\nonumber\\
A^{\mbox{\tiny{$ ++$}}}(s,t) &=& - \frac{\beta_{\mbox{\tiny{$\! +-$}}}}{F_{\pi}^2}\,
\left( s - \frac{4}{3}M_{\pi}^2 \right)\,+\,
\frac{2\alpha_{\mbox{\tiny{$\! +-$}}} M_{\pi^0}^2}{3F_{\pi}^2} ,
\label{AmpTree2}
\end{eqnarray}
together with the corresponding lowest-order partial wave projections,
\begin{eqnarray}
\varphi_0^{\mbox{\tiny{$ +$}}0}(s) &=& \!   -\frac{\beta_{x}}{16\pi F_{\pi}^2}\!
\left[ \frac{\lambda (s)}{2s} + \frac{2}{3} M_{\pi}^2 + \frac{2}{3} M_{\pi^0}^2 \! \right] \! +
\frac{\alpha_{x} M_{\pi^0}^2}{48\pi F_{\pi}^2}  
\nonumber\\
\varphi_1^{\mbox{\tiny{$ +$}}0}(s) &=&    \frac{\beta_{x}}{48\pi F_{\pi}^2}\,
\frac{\lambda (s)}{2s} 
\nonumber\\
\varphi_0^{\mbox{\tiny{$ ++$}}}(s) &=& - \frac{\beta_{\mbox{\tiny{$\! +-$}}}}{16\pi F_{\pi}^2}\,
\left( s  - \frac{4}{3}M_{\pi}^2 \right)\,+\,
\frac{\alpha_{\mbox{\tiny{$\! +-$}}} M_{\pi^0}^2}{24\pi F_{\pi}^2} .
\label{PWTree2}
\end{eqnarray}
Applying the formulae given in sec.~\ref{sec:dispconstpipi}, and recalling that at the one-loop order
$\left\vert f_\ell (s) \right\vert ^2 = [\varphi_\ell(s)]^2 + {\cal O}(E^6)$, one easily obtains
the expressions for the unitarity parts of the various amplitudes, up to a polynomial
ambiguity that can be reabsorbed into the corresponding subtraction polynomials $P(s,t,u)$. For the
amplitudes corresponding to the elastic channels, with either only neutral (\ref{Im_W00}) or only charged pions (\ref{Im_Wpm}),
these expressions read
\begin{eqnarray}
W_0^{00}(s) &=&
\frac{1}{2}\left[16\pi \varphi^{00}_0(s) \right] ^2 {\bar J}_0 (s)
\,+\,
\left[16\pi \varphi^{x}_0(s) \right] ^2\,{\bar J} (s)
\nonumber\\
W^{\mbox{\tiny{$+-$}}}_{0}(s) &=&   
\left[ 16\pi \varphi^{\mbox{\tiny{$+-$}}}_0(s) \right] ^2
{\bar J} (s) 
\,+\,
\frac{1}{2}\,
\left[ 16\pi \varphi^{x}_0(s) \right] ^2 {\bar J}_0 (s)
\nonumber\\
W^{\mbox{\tiny{$+-$}}}_{1}(s) &=&   
\frac{\beta_{\mbox{\tiny{$\! +-$}}}^2}{36 F_{\pi}^4}
\,\left( s  - {4}M_{\pi}^2 \right){\bar J} (s)
\nonumber\\
W^{\mbox{\tiny{$++$}}}_{0}(s) &=& 
\frac{1}{2}\, 
\left[ 16\pi \varphi^{\mbox{\tiny{$++$}}}_0(s) \right] ^2 {\bar J} (s) .
\end{eqnarray}
For the amplitudes corresponding to the processes involving both charged and neutral pions (\ref{Im_Wx}),
one obtains
\begin{eqnarray}
W^{x}_{0}(s) &=& - \frac{1}{2}\,(16\pi)^2
\varphi^{x}_0(s)\,
\varphi^{00}_0(s)  \,{\bar J}_0 (s)
 \,-\,(16\pi)^2
\varphi^{\mbox{\tiny{$+-$}}}_0(s)\, 
\varphi^{x}_0(s)\,{\bar J} (s)
\nonumber\\
W^{\mbox{\tiny{$+$}} 0}_{0}(s)&=& \bigg\{
\frac{\beta_{x}^2}{12 F_{\pi}^4}\,\frac{(M_{\pi}^2 - M_{\pi^0}^2)^2}{s}\,
\left[
s \,-\, 6(M_{\pi}^2 + M_{\pi^0}^2)
\right]
\nonumber\\
&&
+\,\frac{\beta_{x}^2}{F_{\pi}^4}\left[
\frac{s^2}{4} - \frac{s}{3}(M_{\pi}^2 + M_{\pi^0}^2) +
\frac{11 M_{\pi}^4 - 14 M_{\pi}^2 M_{\pi^0}^2 + 11 M_{\pi^0}^4}{18}
\right]
\nonumber\\
&&
-\,\frac{\beta_{x} \alpha_{x} M_{\pi^0}^2}{3 F_{\pi}^4}
\left[
s - \frac{2}{3} (M_{\pi}^2 + M_{\pi^0}^2) + \frac{(M_{\pi}^2 - M_{\pi^0}^2)^2}{s}
\right]
\,+\, \frac{\alpha_{x}^2 M_{\pi^0}^4}{9 F_{\pi}^4}\bigg\}\,
{\bar J}_{\mbox{\tiny{$\! + $}} 0} (s)
\nonumber\\
&& 
+\,\frac{\beta_{x}^2}{3 F_{\pi}^4}\,\frac{(M_{\pi}^2 - M_{\pi^0}^2)^4}{s^2}\,\,
{\bar{\!\!{\bar J}}}_{\mbox{\tiny{$+ $}} 0} (s)
\nonumber\\
W^{\mbox{\tiny{$+$}} 0}_{1}(s)&=& \frac{\beta_{x}^2}{36 F_{\pi}^4}\,
\left[
s \,-\, 2(M_{\pi}^2 + M_{\pi^0}^2)\,+\,\frac{(M_{\pi}^2 - M_{\pi^0}^2)^2}{s}
\right]\,{\bar J}_{\mbox{\tiny{$+ $}} 0}(s) .
\end{eqnarray}
In these last expressions, we have introduced another one-loop integral
[$s_{\mbox{\tiny{$\! + $}} 0} \equiv (M_{\pi}  + M_{\pi^0})^2$],
\begin{equation}
{\bar J}_{\mbox{\tiny{$\!+ $}} 0} (s) = 
 \frac{s}{16\pi^2} \! \int_{s_{\mbox{\tiny{$\!+ $}} 0}}^{\infty}
 \!\frac{dx}{x}\,\frac{1}{x-s-i0}\,\frac{\lambda^{1/2} (x)}{x} ,
 \label{Jbar_+0_Disp}
\end{equation}
together with the subtracted integral $\ {\bar{\!\!{\bar J}}}_{\mbox{\tiny{$\! + $}} 0} (s) = {\bar J}_{\mbox{\tiny{$\! + $}} 0} (s)
- s {\bar J}_{\mbox{\tiny{$\! + $}} 0}^\prime (0)$, i.e.
\begin{equation}
{\bar{\!\!{\bar J}}}_{\mbox{\tiny{$\! +$}} 0} (s) = 
 \frac{s^2}{16\pi^2} \! \int_{s_{\mbox{\tiny{$\!\pm $}} 0}}^{\infty}
 \!\frac{dx}{x^2}\,\frac{1}{x-s-i0}\,\frac{\lambda^{1/2} (x)}{x} .
\end{equation}
The expression for ${\bar J}_{\mbox{\tiny{$\! + $}} 0} (s)$ 
can again be brought into the more familiar form of an integral over a Feynman parameter,
\begin{eqnarray}
\!\!\!\!\!
{\bar J}_{\mbox{\tiny{$\! + $}} 0} (s) = \frac{-1}{16\pi^2}\!
\int_0^1 \!\! dx \ln\left[1 - \frac{x(1-x)s}{M_{\pi}^2 - x (M_{\pi}^2 - M_{\pi^0}^2)}\right] \!,
 \label{Jbar_+0_Loop} 
\end{eqnarray}
through an appropriate change of variable.

The functions $W_0^{00}(s)$, $W^{\mbox{\tiny{$+-$}}}_{0}(s)$, etc. are defined by their
discontinuities up to polynomial ambiguities. At one-loop order, these polynomials need
only be of at most second order in the variables $s,t,u$. Taking into account the
symmetry properties of the corresponding amplitudes, they may therefore be written as
[see also eq. (\ref{polynomials_P}) and the discussion following it]
\begin{eqnarray}
 P^{00}(s,t,u) &=& \frac{\alpha_{00} M_{\pi^0}^2}{F_{\pi}^2}
\,+\,
\frac{3\lambda_{00}^{(1)}}{F_{\pi}^4}\left[
(s-2M_{\pi^0}^2)^2 + (t-2M_{\pi^0}^2)^2 + (u-2M_{\pi^0}^2)^2
\right]
\nonumber\\
P^{x}(s,t,u) &=& \frac{\beta_{x}}{F_{\pi}^2}\,
\left( s - \frac{2}{3}M_{\pi}^2 - \frac{2}{3}M_{\pi^0}^2\right)\,+\,
\frac{\alpha_{x} M_{\pi^0}^2}{3F_{\pi}^2}    
\nonumber\\
&& +\,
\frac{\lambda_{x}^{(1)}}{F_{\pi}^4}(s-2M_{\pi^0}^2)(s-2M_{\pi}^2)  
\,+\,
\frac{\lambda_{x}^{(2)}}{F_{\pi}^4}\left[
(t - M_{\pi}^2 - M_{\pi^0}^2)^2 + (u - M_{\pi}^2 - M_{\pi^0}^2)^2
\right]   
\nonumber\\
P^{\mbox{\tiny{$ +-$}}}(s,t,u) &=& \frac{\beta_{\mbox{\tiny{$ +-$}}}}{F_{\pi}^2}\,
\left( s + t - \frac{8}{3}M_{\pi}^2 \right)\,+\,
\frac{2\alpha_{\mbox{\tiny{$ +-$}}} M_{\pi^0}^2}{3F_{\pi}^2}
\nonumber\\
&& +\,
\frac{\lambda_{\mbox{\tiny{$ +-$}}}^{(1)} + \lambda_{\mbox{\tiny{$ +-$}}}^{(2)}}{F_{\pi}^4}\left[
(s- 2M_{\pi}^2 )^2 + (t - 2M_{\pi}^2)^2
\right]
\,+\,
\frac{2 \lambda_{\mbox{\tiny{$ +-$}}}^{(2)}}{F_{\pi}^4}
(u- 2M_{\pi}^2 )^2   
\label{poly_1loop_amp}
\end{eqnarray}
The five subtraction constants $\lambda_{00}^{(1)}$, $\lambda_{x}^{(i)}$, $\lambda_{\mbox{\tiny{$ +-$}}}^{(i)}$,
$i = 1,2$, are new free parameters. In the isospin limit, they are given by
\begin{eqnarray}
\lambda_{x}^{(i)} \ \to\ \lambda_{i}\,,\quad \lambda_{\mbox{\tiny{$ +-$}}}^{(i)}\ \to\ \lambda_{i}\,,\quad 
\lambda_{00}^{(1)} \ \to\ \frac{\lambda_{1} + 2\lambda_{2}}{3} ,
\label{lambdas_iso}
\end{eqnarray}
where, at this order, $\lambda_{1,2}$ can be expressed in terms of the low-energy constants ${\bar \ell}_1$ and ${\bar \ell}_2$
of \cite{Gasser:1983yg}, cf. \cite{KMSF95} and equation (\ref{lambdas_l1_l2}) below.

The expressions for the $\pi\pi$ scattering amplitudes that follow from the above results agree with those derived 
in refs. \cite{Knecht:1997jw,Meissner:1997fa,KnechtNehme02} from a Feynman graph calculation based 
on the low-energy effective lagrangian for QCD+QED, provided that one drops the contributions
coming from virtual photons, while keeping the difference between charged and neutral pion masses
[this procedure is discussed in greater detail in subsection \ref{sub_csts_at_NLO} and in appendix \ref{app:subtraction}].

\subsection{Infrared behaviour of one-loop amplitudes and form factors}\label{Mto0_1loop}

In the isospin-symmetric situation, the $\pi\pi$ scattering amplitude $A(s\vert t,u)$, the vector
form factor $F_V^\pi(s)$ and the scalar form factor $F_S^\pi (s)$ are
well behaved in the limit where the pion mass $M_\pi$ vanishes. For the pion form factors, this
has been shown explicitly at the two loop level in \cite{GasserMeissner91}. 
This good behaviour in the chiral limit requires that some of the parameters that appear in these
quantities develop themselves logarithmically singular terms in the limit whare the pion mass
vanishes. This is necessary in order to compensate for the singularities coming from the unitarity part,
for, as $M_\pi \rightarrow 0$,
\begin{equation}
{\bar J}(s)\, _{ \widetilde{_{ M_\pi \rightarrow\, 0\,\, }}}
\,\frac{1}{16\pi^2}\,\ln\left(\frac{M_{\pi}^2}{-s}\right)
\,+\, \frac{1}{8 \pi^2} .
\label{Jbar_IR_limit}
\end{equation}
Taking into account the pion mass difference 
offers a wider range of possibilities from this point of view. Indeed, besides the path just
described, reaching the isospin limit first, then letting the common pion mass vanish, two
additional options might be considered, where one lets, say, the neutral pion mass tend to zero,
keeping the charged pion mass fixed, or the other way around. In the first case, singular contributions
in the unitarity part come from the function ${\bar J}_0(s)$, which behaves as in (\ref{Jbar_IR_limit}),
but with $M_\pi$ replaced by $M_{\pi^0}$. In the second case, eq. (\ref{Jbar_IR_limit}) applies directly.
Notice that the function ${\bar J}_{\mbox{\tiny{$\! + $}} 0} (s)$ remains finite as either of these
two limits is taken. It requires that both pion masses vanish for it to develop an infrared singular behaviour.

Let us first consider the $\pi\pi$ scattering amplitudes obtained in the preceding
sub-section. In order for the one-loop amplitudes $A^{00}$, $A^x$, and $A^{\mbox{\tiny{$ +-$}}}$
to remain finite as $M_{\pi^0} \rightarrow 0$, with $M_\pi$ fixed, we must have
\begin{eqnarray}
\alpha_{00} M_{\pi^0} &\rightarrow & 0
\, ,\quad \alpha_x M_{\pi^0}^2 \ \rightarrow \  {\widehat\alpha}_x M_{\pi}^2
\, ,\ \beta_x  \ \rightarrow \ {\widehat\beta}_x \,,
\end{eqnarray}
and [for the sake of simplicity, we keep the notation $F_\pi$ for the pion decay constant, which has a regular behaviour in
either limit]
\begin{eqnarray}
\alpha_{\mbox{\tiny{$ +-$}}} M_{\pi^0}^2 &\sim & - \frac{M_\pi^4}{96\pi^2 { F}_\pi^2}
\,{\widehat\alpha}_x ( {\widehat\alpha}_x + 4 {\widehat\beta}_x ) \ln M_{\pi^0}^2\,+\,{\mbox{finite}}
\nonumber\\
\beta_{\mbox{\tiny{$ +-$}}} &\sim & - \frac{M_\pi^2}{48\pi^2 { F}_\pi^2}
\,{\widehat\beta}_x ( {\widehat\alpha}_x + 4 {\widehat\beta}_x) \ln M_{\pi^0}^2\,+\,{\mbox{finite}}
\nonumber\\
\lambda_{\mbox{\tiny{$ +-$}}}^{(1)} &\sim & -\frac{1}{32\pi^2}\,{\widehat\beta}_x ^2 \ln M_{\pi^0}^2\,+\,{\mbox{finite}}\,,
\end{eqnarray}
whereas the remaining coefficients, $\lambda_{00}^{(1)}$, $\lambda_{\mbox{\tiny{$ +-$}}}^{(2)}$, $\lambda_x^{(1)}$, and
$\lambda_x^{(2)}$ remain finite in this limit. Likewise, the coefficients ${\widehat\alpha}_x$ and 
${\widehat\beta}_x$ are free of infrared singularities. In order to avoid any possible confusion, we remind the reader  
that the coefficients $\alpha_{00}$, $\alpha_x$, $\alpha_{\mbox{\tiny{$ +-$}}}$ appear 
in the amplitudes multiplied by $M_{\pi^0}$ as a pure matter of convention [see the remark after eq. (\ref{AmpTree1})]. 
Therefore, $\alpha_x M_{\pi^0}^2$ and $\alpha_{\mbox{\tiny{$ +-$}}} M_{\pi^0}^2$
need not vanish as $M_{\pi^0}$ tends to zero with $M_\pi$ fixed. This feature is actually exhibited
already by the lowest-order expressions given in eq. (\ref{alphabetaLO}).
Furthermore, the quantities which appear on the right-hand sides of the above equations 
have to be understood as taking their lowest-order values. We have not distinguished them from their
values at next-to-leading order, which appear on the left-hand sides, in order not to overburden
the notation. The appearance of infrared singular behaviours is a loop effect, and higher-order
corrections will induce new singularities. At the next order, these may involve log-squared terms \cite{GasserMeissner91},
with an additional $1/(4\pi F_\pi)^2$ loop suppression factor. Concretely, from eq. (\ref{alphabetaLO}) one obtains
the lowest-order values
\begin{equation}
{\widehat\alpha}_x \,=\, 2 \beta \, , \ {\widehat\beta}_x \,=\, \beta .
\label{alpha_beta_hat}
\end{equation}
Taking now the second limit, $M_{\pi} \rightarrow 0$ with $M_{\pi^0}$ fixed, we find that the
finiteness of the one-loop amplitudes $A^{00}$, $A^x$, and $A^{\mbox{\tiny{$ +-$}}}$ makes
the various coefficients [we denote their lowest-order, infrared finite, limiting values with a 
tilde on top of them, except for $F_\pi$] 
behave as follows:
\begin{eqnarray}
\alpha_{00} &\sim & - \frac{M_{\pi^0}^2}{48\pi^2 { F}_\pi^2}
\,{\widetilde\alpha}_x ( {\widetilde\alpha}_x + 4 {\widetilde\beta}_x) \ln M_{\pi}^2\,+\,{\mbox{finite}}
\nonumber\\
\lambda_{00}^{(1)} &\sim & -\frac{1}{48\pi^2}\,{\widetilde\beta}_x ^2 \ln M_{\pi}^2\,+\,{\mbox{finite}}
\nonumber\\
\alpha_x  &\sim & - \frac{1}{48\pi^2}\, \frac{M_{\pi^0}^2}{{ F}_\pi^2}
\left[
2 {\widetilde\alpha}_{\mbox{\tiny{$ +-$}}} {\widetilde\alpha}_x  + {\widetilde\beta}_{\mbox{\tiny{$ +-$}}} 
({\widetilde\alpha}_x + 4 {\widetilde\beta}_x ) \right] \ln M_{\pi}^2\,+\,{\mbox{finite}}
\nonumber\\
\beta_x  &\sim & - \frac{1}{96\pi^2}\, \frac{M_{\pi^0}^2}{{ F}_\pi^2}
\left[
4 {\widetilde\alpha}_{\mbox{\tiny{$ +-$}}} {\widetilde\beta}_x  + {\widetilde\beta}_{\mbox{\tiny{$ +-$}}} 
({\widetilde\alpha}_x + 4 {\widetilde\beta}_x ) \right] \ln M_{\pi}^2\,+\,{\mbox{finite}}
\nonumber\\
\lambda_x^{(1)}  &\sim & - \frac{1}{32\pi^2}\, {\widetilde\beta}_x {\widetilde\beta}_{\mbox{\tiny{$ +-$}}}
\ln M_{\pi}^2\,+\,{\mbox{finite}}
\nonumber\\
\alpha_{\mbox{\tiny{$ +-$}}}  &\sim & - \frac{5}{48\pi^2} \frac{M_{\pi^0}^2}{{ F}_\pi^2}
\,{\widetilde\beta}_{\mbox{\tiny{$ +-$}}}^2 \ln M_{\pi}^2\,+\,{\mbox{finite}}
\nonumber\\
\beta_{\mbox{\tiny{$ +-$}}} &\sim & - \frac{1}{12\pi^2} \frac{M_{\pi^0}^2}{{ F}_\pi^2}
\,{\widetilde\alpha}_{\mbox{\tiny{$ +-$}}} {\widetilde\beta}_{\mbox{\tiny{$ +-$}}} \ln M_{\pi}^2\,+\,{\mbox{finite}}
\nonumber\\
\lambda_{\mbox{\tiny{$ +-$}}}^{(1)} &\sim & \frac{1}{96\pi^2} 
\,{\widetilde\beta}_{\mbox{\tiny{$ +-$}}}^2\ln M_{\pi}^2\,+\,{\mbox{finite}}
\nonumber\\
\lambda_{\mbox{\tiny{$ +-$}}}^{(2)} &\sim & - \frac{1}{48\pi^2} 
\,{\widetilde\beta}_{\mbox{\tiny{$ +-$}}}^2\ln M_{\pi}^2\,+\,{\mbox{finite}}
\,.
\end{eqnarray}
Now the lowest-order values inferred from eq. (\ref{alphabetaLO}) read:
\begin{equation}
{\widetilde\alpha}_x \,=\, \alpha - 2 \beta \, , \ {\widetilde\alpha}_{\mbox{\tiny{$ +-$}}} \,=\, \alpha - 4 \beta \, ,
\ {\widetilde\beta}_x \,=\, {\widetilde\beta}_{\mbox{\tiny{$ +-$}}} \,=\, \beta .
\end{equation}
In appendix \ref{app:subtraction} we determine the various subtraction constants that
appear in the expressions of the $\pi\pi$ scattering amplitudes obtained after the first
iteration from the corresponding expressions obtained from a one-loop calculation.
One may check that the formulae given in appendix \ref{app:subtraction} 
indeed exhibit the expected infrared behaviour.

Let us now turn towards the form factors, and consider the same two limits.
As $M_{\pi^0}$ vanishes while $M_\pi$ is kept fixed, the form factors remain free
of infrared singularities provided
\begin{eqnarray}
a_S^{\pi} &\sim &  - \frac{1}{32\pi^2 F_\pi^2}  
\, \frac{{\widehat F}_S^{\pi^0}(0)}{{\widehat F}_S^{\pi}(0)}
\,{\widehat\beta}_x \ln M_{\pi^0}^2\,+\,{\mbox{finite}}
\nonumber\\
F_S^{\pi}(0) &\sim &  - \frac{1}{96\pi^2} \frac{M_{\pi}^2}{{ F}_\pi^2}
\,{\widehat F}_S^{\pi^0}(0) ( {\widehat\alpha}_x - 2 {\widehat\beta}_x)\ln M_{\pi^0}^2\,+\,{\mbox{finite}}\, ,
\end{eqnarray}
while $a_S^{\pi^0}$ and $F_S^{\pi^0}(0)$ remain finite [at this order]. Actually, in view
of (\ref{alpha_beta_hat}), this is also true for $F_S^{\pi}(0)$. In the case
of the second limit, $M_\pi \rightarrow 0$ and $M_{\pi^0}$ fixed, infrared finite form factors require that
\begin{eqnarray}
a_V^\pi &\sim & - \frac{1}{96\pi^2 F_\pi^2} 
\,{\widetilde\beta}_{\mbox{\tiny{$ +-$}}} \ln M_{\pi}^2\,+\,{\mbox{finite}}
\nonumber\\
a_S^{\pi^0} &\sim &  - \frac{1}{16\pi^2 F_\pi^2} 
\, \frac{{\widetilde F}_S^{\pi}(0)}{{\widetilde F}_S^{\pi^0}(0)}
\,{\widetilde\beta}_x \ln M_{\pi}^2\,+\,{\mbox{finite}}
\nonumber\\
F_S^{\pi^0}(0) &\sim &  - \frac{1}{48\pi^2} \frac{M_{\pi^0}^2}{{ F}_\pi^2}
\,{\widetilde F}_S^\pi (0) ( {\widetilde\alpha}_x - 2 {\widetilde\beta}_x) \ln M_{\pi}^2\,+\,{\mbox{finite}}
\nonumber\\
a_S^{\pi} &\sim &  - \frac{1}{32\pi^2 F_\pi^2} 
\,{\widetilde\beta}_{\mbox{\tiny{$ +-$}}} \ln M_{\pi}^2\,+\,{\mbox{finite}}
\nonumber\\
F_S^{\pi}(0) &\sim &  - \frac{1}{24\pi^2} \frac{M_{\pi^0}^2}{{ F}_\pi^2}
\, {\widetilde F}_S^\pi (0) {\widetilde\alpha}_{\mbox{\tiny{$ +-$}}} \ln M_{\pi}^2\,+\,{\mbox{finite}}\,.
\end{eqnarray}
Again, one may check that the explicit one-loop expressions given in appendix \ref{app:subtraction}
reproduce the infrared behaviour obtained here from quite general arguments.

To conclude this short discussion, we may observe that many of these infrared
singularities disappear once the second mass also tends to zero, thus restoring the
infrared features of the isospin-symmetric chiral limit.
Upon studying the expressions (\ref{1loop_radii}) and (\ref{1loop_curvatures}) in the light of the present discussion,
a similar worsening of the infrared behaviour, as compared to the isospin-symmetric limit, can also be brought forward in the 
radii and curvatures of the scalar form factors. 
We thus can conclude that, to a certain extent, isospin symmetry tames the infrared behaviour of the soft pion clouds.

\section{SECOND ITERATION: TWO-LOOP EXPRESSIONS} \label{2ndIteration}
\setcounter{equation}{0}

So far, we have completed the first cycle of the procedure sketched in figure~\ref{iterconst}.
The one-loop expressions of the form factors and scattering amplitudes obtained in the preceding section
can now be used in order to construct the two-loop representations of the form factors. At next-to-leading 
order in the chiral expansion, the structure of the form factors is now as follows
\begin{eqnarray}
{\mbox{Re}} F_S^{\pi}(s) &=& F_S^{\pi}(0) \left[
1 \,+\,\Gamma_S^{\pi}(s) \right] \,+\,{\cal O}(E^6) 
\nonumber\\
{\mbox{Re}} F_S^{\pi^0}(s) &=& F_S^{\pi^0}(0) \left[
1 \,+\,\Gamma_S^{\pi^0}(s) \right] \,+\,{\cal O}(E^6) , \quad{ }
\end{eqnarray}
and
\begin{equation}
{\mbox{Re}} F_V^{\pi}(s)  \,=\, 1 + \Gamma_V^{\pi}(s) \,+\,{\cal O}(E^4).
\end{equation}
The one-loop corrections $\Gamma_S^{\pi^0}(s)$, $\Gamma_S^{\pi}(s)$, 
and $\Gamma_V^{\pi}(s)$ are easy to extract from the expressions
of the form factors obtained in the preceding section,
\begin{eqnarray}
\Gamma_S^{\pi^0}(s) &=& 
a_S^{\pi^0}\!s  \,+\, 
16\pi\,\frac{1}{2}\,\varphi_0^{00}(s)\,{\mbox{Re}}\,  {\bar J}_0 (s)
\, -\,16\pi \varphi_0^{x}(s)\,\frac{F_S^{\pi}(0)}{F_S^{\pi^0}(0)}\,{\mbox{Re}}\, {\bar J} (s)
\nonumber\\
\Gamma_S^{\pi}(s) &=& 
a_S^{\pi} s  \,+\, 
16\pi \varphi_0^{\mbox{\tiny{$+-$}}}(s) \,{\mbox{Re}}\,  {\bar J} (s)
\, -\,16\pi\,\frac{1}{2}\,\varphi_0^{x}(s)\,
\frac{F_S^{\pi^0}(0)}{F_S^{\pi}(0)} \,{\mbox{Re}}\, {\bar J}_0 (s)
\nonumber\\
\Gamma_V^{\pi}(s) &=& 
a_V^{\pi} s  \,+\, 
16\pi \varphi_1^{\mbox{\tiny{$+-$}}}(s) \,{\mbox{Re}}\,  {\bar J} (s)
 .
\label{eq:gammaoneloop}
\end{eqnarray}

As far as the discontinuities are concerned, they start at ${\cal O}(E^4)$ for $F_S^{\pi}(s)$ and 
$F_S^{\pi^0}(s)$, and at ${\cal O}(E^2)$ for $F_V^{\pi}(s)$.
Using this power counting and eq.~(\ref{partialwavesamplitude}), one obtains the discontinuities of 
the form factors at next-to-next-to-leading order 
\begin{eqnarray}
{\mbox{Im}}F_S^{\pi^0}(s) &=& 
\frac{1}{2}\,\sigma_{0}(s)F_S^{\pi^0}(0) \left\{ 
\varphi_0^{00}(s)\left[
1 \,+\,\Gamma_S^{\pi^0}(s) \right] \,+\,\psi_0^{00}(s) 
\right\}
\theta(s-4M_{\pi^0}^2)
\nonumber\\
&& \, -\,
\sigma(s)F_S^{\pi}(0)\left\{
\varphi_0^{x}(s)\left[
1 \,+\,\Gamma_S^{\pi}(s) \right] \,+\,\psi_0^{x}(s) 
\right\}\theta(s-4M_{\pi}^2)
\,+\,{\cal O}(E^6)
\nonumber\\
{\mbox{Im}}F_S^{\pi}(s) &=&  
\sigma(s)F_S^{\pi}(0)\left\{
\varphi_0^{\mbox{\tiny{$+-$}}}(s)\left[
1 \,+\,\Gamma_S^{\pi}(s) \right] \,+\,\psi_0^{\mbox{\tiny{$+-$}}}(s) 
\right\}\theta(s-4M_{\pi}^2)
\nonumber\\
&& \, -\,
\frac{1}{2}\,\sigma_{0}(s)F_S^{\pi^0}(0)\left\{
\varphi_0^{x}(s)\left[
1 \,+\,\Gamma_S^{\pi^0}(s) \right] \,+\,\psi_0^{x}(s) 
\right\}\theta(s-4M_{\pi^0}^2)
\,+\,{\cal O}(E^6)
\nonumber\\
{\mbox{Im}}F_V^{\pi}(s) &=&  
\sigma (s)\left\{
\varphi_1^{\mbox{\tiny{$+-$}}}(s) \left[
1 \,+\,\Gamma_V^{\pi}(s) \right] \,+\,\psi_1^{\mbox{\tiny{$+-$}}}(s) 
\right\}\theta(s-4M_{\pi}^2) \,+\,{\cal O}(E^4) .
\label{ImF_NLO}
\end{eqnarray}

\subsection{Partial-wave projections from the one-loop amplitudes}\label{PW_1_loop}

The computation of the one-loop corrections 
$\psi_0^{00}(s)$, $\psi_0^{x}(s)$, $\psi_0^{\mbox{\tiny{$+-$}}}(s)$, $\psi_1^{\mbox{\tiny{$+-$}}}(s)$ 
to the $\pi\pi$ $S$ and $P$ partial wave projections from the one-loop scattering amplitudes is less straightforward, 
and represents the next issue to be addressed.
In order to illustrate how this difficulty can be handled, let us start with the elastic scattering of neutral pions,
i.e. the quantity $\psi^{00}_0(s)$
describing the next-to-leading-order correction to the real part of the $S$-wave projection for $A^{00}(s,t)$
in the range $s\ge 4M_{\pi^0}^2$.
Trading the integration over the scattering angle for an integration over the variable $t$,
with $t_{\mbox{\tiny{$-$}}}(s)\equiv -(s - 4 M_{\pi^0}^2) \le t \le 0$, we obtain (the functions ${\bar J}(t)$ and 
${\bar J}_0(t)$ are real for $t\le 0$)
\begin{eqnarray}
\psi^{00}_0(s) &=&
\frac{\lambda_{0 0}^{(1)} }{16\pi F_{\pi}^4}
(5 s^2 - 16  s M_{\pi^0}^2 + 20 M_{\pi^0}^4)
+\,
\frac{1}{32\pi}\left[ 16\pi \varphi_0^{00}(s)\right]^2\, 
{\rm{Re}}\,{\bar J}_0(s) 
\,+\,
\frac{1}{16\pi}\left[ 16\pi \varphi_0^{x}(s)\right]^2\,
{\rm{Re}}\,{\bar J} (s)  
\nonumber\\
&&\!\!\!\!\!
+\,
\frac{1}{16\pi}\,\frac{1}{s - 4 M_{\pi^0}^2}\,\frac{\alpha_{0 0}^2 M_{\pi^0}^4}{F_{\pi}^4}\!
\int_{t_{-}(s)}^{0}\!\! dt\,{\bar J}_0(t)
+\,
\frac{1}{8\pi}\,\frac{1}{s - 4 M_{\pi^0}^2}\!
\int_{t_{-}(s)}^{0} \! dt \left[ 16\pi \varphi_0^{x}(t)\right]^2
\!\!{\bar J} (t) . \quad\ \,{ }
\label{psi_00_proj}
\end{eqnarray}
It turns out that the remaining integrals can be performed analytically.
The relevant formulae can be found in appendix~\ref{app:integJbar}. The resulting expression
can be written as
\begin{eqnarray}
16\pi \psi^{00}_0(s) &=&  2\,\frac{M_{\pi}^4}{F_\pi^4}\,
\sqrt{\frac{s}{s - 4M_{\pi^0}^2}}\,\Bigg\{
\xi_{00}^{(0)}(s) \sigma_{0}(s) \,+\, 2\xi^{(1;0)}_{00}(s) 
L_{0}(s) 
\,+\, 2\xi^{(1;{\mbox{\tiny$\nabla$}})}_{00}(s)\,\frac{\sigma_{0}(s)}{\sigma_{\mbox{\tiny$\nabla$}}(s)}
\, L_{\mbox{\tiny$\nabla$}}(s)
\nonumber\\
&&
+\,2\xi^{(2;\pm)}_{00}(s) \sigma (s) \sigma_0(s) L(s) \,+\,
2\xi^{(2;0)}_{00}(s) \left(1 - \frac{4 M_{\pi^0}^2}{s}\right) L_0(s)
\nonumber\\
&&
+\, 3 \xi^{(3;0)}_{00}(s)\,\sigma_0 (s)\,
\frac{M_{\pi^0}^2}{s - 4M_{\pi^0}^2}\,L_{0}^2(s) 
\,+\,
3 \xi^{(3;{\mbox{\tiny$\nabla$}})}_{00}(s)\,\sigma_{0}(s)
\frac{M_{\pi}^2}{s - 4M_{\pi^0}^2}
\,L_{\mbox{\tiny$\nabla$}}^2(s)
\Bigg\} ,
\label{psi_00}
\end{eqnarray}
with $\sigma_{\mbox{\tiny $\nabla $}}(s) = \sigma (s-4M_{\pi^0}^2+4M_{\pi}^2)$
and $L_{\mbox{\tiny $\nabla $}}(s) = L(s-4M_{\pi^0}^2+4M_{\pi}^2)$.
The various functions $\xi_{00}^{(0)}(s)$, $\xi^{(1;0)}_{00}(s)$, etc. that enter this expression of $\psi^{00}_0(s)$ 
are polynomials in $s$ and in the subthreshold parameters, which
are given in appendix~\ref{app:polynomials}.
We have written the result in a way that allows for a straightforward connection
with the similar expressions for the isospin-symmetric case, as displayed in
ref. \cite{KMSF95}. Indeed in the limit $M_{\pi^0} \to M_{\pi}$
(and $\alpha_{00}, \alpha_x, \alpha_{\mbox{\tiny{$+-$}}} \to \alpha$, $\beta_x, \beta_{\mbox{\tiny{$+-$}}} \to \beta$) 
one obtains the expected combination of $I=0$ and $I=2$ contributions,
weighted by the corresponding $SU(2)$ Clebsch-Gordan coefficients,
\begin{equation}
\psi^{00}_0(s)\rightarrow \ \stackrel{{ }_{\mbox{\scriptsize{$o$}}}}{\psi} \stackrel{00}{_{\!\! 0}} \!\! (s)
\,\equiv\, 2\,\frac{M_\pi^4}{F_\pi^4}\,
\sqrt{\frac{s}{s-4M_\pi^2}}\,
\sum_{n=0}^3\left[\frac{2}{3}\,\xi^{(n)}_2(s)\,+\,\frac{1}{3}\,\xi^{(n)}_0(s)\right]
k_n(s) ,
\nonumber
\label{lim_psi_00} 
\end{equation}
where, for the reader's convenience, we reproduce the expressions of the functions $k_n(s)$
of ref. \cite{KMSF95} (the function $k_4(s)$ appears only in the $P$ wave component
$\psi^{\mbox{\tiny{$+-$}}}_1(s)$, to be discussed below),
\begin{eqnarray}
&& k_0(s) \,=\, \frac{1}{16\pi}\,\sqrt{\frac{s - 4 M_\pi^2}{s}}\,,
\qquad
k_1(s) \,=\, \frac{1}{8\pi}\,L(s)\,,
\nonumber\\
&& k_2(s) \,=\,  \frac{1}{8\pi}\,\left(1\,-\,\frac{4 M_\pi^2}{s}\right)\,L(s)\,,
\qquad
k_3(s) \,=\, \frac{3}{16\pi}\,\frac{M_\pi^2}{\sqrt{s(s - 4 M_\pi^2)}}\,L^2(s)\,,
\nonumber\\
&& k_4(s) \,=\,\frac{1}{16\pi}\,\frac{M_\pi^2}{\sqrt{s(s - 4 M_\pi^2)}}\,
\left\{
1\,+\,\sqrt{\frac{s}{s - 4 M_\pi^2}}\,L(s)\,+\,\frac{M_\pi^2}{s - 4 M_\pi^2}\,L^2(s)
\right\}
.
\label{k_n}
\end{eqnarray}
For the remaining notation we refer the reader to \cite{KMSF95} (see in particular eqs. (3.36), (3.37) and the Appendix B therein).
Let us, however, point out that the factor of 2 in (\ref{lim_psi_00}) takes care of the difference in normalization in the partial 
waves as compared to that reference, see eq. (\ref{PWdecomp}) and the comment preceding it. 

For the elastic scattering of charged pions, the computation of $\psi^{\mbox{\tiny{$+-$}}}_0 (s)$ 
and of $\psi^{\mbox{\tiny{$+-$}}}_1 (s)$, now in the range $s\ge M_{\pi}^2$, proceeds
along similar lines.
The starting point is provided by the following formulae,
\begin{eqnarray}
\psi^{\mbox{\tiny{$+-$}}}_0(s) &=&
\frac{\lambda_{\mbox{\tiny{$+-$}}}^{(1)} + \lambda_{\mbox{\tiny{$+-$}}}^{(2)}}{ F_{\pi}^4}\,\left(s-2 M_{\pi}^2\right)^2 
\,+\,
\frac{\lambda_{\mbox{\tiny{$+-$}}}^{(1)} + 3\lambda_{\mbox{\tiny{$+-$}}}^{(2)} }{3 F_{\pi}^4}\,
\left(s^2 - 2 s M_{\pi} + 4  M_{\pi}^4
\right)  
\,+\,
\frac{1}{32\pi}\,\left[ 16\pi \varphi_0^{x}(s)\right]^2 {\rm{Re}}\,{\bar J}_0(s) 
\nonumber\\
&&
+\,
\frac{1}{16\pi}\,\left[ 16\pi \varphi_0^{\mbox{\tiny{$+-$}}}(s)\right]^2 {\rm{Re}}\,{\bar J}(s) 
\,+\,
\frac{1}{32\pi}\,\frac{1}{s - 4 M_{\pi}^2}\! 
\int_{t_-(s)}^{0} \! dt \left[ 16\pi \varphi_0^{x}(t)\right]^2{\bar J}_0(t)
\nonumber\\
&&
+\,
\frac{1}{32\pi}\,\frac{1}{s - 4 M_{\pi}^2}\,
\int_{t_-(s)}^{0} dt\,\left\{
2\left[ 16\pi \varphi_0^{\mbox{\tiny{$+-$}}}(t)\right]^2 + \left[ 16\pi \varphi_0^{\mbox{\tiny{$++$}}}(t)\right]^2
\right\}
{\bar J} (t)
\nonumber\\
&&
+\,
\frac{1}{16\pi}\,\frac{1}{s - 4 M_{\pi}^2}\,
\int_{t_-(s)}^{0} dt\,
\frac{\beta_{\mbox{\tiny{$\! +-$}}}^2}{12F_\pi^4}\,(t - 4 M_{\pi}^2)(2s + t - 4 M_{\pi}^2)
{\bar J} (t)
,
\nonumber
\end{eqnarray}
and
\begin{eqnarray}
\psi^{\mbox{\tiny{$+-$}}}_1(s) &=&
-\,\frac{\lambda_{\mbox{\tiny{$+-$}}}^{(1)} - \lambda_{\mbox{\tiny{$+-$}}}^{(2)}}{96\pi F_{\pi}^4}\,s(s - 4  M_{\pi}^2) 
\,+\,
\frac{1}{16\pi}\,\frac{\beta_{\mbox{\tiny{$\! +-$}}}^2}{36 F_{\pi}^4}\,(s - 4  M_{\pi}^2)^2
\,{\rm{Re}}\,{\bar J}(s) 
\nonumber\\ 
&&
 +\,
\frac{1}{32\pi}\,\frac{1}{s - 4 M_{\pi}^2}\! 
\int_{t_-(s)}^{0} \! dt 
\left[ 16\pi \varphi_0^{x}(t)\right]^2 \left( 1 + \frac{2t}{s - 4 M_{\pi}^2}\right) {\bar J}_0(t)
\nonumber\\
&&
 +\,
\frac{1}{32\pi}\,\frac{1}{s - 4 M_{\pi}^2}\!
\int_{t_-(s)}^{0} \! dt
\left\{
2\left[ 16\pi \varphi_0^{\mbox{\tiny{$+-$}}}(t)\right]^2 - \left[ 16\pi \varphi_0^{\mbox{\tiny{$++$}}}(t)\right]^2
\right\}
\left( 1 + \frac{2t}{s - 4 M_{\pi}^2}\right){\bar J}(t)
\nonumber\\
&&
+\,
\frac{1}{16\pi}\,\frac{1}{s - 4 M_{\pi}^2}\,
\int_{t_-(s)}^{0} dt\,
\frac{\beta_{\mbox{\tiny{$\! +-$}}}^2}{12F_\pi^4}\,(t - 4 M_{\pi}^2)(2s + t - 4 M_{\pi}^2)
\left( 1 + \frac{2t}{s - 4 M_{\pi}^2}\right){\bar J} (t)
,
\end{eqnarray}
with now $t_{\mbox{\tiny{$-$}}}(s) = -(s - 4 M_{\pi}^2)$. Performing the remaining integrations
with the help of the formulae given in appendix~\ref{app:integJbar} leads then to
\begin{eqnarray}
16\pi \psi^{\mbox{\tiny{$+-$}}}_0(s) &=&  2\,\frac{M_{\pi}^4}{F_\pi^4}\,
\sqrt{\frac{s}{s - 4M_{\pi}^2}}\,\Bigg\{
\xi_{{\mbox{\tiny{$+-$}}};S}^{(0)}(s) \sigma (s) \,+\, 2\xi^{(1;{\mbox{\tiny{$\pm$}}})}_{{\mbox{\tiny{$+-$}}};S}(s) L (s) 
\,+\, 2\xi^{(1;{\mbox{\tiny$\Delta$}})}_{{\mbox{\tiny{$+-$}}}; S}(s)\,\frac{\sigma (s)}{\sigma_{\mbox{\tiny$\Delta$}}(s)}
\, L_{\mbox{\tiny$\Delta$}}(s)
\nonumber\\
&&
+\,2\xi^{(2;{\mbox{\tiny{$\pm$}}})}_{{\mbox{\tiny{$+-$}}};S}(s) \left(1 - \frac{4 M_{\pi}^2}{s}\right) L (s) \,+\,
2\xi^{(2;0)}_{{\mbox{\tiny{$+-$}}};S}(s) \sigma (s) \sigma_0(s) L_0(s)
\nonumber\\
&&
+\, 3 \xi^{(3;{\mbox{\tiny{$\pm$}}})}_{{\mbox{\tiny{$+-$}}};S}(s)\,\frac{M_{\pi}^2}{\sqrt{s(s - 4M_{\pi}^2)}}\,L^2(s) \,+\,
3 \xi^{(3;{\mbox{\tiny$\Delta$}})}_{{\mbox{\tiny{$+-$}}};S}(s)\,\frac{M_{\pi^0}^2}{\sqrt{s(s - 4M_{\pi}^2)}}
\,L_{\mbox{\tiny$\Delta$}}^2(s)
\Bigg\}
,
\label{psi_+-_0} 
\end{eqnarray}
\begin{eqnarray}
16\pi \psi^{\mbox{\tiny{$+-$}}}_1(s) &=&  2\,\frac{M_{\pi}^4}{F_\pi^4}\,
\sqrt{\frac{s}{s - 4M_{\pi}^2}}\,\Bigg\{
\xi_{{\mbox{\tiny{$+-$}}};P}^{(0)}(s) \sigma (s) \,+\, 2\xi^{(1;{\mbox{\tiny{$\pm$}}})}_{{\mbox{\tiny{$+-$}}};P}(s) L (s) 
\,+\, 2\xi^{(1;{\mbox{\tiny$\Delta$}})}_{{\mbox{\tiny{$+-$}}}; P}(s)\,\frac{\sigma (s)}{\sigma_{\mbox{\tiny$\Delta$}}(s)}
\, L_{\mbox{\tiny$\Delta$}}(s)
\nonumber\\
&&
+\,
2\xi^{(2;{\mbox{\tiny{$\pm$}}})}_{{\mbox{\tiny{$+-$}}};P}(s) \left(1 - \frac{4 M_{\pi}^2}{s}\right) L (s) 
\nonumber\\
&&
+\, 3 \xi^{(3;{\mbox{\tiny{$\pm$}}})}_{{\mbox{\tiny{$+-$}}};P}(s)\,\frac{M_{\pi}^2}{\sqrt{s(s - 4M_{\pi}^2)}}\,L^2(s) \,+\,
3 \xi^{(3;{\mbox{\tiny$\Delta$}})}_{{\mbox{\tiny{$+-$}}};P}(s)\,\frac{M_{\pi^0}^2}{\sqrt{s(s - 4M_{\pi}^2)}}
\,L_{\mbox{\tiny$\Delta$}}^2(s)
\nonumber\\
&&
+\,\xi^{(4;{\mbox{\tiny{$\pm$}}})}_{{\mbox{\tiny{$+-$}}};P}(s)\,\frac{M_{\pi}^2}{\sqrt{s(s-4 M_{\pi}^2)}}\,
\left[
1\,+\,\frac{1}{\sigma (s)}L(s)\,+\,\frac{M_{\pi}^2}{s-4 M_{\pi}^2}\,L^2(s)
\right] 
\nonumber\\
&&
+\,\xi^{(4;{\mbox{\tiny$\Delta$}})}_{{\mbox{\tiny{$+-$}}};P}(s)\,\frac{M_{\pi^0}^2}{\sqrt{s(s-4 M_{\pi}^2)}}\,
\left[
1\,+\,\frac{1}{\sigma_{\mbox{\tiny$\Delta$}}(s)}\,L_{\mbox{\tiny$\Delta$}}(s)
\,+\,\frac{M_{\pi^0}^2}{s-4 M_{\pi}^2}\,L_{\mbox{\tiny$\Delta$}}^2(s)  
\right]
\Bigg\}
, 
\label{psi_+-_1} 
\end{eqnarray}
with $\sigma_{\mbox{\tiny$\Delta $}}(s) = \sigma_0 (s+4M_{\pi^0}^2-4M_{\pi}^2)$
and $L_{\mbox{\tiny$\Delta $}}(s) = L_0(s+4M_{\pi^0}^2-4M_{\pi}^2)$.
The various polynomials that enter the expression of $\psi^{\mbox{\tiny{$+-$}}}_0(s)$ and 
$\psi^{\mbox{\tiny{$+-$}}}_1(s)$ have been gathered in appendix~\ref{app:polynomials}.
Taking the isospin limit, as described above in the case of $\psi^{00}_0(s)$, one
recovers, for the $S$ wave, the expected combination of $I=0$ and $I=2$ contributions,
and, for the $P$ wave, the corresponding $I=1$ component,
\begin{eqnarray}
&&\psi^{\mbox{\tiny{$+-$}}}_0(s)\rightarrow 
\ \stackrel{{ }_{\mbox{\scriptsize{$o$}}}}{\psi} \stackrel{{\mbox{\tiny{$+-$}}}}{_{\!\!\!\! 0}} \! \! (s)
\,\equiv\, 2\,\frac{M_\pi^4}{F_\pi^4}\,
\sqrt{\frac{s}{s-4M_\pi^2}}\,
\sum_{n=0}^3\left[\frac{1}{6}\,\xi^{(n)}_2(s)\,+\,\frac{1}{3}\,\xi^{(n)}_0(s)\right]
k_n(s) ,
\nonumber\\
&&\psi^{\mbox{\tiny{$+-$}}}_1(s)\rightarrow
\ \stackrel{{ }_{\mbox{\scriptsize{$o$}}}}{\psi} \stackrel{{\mbox{\tiny{$+-$}}}}{_{\!\!\!\! 1}} \! \! (s) 
\,\equiv\, 2\,\frac{M_\pi^4}{F_\pi^4}\,
\sqrt{\frac{s}{s-4M_\pi^2}}\,
\sum_{n=0}^4 \frac{1}{2}\,\xi^{(n)}_1(s) k_n(s) ,
\label{lim_psi_+-}
\end{eqnarray}
in full agreement with the results of \cite{KMSF95}.

Turning eventually towards the inelastic scattering $\pi^+ \pi^- \rightarrow \pi^0 \pi^0$, 
i.e. $\psi^{x}_0(s)$, the range of integration corresponding to $-1\le z\equiv\cos\theta\le +1$ is 
$t_{\mbox{\tiny{$-$}}}(s) \le t \le t_{\mbox{\tiny{$+$}}}(s)$, with 
\begin{equation}
t_{\mbox{\tiny{$\pm$}}}(s) \,=\, -\frac{1}{2} (s - 2 M_{\pi}^2 - 2 M_{\pi^0}^2)
\,\pm\,
\frac{1}{2}\,\sqrt{(s- 4 M_{\pi}^2)(s - 4 M_{\pi^0}^2)} .
\end{equation}
For $s\ge 4  M_{\pi}^2$, one has $t\le 0$ and $u\le 0$. In terms of an integration over $t$, one thus obtains
\begin{eqnarray}
\psi^{x}_0(s) 
&=&-\,\frac{\lambda_{x}^{(1)}}{16\pi F_{\pi}^4}(s-2M_{\pi^0}^2)(s-2M_{\pi}^2)  
\,-\,
\frac{\lambda_{x}^{(2)}}{24\pi F_{\pi}^4}\left[
s^2 - s (M_{\pi}^2 + M_{\pi^0}^2) + 4 M_{\pi}^2 M_{\pi^0}^2
\right]   
\nonumber\\ 
&&
+\,
 \varphi_0^{x}(s) \varphi_0^{00}(s) \,\,
\frac{1}{2\pi}\,\left[2\,+\,\sigma_0(s)\,L_0(s)\right] 
+\,
\varphi_0^{\mbox{\tiny{$+-$}}}(s) \varphi_0^{x}(s)\, \,
\frac{1}{\pi}\,\left[2\,+\,\sigma (s)\,L (s)\right] 
\nonumber\\ 
&&
-\,
\frac{1}{8\pi F_\pi^4}\,\frac{1}{\sqrt{(s- 4 M_{\pi}^2)(s - 4 M_{\pi^0}^2)}}\,
\int_{t_-(s)}^{t_+(s)} dt\,\left[
\frac{\beta_{x}}{2}\,\left(t - \frac{2}{3}M_{\pi}^2 - \frac{2}{3}M_{\pi^0}^2\right)\,-\,
\frac{\alpha_{x} M_{\pi^0}^2}{3}
\right]^2 {\bar J}_{\mbox{\tiny{$\! + $}} 0}(t)
\nonumber\\ 
&&
-\,
\frac{1}{24\pi F_{\pi}^4}\,
\frac{\beta_{x}(M_{\pi}^2 - M_{\pi^0}^2)^2}{\sqrt{(s- 4 M_{\pi}^2)(s - 4 M_{\pi^0}^2)}}\,
\int_{t_-(s)}^{t_+(s)} \frac{dt}{t}\,
\left[
\frac{\beta_{x}}{4}\,\left(7t - 6 M_{\pi}^2 - 6 M_{\pi^0}^2\right)\,-\,
{\alpha_{x} M_{\pi^0}^2}
\right] {\bar J}_{\mbox{\tiny{$\! + $}} 0}(t)
\nonumber\\ 
&&
-\,
\frac{1}{24\pi F_{\pi}^4}\,
\frac{\beta_{x}^2(M_{\pi}^2 - M_{\pi^0}^2)^4}{\sqrt{(s- 4 M_{\pi}^2)(s - 4 M_{\pi^0}^2)}}\,
\int_{t_-(s)}^{t_+(s)} \frac{dt}{t^2}\,
\ {\bar{\!\!{\bar J}}}_{\mbox{\tiny{$\! + $}} 0}(t)
\nonumber\\ 
&&
-\,
\frac{1}{96\pi F_{\pi}^4}\,
\frac{\beta_{x}^2}{\sqrt{(s- 4 M_{\pi}^2)(s - 4 M_{\pi^0}^2)}}\,
\int_{t_-(s)}^{t_+(s)} dt (2s + t - 2M_{\pi}^2 - 2M_{\pi^0}^2)
\nonumber\\
&&\qquad\qquad\qquad\qquad\qquad\qquad\qquad\qquad\times
\left[
t - 2 M_{\pi}^2 - 2 M_{\pi^0}^2 \,+\,\frac{(M_{\pi}^2 - M_{\pi^0}^2)^2}{t}\,
\right] {\bar J}_{\mbox{\tiny{$\! + $}} 0}(t)
.
\label{psi_x_proj}
\end{eqnarray}
With the help of the formulae displayed in appendix~\ref{app:integJbar}, one finds
\begin{eqnarray}
16\pi \psi^{x}_0(s) &=&  2\,\frac{M_{\pi}^4}{F_\pi^4}\,
\sqrt{\frac{s}{s - 4M_{\pi}^2}}\,\Bigg\{
\xi_{x}^{(0)}(s) \sigma (s) \,+\, 2\xi^{(1)}_{x} (s) \,\frac{1}{\sqrt{s(s - 4M_{\pi^0}^2)}}
\left[ 
\lambda^{1/2}(t_{\mbox{\tiny{$-$}}}(s)){\cal L}_{\mbox{\tiny{$-$}}} (s) 
-
\lambda^{1/2}(t_{\mbox{\tiny{$+$}}}(s)){\cal L}_{\mbox{\tiny{$+$}}} (s) 
\right]
\nonumber\\
&&
+\, 2\xi^{(2;{\mbox{\tiny{$\pm$}}})}_{x}(s) \left(1 - \frac{4 M_{\pi}^2}{s}\right) L(s)
+\, 2\xi^{(2;0)}_{x}(s) \sigma (s) \sigma_0(s) L_0(s) 
\nonumber\\
&&
+\,
3 \xi^{(3)}_{x}(s)\,\frac{M_{\pi}^2}{\sqrt{s(s - 4M_{\pi^0}^2)}}
\left[
{\cal L}^2_{\mbox{\tiny{$-$}}} (s)  -  {\cal L}^2_{\mbox{\tiny{$+$}}} (s)
\right]
\Bigg\}
\,+\, 16\pi \Delta_1 \psi^{x}_0(s) \,+\, 16\pi \Delta_2 \psi^{x}_0(s) .
\label{psi_x_0} 
\end{eqnarray}

\begin{figure}[t]
\center\epsfig{figure=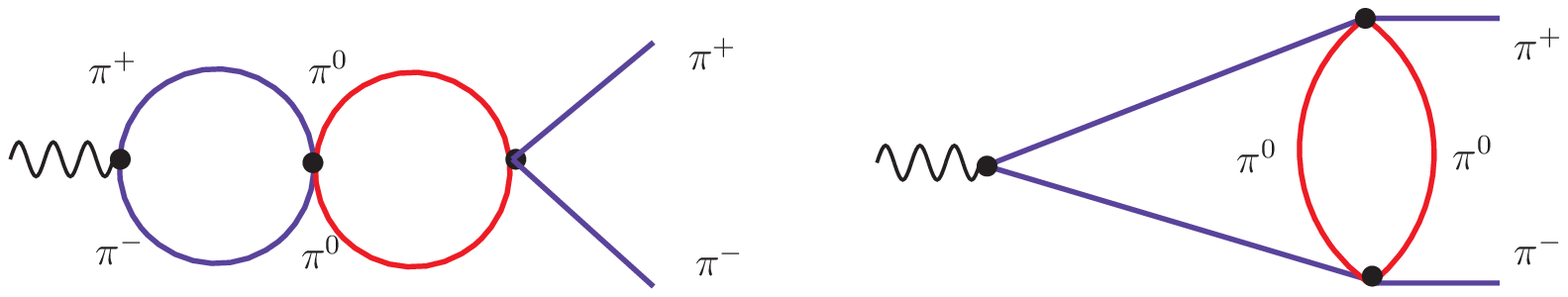,height=2.8cm}
\caption{Examples of two-loop graphs of the factorizing (left) and of the non-factorizing (right) type
involving pions with different masses.\label{fig3}}
\end{figure}

\noindent
Here $\Delta_1 \psi^{x}_0(s)$ and $\Delta_2 \psi^{x}_0(s)$ represent two contributions
that behave as ${\cal O}(M_{\pi}^2 - M_{\pi^0}^2)$ and as ${\cal O}\left( (M_{\pi}^2 - M_{\pi^0}^2)^2\right)$,
respectively, as one approaches the isospin limit. Their expressions, together with those of the remaining
polynomials $\xi_{x}^{(0)}(s)$, etc. are given in appendix~\ref{app:polynomials}. This expression of
$\psi^{x}_0(s)$ involves the function ${\cal L}_{\mbox{\tiny{$\pm $}}}(s)$, defined as
\begin{equation}
{\cal L}_{\mbox{\tiny{$\pm $}}} (s) \,=\, \ln\left[ \chi(t_{\mbox{\tiny{$\pm $}}}(s)) \right] ,
\end{equation}
with
\begin{equation}
\chi(t) \,=\, 
\frac{\sqrt{(M_{\pi} + M_{\pi^0})^2 - t} - \sqrt{(M_{\pi} - M_{\pi^0})^2 - t}}{\sqrt{(M_{\pi} + M_{\pi^0})^2 - t} + \sqrt{(M_{\pi} - M_{\pi^0})^2 - t}}\, .
\label{chi_def}
\end{equation}
Let us point out that
the expression (\ref{psi_x_0}) for $\psi^{x}_0(s)$ holds in the range $s\ge 4 M_{\pi}^2$, where the
functions $t_{\mbox{\tiny{$\pm $}}}(s)$ are real. An analytical continuation is necessary in order to
describe $\psi^{x}_0(s)$ in, say, the range $ 4 M_{\pi^0}^2 \le s \le 4 M_{\pi}^2$, as required, for instance,
for ${\mbox{Im}}F_S^{\pi}(s)$, cf. eq. (\ref{ImF_NLO}). For the applications that will be discussed in 
the following sections, we need not deal with this aspect, and the expression (\ref{psi_x_0})
is sufficient. 
It is also useful to notice that in the isospin limit 
$t_{\mbox{\tiny{$+$}}} (s)$,
$\lambda^{1/2}(t_{\mbox{\tiny{$+$}}}(s)){\cal L}_{\mbox{\tiny{$+$}}} (s)$, and ${\cal L}_{\mbox{\tiny{$\pm $}}}^2 (s)$
all behave as ${\cal O}\left( (M_{\pi}^2 - M_{\pi^0}^2)^2\right)$.
We keep however the contributions involving ${\cal L}_{\mbox{\tiny{$+$}}} (s)$ as indicated in equation (\ref{psi_x_0}),
so that each of the three pieces, when taken separately, displays a regular behaviour as $s$ approaches $4 M_\pi^2$ (from above).
Finally,
in the limit where the value of the mass $M_{\pi^0}$ tends
to $M_\pi$, one recovers the result of \cite{KMSF95},
\begin{eqnarray}
&&\psi^{x}_0(s)\rightarrow \ \stackrel{{ }_{\mbox{\scriptsize{$o$}}}}{\psi} \stackrel{x}{_{0}} \! \! (s)
\,\equiv\, 2\,\frac{M_\pi^4}{F_\pi^4}\,
\sqrt{\frac{s}{s-4M_\pi^2}}\,
\sum_{n=0}^3\left[\frac{1}{3}\,\xi^{(n)}_2(s)\,-\,\frac{1}{3}\,\xi^{(n)}_0(s)\right]
k_n(s) .
\label{lim_psi_x}
\end{eqnarray}

\subsection{Two-loop representation of the form factors and scattering amplitudes}

Having obtained the partial wave projections in the $S$ and $P$ waves from the
relevant one-loop $\pi\pi$ scattering amplitudes, we may now proceed towards
obtaining the two-loop expressions of the form factors and scattering amplitudes.
This requires one to evaluate the dispersive integrals in terms of which they are expressed.
As mentioned previously, we will not be able to work out analytical expressions
for all the integrals involved. Closed expressions will be obtained only for the 
contributions corresponding to so-called factorizing two-loop diagrams, see fig. \ref{fig3}. They will 
involve the functions ${\bar K}_n (s)$, defined as \cite{KMSF95}
\begin{equation}
{\bar K}_n(s) \,=\, \frac{s}{\pi} \int_{4 M_\pi^2}^\infty \frac{dx}{x}\,\frac{k_n(s)}{x - s - i0} ,
\end{equation}
with the functions $k_n(s)$ given in eq. (\ref{k_n}), and the understanding that $K_0(s)$
remains denoted by ${\bar J}(s)$. Explicit expressions of the functions ${\bar K}_n(s)$ in
terms of ${\bar J}(s)$ can be found in ref. \cite{KMSF95}.
There will also appear functions ${\bar K}_n^0(s)$, which are defined in the same way in terms
of functions $k_n^0(s)$, identical to the functions (\ref{k_n}), but with
the charged pion mass $M_\pi$ replaced by $M_{\pi^0}$. Similarly, we will keep the notation ${\bar J}_0(s)$ for 
${\bar K}^0_0(s)$.

Starting with the form factors, we obtain the two-loop representations
\begin{eqnarray}
F_S^{\pi^0}(s) &=& F_S^{\pi^0}\! (0)\! \left(
1 + a_S^{\pi^0} \! s + b_S^{\pi^0} \! s^2 
\right)
\nonumber\\
&&
+\, 8\pi F_S^{\pi^0}\! (0) \varphi_0^{00}(s) 
\left[
1 + a_S^{\pi^0} \! s + \frac{1}{\pi} \varphi_0^{00}(s) - \frac{2}{\pi} \frac{F_S^{\pi}(0)}{F_S^{\pi^0}(0)}\, \varphi_0^{x}(s)
\right] \! {\bar J}_0 (s) 
 \nonumber\\
&&
-\, 16\pi {F_S^{\pi}(0)}\, \varphi_0^{x}(s)
\left[
1 +  a_S^{\pi}\,s + \frac{2}{\pi} \varphi_0^{\mbox{\tiny{$ +-$}}}(s) 
- \frac{1}{\pi} \frac{F_S^{\pi^0}\! (0)}{F_S^{\pi}(0)}\, \varphi_0^{x}(s)
\right] \! {\bar J} (s)
\nonumber\\
&&
+\, \frac{M_{\pi}^4}{F_{\pi}^4}\,F_S^{\pi^0}\! (0)\! \left [
\xi^{(0)}_{00}(s) {\bar J}_0 (s) +
\xi^{(1;0)}_{00} (s) {\bar K}_1^0(s) +
2 \xi^{(2;0)}_{00} (s) {\bar K}_2^0(s) +
\xi^{(3;0)}_{00} (s) {\bar K}_3^0(s)
\right]
\nonumber\\
&&
-\, \,\frac{M_{\pi}^4}{F_{\pi}^4}\,F_S^{\pi}(0) \left[
2 \xi^{(0)}_{x} (s) {\bar J} (s) + 4 \xi^{(2;{\mbox{\tiny{$\!\pm $}}})}_{x} (s) {\bar K}_2(s)
\right]
\nonumber\\
&&
+\, 2 \,\frac{M_{\pi}^4}{F_{\pi}^4} 
\left[ F_S^{\pi^0}\! (0)\xi^{(2;{\mbox{\tiny{$\!\pm $}}})}_{00} (s) - 2 F_S^{\pi}(0) \xi^{(2;0)}_{x} (s) \right]
 \left[ 16 \pi^2 {\bar J} (s) {\bar J}_0 (s) - 2 {\bar J} (s) - 2 {\bar J}_0 (s) \right]
\nonumber\\
&&
+\,  \Delta_{\mbox{\tiny NF}} F_S^{\pi^0}(s) 
\,+\, {\cal O}(E^8) ,
\end{eqnarray}
and
\begin{eqnarray}
F_S^{\pi}(s) &=& F_S^{\pi} (0)\! \left(
1 + a_S^{\pi}  s + b_S^{\pi}  s^2 
\right)
\nonumber\\
&&
+\, 16\pi F_S^{\pi} (0) \varphi_0^{\mbox{\tiny{$ +-$}}}(s)
\left[
1 + a_S^{\pi} s + \frac{2}{\pi} \varphi_0^{\mbox{\tiny{$ +-$}}}(s) 
- \frac{1}{\pi} \frac{F_S^{\pi^0}\!(0)}{F_S^{\pi}(0)}\,\varphi_0^{x}(s)
\right] \! {\bar J} (s) 
 \nonumber\\
&&
-\, 8\pi {F_S^{\pi^0}\! (0)}\, \varphi_0^{x}(s)
\left[
1 +  a_S^{\pi^0}\! s + \frac{1}{\pi} \varphi_0^{00}(s) 
- \frac{2}{\pi} \frac{F_S^{\pi} (0)}{F_S^{\pi^0}\! (0)}\, \varphi_0^{x}(s)
\right] \! {\bar J}_0 (s)
\nonumber\\
&&
+\, 2\,\frac{M_{\pi}^4}{F_{\pi}^4}\,F_S^{\pi} (0) \left [
 \xi_{{\mbox{\tiny{$+-$}}};S}^{(0)}(s) {\bar J} (s) +
 \xi_{{\mbox{\tiny{$+-$}}};S}^{(1;{\mbox{\tiny{$\!\pm $}}})}(s) {\bar K}_1(s) +
2 \xi_{{\mbox{\tiny{$+-$}}};S}^{(2;{\mbox{\tiny{$\!\pm $}}})}(s) {\bar K}_2(s) +
 \xi_{{\mbox{\tiny{$+-$}}};S}^{(3;{\mbox{\tiny{$\!\pm $}}})}(s) {\bar K}_3(s)
\right]
\nonumber\\
&&
-\, \frac{M_{\pi}^4}{F_{\pi}^4}\,F_S^{\pi^0}\! (0) \left[
\xi^{(0)}_{x} (s) {\bar J}_0 (s) + 2\xi^{(2;0)}_{x} (s) {\bar K}_2^0(s)
\right]
\nonumber\\
&&
+\, 2 \,\frac{M_{\pi}^4}{F_{\pi}^4} 
\left[2 F_S^{\pi} (0) \xi_{{\mbox{\tiny{$+-$}}};S}^{(2;0)} (s) - F_S^{\pi^0}\! (0) \xi^{(2;{\mbox{\tiny{$\!\pm $}}})}_{x} (s) \right]
 \left[ 16 \pi^2 {\bar J} (s) {\bar J}_0 (s) - 2 {\bar J} (s) - 2 {\bar J}_0 (s) \right]
\nonumber\\
&&
+\,  \Delta_{\mbox{\tiny NF}} F_S^{\pi}(s)  
\,+\, {\cal O}(E^8) , 
\end{eqnarray}
for the scalar form factors, whereas the vector form factor reads
\begin{eqnarray}
F_V^{\pi}(s) &=&  
1 + a_V^{\pi} \! s + b_V^{\pi} \! s^2 
\nonumber\\
&&
+\, 16 \pi \varphi_1^{\mbox{\tiny{$+-$}}} (s) 
\left[
1 + a_V^{\pi} \! s + \frac{2}{\pi} \varphi_1^{\mbox{\tiny{$+-$}}}(s) 
\right] \! {\bar J} (s) 
 \nonumber\\
&&
+\, 2 \frac{M_{\pi}^4}{F_{\pi}^4} \left [
\xi^{(0)}_{{\mbox{\tiny{$+-$}}};P} (s) {\bar J} (s) +
\xi^{(1;{\mbox{\tiny{$\!\pm $}}})}_{{\mbox{\tiny{$+-$}}};P} (s) {\bar K}_1(s) +
2 \xi^{(2)}_{{\mbox{\tiny{$+-$}}};P} (s) {\bar K}_2(s) +
\xi^{(3;{\mbox{\tiny{$\!\pm $}}})}_{{\mbox{\tiny{$+-$}}};P} (s) {\bar K}_3(s) +
\xi^{(4;{\mbox{\tiny{$\!\pm $}}})}_{{\mbox{\tiny{$+-$}}};P} (s) {\bar K}_4(s)
\right]
\nonumber\\
&&
+\, \Delta_{\mbox{\tiny NF}} F_V^{\pi}(s) 
\,+\, {\cal O}(E^6) . 
\end{eqnarray}
The contributions $\Delta_{\mbox{\tiny NF}} F_S^{\pi^0}(s)$, $\Delta_{\mbox{\tiny NF}} F_S^{\pi}(s)$, and
$\Delta_{\mbox{\tiny NF}} F_V^{\pi}(s)$ stemming from non-factorizing two-loop graphs, see fig. \ref{fig3},
are expressed as dispersive integrals,
\begin{eqnarray}
\Delta_{\mbox{\tiny NF}} F_S^{\pi^0}(s) &=&
\frac{M_{\pi}^4}{F_{\pi}^4}\,F_S^{\pi^0}\! (0)
\frac{s}{\pi} \int_{4 M_{\pi^0}^2}^{\infty}
\frac{dx}{x}\,\frac{1}{x - s - i0} \!
\left[
\xi^{(1;{\mbox{\tiny$\nabla$}})}_{00}(s)\,\frac{1}{8 \pi}\,\frac{\sigma_{0}(x)}{\sigma_{\mbox{\tiny$\nabla$}}(x)}
\, L_{\mbox{\tiny$\nabla$}}(x) \,+\,
\xi^{(3;{\mbox{\tiny$\nabla$}})}_{00}(s)\,\frac{3}{16 \pi}\,\frac{M_{\pi}^2}{x\sigma_{0}(x)}
\, L_{\mbox{\tiny$\nabla$}}^2(x)
\right]
\nonumber\\
&&\!\!\!\!\!
-\,2 \frac{M_{\pi}^4}{F_{\pi}^4}\,F_S^{\pi} (0)
\frac{s}{\pi} \int_{4 M_{\pi}^2}^{\infty}
\frac{dx}{x}\,\frac{1}{x - s - i0}
\left[
\xi^{(1)}_{x} (s) \,\frac{1}{8 \pi}\,\frac{\lambda^{1/2}(t_{\mbox{\tiny $-$}}(x))}{x \sigma_0 (x)}\, {\cal L} (x) \,+\,
\xi^{(3)}_{x}(s)\,\frac{3}{16 \pi}\,\frac{M_{\pi}^2}{x \sigma_0 (x)}
\,{\cal L}^2(x)
\right]
\nonumber\\
&&\!\!\!\!\!
-\, F_S^{\pi} (0)
\frac{s}{\pi} \int_{4 M_{\pi}^2}^{\infty}
\frac{dx}{x}\,\frac{1}{x - s - i0}\,\sigma(x)
\left[
\Delta_1 \psi^{x}_0(x) \,+\,  \Delta_2 \psi^{x}_0(x) 
\right] 
\end{eqnarray}
\begin{eqnarray}
\Delta_{\mbox{\tiny NF}} F_S^{\pi}(s) &=&
2 \frac{M_{\pi}^4}{F_{\pi}^4}\,F_S^{\pi} (0)
\frac{s}{\pi} \int_{4 M_{\pi}^2}^{\infty}
\frac{dx}{x}\,\frac{1}{x - s - i0}
\left[
\xi^{(1;{\mbox{\tiny$\Delta$}})}_{{\mbox{\tiny{$+-$}}}; S}(s) \,\frac{1}{8 \pi}\,\frac{\sigma (x)}{\sigma_{\mbox{\tiny$\Delta$}}(x)}
\, L_{\mbox{\tiny$\Delta$}}(x)
\,+\,
\xi^{(3;{\mbox{\tiny$\Delta$}})}_{{\mbox{\tiny{$+-$}}};S}(s) \,\frac{3}{16 \pi}\,
\frac{M_{\pi^0}^2}{x\sigma(x)}\,L_{\mbox{\tiny$\Delta$}}^2(x)
\right]
\nonumber\\
&&\!\!\!\!\!
-\, \frac{M_{\pi}^4}{F_{\pi}^4}\,F_S^{\pi^0}\! (0)
\frac{s}{\pi} \int_{4 M_{\pi^0}^2}^{\infty}
\frac{dx}{x}\,\frac{1}{x - s - i0}
\left[
\xi^{(1)}_{x} (s) \,\frac{1}{8 \pi}\,\frac{\lambda^{1/2}(t_{\mbox{\tiny $-$}}(x))}{x \sigma (x)}\, {\cal L} (x) \,+\,
\xi^{(3)}_{x}(s)\,\frac{3}{16 \pi}\,\frac{M_{\pi}^2}{x \sigma (x)}
\,{\cal L}^2(x)
\right]
\nonumber\\
&&\!\!\!\!\!
-\,\frac{1}{2}\, F_S^{\pi^0} (0)
\frac{s}{\pi} \int_{4 M_{\pi^0}^2}^{\infty}
\frac{dx}{x}\,\frac{1}{x - s - i0}\,\sigma_0(x)
\left[
\Delta_1 \psi^{x}_0(x) \,+\,  \Delta_2 \psi^{x}_0(x) 
\right] 
\end{eqnarray}
\begin{eqnarray}
\Delta_{\mbox{\tiny NF}} F_V^{\pi}(s) &=&
2 \frac{M_{\pi}^4}{F_{\pi}^4}
\frac{s}{\pi} \int_{4 M_{\pi}^2}^{\infty}
\frac{dx}{x}\,\frac{1}{x - s - i0}
\left\{
\xi^{(1;{\mbox{\tiny$\Delta$}})}_{{\mbox{\tiny{$+-$}}}; P}(s)
\,\frac{1}{8 \pi}\,\frac{\sigma (x)}{\sigma_{\mbox{\tiny$\Delta$}}(x)}
\, L_{\mbox{\tiny$\Delta$}}(x) \,+\,
\xi^{(3;{\mbox{\tiny$\Delta$}})}_{{\mbox{\tiny{$+-$}}};P}(s) \,\frac{3}{16 \pi}\,\frac{M_{\pi^0}^2}{x\sigma (x)}
\,L_{\mbox{\tiny$\Delta$}}^2(x)
\right.
\nonumber\\
&&
\left.
+\,\xi^{(4;{\mbox{\tiny$\Delta$}})}_{{\mbox{\tiny{$+-$}}};P}(s)\,\frac{1}{16 \pi}\,
\frac{M_{\pi^0}^2}{\sqrt{x(x-4 M_{\pi}^2)}}\,
\left[
1\,+\,\frac{1}{\sigma_{\mbox{\tiny$\Delta$}}(x)}\,L_{\mbox{\tiny$\Delta$}}(x)
\,+\,\frac{M_{\pi^0}^2}{x-4 M_{\pi}^2}\,L_{\mbox{\tiny$\Delta$}}^2(x)  
\right]
\right\} .
\end{eqnarray}
Their evaluation has to be performed numerically. Notice, however, that
these representations are not necessarily best suited for this purpose,
due to possible numerical stability problems. These functions are actually
often expressed as two-dimensional integrals, which can be computed more
efficiently.
Examples of such representations can be found in the articles quoted under \cite{2loops},
and we refer the reader to them and to the papers quoted therein for further discussions on these aspects.

We could proceed in a similar way in order to write down two-loop representations
for the $\pi\pi$ amplitudes, but this is not very useful for the applications of interest here. 
For the sake of illustration, let us consider the function $W_0^{00}(s)$, involved
in the amplitude for elastic scattering of two neutral pions, as an example.
With the results already at our disposal, we obtain
\begin{eqnarray}
W_0^{00}(s) &=&
\frac{1}{2}\left[16\pi \varphi^{00}_0(s) \right] ^2 {\bar J}_0 (s)
\,+\,
\left[16\pi \varphi^{x}_0(s) \right] ^2\,{\bar J} (s)
\nonumber\\
&&
+ 32 \pi \,\frac{M_{\pi}^4}{F_{\pi}^4}\,\varphi_0^{00}(s)\left [
\xi^{(0)}_{00}(s) {\bar J}_0 (s) +
\xi^{(1;0)}_{00} (s) {\bar K}_1^0(s) +
\xi^{(2;0)}_{00} (s) {\bar K}_2^0(s) +
\xi^{(3;0)}_{00} (s) {\bar K}_3^0(s)
\right]
\nonumber\\
&&
+ 64 \pi \frac{M_{\pi}^4}{F_{\pi}^4}\,\varphi^{x}_0(s) \left[
 \xi^{(0)}_{x} (s) {\bar J} (s) + \xi^{(2;{\mbox{\tiny{$\!\pm $}}})}_{x} (s) {\bar K}_2(s)
\right]
\nonumber\\
&&
+ 64 \pi \,\frac{M_{\pi}^4}{F_{\pi}^4}  
\varphi_0^{00}(s) \xi^{(2;{\mbox{\tiny{$\!\pm $}}})}_{00} (s)
 \left[  16 \pi^2 {\bar J} (s) {\bar J}_0 (s) - 2 {\bar J} (s) - 2 {\bar J}_0 (s)  \right]
\nonumber\\
&&
+\,  
\Delta_{\mbox{\tiny NF}} W_0^{00}(s) ,
\end{eqnarray}
with
\begin{eqnarray}
\Delta_{\mbox{\tiny NF}} W_0^{00}(s) &=&
32 \pi \,\frac{M_{\pi}^4}{F_{\pi}^4}\,\varphi_0^{00}(s)
\frac{s}{\pi} \int_{4 M_{\pi^0}^2}^{\infty}\!\!
\frac{dx}{x}\,\frac{1}{x - s - i0}
\left[
\xi^{(1;{\mbox{\tiny$\nabla$}})}_{00}(s)\,\frac{1}{8 \pi}\,\frac{\sigma_{0}(x)}{\sigma_{\mbox{\tiny$\nabla$}}(x)}
\, L_{\mbox{\tiny$\nabla$}}(x) +
\xi^{(3;{\mbox{\tiny$\nabla$}})}_{00}(s)\,\frac{3}{16 \pi}\,\frac{M_{\pi}^2}{x\sigma_{0}(x)}
\, L_{\mbox{\tiny$\nabla$}}^2(x)
\right]
\nonumber\\
&&\!\!\!\!\!
+\, 64 \pi \frac{M_{\pi}^4}{F_{\pi}^4}\,\varphi^{x}_0(s)
\frac{s}{\pi} \int_{4 M_{\pi}^2}^{\infty}
\frac{dx}{x}\,\frac{1}{x - s - i0}
\left[
\xi^{(1)}_{x} (s) \,\frac{1}{8 \pi}\,\frac{\lambda^{1/2}(t_{\mbox{\tiny $-$}}(x))}{x \sigma_0 (x)}\, {\cal L} (x)
\,+\,
\xi^{(3)}_{x}(s)\,\frac{3}{16 \pi}\,\frac{M_{\pi}^2}{x \sigma_0 (x)}
\,{\cal L}^2(x)
\right]
\nonumber\\
&&\!\!\!\!\!
+\, 32 \pi \frac{M_{\pi}^4}{F_{\pi}^4}\,\varphi^{x}_0(s)
\frac{s}{\pi} \int_{4 M_{\pi}^2}^{\infty}
\frac{dx}{x}\,\frac{1}{x - s - i0}\,\sigma (x)
\left[
\Delta_1 \psi^{x}_0(x) \,+\,  \Delta_2 \psi^{x}_0(x)
\right] .
\end{eqnarray}
The remaining $W$-functions can be handled in a similar way, but we will not pursue
the matter further here. As for the corresponding subtraction polynomials, they were
already given in eq. (\ref{polynomials_P}).

\section{Isospin breaking in phase shifts} \label{IB_in_phases}
\setcounter{equation}{0}

In this section we turn to the issue of isospin breaking in the phases of the form factors,
making use of the results obtained so far. We begin with a discussion of some general
aspects of the phases of the form factors in the low-energy regime, 
and consider the lowest-order isospin-breaking corrections. The corrections at next order
are then discussed in a framework where only contributions of first order in the difference
of the pion masses are kept.

\subsection{General discussion and leading-order results}\label{IB_in_phases_LO}

The phases of the form factors are defined generically as
\begin{equation}
F(s+i0) = e^{2i\delta (s)}  F(s-i0)
,
\end{equation}
where the phases will be denoted $\delta_0^{\pi^0}(s)$, $\delta_0^{\pi}(s)$, and $\delta_1^{\pi}(s)$
for the form factors $F_S^{\pi^0}(s)$, $F_S^{\pi}(s)$, and  $F_V^{\pi}(s)$, respectively.
For a discussion of the analyticity properties of the form factors, we refer the reader to
ref. \cite{Colangelo:2008sm}.
Each of these phases $\delta_{\ell}(s)$ has
itself a low-energy expansion, $\delta_{\ell}(s) = \delta_{\ell,2}(s) + \delta_{\ell,4}(s)
+ {\cal O}(E^6)$.
Our aim is to address the issue of isospin-breaking corrections
in the phases order by order in this expansion, i.e. our interest lies in the differences 
\begin{equation}
\Delta \delta_\ell (s) \equiv \delta_\ell (s) - \stackrel{o}{\delta} _\ell \!\!(s)
= \Delta_{2} \delta_\ell (s) + \Delta_{4} \delta_\ell (s) + {\cal O}(E^6) 
\end{equation}
between the phases $\delta_\ell (s)$ in the presence of isospin breaking
and the phases $\stackrel{o}{\delta} _\ell \!\!(s)$ in the isospin limit,
with $\Delta_n \delta_\ell (s) \equiv \delta_{\ell , n} (s) - \stackrel{o}{\delta} _{\ell ,n} \!\!(s)$.
For the cases under consideration, we have
\begin{eqnarray}
\delta_{0}^{\pi^0}(s) &=& 
\frac{1}{2}\,\sigma_{0}(s)\left[
 \varphi_0^{00}(s) \,+\, \psi_0^{00}(s) \right]\theta(s-4M_{\pi^0}^2)
\nonumber\\
&&\!\!\!\!\!
-\,
\sigma (s) \,\frac{F_S^{\pi}(0)}{F_S^{\pi^0}(0)}\left[
\varphi_0^{x}(s) 
\frac{1+\Gamma^{\pi}_S(s)}{1+\Gamma^{\pi^0}_S (s)}\,+\,
\psi_0^{x}(s)
\right]\theta(s-4M_{\pi}^2)
\,+\,{\cal O}(E^6) 
\nonumber\\
\delta_{0}^{\pi}(s) &=& 
\sigma (s)\left[ \varphi_0^{\mbox{\tiny{$+-$}}}(s) \,+\,\psi_0^{\mbox{\tiny{$+-$}}}(s)
\right]\theta(s-4M_{\pi}^2)
\nonumber\\
&&\!\!\!\!\! 
-\,\frac{1}{2}\,\sigma_{0}(s)\,\frac{F_S^{\pi^0}(0)}{F_S^{\pi}(0)}
\left[ \varphi_0^{x}(s)
\frac{1+\Gamma^{\pi^0}_S(s)}{1+\Gamma^{\pi}_S (s)}\,+\,
\psi_0^{x}(s)
\right] \theta(s-4M_{\pi^0}^2)
\,+\,{\cal O}(E^6)
\nonumber\\
\delta_{1}^{\pi}(s) &=& 
\sigma (s)\left[ \varphi_1^{\mbox{\tiny{$+-$}}} (s) \,+\,\psi_1^{\mbox{\tiny{$+-$}}} (s)
\right]\theta(s-4M_{\pi}^2)
\,+\,{\cal O}(E^6) .
\end{eqnarray}
We thus deduce that
\begin{eqnarray}
\delta_{0,2}^{\pi^0}(s) &=& 
\frac{1}{2}\,\sigma_{0}(s) \varphi_0^{00}(s) \theta(s-4M_{\pi^0}^2)
 \, -\,
\sigma (s) \varphi_0^{x}(s) \theta(s-4M_{\pi}^2)
\nonumber\\
\delta_{0,2}^{\pi}(s) &=&  
\sigma (s) \varphi_0^{\mbox{\tiny{$+-$}}}(s) \theta(s-4M_{\pi}^2)
\, -\,
\frac{1}{2}\,\sigma_{0}(s) \varphi_0^{x}(s) \theta(s-4M_{\pi^0}^2)
\nonumber\\
\delta_{1,2}^{\pi}(s) &=&  
\sigma (s)
\varphi_1^{\mbox{\tiny{$+-$}}}(s) \theta(s-4M_{\pi}^2) ,
\label{phases_LO}
\end{eqnarray}
at leading order, while, at next-to-leading order,
\begin{eqnarray}
\delta_{0,4}^{\pi^0}(s) &=& 
\frac{1}{2}\,\sigma_{0}(s) \psi_0^{00}(s) \theta(s-4M_{\pi^0}^2)
 \, -\,\sigma (s) \psi_0^{x}(s) \theta(s-4M_{\pi}^2)
 \nonumber\\
 && \,-\,
\sigma (s)\varphi_0^{x}(s)
\left[\left(\frac{F_S^{\pi}(0)}{F_S^{\pi^0}(0)} - 1\right)
+\left(\Gamma_S^{\pi}(s) - \Gamma_S^{\pi^0}(s)\right)
\right] \theta(s-4M_{\pi}^2)
\nonumber\\
\delta_{0,4}^{\pi}(s) &=&  
\sigma (s) \psi_0^{\mbox{\tiny{$+-$}}}(s) \theta(s-4M_{\pi}^2)
\, -\,
\frac{1}{2}\,\sigma_{0}(s) \psi_0^{x}(s) \theta(s-4M_{\pi^0}^2)
 \nonumber\\
 && \, -\,
\frac{1}{2}\,
\sigma_{0}(s)\varphi_0^{x}(s)
\left[\left(\frac{F_S^{\pi^0}(0)}{F_S^{\pi}(0)} - 1\right)
-\left(\Gamma_S^{\pi}(s) - \Gamma_S^{\pi^0}(s)\right)
\right] \theta(s-4M_{\pi^0}^2)
\nonumber\\
\delta_{1,4}^{\pi}(s) &=&  
\sigma (s)
\psi_1^{\mbox{\tiny{$+-$}}}(s) \theta(s-4M_{\pi}^2) .
\label{phases_NLO}
\end{eqnarray}
%
%
%
\begin{figure}[t]
\center\epsfig{figure=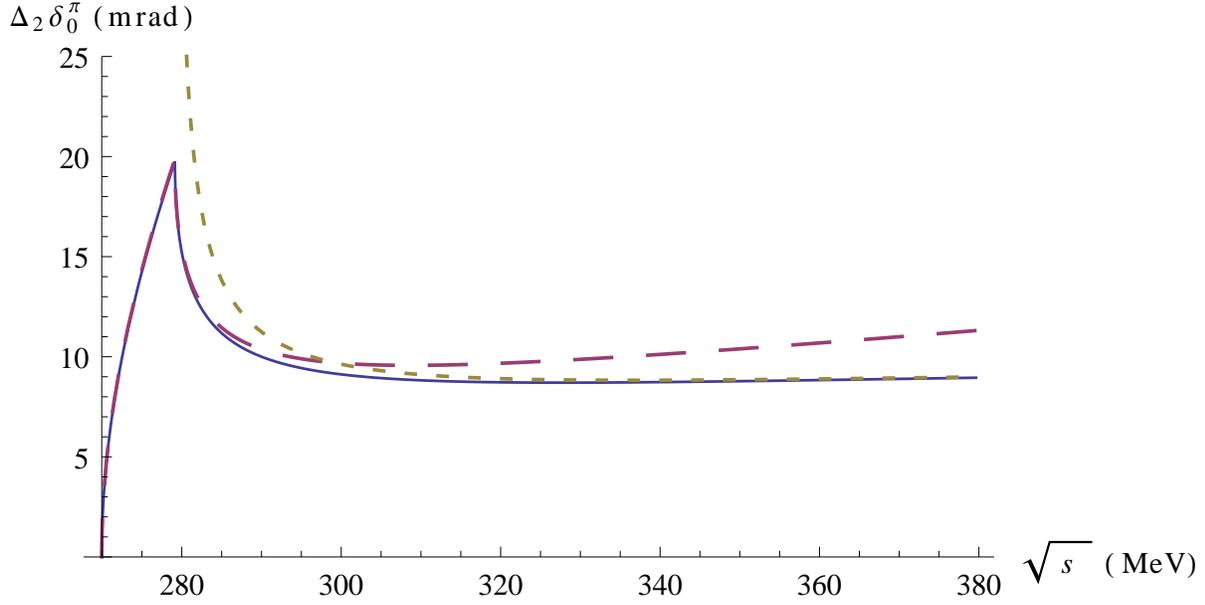,height=8.0cm}
\caption{The lowest-order phase difference $\Delta_2^{\mbox{\scriptsize{LO}}} \delta^\pi_{0}(s)$ (solid line),
defined in eq. (\ref{Delta_2_LO}), as a function of the energy 
$\sqrt{s}$ in MeV, for $\alpha=1.40$ and $\beta =1.08$. Also shown is the quantity $\Delta_2  \delta^\pi_{0}(s)$ (long-dashed line),
defined in eq. (\ref{Delta_2}), as well as the approximation to first order in $\Delta_\pi$
(short-dashed line), cf. eq. (\ref{Delta_2_LO_approx}).\label{fig5}
}
\end{figure}
%
%
Here we have used the property that the quantities $F_S^{\pi^0}(0)/F_S^{\pi}(0) - 1$, 
$\Gamma_S^{\pi}(s)$, and $\Gamma_S^{\pi^0}(s)$ are all of order ${\cal O}(E^2)$. 
At order ${\cal O}(E^n)$, the isospin-symmetric phases $\stackrel{o}{\delta} _{\ell ,n} \!\!(s)$
are obtained from the expressions (\ref{phases_LO}) and (\ref{phases_NLO}) by setting $M_{\pi^0}$ equal to
$M_\pi$, and by replacing the lowest order expressions of the $\pi\pi$ $S$ and $P$ waves,
$\varphi_\ell (s)$ and $\psi_\ell (s)$, by their counterparts in the isospin limit, 
$\stackrel{o}{\varphi} _\ell \!(s)$ and $\stackrel{{ }_{\mbox{\scriptsize{$o$}}}}{\psi} _\ell \!(s)$, 
respectively. In this limit, the differences $F_S^{\pi^0}(0) / F_S^{\pi}(0) - 1$ and 
$\Gamma_S^{\pi}(s) - \Gamma_S^{\pi^0}(s)$ vanish.

Before proceeding with the actual calculation, let us make a few comments. 
First, we should point out an important aspect that emerges from the above expressions,
and that has already been observed in ref. \cite{Colangelo:2008sm}.
At order ${\cal O}(E^2)$, the phases of the form factors are entirely
determined by the $\pi\pi$ scattering data. In the case
of the vector form factor where, due to Bose symmetry, the $\pi^+\pi^-$ 
channel alone contributes, Watson's theorem is still operative: the phase of
$F_V^\pi(s)$ coincides with the phase of the $P$-wave projection of the 
corresponding scattering amplitude $A^{\mbox{\tiny{$+-$}}}(s,t,u)$.
For the scalar form factors, the situation is different, due to the mixing between the two channels 
that contribute to the unitarity sum in the $S$ wave. Nevertheless, the phase
has still a ``universal'' character, in the sense that its expression 
involves only the partial waves of the $\pi\pi$ scattering amplitudes, and makes no explicit reference
to the form factors themselves. This property, however,
rests entirely on the fact that $F_S^{\pi^0}(0) / F_S^{\pi}(0) = 1$ at this order.
Hence, this situation does no longer survive at the next order in the case of the scalar form factors.
In addition to the universal parts, ${\Delta}_4^U \delta_0^{\pi^0}(s)$ and ${\Delta}_4^U \delta_0^{\pi}(s)$,
provided by the ${\cal O}(E^4)$ 
partial waves of the $\pi\pi$ scattering amplitudes, there now appear contributions
${\Delta}_4^F \delta_0^{\pi^0}(s)$ and ${\Delta}_4^F \delta_0^{\pi^0}(s)$ that depends 
explicitly on the form factors considered:
\begin{eqnarray}
{\Delta}_4 \delta_0^{\pi^0}(s) &=& {\Delta}_4^U \delta_0^{\pi^0}(s) \,+\, {\Delta}_4^F \delta_0^{\pi^0}(s) 
\nonumber\\
{\Delta}_4 \delta_0^{\pi}(s) &=& {\Delta}_4^U \delta_0^{\pi}(s) \,+\, {\Delta}_4^F \delta_0^{\pi}(s) 
.
\label{Delta_U_F}
\end{eqnarray}
The universal parts ${\Delta}_4^U \delta_0^{\pi^0}(s)$ and ${\Delta}_4^U \delta_0^{\pi}(s)$ correspond
to the first lines of the expressions of $\delta_{0,4}^{\pi^0}(s)$ and of $\delta_{0,4}^{\pi}(s)$ given
in eq. (\ref{phases_NLO}), respectively, from which the isospin-symmetric contributions are subtracted. 
The second lines in these same expressions correspond to the
form-factor dependent contributions ${\Delta}_4^F \delta_0^{\pi^0}(s)$ and ${\Delta}_4^F \delta_0^{\pi}(s)$.
Finally, we also note that for the scalar form factors
some contributions in the expressions (\ref{phases_LO}) and (\ref{phases_NLO}) 
start at $s=M_{\pi^0}^2$, while others appear only for
$s\ge M_\pi^2$. This is of course the manifestation of the unitarity cusp
in the phases themselves.

\begin{figure}[t]
\center\epsfig{figure=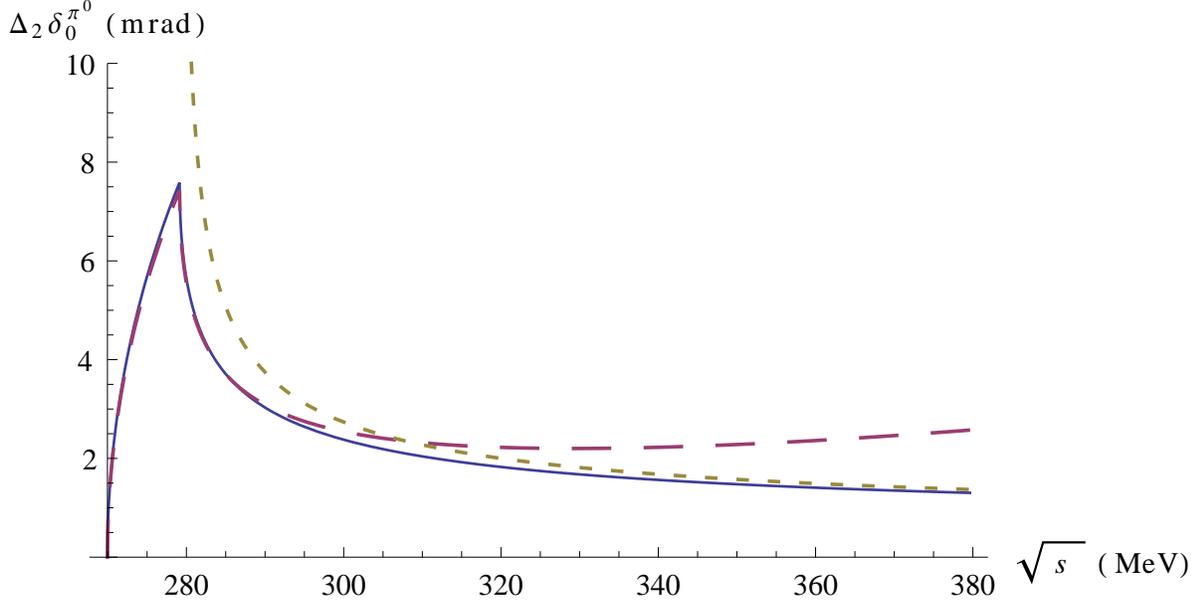,height=8.0cm}
\caption{The lowest-order phase difference $\Delta_2^{\mbox{\scriptsize{LO}}} \delta^{\pi^0}_{0}(s)$ (solid line),
defined in eq. (\ref{Delta_2_LO}), as a function of the energy 
$\sqrt{s}$ in MeV, for $\alpha=1.40$ and $\beta =1.08$. Also shown is the quantity $\Delta_2  \delta^\pi_{0}(s)$ (long-dashed line),
defined in eq. (\ref{Delta_2}), as well as the approximation to first order in $\Delta_\pi$
(short-dashed line), cf. eq. (\ref{Delta_2_LO_approx}).
\label{fig6}
}
\end{figure}

At order ${\cal O}(E^2)$, the isospin-symmetric phases $\stackrel{o}{\delta} _{\ell ,2} \!\!(s)$
are obtained from the expressions (\ref{phases_LO}), putting $M_{\pi^0}$ equal to
$M_\pi$, and replacing the lowest order expressions of the $\pi\pi$ $S$ and $P$ waves
$\varphi_\ell (s)$ by their counterparts in the isospin limit, $\stackrel{o}{\varphi} _\ell \!(s)$, 
which read
\begin{eqnarray}
\stackrel{o}{\varphi} \stackrel{00}{_{\!\! 0}} \!\! (s)
&=& \frac{\alpha M_{\pi}^2}{16\pi F_{\pi}^2}
\nonumber\\
\stackrel{o}{\varphi} \stackrel{x}{_{0}} \! \! (s)
&=&  
-\frac{\beta}{16\pi F_{\pi}^2}\,
\left( s - \frac{4}{3}M_{\pi}^2\right)\,-\,
\frac{\alpha M_{\pi}^2}{48\pi F_{\pi}^2}  
\nonumber\\
\stackrel{o}{\varphi} \stackrel{{\mbox{\tiny{$+-$}}}}{_{\!\!\!\! 0}} \! \! (s)
&=& \frac{\beta}{32\pi F_{\pi}^2}\,
\left( s  - \frac{4}{3}M_{\pi}^2 \right)\,+\,
\frac{\alpha M_{\pi}^2}{24\pi F_{\pi}^2}
\nonumber\\
\stackrel{o}{\varphi} \stackrel{{\mbox{\tiny{$+-$}}}}{_{\!\!\!\! 1}} \! \! (s)
&=& \frac{\beta}{96\pi F_{\pi}^2}\,
\left( s  - {4}M_{\pi}^2 \right) .
\end{eqnarray}
We then obtain
\begin{eqnarray}
 \Delta_2 \delta^{\pi^0}_0 (s)  &=&
\frac{\alpha_{00} M_{\pi^0}^2}{32 \pi F_\pi^2} 
\left[ \sigma_0 (s) \theta (s-4 M_{\pi^0}^2) - \sigma (s) \theta (s-4 M_{\pi}^2) \right]
+\,
\bigg[
\frac{4 \beta_x - 2 \alpha_x - 3 \alpha_{00}}{96 \pi} \frac{\Delta_\pi}{F_\pi^2}
\nonumber\\
&&
+\,
\frac{\beta_{x} - \beta}{16 \pi F_\pi^2} \left( s - \frac{4}{3} M_\pi^2 \right) \,+\,
\frac{\alpha_{x} - \alpha}{48 \pi} \frac{M_{\pi}^2}{F_\pi^2} \,+\,
\frac{\alpha_{00} - \alpha}{32 \pi} \frac{M_{\pi}^2}{F_\pi^2} 
\bigg]
\sigma (s) \theta (s-4 M_{\pi}^2)
\nonumber\\
 \Delta_2 \delta^{\pi}_0 (s)  &=&
\frac{1}{32\pi F_{\pi}^2}
\left[
\beta_x \left( s - \frac{2}{3}M_{\pi^0}^2 - \frac{2}{3}M_{\pi}^2 \right) +\,
\frac{\alpha_x M_{\pi^0}^2}{3}  
\right]
\left[ \sigma_0 (s) \theta (s-4 M_{\pi^0}^2) - \sigma (s) \theta (s-4 M_{\pi}^2) \right]
\nonumber\\
&&
+\,
\bigg[
\frac{2 \beta_x -  \alpha_x - 4 \alpha_{\mbox{\tiny{$+-$}}}}{96 \pi} \,\frac{\Delta_\pi}{F_\pi^2}
\,+\,
\frac{(\beta_{x} - \beta) + (\beta_{\mbox{\tiny{$+-$}}} - \beta )}{32 \pi F_\pi^2} \left( s - \frac{4}{3} M_\pi^2 \right) 
\nonumber\\
&&
+\,
\frac{\alpha_{x} - \alpha}{96 \pi} \frac{M_{\pi}^2}{F_\pi^2} \,+\,
\frac{\alpha_{\mbox{\tiny{$+-$}}} - \alpha}{24 \pi} \frac{M_{\pi}^2}{F_\pi^2} 
\bigg]
\sigma (s) \theta (s-4 M_{\pi}^2)
\nonumber\\
\Delta_2 \delta_1^{\pi}(s) &=&
\frac{1}{96\pi F_{\pi}^2}\,
(\beta_{\mbox{\tiny{$+-$}}} - \beta)\left(s - 4 M_{\pi}^2\right)
\sigma (s) .
\label{Delta_2}
\end{eqnarray}
If we make use of eq. (\ref{alphabetaLO}),
these expressions become
\begin{eqnarray}
 \Delta_2^{\mbox{\scriptsize{LO}}} \delta^{\pi^0}_0 (s)  &=&
\frac{\alpha M_{\pi^0}^2}{32 \pi F_\pi^2} 
\left[ \sigma_0 (s) \theta (s-4 M_{\pi^0}^2) - \sigma (s) \theta (s-4 M_{\pi}^2) \right]
+\,
\frac{8 \beta - 5\alpha}{96 \pi}  \, \frac{\Delta_\pi}{F_\pi^2}\, \sigma (s) \theta (s-4 M_{\pi}^2)
\nonumber\\
 \Delta_2^{\mbox{\scriptsize{LO}}} \delta^{\pi}_0 (s)  &=&
\frac{1}{32\pi F_{\pi}^2}
\left[
\beta \left( s - \frac{4}{3}M_{\pi^0}^2 \right) +\,
\frac{\alpha M_{\pi^0}^2}{3}  
\right]
\left[ \sigma_0 (s) \theta (s-4 M_{\pi^0}^2) - \sigma (s) \theta (s-4 M_{\pi}^2) \right]
\nonumber\\
&&
+\,
\frac{5(4 \beta - \alpha)}{96 \pi}  \, \frac{\Delta_\pi}{F_\pi^2}\, \sigma (s) \theta (s-4 M_{\pi}^2)  ,
\label{Delta_2_LO}
\end{eqnarray}
and $\Delta_2^{\mbox{\scriptsize{LO}}} \delta^{\pi}_1 (s) = 0$. 
Here we have introduced the notation
\begin{equation}
\Delta_\pi \,\equiv\, M_{\pi}^2 - M_{\pi^0}^2 \,.
\end{equation} 
Furthermore, $\Delta_2^{\mbox{\scriptsize{LO}}} \delta_\ell (s)$ represents the leading-order isospin-breaking correction 
when the lowest-order relations (\ref{alphabetaLO}) for the subthreshold parameters $\alpha_{00}$, $\alpha_x\ldots$
have been used. However, the latter also receive higher-order
corrections, that will be discussed below. This means that at next-to-leading order we also need to include a contribution 
\begin{equation}
\Delta_2^{\mbox{\scriptsize{NLO}}} \delta_\ell (s) \,=\, 
\Delta_2 \delta_\ell (s) \,-\,
\Delta_2^{\mbox{\scriptsize{LO}}} \delta_\ell (s)
\label{Delta_2_NLO}
\end{equation}
which takes them into account.
We show, on fig. \ref{fig5}, a plot of $\Delta_2^{\mbox{\scriptsize{LO}}} \delta^{\pi}_0 (s)$ as a function of energy
[the numerical input values we use are given in table \ref{tab:LECs} and in equation (\ref{inputs})].
The curve exhibits the characteristic cusp-type behaviour \cite{cusp61,Cabibbo:2004gq} at the $\sqrt{s} = 2 M_\pi$ threshold, where
$\Delta_2^{\mbox{\scriptsize{LO}}} \delta^{\pi}_0 (s)$ takes its maximal value, close to 20 milliradians. For $\sqrt{s}$ greater than
$\sim 300$ MeV, the value of $\Delta_2^{\mbox{\scriptsize{LO}}} \delta^{\pi}_0 (s)$ stays practically constant, around 10 milliradians.
The higher-order contributions mentioned in eq. (\ref{Delta_2_NLO}) become sizeable only above the cusp.
The curve for $\Delta_2^{\mbox{\scriptsize{LO}}} \delta^{\pi^0}_0 (s)$ follows a similar shape, see figure \ref{fig6}, but its magnitude 
is reduced by roughly a factor of three as compared to $\Delta_2^{\mbox{\scriptsize{LO}}} \delta^{\pi}_0 (s)$. 
In both cases, the difference $\Delta_2^{\mbox{\scriptsize{NLO}}} \delta_\ell (s)$ defined in equation (\ref{Delta_2_NLO}) above
is negligible in the cusp region, but becomes more and more important as the energy increases.
%
\begin{figure}[b]
\center\epsfig{figure=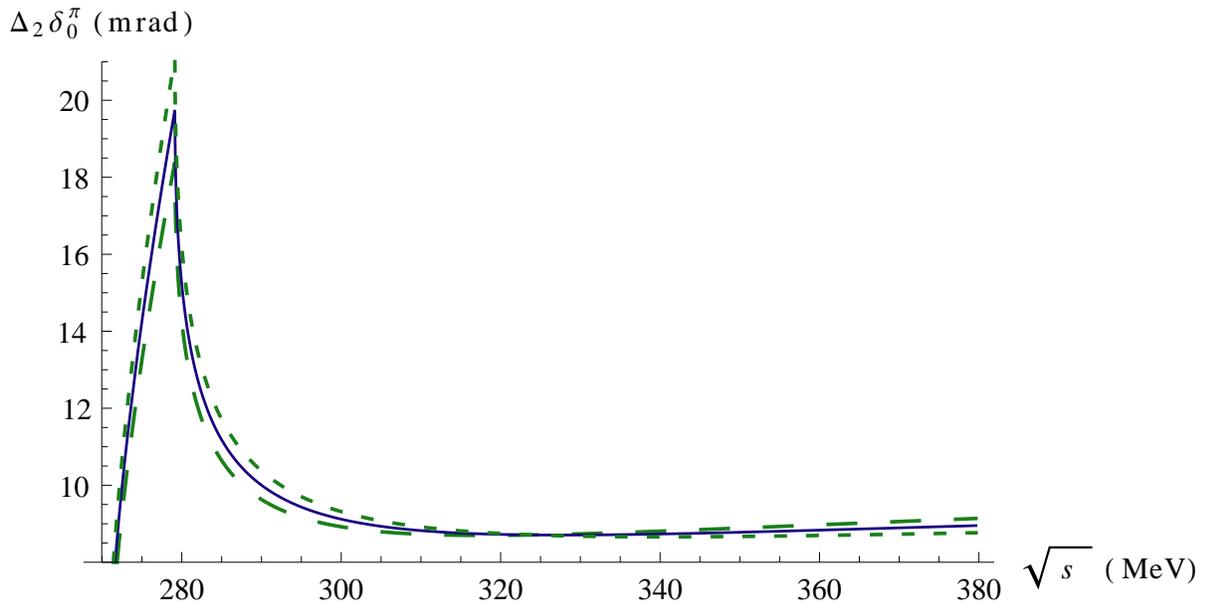,height=8.0cm}
\caption{The lowest-order phase difference $\Delta_2^{\mbox{\scriptsize{LO}}} \delta^{\pi}_{0}(s)$
for different values of $\alpha$ and $\beta$. The solid curve
corresponds to $(\alpha , \beta) = (1.4, 1.08)$, while the cases $(\alpha , \beta) = (1.0, 1.04)$ 
and $(\alpha , \beta) = (1.8, 1.12)$ are represented by
the long-dashed and short-dashed curves, respectively.
\label{fig7}}
\end{figure}

It is also interesting to investigate the sensitivity of the isospin-breaking corrections
with respect to the (unknown) parameters $\alpha$ and $\beta$. This is done in figures
\ref{fig6} and \ref{fig7}. We notice that the dependence is strongest in the vicinity of
the cusp, and is mainly driven by $\alpha$. Indeed, at $s=4 M_\pi^2$, one has
\begin{eqnarray}
 \Delta_2^{\mbox{\scriptsize{LO}}} \delta^{\pi^0}_0 (4 M_\pi^2) &=&
\frac{\alpha M_{\pi^0}^2}{32 \pi F_\pi^2} \, \sigma_0 (4 M_\pi^2)
\nonumber\\
 \Delta_2^{\mbox{\scriptsize{LO}}} \delta^{\pi}_0 (4 M_\pi^2) &=&
\frac{1}{32\pi F_{\pi}^2}
\left[
\beta \left( 4 M_\pi^2 - \frac{4}{3}M_{\pi^0}^2 \right) +\,
\frac{\alpha M_{\pi^0}^2}{3} \right] \sigma_0 (4 M_\pi^2) \, .
\label{Delta_2_LO_cusp}
\end{eqnarray}
The value at the cusp of $\Delta_2^{\mbox{\scriptsize{LO}}} \delta^{\pi^0}_0$ is directly proportional to $\alpha$.
In the case of $\Delta_2^{\mbox{\scriptsize{LO}}} \delta^{\pi}_0$, the value is driven by the contribution
proportional to $\beta$, the contribution proportional to $\alpha$ being suppressed by a factor of three
as compared to $\Delta_2^{\mbox{\scriptsize{LO}}} \delta^{\pi^0}_0$. However, the relative variation in $\beta$
only covers a restricted range, approximatively $\pm 5\%$, so that the variations in $\alpha$ account for the largest part of 
the effect. As one leaves the cusp region towards larger values of $s$, the contribution proportional to $\alpha$
looses weight, and the variations are less ample. This behaviour is conveyed in a simple manner by the high-energy
asymptotic expressions of 
$\Delta_2^{\mbox{\scriptsize{LO}}} \delta^{\pi^0}_0 (s)$, and, even more strongly, of $ \Delta_2^{\mbox{\scriptsize{LO}}} \delta^{\pi}_0 (s)$:
\begin{eqnarray}
\Delta_2^{\mbox{\scriptsize{LO}}} \delta^{\pi^0}_0 (s)  &\sim&
\frac{8 \beta - 5\alpha}{96 \pi}  \,\frac{\Delta_\pi}{F_\pi^2}\,
+\,\frac{\alpha - \beta }{6 \pi}  \,\frac{\Delta_\pi}{F_\pi^2}\,\frac{M_\pi^2}{s} \,+ \dots
\nonumber\\
 \Delta_2^{\mbox{\scriptsize{LO}}} \delta^{\pi}_0 (s)  &\sim&
\frac{26 \beta - 5\alpha}{96 \pi}  \,\frac{\Delta_\pi}{F_\pi^2}\,
+\,\frac{\alpha - 4 \beta }{8 \pi}  \,\frac{\Delta_\pi}{F_\pi^2}\,\frac{M_\pi^2}{s} \,+ \dots 
\label{Delta_2_LO_asym}
\end{eqnarray}

\begin{figure}[t]
\center\epsfig{figure=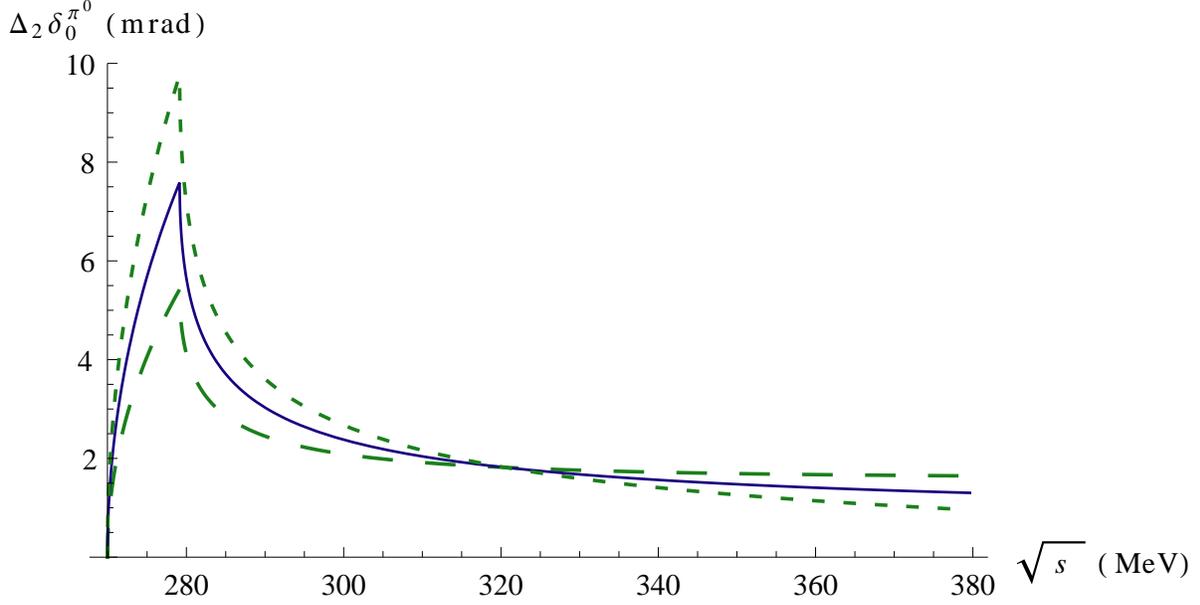,height=8.0cm}
\caption{The lowest-order phase difference $\Delta_2^{\mbox{\scriptsize{LO}}} \delta^{\pi^0}_{0}(s)$
for different values of $\alpha$ and $\beta$. The solid curve
corresponds to $(\alpha , \beta) = (1.4, 1.08)$, while the cases $(\alpha , \beta) = (1.0, 1.04)$ 
and $(\alpha , \beta) = (1.8, 1.12)$ are represented by
the long-dashed and short-dashed curves, respectively.
\label{fig8}}
\end{figure}

\subsection{Isospin-breaking corrections at next-to-leading order}\label{IB_in_phases_NLO}

The evaluation of $\Delta_4 \delta_\ell (s)$,
the isospin-breaking effects in the phases
at next-to-leading order, relies on the results obtained in the
preceding sections. The corresponding numerical analysis will be the
subject of section \ref{numerics} below. Here, we wish to proceed for a while
at the analytical level, but for simplicity, and since the next-to-leading order isospin-breaking corrections
are expected to be small, we will  restrict ourselves to the first order in $\Delta_\pi$. For this purpose, we expand
the various quantities of interest with respect to $\Delta_\pi$, and 
neglect contributions beyond the linear terms. In the case of the phase-space factor for the neutral two-pion state, 
an expansion like:
\begin{equation}
\sigma_0 (s) \,=\, \sigma (s) \left[ 1 \,+\, \frac{2}{s - 4 M_\pi^2}\,\Delta_\pi \,+\, {\cal O}(\Delta_\pi^2) \right] ,   
\label{sigma_0_expand}
\end{equation} 
will not make sense when $s$ remains close to $4 M_\pi^2$. 
This means that our expansion to first order in isospin breaking will only
provide an adequate description in regions of phase space sufficiently away from the $\pi^0 \pi^0$
and $\pi^+ \pi^-$ thresholds. From an experimental
point of view, this needs not constitute a serious drawback, since the vicinity of the two-pion
thresholds is usually part of the regions of phase space where the acceptance is low,
as can be seen, for instance, from \cite{Batley:2007zz,Batley:2010zza} in
the case of the $K^+_{e4}$ decay (see, however, the discussion on
the $K^\pm \rightarrow \pi^0 \pi^0 e^\pm \nu_e$ decay mode in \cite{NA48-2_Ke4_cusp}). 
From a practical point of view, we gain the advantage of having to deal with
expressions which remain tractable. In the rest of this section, we will therefore
remain within the framework set up by these two conditions -- staying away from the two-pion thresholds,
and considering only first-order isospin-violating effects. For illustration, let us
consider the lowest-order corrections in the $S$-wave phases,
$\Delta_2^{\mbox{\scriptsize{LO}}} \delta_0^{\pi^0}(s)$ and 
$\Delta_2^{\mbox{\scriptsize{LO}}} \delta_0^{\pi^0}(s)$ that we
have discussed in the preceding subsection. Applying the procedure that
we have just described to the expressions (\ref{Delta_2_LO}) yields, for $s > 4 M_\pi^2$,
\begin{eqnarray}
\Delta_2^{\mbox{\scriptsize{LO}}} \delta_0^{\pi^0}(s) &=& \sigma (s)\left[
\frac{\alpha}{16\pi}\,\frac{M_{\pi}^2}{s - 4 M_{\pi}^2}
\,+\,\frac{8 \beta - 5 \alpha}{96\pi}
\right] 
\frac{\Delta_\pi}{F_{\pi}^2}
\,+\,{\cal O}(\Delta_\pi^2)
\nonumber\\
\Delta_2^{\mbox{\scriptsize{LO}}} \delta_0^{\pi}(s) &=&\sigma (s)\left[
\frac{8 \beta + \alpha}{48\pi}\,\frac{M_{\pi}^2}{s - 4 M_{\pi}^2}
\,+\,\frac{26 \beta - 5 \alpha}{96\pi}
\right] 
\frac{\Delta_\pi}{F_{\pi}^2}
\,+\,{\cal O}(\Delta_\pi^2) .
\label{Delta_2_LO_approx}
\end{eqnarray}
On figures \ref{fig5} and \ref{fig6}, respectively, these expressions for $\Delta_2^{\mbox{\scriptsize{LO}}} \delta_0^{\pi}(s)$
and $\Delta_2^{\mbox{\scriptsize{LO}}} \delta_0^{\pi^0}(s)$
are shown together with the exact expressions given in equations (\ref{Delta_2}) and  (\ref{Delta_2_LO}).
We see that the curve corresponding to the approximate expression overshoots
the exact formula by about 25\% at $\sqrt{s} = 285$ MeV,
while the difference drops to a level around 5\% for $\sqrt{s}$ above $\sim325$ MeV.

The expressions of the next-to-leading partial waves 
$\stackrel{{ }_{\mbox{\scriptsize{$o$}}}}{\psi} \stackrel{00}{_{\!\! 0}} \!\! (s)$,
$\stackrel{{ }_{\mbox{\scriptsize{$o$}}}}{\psi} \stackrel{{\mbox{\tiny{$+-$}}}}{_{ 0,1}} \! \! (s)$,
and $\stackrel{{ }_{\mbox{\scriptsize{$o$}}}}{\psi} \stackrel{x}{_{0}} \! \! (s)$ in the isospin limit
have been given in eqs. (\ref{lim_psi_00}), (\ref{lim_psi_+-}), and (\ref{lim_psi_x}), respectively.
In order to evaluate $\Delta_4 \delta_\ell (s)$, we will proceed as follows.
First, we replace, in the polynomials $\xi^{(n)}_{00}(s)$, $\xi^{(n)}_{{\mbox{\tiny{$+-$}}};S,P}$, and
$\xi^{(n)}_{x}(s)$, each occurrence of $M_{\pi^0}^2$ by $M_{\pi}^2 - \Delta_\pi$. Keeping
only the terms of first order in $\Delta_\pi$, we thus obtain
\begin{equation}
\xi^{(n)}_{00}(s)\,=\,{\overline\xi}^{(n)}_{00}(s)\,+\,
\frac{\Delta_\pi}{M_\pi^2}\,\delta\xi^{(n)}_{00}(s)\,+\,
{\cal O}(\Delta_\pi^2) \, , 
\label{delta_xi}
\end{equation}
and so on. Next, we have to expand
the functions that multiply the polynomials $\xi^{(n)}_{00}(s)$, $\xi^{(n)}_{{\mbox{\tiny{$+-$}}};S}$,
$\xi^{(n)}_{{\mbox{\tiny{$+-$}}};P}$ and $\xi^{(n)}_{x}(s)$, see eqs. (\ref{psi_00}), (\ref{psi_+-_0}), (\ref{psi_+-_1}),
and (\ref{psi_x_0}), to first order in $\Delta_\pi$. 
Finally, when subtracting from the functions ${\overline\xi}^{(n)}_{00}(s)$, ${\overline\xi}^{(n)}_{{\mbox{\tiny{$+-$}}};S}$, 
${\overline\xi}^{(n)}_{{\mbox{\tiny{$+-$}}};P}$, ${\overline\xi}^{(n)}_{x}(s)$ the corresponding combinations of polynomials 
$\xi^{(n)}_\ell (s)$ arising in the isospin limit, as given in
(\ref{lim_psi_00}), (\ref{lim_psi_+-}), and (\ref{lim_psi_x}), isospin breaking only occurs through differences
like $\alpha_x - \alpha$, or $\beta_{\mbox{\tiny{$+-$}}} - \beta$, for instance.
Collecting these three contributions provides us with an expression for $\Delta_4 \delta_\ell (s)$
accurate at first order in $\Delta_\pi$.

As an illustration, consider the case of $\psi^{00}_0(s)$, cf. eq. (\ref{psi_00}). 
The first step is a straightforward algebraic exercise. It produces functions ${\overline\xi}^{(n)}_{00}(s)$,
$n=0,1,2,3$, with
\begin{equation}
{\overline\xi}^{\,(1)}_{00}(s) \,=\, {\overline\xi}^{\,(1;0)}_{00}(s) \,+\,
{\overline\xi}^{\,(1;{\mbox{\tiny$\nabla$}})}_{00}(s)\,,
\ {\overline\xi}^{\,(2)}_{00}(s) \,=\, {\overline\xi}^{\,(2;0)}_{00}(s) \,+\,
{\overline\xi}^{\,(2;\pm)}_{00}(s)\,,
\ {\overline\xi}^{\,(3)}_{00}(s) \,=\, {\overline\xi}^{\,(3;0)}_{00}(s) \,+\,
{\overline\xi}^{\,(3;{\mbox{\tiny$\nabla$}})}_{00}(s) .
\end{equation}
For the second step,
we expand the various functions that appear in the expression for $\psi^{00}_0(s)$ to first order in $\Delta_\pi$
[the functions $k_n(s)$ are defined in equation (\ref{k_n})]:
\begin{eqnarray}
2 \frac{\sigma(s)}{\sigma_{0}(s)}\,L_{0}(s) &=& 16\pi k_1(s)\,-\,
32\pi\,\frac{\Delta_\pi}{M_{\pi}^2}\left\{
k_0(s)\,+\,
\frac{M_{\pi}^2}{s-4 M_{\pi}^2}\,
\left[4 k_0(s) + k_1(s)\right]\right\}
\,+\, {\cal O}(\Delta_\pi^2)
\nonumber\\ 
2 \frac{\sigma (s)}{\sigma_{\mbox{\tiny$\nabla$}}(s)}\, L_{\mbox{\tiny$\nabla$}}(s) &=&
16\pi k_1(s) + 8\pi\,\frac{\Delta_\pi}{M_{\pi}^2}\left\{
 k_1(s) - k_2(s)\,-\,
\frac{4 M_{\pi}^2}{s-4 M_{\pi}^2}\,
\left[4 k_0(s) + k_1(s)\right]\right\}\,+\,{\cal O}(\Delta_\pi^2)
\nonumber\\ 
2 \sigma (s) \sigma_0(s) L_0(s) &=& 16\pi k_2(s) \,+\,8\pi\frac{\Delta_\pi}{M_{\pi}^2}\,
\left[k_1(s) - k_2(s) - 4 k_0(s)\right]\,+\,
{\cal O}(\Delta_\pi^2)
\nonumber\\ 
\frac{3 M_{\pi^0}^2 \sigma (s)}{s - 4M_{\pi^0}^2}\,L_{0}^2(s) &=& 16\pi k_3(s) \,-\,16\pi\,
\frac{\Delta_\pi}{M_{\pi}^2}\left\{
k_3(s) + \frac{M_{\pi}^2}{s-4 M_{\pi}^2}\,
\left[3 k_1(s) + 4 k_3(s)\right]\right\}\,+\,
{\cal O}(\Delta_\pi^2)
\nonumber\\ 
\frac{3 M_{\pi}^2 \sigma (s)}{s - 4M_{\pi^0}^2}\,L_{\mbox{\tiny$\nabla$}}^2(s) &=& 16\pi k_3(s) \,+\,
16\pi\,
\frac{\Delta_\pi}{M_{\pi}^2}\left\{\frac{3}{4}\,
\left[k_1(s) - k_2(s)\right] \,-\, \frac{M_{\pi}^2}{s-4 M_{\pi}^2}\,
\left[3 k_1(s) + 4 k_3(s) \right] \right\} 
+\,
{\cal O}(\Delta_\pi^2) .\ \qquad{ }
.
\label{expansion_k_n}
\end{eqnarray}
Collecting the two contributions allows us to rewrite eq. (\ref{psi_00}) as
\begin{equation}
\psi^{00}_0(s)
\,=\, 2\,\frac{M_\pi^4}{F_\pi^4}\,
\sqrt{\frac{s}{s-4M_\pi^2}}\,
\sum_{n=0}^3\left[{\overline\xi}^{(n)}_{00}(s) \,+\, \frac{\Delta_\pi}{M_\pi^2}\,\Delta\xi^{(n)}_{00}(s) \right]
k_n(s) \,+\,
{\cal O}(\Delta_\pi^2) ,
\label{psi_00_exp}
\end{equation}
where $\Delta\xi^{(n)}_{00}(s)$ is the sum of $\delta\xi^{(n)}_{00}(s)$ in eq. (\ref{delta_xi}) and of
the contribution generated by the expansions (\ref{expansion_k_n}) to first order in $\Delta_\pi$.
The expressions of the functions $\Delta\xi^{(n)}_{00}(s)$ are displayed in appendix \ref{app:delta_xi}.
Let us briefly comment on the appearance of contributions involving the factor $M_\pi^2/(s - 4 M_\pi^2)$
in equation (\ref{expansion_k_n}).
Upon closer inspection, one finds that the combinations $\left[4 k_0(s) + k_1(s)\right]/(s - 4 M_\pi^2)$,
$\left[3 k_1(s) + 4 k_3(s)\right]/(s - 4 M_\pi^2)$, and $k_2(s)/(s - 4 M_\pi^2)$, become actually 
proportional to $\sigma(s)$ as $s$ approaches $4 M_\pi^2$ from above.

The extraction of the first-order isospin-breaking contributions from the remaining one-loop partial 
waves proceeds along similar lines, and we merely quote the resulting formulae:
\begin{eqnarray}
\psi^{\mbox{\tiny{$+-$}}}_0(s) &=&  2\,\frac{M_{\pi}^4}{F_\pi^4}\,
\sqrt{\frac{s}{s-4M_{\pi}^2}}\,
\sum_{n=0}^3\left[{\overline\xi}^{(n)}_{{\mbox{\tiny{$+-$}}};S} (s)\,+\,
\frac{\Delta_\pi}{M_{\pi}^2}\,\Delta\xi^{(n)}_{{\mbox{\tiny{$+-$}}};S} (s)\right]
k_n(s) \,+\,
{\cal O}(\Delta_\pi^2) 
\nonumber\\
\psi^{\mbox{\tiny{$+-$}}}_1(s) &=&  2\,\frac{M_{\pi}^4}{F_\pi^4}\,
\sqrt{\frac{s}{s-4M_{\pi}^2}}\,
\sum_{n=0}^4\left[{\overline\xi}^{(n)}_{{\mbox{\tiny{$+-$}}};P} (s)\,+\,
\frac{\Delta_\pi}{M_{\pi}^2}\,\Delta\xi^{(n)}_{{\mbox{\tiny{$+-$}}};P} (s)\right]
k_n(s) \,+\,
{\cal O}(\Delta_\pi^2) 
\nonumber\\
\psi^{x}_0(s) &=&  2\,\frac{M_{\pi}^4}{F_\pi^4}\,
\sqrt{\frac{s}{s-4M_{\pi}^2}}\,
\sum_{n=0}^3\left[{\overline\xi}^{(n)}_{x}(s)\,+\,
\frac{\Delta_\pi}{M_{\pi}^2}\,\Delta\xi_{x}^{(n)}(s)\right]
k_n(s) \,+\,
{\cal O}(\Delta_\pi^2) .
\label{psi_+-_exp_and_psi_x_exp}
\end{eqnarray}
More details, as well as expressions of the functions $\Delta\xi^{(n)}_{{\mbox{\tiny{$+-$}}};S} (s)$,
$\Delta\xi^{(n)}_{{\mbox{\tiny{$+-$}}};P} (s)$, and $\Delta\xi_{x}^{(n)}(s)$ are given in
appendix \ref{app:delta_xi}. Working at first order in $\Delta_\pi$ has allowed us
to cast the functions $\psi^{00}_0(s)$, $\psi^{\mbox{\tiny{$+-$}}}_{0,1}(s)$, and $\psi^{x}_0(s)$
into a form that makes the comparison with their expressions in the isospin limit straightforward.

At next-to-leading order, the isospin-breaking contributions to the $P$-wave phase
are thus simply proportional to the difference $\psi^{\mbox{\tiny{$+-$}}}_1 (s)-
\stackrel{{ }_{\mbox{\scriptsize{$o$}}}}{\psi} \stackrel{{\mbox{\tiny{$+-$}}}}{_{\!\!\!\! 1}} \! \! (s) $:
\begin{eqnarray}
 {\Delta}_4 \delta_1^{\pi}(s) &=&
2\,\frac{M_\pi^4}{F_\pi^4}\,
\sum_{n=0}^4\left[ 
{\overline\xi}^{(n)}_{{\mbox{\tiny{$+-$}}};P} (s)
\,-\,\frac{1}{2}\xi_1^{(n)}(s)
\,+\,\frac{\Delta_\pi}{M_{\pi}^2}\,\Delta\xi^{(n)}_{{\mbox{\tiny{$+-$}}};P} (s)
\right]
k_n(s) \,+\,
{\cal O}(\Delta_\pi^2) .
\end{eqnarray}
In the case of the $S$-wave phases, the corresponding corrections are naturally split into
a universal contribution ${\Delta}_4^U \delta_0(s)$ and a form-factor dependent piece
${\Delta}_4^F \delta_0(s)$, cf. eq. (\ref{Delta_U_F}).
Keeping only the first-order isospin-breaking contributions, the two
universal pieces read, again for $s > 4 M_\pi^2$,
\begin{eqnarray}
{\Delta}_4^U \delta_0^{\pi^0}(s) &=& 
\sigma(s)\left\{
\frac{1}{2} \psi_0^{00}(s) -  \frac{1}{2} \stackrel{{ }_{\mbox{\scriptsize{$o$}}}}{\psi} \stackrel{00}{_{\!\! 0}} \!\! (s)
 \, -\, \psi_0^{x}(s) \theta(s-4M_{\pi}^2) + \stackrel{{ }_{\mbox{\scriptsize{$o$}}}}{\psi} \stackrel{x}{_{0}} \! \! (s)
 \,+\,\frac{1}{2}\left( \frac{\sigma_{0}(s)}{\sigma(s)} - 1\right) \psi_0^{00}(s)
\right\}
 \nonumber\\
 &=&
2\,\frac{M_\pi^4}{F_\pi^4}\,
\sum_{n=0}^3\left\{ \frac{1}{2}{\overline\xi}^{\,(n)}_{00}(s)\,-\,
{\overline\xi}^{\,(n)}_{x}(s)
\,-\,\frac{1}{2}\xi_0^{(n)}(s)
\right.
\nonumber\\
&& \left. \!\!\!\!\!\!\!\!
+\,\frac{\Delta_\pi}{M_\pi^2}\,\left[
\frac{1}{2} \Delta\xi_{00}^{(n)}(s)\,-\,\Delta\xi_{x}^{(n)}(s)
\,+\,\frac{1}{3}\,\frac{M_\pi^2}{s - M_\pi^4}
\left( 2 \xi_2^{(n)}(s) + \xi_0^{(n)}(s) \right)\right]
\right\}
k_n(s) \,+\,
{\cal O}(\Delta_\pi^2) ,
\nonumber\\
\nonumber\\
{\Delta}_4^U \delta_0^{\pi}(s) &=&
\sigma(s)\left\{
\psi^{\mbox{\tiny{$+-$}}}_0 (s)-
\stackrel{{ }_{\mbox{\scriptsize{$o$}}}}{\psi} \stackrel{{\mbox{\tiny{$+-$}}}}{_{\!\!\!\! 0}} \! \! (s)
\,-\,
\frac{1}{2} \psi^{x}_0(s)\,+\,
\frac{1}{2} \stackrel{{ }_{\mbox{\scriptsize{$o$}}}}{\psi} \stackrel{x}{_{0}} \! \! (s)
\,-\,\frac{1}{2}\left( \frac{\sigma_{0}(s)}{\sigma(s)} - 1\right) \psi^{x}_0(s)
\right\}
\nonumber\\
&=&
2\,\frac{M_\pi^4}{F_\pi^4}\,
\sum_{n=0}^3\left\{ {\overline\xi}^{\,(n)}_{{\mbox{\tiny{$+-$}}};S}(s)\,-\,\frac{1}{2}
{\overline\xi}^{\,(n)}_{x}(s)
\,-\,\frac{1}{2}\xi_0^{(n)}(s)
\right.
\nonumber\\
&& \left. \!\!\!\!\!\!\!\!
+\,\frac{\Delta_\pi}{M_\pi^2}\,\left[
\Delta\xi_{{\mbox{\tiny{$+-$}}};S}^{(n)}(s)\,-\,\frac{1}{2} \Delta\xi_{x}^{(n)}(s)
\,-\,\frac{1}{3}\,\frac{M_\pi^2}{s - M_\pi^4}
\left(\xi_2^{(n)}(s) - \xi_0^{(n)}(s) \right)\right]
\right\}
k_n(s) \,+\,
{\cal O}(\Delta_\pi^2) 
.
\end{eqnarray}
Concerning the form-factor dependent parts, the
one-loop result (\ref{eq:gammaoneloop}) gives,
at the same level of accuracy,
\begin{eqnarray}
\Gamma_S^{\pi}(s) \,-\,\Gamma_S^{\pi^0}(s) &=&
s(a_S^{\pi} - a_S^{\pi^0})\,+\,
\frac{\Delta_\pi}{32\pi^2F_\pi^2}\left[
-\beta\frac{s}{M_\pi^2}\,+\,\frac{28 \beta + 2 \alpha}{3}\right]
\,+\,
\frac{\Delta_\pi}{96\pi^2F_\pi^2} (4 \beta - \alpha)\,\frac{L(s)}{\sigma(s)}
\nonumber\\
&&\!\!\!\!\!
+\,
\frac{\Delta_\pi}{96\pi^2F_\pi^2} (14 \beta + \alpha)\,{\sigma(s)}L(s)
\,+\,{\cal O}(\Delta_\pi^2) .
\end{eqnarray}

\section{Numerical evaluation} \label{numerics}
\setcounter{equation}{0}

This section is devoted to the numerical evaluation of the isospin breaking
corrections $\Delta_4 \delta_\ell (s)$ keeping the full dependence on $\Delta_\pi$.
For this, we first need to know how the subtraction parameters that appear in the 
amplitudes and form factors after the first iteration are related to the corresponding
ones in the isospin limit.

\subsection{Determination of the subtraction parameters}\label{sub_csts_at_NLO}

From the dispersive representations of the form factors and scattering amplitudes, we have obtained the isospin-breaking corrections 
in the phases of the pion form factors beyond leading order. These expressions involve the normalizations
$F_S^\pi (0)$ and $F_S^{\pi^0} (0)$ and the two subtraction parameters $a_S^\pi$ and $a_S^{\pi^0}$ in the one-loop expressions 
of the form factors, and only a subset of the 
15 subtraction constants that appear in the $\pi\pi$ amplitudes, namely 
$\alpha_{00}$, $\alpha_x$, $\alpha_{{\mbox{\tiny{$+-$}}}}$, $\beta_{x}$, and $\beta_{{\mbox{\tiny{$+-$}}}}$
on the one hand, $\lambda^{(1)}_{00}$, $\lambda^{(i)}_x$ and $\lambda_{{\mbox{\tiny{$+-$}}}}^{(i)}$, $i = 1,2$, on
the other hand. In the isospin limit, the latter are given, as indicated in equation (\ref{lambdas_iso}), 
in terms of the constants $\lambda_i$ discussed and evaluated in refs. \cite{KMSF95,KMSF96}. More
accurate determinations have appeared since then \cite{Colangelo:2001df,DescotesGenon:2001tn}, see below.
We thus merely need to evaluate the size of the isospin-breaking deviations like, say, $\lambda^{(i)}_x - \lambda_i$.
The subset $\alpha_{00}$...$\beta_{{\mbox{\tiny{$+-$}}}}$ is likewise related to the subthreshold parameters
$\alpha$ and $\beta$ in the isospin limit. At lowest order, these relations were given in eq. (\ref{alphabetaLO}),
but in order to evaluate the phases at next-to-leading order, it is necessary to go beyond this approximation.
Again, we only need to know the size of the deviations from the isospin-limit quantities $\alpha$ and $\beta$. 
As discussed at the end of subsection
\ref{FF_and_Amp_tree}, $\alpha$ and $\beta$ represent the observables that we eventually would like to pin down from a 
phenomenological analysis of experimental data, so we have to trace down the dependence on these parameters beyond the 
lowest-order expressions.

Let us now explain how we proceed with these tasks. For this purpose, we briefly come back to the discussion in 
subsection \ref{sec:em_corr}. The framework that we have presented there can be described by an ``effective'' lagrangian,
whose form is similar to the chiral lagrangian used to treat electromagnetic corrections, 
but without including photons as dynamical degrees of freedom, as their effect is supposed to be treated by other means 
or otherwise to be negligible. The leading-order (strong) lagrangian ${\mathcal{L}}_2$ is then supplemented with a contribution 
of the form \cite{Colangelo:2008sm}:
\begin{equation}
{\mathcal{L}}_2 \to {\mathcal{L}}_2 + \widehat{C}\langle QUQU^\dag \rangle \,,\ Q={\rm diag}(2e/3,-e/3)\,,
\label{LO_eff_lag}
\end{equation}
where $\widehat{C}$ is a low-energy constant that breaks isospin symmetry among the pion masses. The
last term, through its transformation properties under chiral symmetry, encodes the information about
the electromagnetic origin of the pion mass difference. Although we could have absorbed it into the definition
of $\widehat{C}$, we have left the electric charge $e$ apparent, in order to make the comparison with
the usual effective theory in presence of electromagnetic interactions more convenient. We call (\ref{LO_eff_lag})
an ``effective" lagrangian since there is no identifiable fundamental theory of which it would constitute
the effective theory in the usual sense \footnote{One of us (M. K.) wishes to thank J. Gasser for numerous interesting discussions
on this and on related issues.}, the quotation marks serving as a reminder of this limitation. Nevertheless, eq. (\ref{LO_eff_lag}) constitutes 
a suitable starting point for a low-energy expansion,
with a well-defined and consistent power counting, which reproduces the features of the framework adopted here
as far as isospin-violating corrections are concerned. 
Thus, the ``effective" lagrangian 
at next-to-leading order is supplemented with the terms described in ref.~\cite{Knecht:1997jw}, but with
the corresponding low-energy constants denoted with a hat, to distinguish them from 
those obtained in the theory with virtual photons included. Indeed, 
the absence of virtual photons modifies the structure of the one-loop divergences, and
the scale dependence of the renormalized low-energy constants
${\widehat k}^r_i(\mu)$ is given by
\begin{equation}
e^2 \mu \frac{d}{d\mu} {\widehat k}^r_i(\mu)\,=\,
- \frac{1}{16\pi^2}\,{\widehat\sigma}_i
,
\end{equation}
with (for the low-energy constants of interest in our case)
\begin{eqnarray}
{\widehat\sigma}_1 &=& {\widehat\sigma}_5 \,=\, -\frac{1}{10}\,\frac{\Delta_\pi}{F^2}
\,,\quad
{\widehat\sigma}_8 \,=\, -\frac{1}{2}\,\frac{\Delta_\pi}{F^2}
\nonumber\\
{\widehat\sigma}_2 &=& {\widehat\sigma}_4 \,=\,{\widehat\sigma}_6\,=\,
\frac{\Delta_\pi^2}{F^2} 
\,,\quad {\widehat\sigma}_3 \,=\, {\widehat\sigma}_7 \,=\, 0
.
\label{sigma_hat}
\end{eqnarray}
One has also a contribution quadratic in the difference $M_\pi^2 - M_{\pi^0}^2$ from the low-energy 
constant ${\widehat k}^r_{14}(\mu)$:
\begin{equation} 
e^4 \mu \frac{d}{d\mu} {\widehat k}^r_{14}(\mu)\,=\,
- \frac{3}{16\pi^2}\,\frac{\Delta_\pi^2}{F^4}
.
\end{equation}
We emphasize that these scale dependences are different from the ones of the equivalent low-energy counterterms discussed in 
ref.~\cite{Knecht:1997jw}, 
since we have considered a theory where no virtual photons are included. Furthermore, they follow in a straightforward manner from the
expressions given in eqs. (3.9)-(3.11) of~\cite{Knecht:1997jw} upon dropping the terms that do not contain the parameter $Z=C/F^4$,
with $M_\pi^2 - M_{\pi^0}^2 = 2 e^2 F^2 Z$ at lowest order, in the notation of that reference [in the case of ${\widehat\sigma}_{14}$,
only the terms in $Z^2$ must be retained, the constant ${\widehat k}^r_{14}$ being multiplied by $e^4$].

The relevant subtraction constants are then obtained upon matching the one-loop expressions 
obtained within the framework we have just described with
the representations obtained in section \ref{1stIteration}. The outcome of this
exercise is displayed in appendix \ref{app:subtraction}. Let us recall here that
there exist explicit one-loop calculations
of the various $\pi\pi$ amplitudes \cite{Knecht:1997jw,KnechtNehme02,Meissner:1997fa}
and form factors \cite{Kubis:1999db} considered here, obtained within the full
QCD+QED effective theory \cite{Urech:1994hd,Neufeld:1994eg,Neufeld:1995mu,Knecht:1997jw,Meissner:1997fa,Schweizer:2002ft}.
As mentioned in subsection \ref{sec:em_corr}, these calculations also include isospin-violating contributions
arising from the exchanges of virtual low-energy photons, which are
however not considered here.  
In order to make a comparison with these one-loop calculations, we must
therefore remove the contributions of virtual photons from the expressions given in these references,
and only keep the effects due to the difference of the pion masses \cite{Colangelo:2008sm}. 
From a practical point of view, this can be done 
as described above: contributions proportional to $e^2$, but without the appropriate $Z$ factor, arise from
the exchange of virtual photons and are discarded, while at the same time the low-energy constants $k^r_i(\mu)$
are replaced by ${\widehat k}^r_i(\mu)$. Furthermore, since we want to display the dependence on the two independent
parameters $\alpha$ and $\beta$, we have, in the computation of the loop contributions, explicitly
kept the quantities $F$ and ${\widehat m} B$ defining the leading-order amplitude (\ref{pipiAmp_LO}), 
for which we have then substituted the lowest-order expressions given in (\ref{alpha_beta_LO}).
This brings in another difference with the one-loop calculations available in the literature.

\subsection{Numerical input values}\label{num_input}

We wish to investigate the size of the isospin-breaking corrections as functions of $\alpha$ and $\beta$, 
for fixed values of $\lambda_{1,2}$ and of the ${\widehat k}^r_i$'s. As mentioned above, the former parameters
have been evaluated before \cite{KMSF95,KMSF96,Colangelo:2001df,DescotesGenon:2001tn} using sum rules and
medium-energy $\pi\pi$ data. For the numerical evaluations below, we use the values from the ``extended fit'' 
of \cite{DescotesGenon:2001tn}:
\begin{equation}
\lambda_1 \,=\, ( -4.18 \pm 0.63 ) \cdot 10^{-3} \ ,\quad \lambda_2 \,=\, ( 8.96 \pm 0.12 ) \cdot 10^{-3} .
\label{lambda_1_2_values}
\end{equation}
Let us notice that the sum rules that lead to this determinations of $\lambda_1$ and $\lambda_2$
also exhibit a mild dependence with respect to $\alpha$ and $\beta$ \cite{KMSF95,KMSF96}. This dependence is, 
however, covered by the quoted uncertainties, and we will therefore not consider it further.

As far as the constants ${\widehat k}^r_i$ are concerned, we will assume that they take the same numerical values
as the low-energy constants ${k}^r_i$. Even though this identification 
constitutes an approximation whose precision is difficult to assess, it is not obvious to consider simple alternatives to this 
choice at the time being. Incidentally, this is also the option that was retained in ref.~\cite{Gasser:2007de}.

To obtain numerical estimates of the ${k}^r_i$s, we proceed in several step. First, we make use of the relation between
these two-flavour constants and their three-flavour counterparts $K_i^r$ \cite{Urech:1994hd}, as worked out at one-loop level 
in ref. \cite{Haefeli:2007ey}. For the constants $K_1^r$...$K_6^r$ we then use the evaluation of ref. \cite{Ananthanarayan:2004qk},
as given by the last line of Table 1 in this reference. These determinations rely on a set of sum-rules \cite{Moussallam:1997xx}
involving QCD four-point functions, that are saturated by the lowest-mass resonances in the corresponding channels. This kind
of minimal hadronic ansatz, which finds some justification in the limit of a large number of colors $N_C$ 
\cite{'tHooft:1973jz,Witten:1979kh}, is usually a good approximation \cite{Eckeretal89}. We therefore endow the numbers of 
ref. \cite{Ananthanarayan:2004qk} with a  relative error of 33\% (1/$N_C$ for $N_C = 3$) accounting for neglected
subleading effects in the $1/N_C$ expansion. We assign the same relative error to the constants
$K_{10}^r$ and $K_{11}^r$ that were estimated along similar lines in ref. \cite{Moussallam:1997xx}. 
Next, the relation between $k_8^r$ and $K_{11}^r$ also involves
\cite{Haefeli:2007ey} the $SU(3)$ low-energy constants \cite{Gasser:1984gg} $L_4^r$ and $L_5^r$. For the latter we have
taken the ${\cal O}(p^4)$ determination $10^3 \cdot L_5^r(M_\rho) = 1.46 \pm 0.15$ from ref. \cite{Amoros:2001cp}.
$L_4^r$ is not so well determined, and can induce significant differences between the patterns of chiral symmetry breaking
for two and three massless flavours \cite{DescotesGenon:1999uh,DescotesGenon:2000ct,DescotesGenon:2000di,DescotesGenon:2003cg}. 
For our present purposes, we take $10^3 \cdot L_4^r(M_\rho) = 0 \pm 0.5$,
a value which was advocated on the basis of the Zweig rule \cite{Gasser:1984gg,Amoros:2001cp}, even though later fits and discussions
favour larger values \cite{DescotesGenon:2007ta,Bernard:2010ex,Bijnens:2011tb}. 
In any case, it turns out that $k_8$ plays only a minor role in the numerical evaluation of 
isospin breaking in the phases.
Finally, the relation between the ${k}^r_i$s of interest to us and the $K_i^r$s also involve four constants, $K_{7,8,9,15}^r$
for which there exist no reliable determinations. We have assigned an overall uncertainty of $\pm 1/(16 \pi^2) = \pm 6.3 \cdot 10^{-3}$,
based on naive dimensional analysis, to the values of $k_{5,6,7,14}^r$ where these constants occur. The results
of this analysis are displayed in table \ref{tab:LECs}. Our values reproduce those given in ref. \cite{Haefeli:2007ey},
where however an overall uncertainty of $\pm 6.3 \cdot 10^{-3}$ was assigned uniformly to all the constants $k_i^r(M_\rho)$.

For the sake of illustration, we let the parameters $\alpha$ and $\beta$
vary within the intervals $1.0 \le \alpha \le 1.8$, $1.04 \le \beta \le 1.12$, suggested by the analysis
of ref. \cite{DescotesGenon:2001tn}. These intervals cover a reasonable range of possibilities, but of course,
if necessary, other values can be considered.

To summarize, for the numerical analysis that follows we thus use as inputs the values of the constants 
$\lambda_{1,2}$ in (\ref{lambda_1_2_values}),
the values of the constants ${\widehat k}^r_i(\mu)$ as given in table \ref{tab:LECs}, together with \cite{pdg2010}
\begin{equation}
M_\pi \,=\, 139.57\ {\mbox{MeV}}\, , \ M_{\pi^0} \,=\, 134.98\ {\mbox{MeV}}\, , \ F_\pi \,=\,  92.2\ {\mbox{MeV}} .
\label{inputs}
\end{equation} 
\indent

\begin{table}[t!]
\begin{center}
\begin{tabular}{|c||c|c|c|c|c|c|c|c|c|}
 \hline
i & 1 & 2 & 3 & 4 & 5 & 6 & 7 & 8 & 14 \\
\hline 
$\stackrel{ }{\widehat{k}_i^r}(M_\rho) \cdot 10^3$ & $8.4 \pm 2.8$ & $3.4 \pm 1.2$ & $2.7 \pm 0.9$ & $1.4 \pm 0.5$ & $-0.8 \pm 6.3$ & $3.9 \pm 6.3$ & $3.7 \pm 6.3$ & $-1.3 \pm 2.5$ & $-0.4 \pm 6.3$ \\
\hline
\end{tabular}
\caption{Values of the low-energy constants ${\widehat{k}_i^r}$ 
used for the estimate of the subtraction constants. 
The values of the $\widehat{k}_i^r$ correspond to the renormalized constants at the scale $\mu = M_\rho = 770$ MeV.
The constants $\stackrel{ }{\widehat{k}_{3}}$ and $\stackrel{ }{\widehat{k}_{7}}$ do not depend on the renormalization
scale, cf. eq. (\ref{sigma_hat}).
\label{tab:LECs}}
\end{center}
\end{table}

\subsection{Size of isospin corrections to the phases}

\begin{figure}[b!]
\center\epsfig{figure=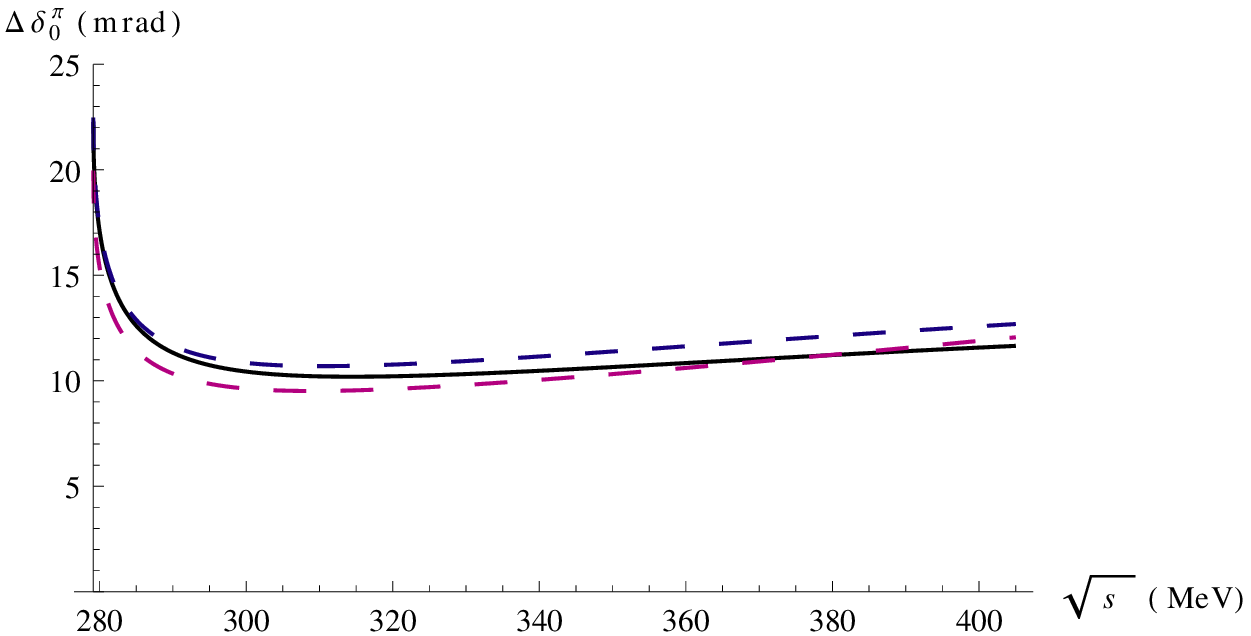,height=8.0cm}
\caption{The total phase difference $\Delta\delta^{\pi}_{0}(s)$ at next-to-leading order 
(solid line) for $\alpha = 1.40$ and $\beta = 1.08$, as a function of energy (in MeV). 
Also shown are the lowest-order
contribution $\Delta_2 \delta^{\pi}_{0}(s)$ (lower dashed curve), and the combination
$\Delta_2 \delta^{\pi}_{0}(s) + \Delta_4^U \delta^{\pi}_{0}(s) = \Delta \delta^{\pi}_{0}(s) - \Delta_4^F \delta^{\pi}_{0}(s)$ (upper dashed curve).
\label{fig9}}
\end{figure}

Let us start with the phase of $F^\pi_S(s)$. We will from now on restrict ourselves to the range 
of energies above the cusp. On figure \ref{fig9} we show the isospin-violating correction 
$\Delta\delta_0^\pi (s) = \Delta_2 \delta_0^\pi (s) + \Delta_4 \delta_0^\pi (s)$, defined as explained
after equation (\ref{phases_NLO}), i.e., for $s > 4M_\pi^2$,
\begin{eqnarray}
\Delta\delta_{0}^{\pi}(s) &=&   
\sigma (s) \left[
\varphi_0^{\mbox{\tiny{$+-$}}}(s) - \stackrel{{ }_{\mbox{\scriptsize{$o$}}}}{\varphi} \stackrel{{\mbox{\tiny{$+-$}}}}{_{\!\!\!\! 0}} \! \! (s)
\right]
\, -\,
\frac{1}{2}\,\sigma_{0}(s) \left[
\psi_0^{x}(s) - \stackrel{{ }_{\mbox{\scriptsize{$o$}}}}{\psi} \stackrel{x}{_{0}} (s)
\right]
 \nonumber\\
&& \!\!\!\!\! + \,
\sigma (s) \left[
\psi_0^{\mbox{\tiny{$+-$}}}(s) - \stackrel{{ }_{\mbox{\scriptsize{$o$}}}}{\psi} \stackrel{{\mbox{\tiny{$+-$}}}}{_{\!\!\!\! 0}} \! \! (s)
\right]
\, -\,
\frac{1}{2}\,\sigma_{0}(s) \left[
\psi_0^{x}(s) - \stackrel{{ }_{\mbox{\scriptsize{$o$}}}}{\psi} \stackrel{x}{_{0}} (s)
\right]
 \nonumber\\
 && \!\!\!\!\! - \,
\frac{1}{2}\,
\sigma_{0}(s)\varphi_0^{x}(s)
\left[\left(\frac{F_S^{\pi^0}(0)}{F_S^{\pi}(0)} - 1\right)
-\left(\Gamma_S^{\pi}(s) - \Gamma_S^{\pi^0}(s)\right)
\right] 
\nonumber\\
&\equiv&
\Delta_2 \delta_0^\pi (s) \,+\,\Delta_4^U \delta_0^\pi (s) \,+\,\Delta_4^F \delta_0^\pi (s) 
.
\end{eqnarray}
The three terms of the decomposition in the second equality correspond, in succession, to the three lines of the first one, see the discussion after
equation (\ref{Delta_U_F}).
In the cusp region, $\Delta\delta_{0}^{\pi}(s) $ is rather well described by $\Delta_2 \delta_0^\pi (s) \,+\,\Delta_4^U \delta_0^\pi (s)$,
the contribution of $\Delta_4^F \delta_0^\pi (s)$ is only marginal. At energies above $\sim 300$ MeV, $\Delta_4^F \delta_0^\pi (s)$
starts to provide a sizeable negative contribution that more and more compensates for $\Delta_4^U \delta_0^\pi (s)$, so that eventually
$\Delta \delta_0^\pi (s) \sim \Delta_2 \delta_0^\pi (s)$. This situation is reproduced if we take values of $\alpha$ and $\beta$ different
from the ones adopted for figure \ref{fig9}, but in the range considered here. We stress that it is $\Delta_2 \delta_0^\pi (s)$, and not
$\Delta_2^{\mbox{\scriptsize{LO}}} \delta_0^\pi (s)$, that provides a good description of the total effect in the region of higher energies.
In this region $\Delta_2 \delta_0^\pi (s)$ and $\Delta_2^{\mbox{\scriptsize{LO}}} \delta_0^\pi (s)$ differ by more than 2 milliradians, 
see figure \ref{fig5}.

\begin{figure}[t]
\center\epsfig{figure=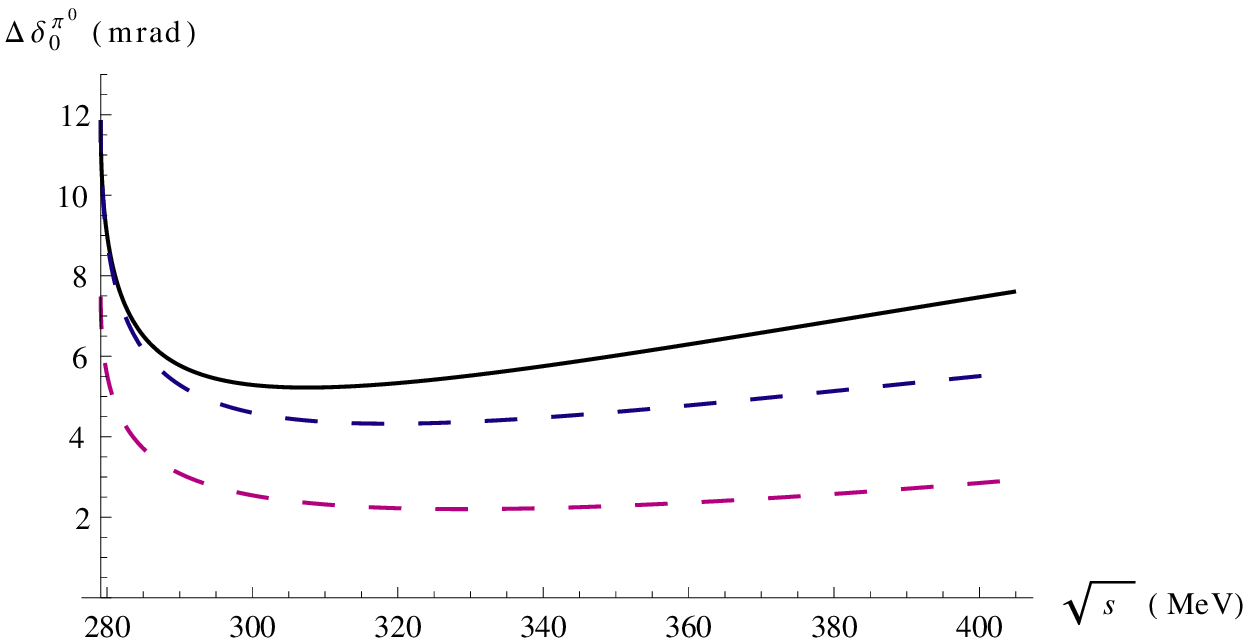,height=8.0cm}
\caption{The total phase difference $\Delta \delta^{\pi^0}_{0}(s)$ at next-to-leading order
(solid line) for $\alpha = 1.40$ and $\beta = 1.08$, as a function of energy (in MeV). 
Also shown are the lowest-order
contribution $\Delta_2 \delta^{\pi^0}_{0}(s)$ (lower dashed curve), and the combination
$\Delta_2 \delta^{\pi^0}_{0}(s) + \Delta_4^U \delta^{\pi^0}_{0}(s) = \Delta \delta^{\pi^0}_{0}(s) - \Delta_4^F \delta^{\pi^0}_{0}(s)$
 (upper dashed curve).
\label{fig10}}
\end{figure}

Turning next to the phase of $F^{\pi^0}_S (s)$,
the definition of the different relevant quantities reads
\begin{eqnarray}
\Delta \delta_{0}^{\pi^0}(s) &=&   
\frac{1}{2}\,\sigma_0 (s) \left[
\varphi_0^{00}(s) - \stackrel{{ }_{\mbox{\scriptsize{$o$}}}}{\varphi} \stackrel{00}{_{\!\! 0}} \!  (s)
\right]
\, -\,
\sigma (s) \left[
\varphi_0^{x}(s) - \stackrel{{ }_{\mbox{\scriptsize{$o$}}}}{\varphi} \stackrel{x}{_{0}} (s)
\right]
 \nonumber\\
&& \!\!\!\!\! + \,
\frac{1}{2}\,
\sigma_0 (s) \left[
\psi_0^{00}(s) - \stackrel{{ }_{\mbox{\scriptsize{$o$}}}}{\psi} \stackrel{00}{_{\!\! 0}} \! (s)
\right]
\, -\,
\sigma (s) \left[
\psi_0^{x}(s) - \stackrel{{ }_{\mbox{\scriptsize{$o$}}}}{\psi} \stackrel{x}{_{0}} (s)
\right]
 \nonumber\\
 && \!\!\!\!\! -\,
\sigma (s)\varphi_0^{x}(s)
\left[\left(\frac{F_S^{\pi}(0)}{F_S^{\pi^0}(0)} - 1\right)
+\left(\Gamma_S^{\pi}(s) - \Gamma_S^{\pi^0}(s)\right)
\right] 
\nonumber\\
&\equiv&
\Delta_2 \delta_0^{\pi^0} (s) \,+\,\Delta_4^U \delta_0^{\pi^0} (s) \,+\,\Delta_4^F \delta_0^{\pi^0} (s) 
.
\end{eqnarray}
We find a rather different situation than in the previous case.
As can be seen on figure \ref{fig10}, the correction $\Delta_4\delta_0^{\pi^0} (s)$ is large, with 
$\Delta_4^U \delta_0^{\pi^0} (s)$ and $\Delta_4^F \delta_0^{\pi^0} (s)$ both contributing in a substantial way.
Here, $\Delta_2 \delta_0^{\pi^0} (s)$, and even less so $\Delta_2^{\mbox{\scriptsize{LO}}} \delta_0^{\pi^0} (s)$, do not provide
a decent representation of the full isospin-violating contribution. Again, the situation shown on figure \ref{fig10}
for specific values of $\alpha$ and $\beta$ is actually generic for all values of these parameters in the ranges considered.

Both $\Delta \delta_{0}^{\pi}(s)$ and $\Delta \delta_{0}^{\pi^0}(s)$ receive contributions from 
$\psi_0^{x}(s)$, which contains a piece $\Delta_2 \psi_0^{x}(s)$ of order ${\cal O}(\Delta_\pi^2)$.
We have checked that its numerical value is indeed tiny, in the range between $-3 \cdot 10^{-3}$ milliradian and 
$-2 \cdot 10^{-3}$ milliradian for all values of $s$ and of the parameters $\alpha$ and $\beta$ considered here.

\begin{figure}[t]
\center\epsfig{figure=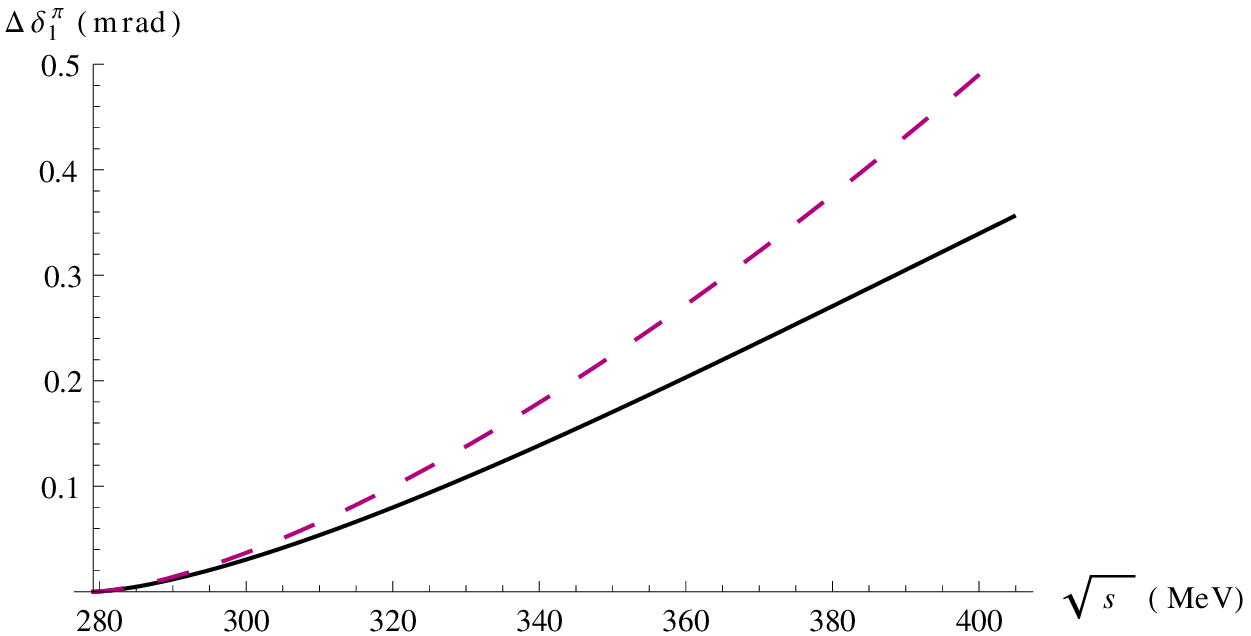,height=8.0cm}
\caption{The total phase difference $\Delta \delta^{\pi}_{1}(s)$ at next-to-leading order
(solid line) for $\alpha = 1.40$ and $\beta = 1.08$, as a function of energy (in MeV). 
The lowest-order
contribution $\Delta_2 \delta^{\pi}_{1}(s)$ (dashed line) is also shown.
\label{fig11}}
\end{figure}

In the case of $F_V^\pi (s)$, the form factor effects are absent from the isospin-violating contribution to the phase,
which reads simply
\begin{eqnarray}
\Delta \delta_{1}^{\pi}(s) &=&  
\sigma (s) \left[
\varphi_1^{\mbox{\tiny{$+-$}}}(s) - \stackrel{{ }_{\mbox{\scriptsize{$o$}}}}{\varphi} \stackrel{{\mbox{\tiny{$+-$}}}}{_{\!\!\!\! 1}} \! \! (s)
\right]
\, -\,
\sigma (s)\left[
\psi_1^{\mbox{\tiny{$+-$}}}(s) - \stackrel{{ }_{\mbox{\scriptsize{$o$}}}}{\psi} \stackrel{{\mbox{\tiny{$+-$}}}}{_{\!\!\!\! 1}} \! \! (s)
\right]
\nonumber\\
&\equiv&
\Delta_2 \delta_1^\pi (s) \,+\,\Delta_4 \delta_1^\pi (s)
.
\end{eqnarray}
As can be inferred from figure \ref{fig11}, the overall effect is small in this case, but 
the unitarity correction gives a substantial decrease of the lowest-order correction.

\begin{table}[b]
\begin{center}
\begin{tabular}{|c||c|c|c|c||c|c|c|c||c|c|c|}
\hline
$\sqrt{s}$ (MeV) & $\Delta_2 \delta_0^{\pi}(s)$ & $\Delta_4^U \delta_0^{\pi}(s)$ &
$\Delta_4^F \delta_0^{\pi}(s)$  & $\Delta \delta_0^{\pi}(s)$ & $\Delta_2 \delta_0^{\pi^0}(s)$ & $\Delta_4^U \delta_0^{\pi^0}(s)$ &
$\Delta_4^F \delta_0^{\pi^0}(s)$  & $\Delta \delta_0^{\pi^0}(s)$ & $\Delta_2 \delta_1^{\pi}(s)$ & $\Delta_4 \delta_1^{\pi}(s)$ & $\Delta \delta_1^{\pi}(s)$
\\ \hline \hline
286.07 & 11.10 & 1.42 & -0.29  & 12.23 & 3.53 & 2.37  & 0.38  & 6.29 & 0.007 & -0.001 & 0.006
\\ \hline
295.97 & 9.80 & 1.24  & -0.38  & 10.66  & 2.70 & 2.08  & 0.61  & 5.40 & 0.027 & -0.005 & 0.022
\\ \hline
304.89 & 9.54  & 1.19  & -0.45  & 10.28 & 2.41 & 2.04  & 0.78  & 5.23 & 0.050 & -0.009 & 0.041
\\ \hline
313.47 & 9.53  & 1.17  & -0.51  & 10.19 & 2.27 & 2.06  & 0.91  & 5.25 & 0.077 & -0.015 & 0.062
\\ \hline
322.02 & 9.64  & 1.15  & -0.57  & 10.23 & 2.21 & 2.11  & 1.04  & 5.36  & 0.107 & -0.022 & 0.085
\\ \hline
330.78 & 9.82  & 1.13  & -0.62  & 10.33 & 2.20 & 2.18  & 1.16  & 5.54 & 0.141 & -0.030 & 0.111
\\ \hline
340.17 & 10.05   & 1.11  & -0.68 & 10.48 & 2.23 & 2.25  & 1.28  & 5.76 & 0.180 & -0.041 & 0.139
\\ \hline
350.92 & 10.34  & 1.06 & -0.74  & 10.67 & 2.29 & 2.34  & 1.41  & 6.04   & 0.228 & -0.055 & 0.173
\\ \hline
364.52 & 10.75   & 1.00   & -0.82 & 10.93 & 2.41  & 2.45  & 1.57  & 6.42   & 0.295 & -0.076 & 0.218 
\\ \hline
389.71 & 11.55 & 0.80   & -0.95  & 11.40 & 2.70 & 2.61  & 1.85 & 7.16 & 0.430 & -0.126 & 0.304
\\
\hline
\end{tabular}
\caption{The isospin-breaking corrections at order ${\cal O}(E^4)$ to the phases (in milliradians) of the scalar
and vector form factors for several values of the energy in the range between the $2M_\pi$ threshold and the kaon mass. 
The break-up into the various contributions discussed in the text is also given. }
\label{table_num}
\end{center}
\end{table}

\begin{figure}[b]
\center\epsfig{figure=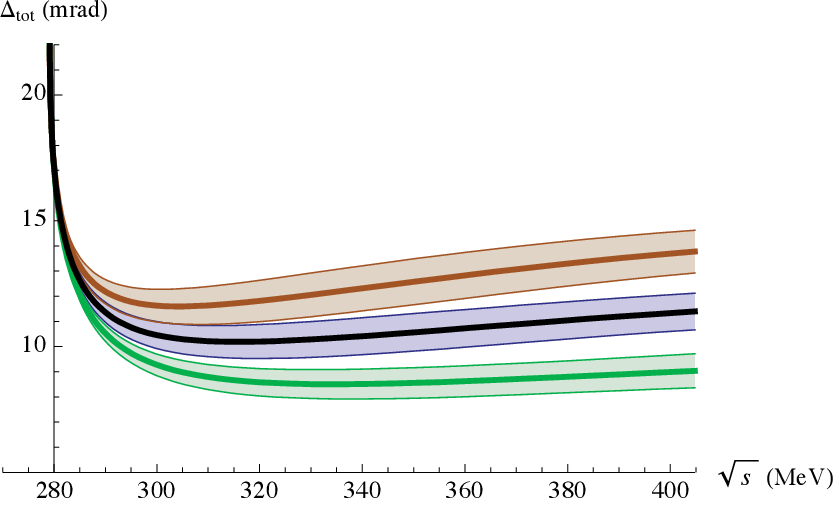,height=8.0cm}
\caption{The isospin-breaking correction $\Delta_{\mbox{\scriptsize{tot}}}^\pi$ as a function
of energy, with the error band corresponding to the uncertainties on the input
parameters, for  $(\alpha,\beta)$=(1.8,1.04) (lower band, green), (1.4,1.08) (middle band, blue), (1.0,1.12) 
(upper band, brown).  }
\label{fig12}
\end{figure}

In table \ref{table_num}, we have also summarized our result in numerical form, for some values
of the energy in the range between the $2 M_\pi$ threshold and the kaon mass.
The presence of the form-factor dependent contributions in $\Delta_4 \delta^{\pi}_{0}(s)$
and in $\Delta_4 \delta^{\pi^0}_{0}(s)$ however precludes a direct application of our results
to the experimentally more interesting case of the $K^+_{e4}$ decay modes. This 
necessitates a dedicated study, which will be reported on elsewhere \cite{wip}.
Nevertheless, it is interesting to investigate the kind of conclusions that such
a study might lead to from the analysis presented so far for the scalar and vector
form factors of the pion. The quantity that comes closest to the observable of
interest in the context of the decay mode $K^\pm \rightarrow \pi^+ \pi^- e^\pm
{\stackrel{(-)}{\nu_e}}$ is the difference between the $S$ and $P$ phases,
$\delta_0^\pi (s) - \delta_1^\pi (s)$, for which the total isospin-breaking correction reads
\begin{eqnarray}
\Delta_{\mbox{\scriptsize{tot}}}^\pi (s) & \equiv & \Delta \delta_0^\pi (s) - \Delta \delta_1^\pi (s)
\nonumber\\
& = & \Delta_2 \delta_0^\pi (s) + \Delta_4^U \delta_0^\pi (s)
+ \Delta_4^F \delta_0(^\pi s) - \Delta_2 \delta_1^{\pi}(s)-\Delta_4 \delta_1^{\pi}(s)
.
\end{eqnarray} 
This correction in the phase difference is shown on figure \ref{fig12}, together with the error
band induced by the uncertainties on the various input parameters, for fixed values of
$\alpha$ and $\beta$. The main contributors
to these error bars are the low-energy constants ${\widehat k}_i$, and in particular
${\widehat k}_{1,2,3,4}$, which enter in the correction $\Delta_4^U \delta_0^\pi (s)$,
through the isospin-breaking differences such as $\alpha_{00} - \alpha$, $\alpha_x - \alpha$, and so on. 
Similar error bands have to be associated to the curves for
$\Delta_4 \delta_0^\pi (s)$ and $\Delta_4 \delta_0^{\pi^0} (s)$ shown on figs. \ref{fig9}
and \ref{fig10}, respectively. Except in the vicinity of the cusp, the correction is
thus relatively constant, around 11 milliradians for $(\alpha,\beta)=(1.4,1.08)$, with an uncertainty that is 
somewhat less than $\pm$1 milliradian. We show the correction, with the associated error band, for three
sets of values for $(\alpha, \beta)$. Despite the uncertainties, there remains a sensitivity with
respect to these parameters.

We have also compared the exact results obtained in this section with the
approximation where only corrections of first order in $\Delta_\pi$ are retained,
as discussed in subsection \ref{IB_in_phases_NLO}. For the range of parameters considered here,
we find that using the approximate expressions
for $\Delta_4 \delta_0^{\pi^0} (s)$, $\Delta_4 \delta_0^\pi (s)$, and $\Delta_4 \delta_1^\pi (s)$
does not modify the values of $\Delta \delta_0^{\pi^0} (s)$, $\Delta \delta_0^\pi (s)$, and $\Delta \delta_1^\pi (s)$
by more than a few percents, as soon as the energy is more than $\sim 20$ MeV higher than the $2 M_\pi$ threshold, i.e.$\sqrt{s}\ge 300$ MeV. 
For practical purposes, one possible option consists therefore 
in keeping the exact expressions for $\Delta_2 \delta_0^{\pi^0} (s)$, $\Delta_2 \delta_0^\pi (s)$, and $\Delta_2 \delta_1^\pi (s)$,
and in using the expressions truncated at first order in $\Delta_\pi$ for the next-to-leading contributions.

\section{Summary and conclusions} \label{conclusion}
\setcounter{equation}{0}

In this article, we have addressed the issue of isospin breaking due to the difference between
the charged and neutral pion masses in the pion form factors and 
$\pi\pi$ scattering amplitudes in the low-energy domain. We have implemented a dispersive approach to 
obtain representations of the various $\pi\pi$ scattering amplitudes and pion form factors that
are valid at next-to-next-to-leading order in the low-energy expansion. These representations
rely on general properties such as relativistic invariance, analyticity, crossing, and unitarity, combined with the chiral 
counting for the form factors and partial waves, and provide an extension of the general frameworks 
developped previously in the isospin-symmetric case to the situation where
isospin-breaking effects are taken into account. This construction needs as inputs the lowest-order
expressions of the pion form factors and $\pi\pi$ $S$ and $P$ partial waves, and proceeds through a two-step
iterative process, the partial wave projections obtained from the one-loop representation after
the first step serving as inputs for the second step. At the two-loop level, we have obtained partially
analytical expressions only, due to the difficulty of performing the dispersive integrals
related to contributions of the non-factorizing type. We have nevertheless shown that in the limit
where the pion mass difference vanishes, we reproduce known two-loop results for the scattering
amplitudes and form factors in the isospin limit. On the other hand, we have obtained explicit
analytical expressions for the phases of the two-loop form factors, on which we have focused. 
We have also provided somewhat more tractable expressions of the
isospin-breaking corrections in the phases valid at first order in the difference $M_\pi^2 - M_{\pi^0}^2$. 
This approximation provides a very good description of isospin-breaking effects at next-to-leading order in the phases, 
for all energies lying between $\sim 300$ MeV and the kaon mass.

The dispersive representations involve a limited number of subtraction constants, which have to be 
fixed from experimental data or theoretical sources. We have related the subtraction constants
involved in the phases of the two-loop form factors to their counterparts in the isospin limit,
and we have provided a numerical evaluation of the isospin-breaking corrections in these subtraction constants.
This has allowed us to perform a quantitative study of the size of isospin-violating effects
in the phases of the form factors.

We have displayed our results in terms of the subthreshold parameters $\alpha$ and $\beta$ of
the pion scattering amplitude in the isospin limit. These parameters represent the unknown quantities
that have to be extracted from low-energy data. Equivalently, one may take the two $S$-wave scattering
lengths $a_0^0$ and $a_0^2$ as unknowns, and we have provided the formulae necessary in order to operate
the translation between the two formulations.

As far as the phases of the form factors are concerned, the main difference with the isospin-symmetric
situation is that Watson's theorem is no longer operative when the various $\pi\pi$ intermediate states that contribute to 
the unitarity sum become distinguishable, as a consequence of the explicit breaking of isospin symmetry through 
the pion mass-difference. The phase is still given by a universal contribution, expressed entirely in terms of data
related to the $\pi\pi$ scattering amplitudes, but at next-to-leading order there appears a second contribution, 
that explicitly depends on the form factors under consideration. The numerical size of this form-factor
dependent part is relatively small in the case of the $\pi^+\pi^-$ scalar form factor, but definitely
more important in the $\pi^0\pi^0$ case. 

We have also investigated the sensitivity of the isospin-breaking correction to two
subthreshold parameters $\alpha$ and $\beta$, or equivalently to the two $S$-wave scattering lengths $a_0^0$ and $a_0^2$,
which represent the quantities that have to be determined from data. Despite the uncertainties
induced by the various other parameters, we find that the correction remains
sensitive to the values of the subthreshold parameters. For the range of values that we have considered,
and depending on the value of the energy, the effect in the total correction can represent up to 5 milliradians.
The issue raised in the introduction about the possibility of a bias if isospin-breaking corrections are 
evaluated for fixed values of these scattering lengths remains therefore, in our opinion, open.
Obviously, the situation that prevails after the present study devoted to the scalar form factors of the pions
need not be representative of the one encountered in the case of the $K_{\ell 4}$ form factors.
To settle the issue, a dedicated study of this experimentally more interesting case is needed.
This will be the subject of a forthcoming article \cite{wip}.

\begin{acknowledgments}
During the long course of this work, we have benefited from numerous discussions and/or from correspondence
with several colleagues, to whom we wish to express our gratitude: V. Bernard,
B.~Bloch-Devaux, J.~Gasser, K. Kampf, B. Kubis, H. Leutwyler, B.~Moussallam, J. Novotn\'y, U.-G. Mei{\ss}ner, A. Rusetsky, and M. Zdr\'ahal.
We are also indebted to V. Bernard for a very careful and critical reading of the manuscript.

\end{acknowledgments}

\appendix

\renewcommand{\theequation}{\Alph{section}.\arabic{equation}}

\section{Indefinite integrals of the scalar loop function} \label{app:integJbar}

In this appendix, we list the integrals involving
the scalar loop function ${\bar J}(t)$ that are used for the computation
of the one-loop partial wave projections in section \ref{2ndIteration}.
For unequal masses $m_1\neq m_2$ [when necessary, we assume that $m_1>m_2$], 
the loop function ${\bar J}(t)$ is given by
\begin{equation}
{\bar J}(t) = \frac{1}{16\pi^2}\,
\bigg\{1 + \frac{m_1^2 - m_2^2}{t}\ln\frac{m_2}{m_1} -
\frac{m_1^2 + m_2^2}{m_1^2 - m_2^2}\ln\frac{m_2}{m_1} +
\frac{\lambda^{1/2}(t)}{2t} 
\ln\frac{[t-\lambda^{1/2}(t)]^2 - (m_1^2 - m_2^2)^2}{[t+\lambda^{1/2}(t)]^2 - (m_1^2 - m_2^2)^2}
\bigg\}
\end{equation}
where $\lambda(t)=t^2-2t(m_1^2 + m_2^2) + (m_1^2 - m_2^2)^2$. The above
expression holds for $t<(m_1 - m_2)^2$. For $m_1 = m_2 = M_\pi$, or
$m_1 = m_2 = M_{\pi^0}$, this expression corresponds to the
functions ${\bar J}(t)$ and ${\bar J}_0(t)$, respectively, defined in
eqs. (\ref{JbarDisp}) and (\ref{Jbar1loop}). Finally, the function ${\bar J}_{\mbox{\tiny{$\! + $}} 0} (s)$
defined in eqs. (\ref{Jbar_+0_Disp}) and (\ref{Jbar_+0_Loop}) corresponds to the
choice $m_1 = M_\pi$ and $m_2 = M_{\pi^0}$.

The partial wave projections of the one-loop $\pi\pi$ amplitudes discussed require the computation of integrals
of the type
\begin{equation}
\int dt t^n {\bar J}(t)
\end{equation}
where $n$ can take negative or positive integer values.
In the present context, the range $-2\le n\le 3$ is particularly relevant.
It proves convenient to define the variable $\chi(t)$ through
\begin{equation}
t\,=\,m_1^2 + m_2^2 - m_1m_2\left(\chi\,+\,\frac{1}{\chi}\right),
\end{equation}
so that $0\le\chi\le1$ when $-\infty < t <(m_1 - m_2)^2$. The expression of
$\chi$ in terms of $t$ is given in eq. (\ref{chi_def}), upon replacing
$M_\pi$ by $m_1$, and $M_{\pi^0}$ by $m_2$,
\begin{equation}
\chi(t) \,=\, 
\frac{\sqrt{(m_1 + m_2)^2 - t} - \sqrt{(m_1 - m_2)^2 - t}}{\sqrt{(m_1 + m_2)^2 - t} + \sqrt{(m_1 - m_2)^2 - t}}\, .
\end{equation}
For strictly positive values of the integer $n$, the corresponding
(indefinite) integrals take a relatively simple form, even when the masses $m_1$
and $m_2$ are not equal  (irrelevant integration
constants have been discarded),
\begin{eqnarray}
16\pi^2 \int dt\, t {\bar J}(t) &=& 
 \left[\frac{3}{2} - \frac{m_1^2 + m_2^2}{m_1^2 - m_2^2}\ln\frac{m_2}{m_1}\right]\times\frac{t^2}{2}
\,+\, \left[(m_1^2 - m_2^2)\ln\frac{m_2}{m_1} - \frac{m_1^2 + m_2^2}{2}\right]\times t
\nonumber\\
&&
-\, \frac{1}{2}(t-m_1^2 - m_2^2)\lambda^{1/2}(t)\ln\chi(t) + m_1^2 m_2^2 \ln^2\chi(t)  
\\
\nonumber\\
16\pi^2 \int dt\, t^2 {\bar J}(t) &=&   
 \left[\frac{4}{3} - \frac{m_1^2 + m_2^2}{m_1^2 - m_2^2}\ln\frac{m_2}{m_1}\right]\times\frac{t^3}{3}
\,+\, \left[(m_1^2 - m_2^2)\ln\frac{m_2}{m_1} - \frac{m_1^2 + m_2^2}{6}\right]\times\frac{t^2}{2}
\nonumber\\
&&
-\,\frac{t}{6}\,(m_1^4 + m_2^4 + 10 m_1^2 m_2^2)
\nonumber\\
&&
-\, \frac{1}{6}\left[
2t^2 - (m_1^2 + m_2^2)t - (m_1^4 + m_2^4 + 10 m_1^2 m_2^2)
\right] \lambda^{1/2}(t)\ln\chi(t)
\nonumber\\
&&
 +\, m_1^2 m_2^2 (m_1^2 + m_2^2) \ln^2\chi(t)  
\\
\nonumber\\
16\pi^2 \int dt\, t^3 {\bar J}(t) &=&    
 \left[\frac{5}{4} - \frac{m_1^2 + m_2^2}{m_1^2 - m_2^2}\ln\frac{m_2}{m_1}\right]\times\frac{t^4}{4}
\,+\, \left[(m_1^2 - m_2^2)\ln\frac{m_2}{m_1} - \frac{m_1^2 + m_2^2}{12}\right]\times\frac{t^3}{3}
\nonumber\\
&&
-\,\frac{t^2}{24}\,(m_1^4 + m_2^4 + 8 m_1^2 m_2^2)\,-\,
\frac{t}{12}\,(m_1^2 + m_2^2)( m_1^4 +  m_2^4 + 28 m_1^2 m_2^2)
\nonumber\\
&&
-\, \frac{1}{12}\left[
3t^3 - (m_1^2 + m_2^2)t^2 - (m_1^4 + m_2^4 + 8 m_1^2 m_2^2)t
\right.
\nonumber\\
&&\qquad\qquad\qquad\left.
- (m_1^2 + m_2^2)(m_1^4 + m_2^4 + 28 m_1^2 m_2^2)
\right] \lambda^{1/2}(t)\ln\chi(t)
\nonumber\\
&&
 +\, m_1^2 m_2^2  (m_1^4 + m_2^4 + 3 m_1^2 m_2^2) \ln^2\chi(t)  
.
\end{eqnarray}
For $n\le 0$, the corresponding expressions are more complicated, at least when
the masses are different, but can be given a simpler form when expressed in terms 
of the function $H_{1,0} (x) \,=\, - {\mbox{Li}}_2 (x) - \ln x \ln (1-x)$,
which belongs to the family of harmonic polylogarithms \cite{HPL}:
\begin{eqnarray}
16\pi^2 \int dt\, {\bar J}(t) &=& 
\left[2 - \frac{m_1^2 + m_2^2}{m_1^2 - m_2^2}\ln\frac{m_2}{m_1}\right]\times t
 \,-\, \lambda^{1/2}(t)\ln\chi(t) \,+\, \frac{m_1^2 + m_2^2}{2} \ln^2\chi(t) 
\nonumber\\
&&
+\, (m_1^2 - m_2^2)
\left[
H_{1,0} \left(\frac{m_1}{m_2}\chi(t)\right) \,-\, H_{1,0} \left(\frac{m_2}{m_1}\chi(t)\right)
\,+\,\ln\frac{m_1}{m_2} \, \ln\chi(t)
\right]
\\
\nonumber\\
16\pi^2 \int \frac{dt}{t}\, {\bar J}(t) &=& 
 \left[m_1^2 + m_2^2 - 2m_1m_2\chi(t)\right]\frac{\ln\chi(t)}{t} - \frac{1}{2}\ln^2\chi(t)
- \ln \chi (t)
- (m_1^2 - m_2^2)\ln\frac{m_2}{m_1}\times\frac{1}{t} 
\nonumber\\
&&
-\, \frac{m_1^2 + m_2^2}{m_1^2 - m_2^2}
\left[
H_{1,0} \left(\frac{m_1}{m_2}\chi(t)\right) \,-\, H_{1,0} \left(\frac{m_2}{m_1}\chi(t)\right)
\,+\,\ln\frac{m_1}{m_2} \, \ln\chi(t)
\right]
\\
\nonumber\\
16\pi^2 \int \frac{dt}{t^2}\,\, {\bar{\!\!{\bar J}}}(t) &=&
- \left[\frac{1}{2} - \frac{m_1^2 + m_2^2}{m_1^2 - m_2^2}\ln\frac{m_2}{m_1}\right]\times\frac{1}{t}
\,-\, (m_1^2 - m_2^2)\ln\frac{m_2}{m_1}\times\frac{1}{2t^2} 
\nonumber\\
&&
-\,
\frac{\lambda^{1/2}(t)}{2 t^2}
\left[ \frac{m_1^2 + m_2^2}{(m_1^2 - m_2^2)^2} t\,-\,1 \right]
\ln\chi(t)
\nonumber\\
&&
-\,  2\frac{m_1^2  m_2^2}{(m_1^2 - m_2^2)^3}
\left[
H_{1,0} \left(\frac{m_1}{m_2}\chi(t)\right) \,-\, H_{1,0} \left(\frac{m_2}{m_1}\chi(t)\right)
\,+\,\ln\frac{m_1}{m_2} \, \ln\chi(t)
\right]
\end{eqnarray}
The range of integration, $t_{\mbox{\tiny $-$}} (s)\le t\le t_{\mbox{\tiny $+$}} (s)$, with 
\begin{equation}
t_{\mbox{\tiny $\pm$}}(s) \,=\, - \frac{1}{2}\,(s - 2 m_1^2 - 2 m_2^2) \,\pm\, 
\frac{1}{2}\,\sqrt{(s - 4 m_1^2)(s - 4 m_2^2)} ,
\end{equation} 
will depend on the process under consideration.
As the masses become equal, $m_2 \rightarrow m_1$, one has [$\Delta_{12}\equiv m_1^2 - m_2^2$]
\begin{eqnarray}
t_{\mbox{\tiny $-$}} (s) & = & - (s - 4 m_1^2) - 2 \Delta_{12} + {\cal O}(\Delta_{12}^2)
\nonumber\\
t_{\mbox{\tiny $+$}} (s) & = & - \frac{\Delta_{12}^2}{s - 4 m_1^2} \,+\, {\cal O}(\Delta_{12}^3) .
\end{eqnarray}
Then
\begin{eqnarray}
\chi_{\mbox{\tiny $-$}} (s) &=& \frac{1 - \sigma (s)}{1 + \sigma (s)}\,+\, {\cal O}(\Delta_{12})
\nonumber\\
\chi_{\mbox{\tiny $+$}} (s) &=& 1 \,-\, \frac{1}{2} \frac{\Delta_{12}}{m_1^2} \frac{1}{\sigma (s)} \,+\, {\cal O}(\Delta_{12}^2)
\nonumber\\
\lambda( t_{\mbox{\tiny $-$}} (s) ) &=& s (s - 4 m_1^2)\,+\, {\cal O}(\Delta_{12})
\nonumber\\
\lambda( t_{\mbox{\tiny $+$}} (s) ) &=& \frac{s}{s - 4 m_1^2}\,\Delta_{12}^2 \,+\, {\cal O}(\Delta_{12}^3) ,
\label{m1_m2_limit}
\end{eqnarray}
where $\chi_{\mbox{\tiny $\pm $}} (s) \equiv \chi (t_{\mbox{\tiny $\pm $}} (s) )$.

\section{Polynomials of the next-to-leading-order $\pi\pi$ partial waves} \label{app:polynomials}

The expressions of the one-loop the partial-wave projections displayed in eqs. (\ref{psi_00}),
(\ref{psi_+-_0}), (\ref{psi_+-_1}), and (\ref{psi_x_0}) involve a certain number of
polynomials whose expressions are given in this appendix. In the case of $\psi^{00}_0(s)$,
these polynomials read
\begin{eqnarray}
\xi_{00}^{(0)}(s) &=& 
\frac{\lambda_{00}^{(1)}}{2 M_\pi^4}\,
\left({5}s^2\,-\,{16}M_{\pi^0}^2 s\,+\,20M_{\pi^0}^4\right)
\nonumber\\
&+&
\frac{1}{16\pi^2 M_\pi^4}\bigg\{
\frac{16}{9}\beta_{x}^2 s^2 \,+\,
\frac{1}{18}\beta_{x} s\left[\beta_{x}\left(
9 M_{\pi}^2 - 106 M_{\pi^0}^2\right) \,-\,3\alpha_{x}M_{\pi^0}^2\right]
\nonumber\\
&&
+\,\frac{2}{9}M_{\pi^0}^4\left(2 \alpha_{x}^2 + 9 \alpha_{00}^2\right)
\,-\,\frac{2}{9}\beta_{x}\alpha_{x}M_{\pi^0}^2\left(11 M_{\pi}^2 - 7 M_{\pi^0}^2 \right)
\,+\,\frac{2}{9}\beta_{x}^2\left(34 M_{\pi^0}^4 + 5 M_{\pi}^4
-11 M_{\pi^0}^2 M_{\pi}^2 \right)
\bigg\}
\nonumber\\
\xi^{(1;0)}_{00}(s) &=&\frac{\alpha_{00}^2 M_{\pi^0}^4}{64\pi^2 M_\pi^4}
\nonumber\\
\xi^{(1;{\mbox{\tiny$\nabla$}})}_{00}(s) &=&\frac{1}{32\pi^2 M_\pi^4}\bigg\{
\frac{1}{3}\beta_{x}^2 s^2 \,+\,
\frac{1}{3}\beta_{x} s\left[\beta_{x}\left(
3 M_{\pi}^2 - 6 M_{\pi^0}^2\right) \,-\,\alpha_{x}M_{\pi^0}^2\right]
\nonumber\\
&&
+\,\frac{1}{9} \alpha_{x}^2 M_{\pi^0}^4
\,-\,\frac{2}{9}\beta_{x}\alpha_{x}M_{\pi^0}^2\left(5 M_{\pi}^2 - 4 M_{\pi^0}^2 \right)
\,-\,\frac{1}{9}\beta_{x}^2\left( 2 M_{\pi}^4
+ 16 M_{\pi^0}^2 M_{\pi}^2 - 28 M_{\pi^0}^4\right)
\bigg\}
\nonumber\\
\xi^{(2;0)}_{00}(s) &=&\frac{\alpha_{00}^2 M_{\pi^0}^4}{128\pi^2 M_\pi^4}
\nonumber\\
\xi^{(2;{\mbox{\tiny{$\pm$}}})}_{00}(s) 
&=&
\frac{1}{64\pi^2 M_\pi^4}\bigg[
\beta_{x} \left( s - \frac{2}{3} M_{\pi}^2 - \frac{2}{3} M_{\pi^0}^2\right)
\,+\,\frac{1}{3}\alpha_{x}  M_{\pi^0}^2
\bigg]^2
\nonumber\\ 
\xi^{(3;0)}_{00}(s) &=& -\frac{\alpha_{00}^2 M_{\pi^0}^4}{96\pi^2 M_\pi^4}
\nonumber\\
\xi^{(3;{\mbox{\tiny$\nabla$}})}_{00}(s) &=&  -\frac{1}{432\pi^2 M_\pi^4}\,
\left[
2\beta_{x}^2(5 M_{\pi}^4 - 2 M_{\pi}^2 M_{\pi^0}^2 + 2 M_{\pi^0}^4)
\,+\,2\beta_{x}\alpha_{x} M_{\pi^0}^2 (M_{\pi}^2 - 2 M_{\pi^0}^2)
\,+\, \alpha_{x}^2 M_{\pi^0}^4
\right]
.
\end{eqnarray}

\noindent
The polynomials involved in the expression for $\psi^{\mbox{\tiny{$+-$}}}_0 (s)$ read
\begin{eqnarray}
\xi_{{\mbox{\tiny{$+-$}}};S}^{(0)}(s) &=& 
\frac{\lambda_{\mbox{\tiny{$+-$}}}^{(1)} + \lambda_{\mbox{\tiny{$+-$}}}^{(2)}}{2 M_{\pi}^4}\,\left(s-2 M_{\pi}^2\right)^2 
\,+\,
\frac{\lambda_{\mbox{\tiny{$+-$}}}^{(1)} + 3\lambda_{\mbox{\tiny{$+-$}}}^{(2)} }{6 M_{\pi}^4}\,
\left(s^2 - 2 s M_{\pi}^2 + 4  M_{\pi}^4
\right)
\nonumber\\
&&
+\,\frac{1}{864\pi^2 M_\pi^4}\,
\left[
\frac{s^2}{8}\left(203\beta_{\mbox{\tiny{$+-$}}}^2 + 300 \beta_{x}^2\right)\,+\,
s  M_{\pi^0}^2 \left(36 \beta_{\mbox{\tiny{$+-$}}}\alpha_{\mbox{\tiny{$+-$}}} + \frac{27}{4} \beta_{x} \alpha_{x}\right)
 \right.
\nonumber\\
&&\left.
-\,\frac{s}{4} \left( 473 \beta_{\mbox{\tiny{$+-$}}}^2 M_{\pi}^2 + 390 \beta_{x}^2 M_{\pi}^2 + 
45 \beta_{x}^2 M_{\pi^0}^2 \right)
\,+\,172 \beta_{\mbox{\tiny{$+-$}}}^2 M_{\pi}^4 \,+\, 108 \beta_{x}^2 M_{\pi}^4
\,+\, 21 \beta_{x}^2 M_{\pi^0}^4
 \right.
\nonumber\\
&&\left.
-\, 21 \beta_{x}^2 M_{\pi^0}^2 M_{\pi}^2
\,+\,78 \alpha_{\mbox{\tiny{$+-$}}}^2 M_{\pi^0}^4  \,+\, \frac{15}{2} \alpha_{x}^2 M_{\pi^0}^4 
\,-\, M_{\pi^0}^2 \left(
48 \beta_{\mbox{\tiny{$+-$}}} \alpha_{\mbox{\tiny{$+-$}}} M_{\pi}^2 + 39 \beta_{x} \alpha_{x} M_{\pi^0}^2
- 15 \beta_{x} \alpha_{x}  M_{\pi}^2\right)
\right]
\nonumber\\
\xi^{(1;{\mbox{\tiny{$\pm$}}})}_{{\mbox{\tiny{$+-$}}};S}(s)  &=& \frac{1}{576\pi^2 M_\pi^4}\,
\left[\frac{7}{4}\beta_{\mbox{\tiny{$+-$}}}^2 s^2 - 10 \beta_{\mbox{\tiny{$+-$}}}^2 s M_{\pi}^2 + 
15 \beta_{\mbox{\tiny{$+-$}}}^2 M_{\pi}^4 + 6 \alpha_{\mbox{\tiny{$+-$}}}^2 M_{\pi^0}^4 
\right]
\nonumber\\
\xi^{(1;{\mbox{\tiny$\Delta$}})}_{{\mbox{\tiny{$+-$}}}; S}(s) &=& \frac{1}{1152\pi^2 M_\pi^4}\,
\left[ 3\beta_{x}^2 s^2 \,-\,9 \beta_{x}^2 s \left(2 M_{\pi}^2 - M_{\pi^0}^2\right)
\,+\, 2 \beta_{x}^2 \left( 14 M_{\pi}^4 - 8 M_{\pi^0}^2 M_{\pi}^2
 -  M_{\pi^0}^4\right)
 \right.
\nonumber\\
&&\left.
-\, 3 \beta_{x} \alpha_{x} s M_{\pi^0}^2
+ 2 \beta_{x} \alpha_{x} M_{\pi^0}^2 \left( 4 M_{\pi}^2 - 5 M_{\pi^0}^2\right)
\,+\, \alpha_{x}^2 M_{\pi^0}^4
\right]
\nonumber\\
\xi^{(2;0)}_{{\mbox{\tiny{$+-$}}};S}(s) &=&
\frac{1}{128 \pi^2 M_\pi^4}\,
\bigg[
\beta_{x} \left( s - \frac{2}{3} M_{\pi}^2 - \frac{2}{3}  M_{\pi^0}^2\right)
\,+\,\frac{1}{3}\alpha_{x}  M_{\pi^0}^2
\bigg]^2
\nonumber\\
\xi^{(2;{\mbox{\tiny{$\pm$}}})}_{{\mbox{\tiny{$+-$}}};S}(s) &=& 
\frac{1}{64 \pi^2 M_\pi^4}\,
\left[
\frac{\beta_{\mbox{\tiny{$+-$}}}}{2} \left( s - \frac{4}{3} M_{\pi}^2 \right)
\,+\,\frac{2}{3} \alpha_{\mbox{\tiny{$+-$}}} M_{\pi^0}^2
\right]^2
\nonumber\\
\xi^{(3;{\mbox{\tiny{$\pm$}}})}_{{\mbox{\tiny{$+-$}}};S}(s) &=& - \frac{1}{288\pi^2 M_\pi^4}\,
\left[
5 \beta_{\mbox{\tiny{$+-$}}}^2 M_{\pi}^4 \,+\,
2 \alpha_{\mbox{\tiny{$+-$}}}^2 M_{\pi^0}^4\,-\,\frac{3}{2} \beta_{\mbox{\tiny{$+-$}}}^2 s M_{\pi}^2
\right]
\nonumber\\
\xi^{(3;{\mbox{\tiny$\Delta$}})}_{{\mbox{\tiny{$+-$}}};S}(s) &=& \frac{1}{1728 \pi^2 M_\pi^4}\,
\left[
2{\beta_{x}^2} \left(
-2 M_{\pi}^4 + 2 M_{\pi}^2 M_{\pi^0}^2 - 5 M_{\pi^0}^4 \right)
\,+\,
 2 \beta_{x} \alpha_{x} M_{\pi^0}^2 \left(2 M_{\pi}^2 - M_{\pi^0}^2\right)
\,-\,
\alpha_{x}^2\, M_{\pi^0}^4
\right]
,
\end{eqnarray}
while for $\psi^{\mbox{\tiny{$+-$}}}_1 (s)$ we obtain
\begin{eqnarray}
\xi_{{\mbox{\tiny{$+-$}}};P}^{(0)}(s) &=& 
- \frac{\lambda_{\mbox{\tiny{$+-$}}}^{(1)} - \lambda_{\mbox{\tiny{$+-$}}}^{(2)}}{12 M_{\pi}^4}\,s\left(s-4 M_{\pi}^2\right) 
\nonumber\\
&&
+\,\frac{1}{3456\pi^2 M_\pi^4}\,
\left[
\frac{s^2}{4}\left(71\beta_{\mbox{\tiny{$+-$}}}^2 - 75 \beta_{x}^2\right)\,+\,
s  M_{\pi^0}^2 \left(44 \beta_{\mbox{\tiny{$+-$}}}\alpha_{\mbox{\tiny{$+-$}}} + 11 \beta_{x} \alpha_{x}\right)
 \right.
\nonumber\\
&&\left.
-\,s\left( 169 \beta_{\mbox{\tiny{$+-$}}}^2 M_{\pi}^2 - 128 \beta_{x}^2 M_{\pi}^2 + 
19 \beta_{x}^2 M_{\pi^0}^2 \right)
\,+\,341 \beta_{\mbox{\tiny{$+-$}}}^2 M_{\pi}^4 \,-\, 200 \beta_{x}^2 M_{\pi}^4 \,-\, 3 \beta_{x}^2 M_{\pi^0}^4
 \right.
\nonumber\\
&&\left.
\,+\, 148 \beta_{x}^2 M_{\pi^0}^2 M_{\pi}^2
\,+\,12 \alpha_{\mbox{\tiny{$+-$}}}^2 M_{\pi^0}^4  \,+\, 3 \alpha_{x}^2 M_{\pi^0}^4 
\,-\,4 M_{\pi^0}^2 \left(
92 \beta_{\mbox{\tiny{$+-$}}} \alpha_{\mbox{\tiny{$+-$}}} M_{\pi}^2 + 9 \beta_{x} \alpha_{x} M_{\pi^0}^2
+ 14 \beta_{x} \alpha_{x}  M_{\pi}^2\right)
\right]
\nonumber\\
\xi^{(1;{\mbox{\tiny{$\pm$}}})}_{{\mbox{\tiny{$+-$}}};P}(s)  &=& \frac{1}{1152\pi^2 M_\pi^4}\,
\left[\beta_{\mbox{\tiny{$+-$}}}^2 s^2 - 7 \beta_{\mbox{\tiny{$+-$}}}^2 s M_{\pi}^2 + 21 \beta_{\mbox{\tiny{$+-$}}}^2 M_{\pi}^4
+ 4 \beta_{\mbox{\tiny{$+-$}}} \alpha_{\mbox{\tiny{$+-$}}} s M_{\pi^0}^2
- 24 \beta_{\mbox{\tiny{$+-$}}} \alpha_{\mbox{\tiny{$+-$}}} M_{\pi^0}^2 M_{\pi}^2
\right]
\nonumber\\
\xi^{(1;{\mbox{\tiny$\Delta$}})}_{{\mbox{\tiny{$+-$}}}; P}(s) &=& \frac{1}{576\pi^2 M_\pi^4}\,
\left[ - \frac{3}{4}\beta_{x}^2 s^2 \,+\, \beta_{x}^2 s \left(5 M_{\pi}^2 - M_{\pi^0}^2\right)
\,-\, \beta_{x}^2 \left( 8 M_{\pi}^4 - 6 M_{\pi^0}^2 M_{\pi}^2
 - \frac{1}{2} M_{\pi^0}^4\right)
 \right.
\nonumber\\
&&\left.
+\, \frac{1}{2} \beta_{x} \alpha_{x} s M_{\pi^0}^2
- \beta_{x} \alpha_{x} M_{\pi^0}^2 \left( 2 M_{\pi}^2 + M_{\pi^0}^2\right)
\right]
\nonumber\\
\xi^{(2;{\mbox{\tiny{$\pm$}}})}_{{\mbox{\tiny{$+-$}}};P}(s) &=&
\frac{1}{2304\pi^2 M_\pi^4}\,\beta_{\mbox{\tiny{$+-$}}}^2 \left( s - 4 M_{\pi}^2 \right)^2
\nonumber\\
\xi^{(3;{\mbox{\tiny{$\pm$}}})}_{{\mbox{\tiny{$+-$}}};P}(s) &=& \frac{1}{864\pi^2 M_\pi^4}\,
\left[
\beta_{\mbox{\tiny{$+-$}}}^2 M_{\pi}^4 \,+\,
4 \beta_{\mbox{\tiny{$+-$}}} \alpha_{\mbox{\tiny{$+-$}}} M_{\pi}^2 M_{\pi^0}^2\,-\,
2 \alpha_{\mbox{\tiny{$+-$}}}^2 M_{\pi^0}^4\,+\,\frac{9}{2} \beta_{\mbox{\tiny{$+-$}}}^2 s M_{\pi}^2
\right]
\nonumber\\
\xi^{(3;{\mbox{\tiny$\Delta$}})}_{{\mbox{\tiny{$+-$}}}; P}(s) &=& \frac{1}{864\pi^2 M_\pi^4}\,
\left[
{\beta_{x}^2} \left(
-2 M_{\pi}^4 + 2 M_{\pi}^2 M_{\pi^0}^2 - 5 M_{\pi^0}^4 \right)
\,+\,
 \beta_{x} \alpha_{x} M_{\pi^0}^2 \left(2 M_{\pi}^2 - M_{\pi^0}^2\right)
\,-\,
\frac{\alpha_{x}^2}{2}\, M_{\pi^0}^4
\right]
\nonumber\\
\xi^{(4;{\mbox{\tiny{$\pm$}}})}_{{\mbox{\tiny{$+-$}}};P}(s) &=& \frac{1}{288\pi^2 M_\pi^4}\,
\left[
23 \beta_{\mbox{\tiny{$+-$}}}^2 M_{\pi}^4 \,-\,
16 \beta_{\mbox{\tiny{$+-$}}} \alpha_{\mbox{\tiny{$+-$}}} M_{\pi}^2 M_{\pi^0}^2\,-\,
4 \alpha_{\mbox{\tiny{$+-$}}}^2 M_{\pi^0}^4
\right]
\nonumber\\
\xi^{(4;{\mbox{\tiny$\Delta$}})}_{{\mbox{\tiny{$+-$}}}; P}(s) &=& \frac{1}{288\pi^2 M_\pi^4}\,
\left[
{\beta_{x}^2} \left(
-4 M_{\pi}^4 +16 M_{\pi}^2 M_{\pi^0}^2 - 25 M_{\pi^0}^4 \right)
\,-\,
{4} \beta_{x} \alpha_{x} M_{\pi^0}^2 \left( 2 M_{\pi^0}^2 - M_{\pi}^2\right)
\,-\,
\alpha_{x}^2 M_{\pi^0}^4
\right]
.
\end{eqnarray}

\noindent
Finally, for  $\psi^{x}_0(s)$ the polynomials are given by
\begin{eqnarray}
\nonumber\\
\xi_{x}^{(0)}(s) &=& -\,\frac{\lambda_{x}^{(1)}}{2 M_{\pi}^4}\,\left(s-2 M_{\pi}^2\right)\left(s-2 M_{\pi^0}^2\right)
\,-\,
\frac{\lambda_{x}^{(2)} }{3 M_{\pi}^4}\,
\left(s^2 - s M_{\pi}^2 - s M_{\pi^0}^2 + 4  M_{\pi}^2 M_{\pi^0}^2
\right)
\nonumber\\
&& - \frac{1}{864\pi^2 M_\pi^4}\,
\left\{
\frac{s^2}{4}\beta_{x}\left(108\beta_{\mbox{\tiny{$+-$}}} + 11 \beta_{x}\right)\,+\,
s \beta_{x} M_{\pi^0}^2 \left(36 \alpha_{\mbox{\tiny{$+-$}}} + \frac{45}{2} \alpha_{x}
+ 27 \alpha_{00}\right)
 \right.
\nonumber\\
&&\left.
-\,s  \beta_{x}\left[ 18 \beta_{\mbox{\tiny{$+-$}}}\left( 3M_{\pi}^2 + M_{\pi^0}^2\right) 
+ \frac{91}{2} \beta_{x} \left( M_{\pi}^2 + M_{\pi^0}^2\right) \right]  
\,+\,9 s\beta_{\mbox{\tiny{$+-$}}}\alpha_{x} M_{\pi^0}^2
 \right.
\nonumber\\
&&\left.
 +\, \beta_{x}\left[ \beta_{x}\left(144 M_{\pi}^4 + 144 M_{\pi^0}^4
- 112 M_{\pi^0}^2 M_{\pi}^2\right)\,+\,
24 \beta_{\mbox{\tiny{$+-$}}} M_{\pi}^2 \left( M_{\pi}^2 + M_{\pi^0}^2\right) \right]
 \right.
\nonumber\\
&&\left.
-\, 6 \beta_{x} M_{\pi^0}^2 \left(
4 \alpha_{\mbox{\tiny{$+-$}}}   
+ 3 \alpha_{00} \right) \left( M_{\pi}^2 + M_{\pi^0}^2\right)
\,-\, 12\beta_{\mbox{\tiny{$+-$}}}\alpha_{x} M_{\pi^0}^2 M_{\pi}^2
\,+\,3 \alpha_{x} M_{\pi^0}^4 \left(
3 \alpha_{00} + 4 \alpha_{\mbox{\tiny{$+-$}}}  + 6 \alpha_{x} \right)
\right\}
\nonumber\\
\xi_{x}^{(1)}(s) &=&  - \frac{1}{288\pi^2 M_\pi^4}\,
\left\{
\frac{1}{4}\beta_{x}^2 s^2\,-\,\beta_{x} s \left[
\beta_{x}  \left( 3 M_{\pi}^2 + \frac{5}{2} M_{\pi^0}^2\right)  
- \frac{3}{2} \alpha_{x} M_{\pi^0}^2 \right]
\right.
\nonumber\\
&&
\left. 
 \,+\,\alpha_{x}^2 M_{\pi^0}^4 
 \,+\,\beta_{x}\alpha_{x} M_{\pi^0}^2 \left( 2 M_{\pi^0}^2 - M_{\pi}^2\right)
 \,+\,\beta_{x}^2 
 \left( 7 M_{\pi}^4 + 10 M_{\pi^0}^4 - 7 M_{\pi}^2 M_{\pi^0}^2 \right)
 \right\}
\nonumber\\
\nonumber\\
\xi_{x}^{(2;0)}(s) &=& - \frac{1}{128 \pi^2 M_\pi^4}\,
\alpha_{00} M_{\pi^0}^2 \left[\beta_{x} s \,-\,
\frac{2}{3}\beta_{x} \left( M_{\pi}^2 + M_{\pi^0}^2\right)\,+\,
\frac{1}{3}\alpha_{x} M_{\pi^0}^2 \right]
\nonumber\\
\nonumber\\
\xi_{x}^{(2;\pm)}(s) &=& - \frac{1}{128\pi^2 M_\pi^4}
\left[\beta_{x} s \,-\,
\frac{2}{3}\beta_{x} \left( M_{\pi}^2 + M_{\pi^0}^2\right)\,+\,
\frac{1}{3}\alpha_{x} M_{\pi^0}^2 \right]
\left[
\beta_{\mbox{\tiny{$+-$}}} \left( s - \frac{4}{3} M_{\pi}^2 \right)
+ \frac{4}{3} \alpha_{\mbox{\tiny{$+-$}}} M_{\pi^0}^2 
\right]
\nonumber\\
\nonumber\\
\xi_{x}^{(3)}(s) &=& \frac{1}{864 \pi^2 M_\pi^4}\,\left[
-3 \beta_{x}^2 \frac{s}{M_{\pi}^2}\,
\left(M_{\pi}^4 + M_{\pi^0}^4 + M_{\pi}^2 M_{\pi^0}^2\right)\,
\right.
\nonumber\\
&&
\left. +\,
2 \beta_{x}\alpha_{x} M_{\pi^0}^2 \left( M_{\pi}^2 - M_{\pi^0}^2
+ \frac{M_{\pi^0}^4}{M_{\pi}^2}\right) 
\,+\,\alpha_{x}^2 M_{\pi^0}^4 \left( 1 \,+\,\frac{M_{\pi^0}^2}{M_{\pi}^2}\right)
\right.
\nonumber\\
&&
\left. +\,
 10 \beta_{x}^2 \left( 1 \,+\,\frac{M_{\pi^0}^2}{M_{\pi}^2}\right)
 \left( M_{\pi}^4 +  M_{\pi^0}^4 -  M_{\pi}^2 M_{\pi^0}^2 \right)
\right]
\end{eqnarray}
In addition, eq. (\ref{psi_x_0}) involves two other contributions,
\begin{eqnarray}
16\pi \Delta_1 \psi^{x}_0(s) &=&  
 \frac{1}{96\pi^2 F_\pi^4}  \left[
\frac{1}{6} \beta_{x}^2 s \,+\,\beta_{x} \alpha_{x}  M_{\pi^0}^2 
\,+\, \beta_{x}^2  \left( M_{\pi}^2 + M_{\pi^0}^2 \right)
\right]
\nonumber\\
&&
\times
\left[
\left(\sqrt{\frac{s - 4 M_{\pi}^2}{s- 4 M_{\pi^0}^2}}\,-\,1\right)
\lambda^{1/2}(t_{\mbox{\tiny $-$}}(s)) {\cal L}_{\mbox{\tiny $-$}} (s)
\,-\,
\left(\sqrt{\frac{s - 4 M_{\pi}^2}{s- 4 M_{\pi^0}^2}}\,+\,1\right)
\lambda^{1/2}(t_{\mbox{\tiny $+$}}(s)) {\cal L}_{\mbox{\tiny $+$}} (s)
\right]
,\qquad{ }
\end{eqnarray}
and
\begin{eqnarray}
16\pi \Delta_2 \psi^{x}_0(s) &=& \frac{1}{144\pi^2 F_\pi^4}\,\left[
1 \,-\, \frac{M_{\pi}^2 + M_{\pi^0}^2}{M_{\pi}^2 - M_{\pi^0}^2}\,
\ln\frac{M_{\pi}}{M_{\pi^0}} 
\right]\times
\bigg[
\frac{1}{2} \beta_{x}^2 s^2
\,-\, 5\beta_{x}^2 s \left(M_{\pi}^2 + M_{\pi^0}^2\right)
\nonumber\\
&&
\,+\, 2 \beta_{x}^2 \left( 7 M_{\pi}^4 + 7 M_{\pi^0}^4 - 6 M_{\pi}^2 M_{\pi^0}^2\right)
\,+\,3\beta_{x}\alpha_{x}  M_{\pi^0}^2 s \,-\, 
2\beta_{x}\alpha_{x}  M_{\pi^0}^2  \left(M_{\pi}^2 + M_{\pi^0}^2\right) \,
\,+\,2 \alpha_{x}^2 M_{\pi^0}^4
\bigg]
\nonumber\\
&&\, -
\frac{1}{144\pi^2 F_\pi^4}\,
\frac{1}{\sqrt{(s- 4 M_{\pi}^2)(s - 4 M_{\pi^0}^2)}}\,
\bigg\{
\left[4 \beta_{x}^2(5 M_{\pi}^4 - 2  M_{\pi}^2  M_{\pi^0}^2 + 5  M_{\pi^0}^4)
- 6 \beta_{x}^2 s (M_{\pi}^2 + M_{\pi^0}^2)
\right.
\nonumber\\
&&
\left.
\,+\,4  \beta_{x} \alpha_{x} M_{\pi^0}^2 (M_{\pi}^2 + M_{\pi^0}^2)\,+\,
2 \alpha_{x}^2 M_{\pi^0}^4 
\right]
\left[ {\cal F}_{\mbox{\tiny $+$}}(s) - {\cal F}_{\mbox{\tiny $-$}}(s) \right] 
\nonumber\\
&&
+\, 6 \beta_x\left[
2 \beta_{x} (M_{\pi}^2 + M_{\pi^0}^2) + \alpha_{x} M_{\pi^0}^2
- \frac{s}{2} \beta_{x} \right]
\left[ {\cal G}_{\mbox{\tiny $+$}}(s) - {\cal G}_{\mbox{\tiny $-$}}(s) \right] 
\nonumber\\
&&
+\,3 \beta_x^2
\left[ {\cal H}_{\mbox{\tiny $+$}}(s) - {\cal H}_{\mbox{\tiny $-$}}(s) \right]
\bigg\} ,
\end{eqnarray}
where ${\cal F}_{\mbox{\tiny $\pm $}}(s) \equiv {\cal F} (t_{\mbox{\tiny $\pm $}}(s))$,
and similar definitions for ${\cal G}_{\mbox{\tiny $\pm $}}(s)$ and ${\cal H}_{\mbox{\tiny $\pm $}}(s)$,
with
\begin{equation}
t_{\mbox{\tiny{$\pm$}}}(s) \,=\, -\frac{1}{2} (s - 2 M_{\pi}^2 - 2 M_{\pi^0}^2)
\,\pm\,
\frac{1}{2}\,\sqrt{(s- 4 M_{\pi}^2)(s - 4 M_{\pi^0}^2)} ,
\end{equation}
and
\begin{eqnarray}
{\cal F}(t) &=& (M_{\pi}^2 - M_{\pi^0}^2)\left[
H_{1,0} \left(\frac{M_{\pi}}{M_{\pi^0}}\chi(t)\right) \,-\, H_{1,0} \left(\frac{M_{\pi^0}}{M_{\pi}}\chi(t)\right)
\,+\,\ln\frac{M_{\pi}}{M_{\pi^0}} \, \ln\chi(t)
\right]
\nonumber\\
{\cal G}(t) &=&  (M_{\pi}^2 + M_{\pi^0}^2){\cal F}(t)  + (M_{\pi}^2 - M_{\pi^0}^2) t \ln\frac{M_{\pi}}{M_{\pi^0}}
\nonumber\\
&&\!\!\!
 - (M_{\pi}^2 - M_{\pi^0}^2)^2 \left[
\left( M_{\pi}^2 + M_{\pi^0}^2 - 2 M_{\pi} M_{\pi^0} \chi(t) \right) \frac{\ln\chi(t)}{t}
- \frac{1}{2} \ln^2\chi(t) - \ln\chi(t) + \frac{M_{\pi}^2 - M_{\pi^0}^2}{t} \ln\frac{M_{\pi}}{M_{\pi^0}}
\right]
\nonumber\\
{\cal H}(t) &=& - 4 M_{\pi}^2 M_{\pi^0}^2 {\cal F}(t)  
 - (M_{\pi}^2 - M_{\pi^0}^2) t^2 \ln\frac{M_{\pi}}{M_{\pi^0}}
\nonumber\\
&&\!\!\!
 -  (M_{\pi}^2 - M_{\pi^0}^2)^4 \left[
2 \left(
\frac{1}{2} + \frac{M_{\pi}^2 + M_{\pi^0}^2}{M_{\pi}^2 - M_{\pi^0}^2} \ln\frac{M_{\pi}}{M_{\pi^0}}
\right) \frac{1}{t}
- \frac{M_{\pi}^2 - M_{\pi^0}^2}{t^2} \ln\frac{M_{\pi}}{M_{\pi^0}}
\right.
\nonumber\\
&&
\left.
\qquad\qquad
+ \frac{1}{t^2}
\left(
\frac{M_{\pi}^2 + M_{\pi^0}^2}{(M_{\pi}^2 - M_{\pi^0}^2)^2} t - 1
\right) \lambda^{1/2}(t) \ln \chi(t) \right]
.
\end{eqnarray}
Notice that individual terms in $\Delta_2 \psi^{x}_0(s)$ may behave as ${\cal O}[\Delta_\pi^2 \times \ln(\Delta_\pi / M_\pi^2)]$
in the isospin limit, but cancellations occur between these terms, so that overall $\Delta_2 \psi^{x}_0(s) = {\cal O}(\Delta_\pi^2)$.

\section{Two-loop form factors in the isospin limit} \label{app:IsoLimit}   

In this appendix, we discuss the scalar and vector form factors at two loops
in the isospin limit, where the complications due to the mass difference between
neutral and charged pions are absent, and the dispersive integrals can all
be expressed in terms of the known function ${\bar J}(s)$. 
The two form factors $F_S^{\pi^0}$ and
$F_S^{\pi^\pm}$ then become identical, since in both cases the two-pion states
are projected on their $I=0$, S-wave components, with identical Clebsch-Gordan
coefficients. Furthermore, this exercise will provide a check of
the calculation in the general case, which has to reduce to the expressions
to be found below in the limit $M_{\pi^0} \to M_{\pi}$. Let us recall that ref.
\cite{GasserMeissner91} did not give analytical expressions for the form 
factors at two loops. Analytical two-loop expressions for the scalar and the vector 
form factors were given in \cite{Bijnens98}. The analytical expression for
the vector form factor only had also been given earlier in ref. \cite{Colangelo96}.

For the first iteration, the discontinuities of the form factors reduce to
[in this appendix, we omit the superscript $\pi$ most of the time]
\begin{eqnarray}
{\mbox{Im}}F_S(s) &=& 
 \sigma(s)F_S (0)  
\varphi_0(s) 
\theta(s-4M_{\pi}^2)\,+\,{\cal O}(E^6)
\nonumber\\
{\mbox{Im}}F_V (s) &=& 
 \sigma(s)  
\varphi_1(s) 
\theta(s-4M_{\pi}^2)\,+\,{\cal O}(E^4)\,,
\end{eqnarray}
with
\begin{eqnarray}
\varphi_0 (s)&=& \varphi_0^{\mbox{\tiny{$+-$}}}(s) \,-\,\frac{1}{2}\,\varphi_0^{x}(s)
\,=\,
\frac{1}{2}\,\varphi_0^{00}(s) \,-\,  \varphi_0^{x}(s)
\,=\,
\frac{1}{16\pi F_\pi^2}\,\left[
\beta \left( s - \frac{4}{3}\,M_\pi^2 \right) \,+\, \frac{5}{6}\, \alpha M_\pi^2
\right]
\nonumber\\
\varphi_1 (s) &=& \varphi_1^{\mbox{\tiny{$+-$}}}(s)\,=\,\frac{1}{96\pi F_\pi^2}
\,\beta(s - 4M_\pi^2).
\end{eqnarray}
The one-loop expressions of the form factors in the isospin limit read
\begin{eqnarray}
F_S (s) &=& F_S (0) \left[
1 \,+\, \frac{1}{6}\langle r^2\rangle_S^{\pi}\,{s}
\,+\, c_S^{\pi}\,{s^2} 
\,+\, U_S(s)
\right]
\\
\nonumber\\
F_V &=& 1 \,+\, \frac{1}{6}\langle r^2\rangle_V^{\pi}\, {s} 
\,+\, c_V^{\pi}\, {s^2} 
\,+\, U_V(s)
,
\end{eqnarray}
with
\begin{eqnarray}
U_S (s) &=& 16\pi\varphi_0(s){\bar J}(s)\,+\,
\frac{M_\pi^2}{16\pi^2 F_\pi^2}\,\bigg\{
\frac{1}{36}\,\frac{s}{M_\pi^2}\,(8\beta - 5\alpha) 
\,-\,
\frac{1}{360}\,\left(\frac{s}{M_\pi^2}\right)^2\,(52\beta + 5\alpha)
\bigg\}
\\
\nonumber\\
U_V (s) &=& \frac{M_\pi^2}{16\pi^2 F_\pi^2}\,\beta\,\left\{
\frac{1}{9}\,\frac{s}{M_\pi^2}\,\left[1\,+\,24\pi^2\sigma^2(s){\bar J}(s)\right]
\,-\,\frac{1}{60}\,\left(\frac{s}{M_\pi^2}\right)^2
\right\} .
\end{eqnarray}

In order to implement the second iteration, it is necessary to 
include the next-to-leading contributions to the discontinuities
of the form factors,
\begin{eqnarray}
{\mbox{Im}}F_S (s) &=& 
 \sigma(s)F_S (0) \left\{ 
\varphi_0(s)\left[
1 \,+\,\Gamma_S (s) \right] \,+\,\psi_0(s) 
\right\}
\theta(s-4M_{\pi}^2)\,+\,{\cal O}(E^6)
\nonumber\\
{\mbox{Im}}F_V (s) &=& 
 \sigma(s) \left\{ 
\varphi_1(s)\left[
1 \,+\,\Gamma_V (s) \right] \,+\,\psi_1(s) 
\right\}
\theta(s-4M_{\pi}^2)\,+\,{\cal O}(E^6)
.
\nonumber
\end{eqnarray}
The relevant one-loop corrections $\Gamma_S (s)$  
and $\Gamma_V (s)$ to the real parts of the form factors
are easy to obtain from their expressions given above,
\begin{eqnarray}
\Gamma_S (s) &=& \frac{s}{6} \left[
\langle r^2\rangle_S^{\pi}
\,+\,\frac{1}{96\pi^2 F_\pi^2}\,(8\beta - 5\alpha)\right]  
\,+\,\frac{1}{\pi}\,\varphi_0(s)[2 + \sigma(s) L(s)]  
\\
\nonumber\\
\Gamma_V (s) &=&\frac{s}{6} \left[
\langle r^2\rangle_V^{\pi}
\,+\,\frac{1}{24\pi^2 F_\pi^2}\,\beta \right] 
\,+\, \frac{1}{\pi}\,\varphi_1(s)
[2 + \sigma(s) L(s)] 
.
\nonumber
\end{eqnarray}

As for the one-loop contributions to the real parts of the $S$ and $P$
partial waves, they are conveniently expressed as
\begin{eqnarray}
\psi_0 (s)&=& \psi_0^{\mbox{\tiny{$+-$}}}(s) \,-\,\frac{1}{2}\,\psi_0^{x}(s)
\,=\,
\frac{1}{2}\,\psi_0^{00}(s) \,-\,  \psi_0^{x}(s)
\nonumber\\
\psi_1 (s) &=& \psi_1^{\mbox{\tiny{$+-$}}}(s),
\nonumber
\end{eqnarray}
where the expressions for $\psi_0^{00}(s)$, $\psi_{0,1}^{\mbox{\tiny{$+-$}}}(s)$, and 
$\psi_0^{x}(s)$ are given in eqs. (\ref{lim_psi_00}), (\ref{lim_psi_+-}), and (\ref{lim_psi_x}),
respectively. They involve the polynomials $\xi_{a}^{(n)} (s)$ of reference \cite{KMSF95}.
We reproduce them in a slightly different notation, in terms of the variable $s$
instead of the relative momentum $q = \sqrt{s/4M_\pi^2 -1}$ in the center-of-mass frame,
\begin{eqnarray}
\xi_0^{(0)}(s) &=&  \frac{1}{432\pi^2}\,(105\alpha^2 - 120 \alpha\beta + 392 \beta^2)\,+\,
\frac{2}{3}\,(11\lambda_1 + 14\lambda_2)
\nonumber\\
&&
+\,\left\{
\frac{1}{864\pi^2}\,(180\alpha - 617 \beta)\beta \,-\,
\frac{20}{3}\,(\lambda_1 + \lambda_2)
\right\}\,\frac{s}{M_\pi^2}
\nonumber\\
&&
+\,\left\{
\frac{311}{1728\pi^2}\,\beta^2 \,+\, \frac{1}{6}\,(11\lambda_1 + 14\lambda_2)
\right\}\,\left(\frac{s }{M_\pi^2}\right)^2
\nonumber\\
\xi_0^{(1)}(s) &=& \frac{5}{192\pi^2}\,(\alpha^2 + 4 \beta^2)\,-\,
\frac{5}{72\pi^2}\,\beta^2\,\frac{s }{M_\pi^2}\,+\,
\frac{7}{576\pi^2}\,\beta^2\,\left(\frac{s  }{M_\pi^2}\right)^2
\nonumber\\
\xi_0^{(2)}(s) &=& \frac{1}{1152\pi^2}\,(25\alpha^2 - 80 \alpha\beta + 64 \beta^2)\,+\,
\frac{1}{96\pi^2}\,\beta (5\alpha - 8 \beta)\,\frac{s}{M_\pi^2}\,+\,
\frac{1}{32\pi^2}\,\beta^2\,\left(\frac{s  }{M_\pi^2}\right)^2
\nonumber\\
\xi_0^{(3)}(s) &=& -\frac{5}{288\pi^2}\,( \alpha^2 + 4 \beta^2)\,+\,
\frac{1}{48\pi^2}\,\beta^2\,\frac{s  }{M_\pi^2}
\nonumber\\
\nonumber\\
\xi_2^{(0)}(s) &=&  \frac{1}{864 \pi^2}\,(93\alpha^2 + 48 \alpha\beta + 112 \beta^2)\,+\,
\frac{4}{3}\,(\lambda_1 + 4\lambda_2)
\nonumber\\
&&
-\,\left\{
\frac{1}{1728 \pi^2}\,(207\alpha + 256 \beta)\beta \,+\,
\frac{2}{3}\,(\lambda_1 + 7 \lambda_2)
\right\}\,\frac{s}{M_\pi^2}
\nonumber\\
&&
+\,\left\{
\frac{265}{3456 \pi^2}\,\beta^2 \,+\, \frac{1}{3}\,(\lambda_1 + 4\lambda_2)
\right\}\,\left(\frac{s }{M_\pi^2}\right)^2
\nonumber\\
\xi_2^{(1)}(s) &=& \frac{1}{192\pi^2}\,(3\alpha^2 - 2 \alpha \beta)\,-\,
\frac{1}{576\pi^2}\,\beta (9 \alpha + 7 \beta )\,\frac{s }{M_\pi^2}\,+\,
\frac{11}{1152\pi^2}\,\beta^2\,\left(\frac{s  }{M_\pi^2}\right)^2
\nonumber\\
\xi_2^{(2)}(s) &=& \frac{1}{288\pi^2}\,(\alpha^2 + 4 \alpha\beta + 4 \beta^2)\,-\,
\frac{1}{96\pi^2}\,\beta (\alpha + 2 \beta)\,\frac{s}{M_\pi^2}\,+\,
\frac{1}{128\pi^2}\,\beta^2\,\left(\frac{s  }{M_\pi^2}\right)^2
\nonumber\\
\xi_2^{(3)}(s) &=& \frac{1}{288\pi^2}\,( -3\alpha^2 + 2 \alpha \beta)\,-\,
\frac{1}{96\pi^2}\,\beta^2\,\frac{s  }{M_\pi^2}
\nonumber\\
\nonumber\\
\xi_1^{(0)}(s) &=&  \frac{1}{1728\pi^2}\,(15\alpha^2 - 460 \alpha\beta + 286 \beta^2) 
\nonumber\\
&&
+\,\left\{
\frac{5}{1728\pi^2}\,(11\alpha - 12 \beta)\beta \,+\,
\frac{2}{3}\,(\lambda_1 - \lambda_2)
\right\}\,\frac{s}{M_\pi^2}
\nonumber\\
&&
-\,\left\{
\frac{1}{1728\pi^2}\,\beta^2 \,+\, \frac{1}{6}\,(\lambda_1 - \lambda_2)
\right\}\,\left(\frac{s }{M_\pi^2}\right)^2
\nonumber\\
\xi_1^{(1)}(s) &=& -\frac{1}{96\pi^2}\,\beta(5\alpha - 3 \beta)\,+\,
\frac{1}{576\pi^2}\,\beta(5\alpha + \beta)\,\frac{s }{M_\pi^2}\,-\,
\frac{1}{1152\pi^2}\,\beta^2\,\left(\frac{s  }{M_\pi^2}\right)^2
\nonumber\\
\xi_1^{(2)}(s) &=& \frac{1}{72\pi^2}\,\beta^2\,\left(\frac{s - 4M_\pi^2}{4M_\pi^2}\right)^2
\nonumber\\
\xi_1^{(3)}(s) &=& \frac{1}{864\pi^2}\,
(-5\alpha^2 + 10 \alpha\beta - 8 \beta^2) \,+\,
\frac{1}{96\pi^2}\,\beta^2\,\frac{s  }{M_\pi^2}
\nonumber\\
\xi_1^{(4)}(s) &=& - \frac{5}{144\pi^2}\,(\alpha^2 + 4\alpha\beta - 2 \beta^2) 
\,.
\nonumber
\end{eqnarray}
 
Putting all elements together leads to
\begin{eqnarray}
\frac{1}{F_S (0)}\,{\mbox{disc}}\,F_S (s) &=&
\sum_{n=0}^3 {\cal S}_n(s) k_n(s) \times \theta(s-4M_\pi^2 )
\nonumber\\
\\
{\mbox{disc}}\,F_V (s) &=&
\sum_{n=0}^4 {\cal V}_n(s) k_n(s) \times \theta(s-4M_\pi^2 )
\end{eqnarray}
with
\begin{eqnarray}
{\cal S}_n(s) &=&  16\pi\varphi_0(s)\delta_{n,0}
\nonumber\\ 
&& +\,
\left\{
\frac{8\pi s}{3}\left[
\langle r^2\rangle_S^{\pi}
\,+\,\frac{1}{96\pi^2 F_\pi^2}\,(8\beta - 5\alpha)\right]  \,+\,
32 \varphi_0(s)\right\} \varphi_0(s)\delta_{n,0}
\nonumber\\ 
&&
+\,8\left[\varphi_0(s)\right]^2 \delta_{n,2}
  \,+\, \frac{M_\pi^4}{F_\pi^4}\, \xi_0^{(n)}(s) 
\nonumber\\  
{\cal V}_n(s) &=&  16\pi\varphi_1(s)\delta_{n,0}
\nonumber\\ 
&& +\,
\left\{
\frac{8\pi s}{3}\left[
\langle r^2\rangle_V^{\pi}
\,+\,\frac{1}{24\pi^2 F_\pi^2}\,\beta \right]  \,+\,
32 \varphi_1(s)\right\} \varphi_1(s)\delta_{n,0}
\nonumber\\ 
&&
+\,8\left[\varphi_1(s)\right]^2 \delta_{n,2}
  \,+\, \frac{M_\pi^4}{F_\pi^4}\,  \xi_1^{(n)}(s)
  .
\end{eqnarray}
Performing the dispersive integrals gives
\begin{eqnarray}
U_S (s) &=&
\sum_{n=0}^3 {\cal S}_n(s) {\bar K}_n(s)  \,+\, P_S (s)
\nonumber\\
\\
U_V (s) &=&
\sum_{n=0}^4 {\cal V}_n(s) {\bar K}_n(s) \,+\, P_V (s).
\end{eqnarray}
The two second-order polynomials $P_S (s)$ and $P_V (s)$ are
obtained as follows. First write
\begin{equation}
{\bar K}_n(s) \,=\,\kappa_n^{(1)} \frac{s}{M_\pi^2} \,+\, 
\kappa_n^{(2)} \left(\frac{s}{M_\pi^2}\right)^2 \,+\,
\frac{s^3}{\pi}\,\int_{4M_\pi^2}^\infty\,\frac{dx}{x^3}\,\frac{k_n(x)}{x-s-i0},
\end{equation}
and, next, expand the polynomials ${\cal S}_n(s)$ and ${\cal V}_n(s)$,
\begin{eqnarray}
{\cal S}_n(s) &=& {\cal S}_n^{(0)} \,+\, {\cal S}_n^{(1)} \frac{s}{M_\pi^2} \,+\, 
 {\cal S}_n^{(2)} 
\left(\frac{s}{M_\pi^2}\right)^2
\nonumber\\
{\cal V}_n(s) &=& {\cal V}_n^{(0)} \,+\, {\cal V}_n^{(1)} \frac{s}{M_\pi^2} \,+\, 
 {\cal V}_n^{(2)} 
\left(\frac{s}{M_\pi^2}\right)^2\,.
\nonumber
\end{eqnarray}
The polynomials $P_S (s)$ and $P_V (s)$ are then given by
\begin{eqnarray}
P_S (s) &=& -\,\frac{s^2}{M_\pi^4}\,\sum_{n=0}^3
\left[\kappa_n^{(2)}{\cal S}_n^{(0)}\,+\,\kappa_n^{(1)}{\cal S}_n^{(1)}\right]\,-\,
\frac{s}{M_\pi^2}\,\sum_{n=0}^3
\kappa_n^{(1)}{\cal S}_n^{(0)} \,,
\nonumber\\
P_V (s) &=& -\,\frac{s^2}{M_\pi^4}\,\sum_{n=0}^4
\left[\kappa_n^{(2)}{\cal V}_n^{(0)}\,+\,\kappa_n^{(1)}{\cal V}_n^{(1)}\right]\,-\,
\frac{s}{M_\pi^2}\,\sum_{n=0}^4
\kappa_n^{(1)}{\cal V}_n^{(0)}\,.
\end{eqnarray}
Using the coefficients $\kappa_n^{(1)}$ and $\kappa_n^{(2)}$  displayed in the following table:

\indent

\begin{center}
\begin{tabular}{|c|c|c|c|c|c|}
\hline
$n$ & $0$ & $1$ & $2$ & $3$ & $4$ \\
\hline
$\pi^2\kappa_n^{(1)}$ & $\frac{1}{96}$ & $- \frac{1}{16}$ & $- \frac{1}{24}$ & $\frac{\pi^2}{192}\,-\,\frac{1}{32}$ 
& $- \frac{\pi^2}{576}\,+\,\frac{1}{64}$ \\
\hline
$\pi^2\kappa_n^{(2)}$ &  $\frac{1}{960}$ & $ - \frac{1}{192}$ & $- \frac{7}{2880}$ & $\frac{\pi^2}{960}\,-\,\frac{1}{128}$
& $-\frac{\pi^2}{1920}\,+\,\frac{19}{3840}$ \\
\hline
\end{tabular}
\end{center}

\indent

\noindent
we obtain
\begin{eqnarray}
P_S (s) &=& 
\frac{M_\pi^2}{16\pi^2 F_\pi^2}\left[
\frac{8\beta - 5\alpha}{36}\,
\left(\frac{s}{M_\pi^2}\right)\,-\,
\frac{52\beta + 5\alpha}{360}\,
\left(\frac{s}{M_\pi^2}\right)^2
\right]
\nonumber\\
&& + \left(\frac{M_\pi^2}{16\pi^2 F_\pi^2}\right)^2
\left\{
\left(\frac{s}{M_\pi^2}\right)
\left[
- \frac{1}{324}(45\alpha^2 + 232\beta^2)
\,+\,\frac{5\pi^2}{216} (\alpha^2 + 4\beta^2)
\,-\,\frac{16\pi^2}{9} (11\lambda_1 + 14\lambda_2)
\right]
\right.
\nonumber\\
&& +
\left.
\left(\frac{s}{M_\pi^2}\right)^2
\left[
\frac{2\pi^2}{27}(8\beta - 5\alpha)F_\pi^2 \langle r^2\rangle_S^{\pi}
\,-\,\frac{1}{3240}(135\alpha^2 +  362\beta^2)
\right.\right.
\nonumber\\
&&\qquad \qquad \qquad
\left.\left.
\,+\,\frac{\pi^2}{216} (\alpha^2 - 2\beta^2)
\,+\,\frac{8\pi^2}{45} (89\lambda_1 + 86\lambda_2)
\right]
\right\}
,
\nonumber\\
P_V (s) &=& 
\frac{M_\pi^2}{16\pi^2 F_\pi^2}\left[
\frac{\beta}{9}\,
\left(\frac{s}{M_\pi^2}\right)\,-\,
\frac{\beta}{60}\,
\left(\frac{s}{M_\pi^2}\right)^2
\right]
\nonumber\\
&& + \left(\frac{M_\pi^2}{16\pi^2 F_\pi^2}\right)^2
\left\{
\left(\frac{s}{M_\pi^2}\right)
\left[
\frac{1}{648}(45\alpha^2 + 340\alpha\beta - 94\beta^2)
\,-\,\frac{\pi^2}{648} (5\alpha^2 + 50\alpha\beta - 28\beta^2)
\right]
\right.
\nonumber\\
&& +
\left.
\left(\frac{s}{M_\pi^2}\right)^2
\left[
\frac{8\pi^2}{27}\,\beta F_\pi^2 \langle r^2\rangle_V^{\pi}
\,+\,\frac{1}{6480}(195\alpha^2 + 1650\alpha\beta + 230\beta^2)
\right.\right.
\nonumber\\
&&\qquad \qquad \qquad
\left.\left.
\,-\,\frac{\pi^2}{3240} (10\alpha^2 + 70\alpha\beta + 7\beta^2)
\,-\,\frac{16\pi^2}{9} (\lambda_1 - \lambda_2)
\right]
\right\}
.
\end{eqnarray}

\noindent
The relations \cite{KMSF95}
\begin{eqnarray}
{\bar K}_2(s) &=& \left(1\,-\,\frac{4M_\pi^2}{s}\right) {\bar K}_1(s)
\,-\,\frac{1}{4\pi^2}
\nonumber\\
{\bar K}_3(s) &=& 3\,\frac{s - 4 M_\pi^2}{M_\pi^2}\,{\bar K}_4(s)
\,-\, 3{\bar J}(s) \,-\, \frac{3}{2}\,{\bar K}_1(s) \,+\,
\left(\frac{3}{32\pi^2}\,-\,\frac{1}{64}\right)\,\frac{s }{M_\pi^2}\,,
\end{eqnarray}
allow us to eliminate ${\bar K}_2(s)$ from $F_S (s)$, and both
${\bar K}_2(s)$ and ${\bar K}_3(s)$ from $F_V (s)$,
thus leading to the following expressions of the form factors
\begin{eqnarray}
\lefteqn{
F_S(s)/F_S(0) \,=\,
1 \,+\, \frac{1}{6}\langle r^2\rangle_S^{\pi}\, {s} 
\,+\, c_S^{\pi}\,s^2
}
\nonumber\\
&& + \,\frac{M_\pi^2}{16\pi^2 F_\pi^2}
\left\{
16\pi^2 \left(
\frac{s}{M_\pi^2}\beta - \frac{1}{6} (8\beta - 5\alpha)
\right){\bar J}(s)
+ \frac{1}{36} \frac{s}{M_\pi^2} (8\beta - 5\alpha) - 
\frac{1}{360}\left(\frac{s}{M_\pi^2}\right)^2
(52\beta + 5\alpha)
\right\}
\nonumber\\
&& + \, 
\left(\frac{M_\pi^2}{16\pi^2 F_\pi^2}\right)^2 
\left\{
\frac{8\pi^2}{9}\,{\bar J}(s)
\left[
\left(\frac{s}{M_\pi^2}\right)^2
\left(
48\pi^2 (11{\lambda}_1 + 14 {\lambda}_2) + 48\pi^2 F_\pi^2 \beta \langle r^2\rangle_S^{\pi} + 
\frac{1}{6}\beta(551\beta - 15 \alpha)
\right)
\right.\right.
\nonumber\\
&& \left.\left.
\qquad\qquad\qquad\qquad\qquad
\,-\,
\left(\frac{s}{M_\pi^2}\right)
\left(
1920 \pi^2 (\lambda_1 + \lambda_2)
+ 8\pi^2 F_\pi^2 (8\beta - 5\alpha) \langle r^2\rangle_S^{\pi}
\right.\right.\right.
\nonumber\\
&& \left.\left.\left.
\qquad\qquad\qquad\qquad\qquad\qquad\qquad\qquad
+ \frac{1}{12} (3684\beta^2 - 1520\alpha\beta + 25\alpha^2)
\right)
\right.\right.
\nonumber\\
&& \left.\left.
\qquad\qquad\qquad\qquad\qquad\qquad
\,+\,
192\pi^2 (11{\lambda}_1 + 14 {\lambda}_2)  
+ \frac{1}{3} (976\beta^2 - 480\alpha\beta + 285\alpha^2) 
\right]
\right.
\nonumber\\
&&
\left. 
\qquad +\,
12\pi^2 {\bar K}_1(s)
\left[
\frac{43}{27}\left(\frac{s}{M_\pi^2}\right)^2 \beta^2
- \frac{20}{27}\left(\frac{s}{M_\pi^2}\right) \beta (14\beta - 3\alpha)
+ \frac{4}{27} (127 \beta^2 - 80 \alpha\beta + 10 \alpha^2)
\right.\right.
\nonumber\\
&& \left.\left.
\qquad\qquad\qquad\qquad\qquad\qquad
- \frac{4}{27}\left(\frac{M_\pi^2}{s}\right) (64\beta^2 - 80\alpha\beta + 25\alpha^2)
\right]
\right.
\nonumber\\
&&
\left. 
\qquad +\,
\frac{16\pi^2}{3} {\bar K}_3(s)
\left[
\left(\frac{s}{M_\pi^2}\right) \beta^2 - \frac{5}{6} (4\beta^2 + \alpha^2)
\right]
\right.
\nonumber\\
&&
\left. 
\qquad +\,
\left(\frac{s}{M_\pi^2}\right)^2
\left[  \frac{8\pi^2}{45} (89 \lambda_1 + 86 \lambda_2) 
+ \frac{2\pi^2}{27}(8\beta - 5\alpha) F_\pi^2\langle r^2\rangle_S^{\pi} 
- \frac{1}{3240} (13322\beta^2 + 135 \alpha^2)
\right.\right.
\nonumber\\
&& \left.\left.
\qquad\qquad\qquad\qquad
- \frac{\pi^2}{216} (2\beta^2 - \alpha^2)
\right]
\right.
\nonumber\\
&&
\left. 
\qquad +\,
\left(\frac{s}{M_\pi^2}\right)
\left[
- \frac{16\pi^2}{9} (11 \lambda_1 + 14 \lambda_2)
+ \frac{1}{324} (3224\beta^2 - 2160 \alpha\beta - 45\alpha^2)
+ \frac{5\pi^2}{216} (4\beta^2 + \alpha^2)
\right]
\right.
\nonumber\\
&&
\left. 
\qquad  -\,\frac{1}{9} (8\beta - 5\alpha)^2
\right\}  
,
\end{eqnarray}

\begin{eqnarray}
\lefteqn{
F_V(s) \,=\, 
1 \,+\, \frac{1}{6}\langle r^2\rangle_V^{\pi}\, {s}  
\,+\, c_V^{\pi}\, s^2
\,+ \,\frac{M_\pi^2}{16\pi^2 F_\pi^2}
\left\{
\frac{8\pi^2}{3}\,\beta \left(
\frac{s}{M_\pi^2} - 4
\right){\bar J}(s)
+ \frac{1}{9} \beta \frac{s}{M_\pi^2} - 
\frac{1}{60}\,\beta \left(\frac{s}{M_\pi^2}\right)^2
\right\}
}
\nonumber\\
&& + \, 
\left(\frac{M_\pi^2}{16\pi^2 F_\pi^2}\right)^2 
\left\{
\frac{8\pi^2}{9}\,{\bar J}(s)
\left[
\left(\frac{s}{M_\pi^2}\right)^2
\left(
[48\pi^2(\lambda_2 - \lambda_1) - \frac{1}{2}] 
+ \frac{1}{2} [16\pi^2 F_\pi^2 \beta\langle r^2\rangle_V^{\pi} + 1]   + \frac{7}{6}\, \beta^2
\right)
\right.\right.
\nonumber\\
&& \left.\left.
\qquad\qquad\qquad\qquad\qquad
\,-\,
4 \left(\frac{s}{M_\pi^2}\right)
\left(
[48\pi^2(\lambda_2 - \lambda_1) - \frac{1}{2}] 
+ \frac{1}{2} [16\pi^2 F_\pi^2 \beta\langle r^2\rangle_V^{\pi} + 1] 
 + \frac{5}{24}\beta(34\beta - 11\alpha)
\right) 
\right.\right.
\nonumber\\
&& \left.\left.
\qquad\qquad\qquad\qquad\qquad
\,+\,\frac{5}{6}\,(86 \beta^2 - 104 \alpha\beta + 9 \alpha^2)
\right]
\right.
\nonumber\\
&&
\left. 
\qquad +\,
\frac{2\pi^2}{9}\, {\bar K}_1(s)
\left[
\left(\frac{s}{M_\pi^2}\right)^2 \beta^2
- 10 \left(\frac{s}{M_\pi^2}\right) \beta(4\beta - \alpha)
+ 2(74\beta^2 - 40\alpha\beta + 5\alpha^2) - 128\left(\frac{M_\pi^2}{s}\right) \beta^2
\right]
\right.
\nonumber\\
&&
\left. 
\qquad +\,
8\pi^2 {\bar K}_4(s)
\left[
\left(\frac{s}{M_\pi^2}\right)^2 \beta^2 - 
\frac{1}{9}\left(\frac{s}{M_\pi^2}\right) (44\beta^2 - 10\alpha\beta + 5\alpha^2)
 +  \frac{2}{9} (26\beta^2 - 40\alpha\beta + 5\alpha^2)
\right]
\right.
\nonumber\\
&&
\left. 
\qquad +\,
\left(\frac{s}{M_\pi^2}\right)^2
\left[  \frac{16\pi^2}{9} (\lambda_2 - \lambda_1) 
+ \frac{8\pi^2}{27}\beta F_\pi^2\langle r^2\rangle_V^{\pi} 
+ \frac{1}{6480} (1130\beta^2 + 1650 \alpha\beta +195\alpha^2)
\right.\right.
\nonumber\\
&& \left.\left.
\qquad\qquad\qquad\qquad
- \frac{\pi^2}{1620} (71\beta^2 + 35\alpha\beta + 5\alpha^2)
\right]
\right.
\nonumber\\
&&
\left. 
\qquad +\,
\left(\frac{s}{M_\pi^2}\right)
\left[
 \frac{1}{648} (338\beta^2 + 520 \alpha\beta - 45\alpha^2)
+ \frac{\pi^2}{648} (52\beta^2 - 80\alpha\beta + 10\alpha^2)
\right] \,-\,\frac{16}{9} \beta^2
\right\} 
.
\end{eqnarray}

In order to recover the expressions of \cite{Colangelo96} from these formulae,
one simply needs to replace the
various quantities by their
expressions at leading or at next-to-leading
order, as they can be found in refs. \cite{KMSF95,Gasser:1983yg},
\begin{eqnarray}
\alpha &=&
1 \,+\, \frac{1}{32\pi^2}\frac{M_\pi^2}{F_\pi^2}\,(4 {\bar \ell}_4 - 3 {\bar \ell}_3 - 1)
\nonumber\\
\beta &=&
1 \,+\, \frac{1}{8\pi^2}\frac{M_\pi^2}{F_\pi^2}\,( {\bar \ell}_4  - 1)
\nonumber\\
\lambda _1 &=&
\frac{1}{48\pi^2} \,\left( {\bar \ell}_1 - \frac{4}{3} \right)
\nonumber\\
\lambda_2 &=&
\frac{1}{48\pi^2} \,\left( {\bar \ell}_2 - \frac{5}{6} \right)
\nonumber\\
\langle r^2\rangle_S^{\pi} &=&
\frac{3}{8\pi^2 F_\pi^2} \,\left( {\bar \ell}_4 - \frac{13}{12} \right)
\nonumber\\
\langle r^2\rangle_V^{\pi} &=&
\frac{1}{16\pi^2 F_\pi^2} \,( {\bar \ell}_6 - 1)
.
\label{alpha_bet_std}
\end{eqnarray}
At the end of this process, we then obtain perfect agreement with ref. \cite{Colangelo96}.

\section{First-order isospin-breaking corrections to the one-loop partial waves } \label{app:delta_xi}

The purpose of this appendix is to provide the explicit expressions of the functions
that describe the isospin-breaking corrections to the one-loop partial waves, as given
in eqs. (\ref{psi_00_exp}) and (\ref{psi_+-_exp_and_psi_x_exp}).

In the case of $\psi^0_{00} (s)$ the corrections that appear in eq. (\ref{psi_00_exp}) read
\begin{eqnarray}
\Delta\xi_{00}^{(0)}(s) &=& \frac{4}{3}\,(\lambda_1 + 2 \lambda_2)
\left(2\frac{s}{M_{\pi}^2} - 5\right)\,+\,
\frac{1}{576\pi^2}\,\left(
160\beta^2 \frac{s}{M_{\pi}^2} + 6 \alpha\beta \frac{s}{M_{\pi}^2}
+ 24 \alpha\beta -504 \beta^2 - 203 \alpha^2 \right)
\nonumber\\
&& -\,
\frac{1}{72\pi^2}\,\frac{M_{\pi}^2}{s - 4M_{\pi}^2}\,
(44 \beta^2 - 28 \alpha\beta + 11 \alpha^2)
\nonumber\\
\Delta\xi_{00}^{(1)}(s) &=& 
\frac{1}{2304\pi^2}\,\left(12 \beta^2 \frac{s^2}{M_{\pi}^4} +
60 \beta^2 \frac{s}{M_{\pi}^2} + 12 \alpha\beta \frac{s}{M_{\pi}^2}
- 79 \alpha^2 -368 \beta^2 \right)
\nonumber\\
&& -\,
\frac{1}{12\pi^2}\,\frac{M_{\pi}^2}{s - 4M_{\pi}^2}\,
( \beta  -   \alpha)\beta  
\nonumber\\
\Delta\xi_{00}^{(2)}(s) &=& 
\frac{1}{2304\pi^2}\,\left( - 12 \beta^2 \frac{s^2}{M_{\pi}^4} +
84 \beta^2 \frac{s}{M_{\pi}^2} - 12 \alpha\beta \frac{s}{M_{\pi}^2}
- 53 \alpha^2 + 48 \alpha\beta - 64 \beta^2 \right)
\nonumber\\
\Delta\xi_{00}^{(3)}(s) &=& 
\frac{1}{864\pi^2}\,( 8  \beta^2 - 12 \alpha\beta + 31 \alpha^2 )
\,+\,\frac{1}{216\pi^2}\,\frac{M_{\pi}^2}{s - 4M_{\pi}^2}\,
(20 \beta^2 - 4 \alpha\beta + 11 \alpha^2)
.
\end{eqnarray}


In the case of $\pi^+ \pi^-$ scattering, we proceed as described in subsection \ref{IB_in_phases_NLO}:
one first obtains the functions [$X=S,P$]
\begin{eqnarray}
&
{\overline\xi}^{\,(1)}_{{\mbox{\tiny{$+-$}}};X} (s) \,=\, 
{\overline\xi}^{(1;{\mbox{\tiny{$\!\pm $}}})}_{{\mbox{\tiny{$+-$}}};X} (s) \,+\,
{\overline\xi}^{\,(1;{\mbox{\tiny$\Delta$}})}_{{\mbox{\tiny{$+-$}}};X} (s) 
&
\quad
 {\overline\xi}^{\,(2)}_{{\mbox{\tiny{$+-$}}};X} (s) \,=\, 
{\overline\xi}^{(2;{\mbox{\tiny{$\!\pm $}}})}_{{\mbox{\tiny{$+-$}}};X} (s) \,+\,
{\overline\xi}^{\,(2;0)}_{{\mbox{\tiny{$+-$}}};X} (s) 
\nonumber\\
&
{\overline\xi}^{\,(3)}_{{\mbox{\tiny{$+-$}}};X} (s) \,=\, 
{\overline\xi}^{(3;{\mbox{\tiny{$\!\pm $}}})}_{{\mbox{\tiny{$+-$}}};X} (s) \,+\,
{\overline\xi}^{\,(3;{\mbox{\tiny$\Delta$}})}_{{\mbox{\tiny{$+-$}}};X} (s)
&
\quad
 {\overline\xi}^{\,(4)}_{{\mbox{\tiny{$+-$}}};P} (s) \,=\, 
{\overline\xi}^{(4;{\mbox{\tiny{$\!\pm $}}})}_{{\mbox{\tiny{$+-$}}};P} (s) \,+\,
{\overline\xi}^{\,(4;{\mbox{\tiny$\Delta$}})}_{{\mbox{\tiny{$+-$}}};P} (s)
. \qquad { }
\end{eqnarray}
The expansion of the remaining functions gives
\begin{eqnarray}
2 \frac{\sigma (s)}{\sigma_{\mbox{\tiny$\Delta$}}(s)}\, L_{\mbox{\tiny$\Delta$}}(s) &=&
16\pi k_1(s) \,-\,8\pi\,\frac{\Delta_\pi}{M_{\pi}^2}
\,\left[4 k_0(s)  + k_1(s) - k_2(s) \right] \,+\,{\cal O}(\Delta_\pi^2)
\nonumber\\ 
3 \,
\frac{M_{\pi^0}^2}{\sqrt{s(s - 4M_{\pi}^2)}}\,L_{\mbox{\tiny$\Delta$}}^2(s) &=& 16\pi k_3(s)
 \,-\,8\pi\,\frac{\Delta_\pi}{M_{\pi}^2}
\,\left[ \frac{3}{2} k_1(s) -  \frac{3}{2} k_2(s) + 2 k_3(s)\right] \,+\,
{\cal O}(\Delta_\pi^2)
,
\end{eqnarray}
and
\begin{eqnarray}
\!\!\!\!
\frac{M_{\pi^0}^2}{\sqrt{s(s-4 M_{\pi}^2)}}\,
\left[
1\,+\,\frac{1}{\sigma_{\mbox{\tiny$\Delta$}}(s)}\,L_{\mbox{\tiny$\Delta$}}(s)
\,+\,\frac{M_{\pi^0}^2}{s-4 M_{\pi}^2}\,L_{\mbox{\tiny$\Delta$}}^2(s)  
\right]   &=& 16\pi k_4(s) 
\nonumber\\
&&\!\!\!\!\!\!\!
\,-\,8\pi\,\frac{\Delta_\pi}{M_{\pi}^2}
\,\left[4 k_4(s)\,-\,\frac{M_{\pi}^2}{s - 4 M_{\pi}^2}\,k_2(s) \right]\,+\,
{\cal O}(\Delta_\pi^2)
.\quad\qquad{ }
\end{eqnarray}
For the $S$ and $P$ partial-wave projections $\psi^{\mbox{\tiny{$+-$}}}_0(s)$ and $\psi^{\mbox{\tiny{$+-$}}}_1(s)$, 
this then leads to the expression (\ref{psi_+-_exp_and_psi_x_exp}), with
\begin{eqnarray}
 \Delta\xi^{(0)}_{{\mbox{\tiny{$+-$}}};S} (s) &=&  
\frac{1}{576\pi^2}\,\left( -12 \beta^2  \frac{s^2}{M_{\pi}^4} +
\frac{81}{2} \beta^2 \frac{s}{M_{\pi}^2} - \frac{63}{2} \alpha\beta \frac{s}{M_{\pi}^2}
+ 84 \alpha\beta - 40 \beta^2 - 116 \alpha^2 \right) 
\nonumber\\
\Delta\xi^{(1)}_{{\mbox{\tiny{$+-$}}};S} (s) &=& 
\frac{1}{2304\pi^2}\,\left( 6 \beta^2 \frac{s^2}{M_{\pi}^4} -
33 \beta^2 \frac{s}{M_{\pi}^2} + 15 \alpha\beta \frac{s}{M_{\pi}^2}
- 51 \alpha^2 + 16 \alpha\beta + 56 \beta^2 \right)
\nonumber\\
\Delta\xi^{(2)}_{{\mbox{\tiny{$+-$}}};S} (s) &=& 
\frac{1}{2304\pi^2}\,\left( - 6 \beta^2 \frac{s^2}{M_{\pi}^4} +
39 \beta^2 \frac{s}{M_{\pi}^2} - 45 \alpha\beta \frac{s}{M_{\pi}^2}
- 37 \alpha^2 + 64 \alpha\beta - 48 \beta^2 \right)
\nonumber\\
\Delta\xi^{(3)}_{{\mbox{\tiny{$+-$}}};S} (s) &=& 
\frac{1}{1728\pi^2}\,( 26  \beta^2 - 2 \alpha\beta + 27 \alpha^2 ) 
\nonumber\\
 \Delta\xi^{(0)}_{{\mbox{\tiny{$+-$}}};P} (s) &=&
\frac{1}{1728\pi^2}\,\left(\frac{9}{2}  \beta^2  \frac{s^2}{M_{\pi}^4} -
\frac{29}{2} \beta^2 \frac{s}{M_{\pi}^2} - \frac{61}{2} \alpha\beta \frac{s}{M_{\pi}^2}
+ 266 \alpha\beta - 62 \beta^2 - 15 \alpha^2 \right) 
\nonumber\\
\Delta\xi^{(1)}_{{\mbox{\tiny{$+-$}}};P} (s) &=& 
\frac{1}{2304\pi^2}\,\left( \frac{3}{2} \beta^2 \frac{s^2}{M_{\pi}^4} -
4 \beta^2 \frac{s}{M_{\pi}^2} - 11 \alpha\beta \frac{s}{M_{\pi}^2}
+ \alpha^2 + 68 \alpha\beta - 15 \beta^2 \right)
\nonumber\\
\Delta\xi^{(2)}_{{\mbox{\tiny{$+-$}}};P} (s) &=& 
\frac{1}{2304\pi^2}\,\left( - \frac{3}{2} \beta^2 \frac{s^2}{M_{\pi}^4} +
8 \beta^2 \frac{s}{M_{\pi}^2} +  \alpha\beta \frac{s}{M_{\pi}^2}
-  \alpha^2 - 4 \alpha\beta - 13 \beta^2 \right)
\nonumber\\
&& -\,\frac{1}{576\pi^2}\,
\frac{M_{\pi}^2}{s - 4 M_{\pi}^2}\,\left[
\alpha^2 + 4 \alpha\beta + 13 \beta^2 \right]
\nonumber\\
\Delta\xi^{(3)}_{{\mbox{\tiny{$+-$}}};P} (s) &=& 
\frac{1}{1728\pi^2}\,( 26  \beta^2 - 10 \alpha\beta + 11 \alpha^2 ) 
\nonumber\\
\Delta\xi^{(4)}_{{\mbox{\tiny{$+-$}}};P} (s) &=& 
\frac{1}{24\pi^2}\,( 5  \beta^2 + 3 \alpha\beta +  \alpha^2 ) 
.
\end{eqnarray}

Turning eventually to the inelastic $\pi^+ \pi^- \to \pi^0 \pi^0$ channel,
one first rewrites the polynomials in eq. (\ref{psi_x_0}) as $\xi^{(n)}_{x}(s) = {\overline\xi}^{(n)}_{x}(s)\,+\,
(\Delta_\pi/F_\pi^2) \delta\xi^{(n)}_{x}(s) + {\cal O}(\Delta_\pi^2)$,
with ${\overline\xi}^{(2)}_{x}(s) \,=\, {\overline\xi}^{\,(2;{\mbox{\tiny{$\pm $}}})}_{x}(s) \,+\,
{\overline\xi}^{\,(2;0)}_{x}(s)$. Next, one proceeds with the expansion of the remaining functions,
\begin{eqnarray}
2\,\frac{\lambda^{1/2}(t_{\mbox{\tiny{$-$}}}(s))}{\sqrt{s(s - 4 M_{\pi^0}^2)}}\,
{\cal L}_{\mbox{\tiny{$-$}}} (s)   &=& 16\pi k_1(s)\,-\,16\pi\,\frac{\Delta_\pi}{M_{\pi^\pm}^2}\,
\left\{k_0(s)\,+\,\frac{M_{\pi}^2}{s - 4 M_{\pi^\pm}^2}\left[
4 k_0(s)\,+\,k_1(s)\right]\right\}\,+\,{\cal O}(\Delta_\pi^2)
\nonumber\\ 
3 \,\frac{M_{\pi}^2}{\sqrt{s(s - 4M_{\pi^0}^2)}}\,{\cal L}_{\mbox{\tiny{$-$}}}^2(s) &=&
16\pi k_3(s) \,-\,16\pi\,\frac{\Delta_\pi}{s - 4 M_{\pi}^2}\left[
\frac{3}{2} k_1(s)\,+\,2k_3(s)\right] \,+\,
{\cal O}(\Delta_\pi^2) .  
\end{eqnarray}
Similar expressions with $t_{\mbox{\tiny{$-$}}}(s)$ replaced by $t_{\mbox{\tiny{$+$}}}(s)$
are of order ${\cal O}(\Delta_\pi^2)$, cf. equation (\ref{m1_m2_limit}), and thus need not be retained in the present context.
Finally, there are the two additional pieces to consider,
\begin{eqnarray}
\Delta_1 \psi^{x}_0(s) &=& 2\,\frac{M_{\pi}^4}{F_\pi^4}\,
\sqrt{\frac{s}{s - 4M_{\pi}^2}}\times\frac{\Delta_\pi}{M_{\pi}^2}\,
\frac{(-1)}{192\pi^2}\left[
\frac{1}{6}\beta^2\frac{s}{M_{\pi}^2} + \alpha\beta + 2 \beta^2\right] k_1(s) \,+\,
{\cal O}(\Delta_\pi^2) , 
\nonumber\\  
\Delta_2 \psi^{x}_0(s) &=& 
{\cal O}(\Delta_\pi^2) . 
\end{eqnarray}
Putting the various parts together then leads to the expression given in (\ref{psi_+-_exp_and_psi_x_exp}), with
\begin{eqnarray}
\Delta\xi_{x}^{(0)}(s) &=& - \lambda_1 \left(\frac{s}{M_{\pi}^2} - 2\right)\,-\,
\frac{\lambda_2}{3} \left(\frac{s}{M_{\pi}^2} - 4\right)
\nonumber\\
&& +\,
\frac{1}{1728\pi^2}\,\left(\frac{3}{2}\beta^2 \frac{s^2}{M_{\pi}^4} 
- 154 \beta^2  \frac{s}{M_{\pi}^2} + 225 \alpha\beta \frac{s}{M_{\pi}^2}
+ 352 \beta^2 - 270 \alpha\beta + 171 \alpha^2 \right)
\nonumber\\
&& -\,
\frac{1}{72\pi^2}\,\frac{M_{\pi}^2}{s - 4M_{\pi}^2}\,
(8 \beta^2 - 7 \alpha\beta - \alpha^2)
\nonumber\\ 
\Delta\xi_{x}^{(1)}(s) &=& 
\frac{1}{2304\pi^2}\,\left( -20 \beta^2 \frac{s}{M_{\pi}^2} + 3 \alpha\beta \frac{s}{M_{\pi}^2}
+ 80 \beta^2 + 36 \alpha\beta + 13 \alpha^2  \right)
\nonumber\\
&& +\,
\frac{1}{48\pi^2}\,\frac{M_{\pi}^2}{s - 4M_{\pi}^2}\,
\alpha\beta 
\nonumber\\
\Delta\xi_{x}^{(2)}(s) &=& 
\frac{1}{2304\pi^2}\,\left( - 12 \beta^2 \frac{s}{M_{\pi}^2} 
  + 57 \alpha\beta \frac{s}{M_{\pi}^2} + 16 \beta^2 - 104 \alpha\beta
+ 31 \alpha^2 \right)
\nonumber\\
\Delta\xi_{x}^{(3)}(s) &=& 
\frac{1}{864\pi^2}\,( 9 \beta^2 \frac{s}{M_{\pi}^2} - 12  \beta^2 - 4 \alpha\beta - 5 \alpha^2 )
\,+\,\frac{1}{216\pi^2}\,\frac{M_{\pi}^2}{s - 4M_{\pi}^2}\,
(8 \beta^2 -  \alpha\beta -  \alpha^2)
.
\end{eqnarray}

Let us close this appendix with a remark concerning the occurrence of contributions proportional to
${M_{\pi}^2}/(s - 4M_{\pi}^2)$ in the expressions of the functions $\Delta \xi^{(n)}(s)$. When summed 
together into the functions $\psi^{00}_0 (s)$, $\psi^x_0 (s)$, and $\psi_1^{{\mbox{\tiny{$+-$}}}} (s)$ [they are absent in 
$\psi_0^{{\mbox{\tiny{$+-$}}}} (s)$], these singularities combine to give a regular
behaviour as $s \to 4 M_\pi^2$. Just like their lowest-order counterparts $\varphi^{00}_0 (s)$, 
$\varphi^x_0 (s)$, $\varphi_0^{{\mbox{\tiny{$+-$}}}} (s)$, and $\varphi_1^{{\mbox{\tiny{$+-$}}}} (s)$,
the real parts of the partial-wave projections at next-to-leading order are regular at $s=4 M_\pi^2$, 
and the expansion in powers of $\Delta_\pi$ preserves this regularity,
see also the remark following eq. (\ref{psi_00_exp}).

\section{Expressions of the subtraction constants} \label{app:subtraction}

The phases of the form factors discussed in section \ref{IB_in_phases} involve a certain number
of subtraction constants, whose values are not fixed by the general properties underlying the
dispersive relations that form the starting point of our construction.
Two sets of parameters, $\alpha_{00}$, $\alpha_{x}$, and $\alpha_{\mbox{\tiny{$ +-$}}}$ on the one hand,
and $\beta_{x}$ and $\beta_{\mbox{\tiny{$ +-$}}}$ on the other hand, are directly related to the
parameters $\alpha$ and $\beta$ of the isospin-symmetric $\pi\pi$ amplitude, that are themselves
related to the two scattering lengths in the $S$ wave, cf. eq.  (\ref{alpha_beta_a00_a02}). 
They represent the quantities to be extracted from experiment. What we need to know, however,
is what becomes of the relations (\ref{alphabetaLO}) at next-to-leading order. For the remaining
set of parameters, the $\lambda$'s, the isospin-breaking corrections to their values
in the isospin limit, given in eqs. (\ref{lambdas_iso}), 
also need to be worked out. 
In order to obtain this information, we have performed a one-loop calculation of the form factors and scattering 
amplitudes using the ``effective" lagrangian approach described in  subsection \ref{sub_csts_at_NLO}.
The results of this calculation are shown in this appendix.

First, at next-to-leading order, the expressions (\ref{alphabetaLO}) become [the definition of the constants ${\widehat{\cal K}}^{00}_1$
and ${\widehat{\cal K}}^{00}_2$ in terms of low-energy constants introduced in \cite{Knecht:1997jw} 
is given in eq. (\ref{cal_K}) below]
\begin{eqnarray}
\frac{F_\pi^2}{F^2} \left( 4 - 3 \frac{2 {\widehat m}B}{M_{\pi^0}^2} \right) - \alpha
&=&
\frac{\Delta_\pi}{M_\pi^2} (\beta - \alpha) \,+\,
\frac{1}{96\pi^2} \frac{M_{\pi^0}^2}{ F_\pi^2} ( 11 \alpha^2 -8 \beta^2  ) \,+\, 
\frac{1}{48\pi^2} \frac{\Delta_\pi}{F_\pi^2} \beta (\beta + 5 \alpha)  
\nonumber\\
&&
+\,
\frac{1}{32\pi^2} \frac{M_{\pi^0}^2}{F_\pi^2} \left( 4 \alpha^2 - 7 \alpha \beta + 6 \beta^2 \right) L_\pi
\,-\,
3 \beta \frac{e^2}{32\pi^2} \left( {\widehat{\cal K}}^{00}_1 +  {\widehat{\cal K}}^{00}_2 \right) 
\nonumber\\
\frac{F_\pi^2}{F^2} - \beta
&=&
\frac{1}{48\pi^2} \frac{M_\pi^2}{F_\pi^2} \beta (\beta + 5 \alpha)
,
\end{eqnarray}
with $L_\pi\equiv \ln(M_\pi^2/M_{\pi^0}^2)$.
Notice the occurrence of the term $\left( \beta - \alpha \right) \frac{\Delta_\pi}{M_\pi^2}$ in the first expression.
Since $\beta - \alpha \sim {\cal O}(M_\pi^2 \times \ln M_\pi^2)$, this term
reveals a logarithmic singularity (at most) in the chiral limit. Actually, it is finite as $M_{\pi^0} \rightarrow 0$. However,
as $M_{\pi} \rightarrow 0$, it develops an infrared singular behaviour,
\begin{equation}
 \left( \beta - \alpha \right) \frac{\Delta_\pi}{M_\pi^2} \sim \frac{1}{32\pi^2} \frac{M_{\pi^0}^2}{F_\pi^2}
\left( 7 \beta - 4 \alpha \right) \alpha \ln M_\pi^2 .
\label{alpha-beta_IR}
\end{equation}
We then obtain the following identification, at one-loop precision, with the various parameters
involved in the polynomial part of these amplitudes:
\begin{eqnarray}
\alpha_{00} &=&  \alpha \,+\, \left( \beta - \alpha \right) \frac{\Delta_\pi}{M_\pi^2}
\,+\,
\frac{1}{48 \pi^2} \frac{\Delta_\pi}{F_\pi^2} \left( 5 \alpha + \beta \right) \beta
\,+\,
\frac{1}{96 \pi^2} \frac{M_{\pi^0}^2}{F_\pi^2} \left( 2 \beta^2 - 3 \alpha \beta + 10 \alpha^2 \right) L_\pi
\nonumber\\
&&
+\,
 \beta \frac{e^2}{32\pi^2} 
\left( {\widehat{\cal K}}^{00}_1 +  {\widehat{\cal K}}^{00}_2  \right)
\,-\,
\alpha \frac{e^2}{32\pi^2} {\widehat{\cal K}}^{00}_2
\nonumber\\
\alpha_x &=& \alpha \,+\, 2 \beta \frac{\Delta_\pi}{M_{\pi^0}^2}  \,+\, \left( \beta - \alpha \right) \frac{\Delta_\pi}{M_\pi^2}
\,+\, \frac{1}{48 \pi^2} \frac{\Delta_\pi}{F_\pi^2} 
\left[\beta \left(11 - 18 \frac{\Delta_\pi}{M_{\pi^0}^2} \right) - 17 \alpha \right] \beta
\nonumber\\
&&
-\,
\frac{1}{96 \pi^2} \frac{M_{\pi^0}^2}{F_\pi^2} \left[ 6 \beta^2 \left( 9 + 2 \frac{\Delta_\pi}{M_{\pi^0}^2} \right)
- \alpha \beta \left( 47 + 6 \frac{\Delta_\pi}{M_{\pi^0}^2} \right) - 4 \alpha^2 \right] L_\pi
\nonumber\\
&&
-\,
\frac{1}{24 \pi^2} \frac{M_{\pi^0}^2}{ F_\pi^2} 
\left[
\frac{M_{\pi^0}^2}{\Delta_\pi} L_\pi \,-\, 1 \right] \left( \alpha - \beta \right)
\left[
\beta \left( 4 \frac{\Delta_\pi}{M_{\pi^0}^2} + 1 \right) + \alpha \right]
\nonumber\\
&&
+\,
\frac{\Delta_\pi }{M_{\pi^0}^2} \, \beta \frac{e^2}{32\pi^2} {\widehat{\cal K}}^{x}_1 
\,+\,
 \beta \frac{e^2}{32\pi^2} 
\left( 2 {\widehat{\cal K}}^{x}_1 + {\widehat{\cal K}}^{x}_2 
+ 4 {\widehat{\cal K}}^{x}_3 - 3 {\widehat{\cal K}}^{00}_1 - 3  {\widehat{\cal K}}^{00}_2 \right)
\,-\,
 \alpha \frac{e^2}{32\pi^2} {\widehat{\cal K}}^{x}_3
\nonumber\\
\alpha_{\mbox{\tiny{$ +-$}}} &=& \alpha \,+\, 4 \beta \frac{\Delta_\pi}{M_{\pi^0}^2}  
\,+\, \left( \beta - \alpha \right) \frac{\Delta_\pi}{M_\pi^2}
\,+\, \frac{1}{16 \pi^2} \frac{\Delta_\pi}{F_\pi^2} 
\left[ \beta \left(3 - 28 \frac{\Delta_\pi}{M_{\pi^0}^2} \right) - 13 \alpha \right] \beta
\nonumber\\
&&
+\,
\frac{1}{96 \pi^2} \frac{M_{\pi^0}^2}{F_\pi^2} 
\left[ 2 \beta^2 \left( 6 \frac{\Delta_\pi^2}{M_{\pi^0}^4} - 16 \frac{\Delta_\pi}{M_{\pi^0}^2} - 45 \right)
+ \alpha \beta \left( 97 + 8 \frac{\Delta_\pi}{M_{\pi^0}^2} \right) + 2 \alpha^2 \right] L_\pi
\nonumber\\
&&
+\,
\beta \frac{e^2}{32\pi^2} 
\left( {\widehat{\cal K}}^{\mbox{\tiny{$ +-$}}}_1 + {\widehat{\cal K}}^{\mbox{\tiny{$ +-$}}}_2 
+ 4 {\widehat{\cal K}}^{\mbox{\tiny{$ +-$}}}_3 - 3 {\widehat{\cal K}}^{00}_1 - 3  {\widehat{\cal K}}^{00}_2 \right)
\,-\,
\alpha \frac{e^2}{32\pi^2} {\widehat{\cal K}}^{\mbox{\tiny{$ +-$}}}_3
\nonumber\\
&&
+\,
\frac{\Delta_\pi}{M_{\pi^0}^2} \,\beta \frac{e^2}{32\pi^2} 
\left( {\widehat{\cal K}}^{\mbox{\tiny{$ +-$}}}_1 + {\widehat{\cal K}}^{\mbox{\tiny{$ +-$}}}_2 \right)
\,+\,
 \frac{F_\pi^2}{M_{\pi^0}^2}
\left[
24 e^4 {\widehat k}^r_{14}(\mu) - \frac{9}{4\pi^2}\frac{\Delta_\pi^2}{F^4}
\ln\frac{M_\pi^2}{\mu^2}\right]
\nonumber\\
 \beta_x &=& \beta \,+\,
\frac{1}{96 \pi^2} \frac{\Delta_\pi}{ F_\pi^2} \left( 10 \alpha - 19 \beta \right) \beta
\,+\,
\frac{1}{96 \pi^2} \frac{M_{\pi^0}^2}{F_\pi^2} \left( 13 \alpha - 10 \beta \right) \beta L_\pi
\,+\,
\frac{1}{48 \pi^2} \frac{M_{\pi^0}^2}{F_\pi^2}
\left[ \frac{M_{\pi^0}^2}{\Delta_\pi} L_\pi \,-\, 1 \right] \left(4 \beta - \alpha \right) \beta
\nonumber\\
&&
+\, \beta \frac{e^2}{32\pi^2} {\widehat{\cal K}}^{x}_1 
\nonumber\\
 \beta_{\mbox{\tiny{$ +-$}}} &=& \beta \,+\,
\frac{1}{48 \pi^2} \frac{\Delta_\pi}{F_\pi^2}
\left(  5 \alpha - 20 \beta \right) \beta
\,+\,
\frac{1}{24 \pi^2} \frac{M_{\pi^0}^2}{F_\pi^2} 
\left[ \beta \left( 3 \frac{\Delta_\pi}{M_{\pi^0}^2} + 2 \right) - 2 \alpha \right] \beta L_\pi
\,+\, \beta \frac{e^2}{32\pi^2} {\widehat{\cal K}}^{\mbox{\tiny{$ +-$}}}_1
\nonumber\\
\lambda_{00}^{(1)} &=& \frac{1}{3} \left( \lambda_1 + 2 \lambda_2  \right)
\nonumber\\
\lambda_{x}^{(1)} &=& \lambda_1 \,+\, \frac{1}{96\pi^2} \left[
\frac{M_{\pi^0}^2}{\Delta_\pi} L_\pi - 1 \right] \beta^2
\nonumber\\
\lambda_{x}^{(2)} &=& \lambda_2 \,-\,\frac{1}{48\pi^2} \left[
\frac{M_{\pi^0}^2}{\Delta_\pi} L_\pi - 1 \right] \beta^2
\nonumber\\
\lambda_{\mbox{\tiny{$ +-$}}}^{(1)} &=& \lambda_1 \,+\, \frac{1}{32\pi^2}\,L_\pi \beta^2
\nonumber\\
\lambda_{\mbox{\tiny{$ +-$}}}^{(2)} &=& \lambda_2 
,
\end{eqnarray}
\begin{eqnarray}
\frac{e^2}{32\pi^2}\,{\widehat{\cal K}}^{00}_1
&=&
e^2\left[- \frac{20}{9} {\widehat k}^r_1(\mu) - \frac{20}{9} {\widehat k}^r_2(\mu) +
4 {\widehat k}_3 - 2 {\widehat k}^r_4(\mu) \right]
\,+\,
\frac{\beta}{8 \pi^2} \frac{\Delta_\pi}{F_\pi^2}\,\ln\frac{M_\pi^2}{\mu^2}
\nonumber\\
\frac{e^2}{32\pi^2}\,{\widehat{\cal K}}^{00}_2
&=&
e^2\left[ \frac{40}{9}  {\widehat k}^r_1(\mu) + \frac{40}{9} {\widehat k}^r_2(\mu) -
8 {\widehat k}_3 + 4 {\widehat k}^r_4(\mu) -
\frac{20}{9} {\widehat k}^r_5(\mu) - \frac{20}{9} {\widehat k}^r_6(\mu) -
\frac{4}{9} {\widehat k}_7
\right]
\,-\,
\frac{3\beta}{16 \pi^2}\,\frac{\Delta_\pi}{F_\pi^2}\,\ln\frac{M_\pi^2}{\mu^2}
\nonumber\\
\frac{e^2}{32\pi^2} {\widehat{\cal K}}^{x}_1
&=&
e^2\left[- \frac{40}{9} {\widehat k}^r_1(\mu) + \frac{32}{9} {\widehat k}^r_2(\mu) -
8 {\widehat k}_3 + 4 {\widehat k}^r_4(\mu) \right]
\,-\,
\frac{\beta}{4\pi^2} \frac{\Delta_\pi}{F_\pi^2}\,\ln\frac{M_\pi^2}{\mu^2}
\nonumber\\
\frac{e^2}{32\pi^2} {\widehat{\cal K}}^{x}_2
&=&
e^2\left[ -24 {\widehat k}^r_2(\mu)\,+\,
24 {\widehat k}_3 - 12 {\widehat k}^r_4(\mu) \right]
\,+\,
\frac{9\beta}{8 \pi^2} \frac{\Delta_\pi}{F_\pi^2}\,\ln\frac{M_\pi^2}{\mu^2}
\nonumber\\
\frac{e^2}{32\pi^2} {\widehat{\cal K}}^{x}_3
&=&
e^2 \left[ \frac{40}{9}  {\widehat k}^r_1(\mu) + \frac{40}{9} {\widehat k}^r_2(\mu) -
4 {\widehat k}_3 + 2 {\widehat k}^r_4(\mu) -
\frac{20}{9} {\widehat k}^r_5(\mu) + \frac{52}{9} {\widehat k}^r_6(\mu) -
\frac{4}{9} {\widehat k}_7 + 8 {\widehat k}^r_8(\mu)
\! \right]
-
\frac{\beta}{4 \pi^2} \frac{\Delta_\pi}{F_\pi^2} \ln\frac{M_\pi^2}{\mu^2}
\nonumber\\
\frac{e^2}{32\pi^2} {\widehat{\cal K}}^{\mbox{\tiny{$ +-$}}}_1
&=&
e^2\left[- \frac{20}{9} {\widehat k}^r_1(\mu) + \frac{52}{9} {\widehat k}^r_2(\mu) +
12 {\widehat k}_3 + 6 {\widehat k}^r_4(\mu) \right]
\,-\,
\frac{3\beta}{8\pi^2} \frac{\Delta_\pi}{F_\pi^2}\,\ln\frac{M_\pi^2}{\mu^2}
\nonumber\\
\frac{e^2}{32\pi^2} {\widehat{\cal K}}^{\mbox{\tiny{$ +-$}}}_2
&=&
e^2\left[- \frac{20}{3} {\widehat k}^r_1(\mu) - \frac{92}{3} {\widehat k}^r_2(\mu) -
12 {\widehat k}_3 - 6 {\widehat k}^r_4(\mu) \right]
\,+\,
\frac{9\beta}{8\pi^2} \frac{\Delta_\pi}{F_\pi^2}\,\ln\frac{M_\pi^2}{\mu^2}
\nonumber\\
\frac{e^2}{32\pi^2} {\widehat{\cal K}}^{\mbox{\tiny{$ +-$}}}_3
&=&
e^2 \left[ \frac{40}{9}  {\widehat k}^r_1(\mu) + \frac{40}{9} {\widehat k}^r_2(\mu) -
\frac{20}{9} {\widehat k}^r_5(\mu) + \frac{124}{9} {\widehat k}^r_6(\mu) -
\frac{4}{9} {\widehat k}_7 + 16 {\widehat k}^r_8(\mu)
\right]
-
\frac{5\beta}{16 \pi^2} \frac{\Delta_\pi}{F_\pi^2}\,\ln\frac{M_\pi^2}{\mu^2}
.
\label{cal_K}
\end{eqnarray}
From the one-loop expressions of the form factors, we obtain the following
information on the subtraction constants in eq. (\ref{1loopFF}): at this level
of accuracy, $a_S^{\pi^0}$ and $a_V^\pi$ are unchanged as compared to the isospin limit, 
while
\begin{eqnarray}
a_S^{\pi} - a_S^{\pi^0}
&=& \frac{\beta}{32\pi^2 F_\pi^2} \, L_\pi
.
\label{aSpi-aSpi0}
\end{eqnarray}
Finally [we have discarded contributions proportional to $e^2 (m_u - m_d)$ or to $(m_u - m_d)^2$]
\begin{equation}
F_S^{\pi}(0) \,=\, F_S^{\pi^0}(0)\left[
1 \,+\,\frac{e^2}{32\pi^2} \left(
{\widehat{\cal K}}^x_1 + \frac{1}{3} {\widehat{\cal K}}^x_2 + {\widehat{\cal K}}^x_3 - 2 {\widehat{\cal K}}^{00}_1 - {\widehat{\cal K}}^{00}_2
\right)
\,-\,
\frac{1}{8\pi^2} \frac{\Delta_\pi}{F_{\pi}^2} \beta
\,-\, \frac{1}{48\pi^2} \frac{M_{\pi^0}^2}{F_{\pi}^2} \left( 5 \beta + \alpha \right) L_\pi
\right] .
\end{equation}
As far as comparison is possible, we find agreement with the existing results in the
literature quoted at the beginning of this appendix, except in two instances. 
The expressions for charged pion scattering given in ref. \cite{KnechtNehme02} only
included corrections of first order in isospin breaking, with which we agree. The formulae
we give here are not limited to this approximation. Furthermore, we found a slight 
disagreement with the result of \cite{Kubis:1999db}
for $F_V(s)$: the radius $\langle r \rangle_V^\pi$ exhibits an infrared divergence proportional
to $\ln M_{\pi^0}^2$ as $M_{\pi^0}$ goes to zero, whereas we find that $\langle r \rangle_V^\pi$
remains finite in this limit, but diverges as  $\sim\ln M_{\pi}^2$ if we send the charged pion mass
to zero, keeping $M_{\pi^0}$ fixed. This is also what follows from our analysis in subsection \ref{Mto0_1loop}
\footnote{The authors of ref. \cite{Kubis:1999db} have informed us that they agree with us on this point.
We thank B. Kubis for correspondence regarding this issue.}.
In this context, it is important to stress that in the expressions given above, the scale-independent 
low-energy constants ${\bar \ell}_i$ are defined as
\begin{equation}
\ell_i^r(\mu) \,=\,\frac{\gamma_i}{32\pi^2}\left( {\bar \ell}_i + \ln\frac{M_\pi^2}{\mu^2}\right) ,
\end{equation}
i.e. the normalization of the logarithm is provided by the charged pion mass.
We have also checked that the results given in this appendix display infrared
behaviours in agreement with the ones obtained in section \ref{Mto0_1loop}, provided
one takes
\begin{equation}
\lambda_1 \,=\, \frac{1}{48 \pi^2} \left( {\bar \ell}_1 - \frac{4}{3} \right) 
\beta^2 
, \quad 
\lambda_2 \,=\, \frac{1}{48 \pi^2} \left( {\bar \ell}_2 - \frac{5}{6} \right) 
\beta^2 
.
\label{lambdas_l1_l2}
\end{equation}
Notice that these expressions differ from the ones given at the end of appendix \ref{app:IsoLimit}, see eq. (\ref{alpha_bet_std}),
by the factor $\beta^2$. Both are compatible at one-loop order, where one would take $\beta =1$ in the above formula.

\section{Two-loop phases in terms of scattering lengths } \label{app:scatt_lengths}

It is perfectly possible, within the framework adopted in this article,
to write down expressions that involve the scattering lengths instead of the
subthreshold parameters. This is achieved by choosing a parameterization of the lowest-order amplitudes
in terms of the scattering lengths, i.e. the value of the amplitudes at 
their respective thresholds, rather than in terms of their values at the 
center of the Mandelstam triangle, as done in the rest of the present article. The expressions (\ref{AmpTree1})
and (\ref{AmpTree2}) are thus replaced by
\begin{eqnarray}
A^{x}(s,t) &=&  16\pi \left[ a_x 
 \,+\, b_x\, \frac{s - 4 M_{\pi}^2}{F_{\pi}^2}
\right] 
\nonumber\\
A^{\mbox{\tiny{$+-$}}}(s,t) &=& 16\pi \left[ a_{\mbox{\tiny{$+-$}}}   
\,+\, b_{\mbox{\tiny{$+-$}}} \frac{s-4M_{\pi}^2}{F_{\pi}^2}\,+\, c_{\mbox{\tiny{$+-$}}} \frac{t-u}{F_{\pi}^2}
\right] 
\nonumber\\
A^{00}(s,t) &=& 16\pi a_{00}
\nonumber\\
A^{{\mbox{\tiny{$+$}}}0}(s,t) &=& 16\pi \left[ a_{{\mbox{\tiny{$+$}}} 0}
\,+\, b_{{\mbox{\tiny{$+$}}}0}\,\frac{s-(M_{\pi} + M_{\pi^0})^2}{F_{\pi}^2}\,
+\, c_{{\mbox{\tiny{$+$}}} 0}\,\frac{t-u+(M_{\pi} - M_{\pi^0})^2}{F_{\pi}^2}
\right]   
\nonumber\\
A^{\mbox{\tiny{$++$}}}(s,t) &=& 16\pi \left[  a_{\mbox{\tiny{$++$}}}  
\,+\, b_{\mbox{\tiny{$++$}}} \frac{s - 4 M_{\pi}^2}{F_{\pi}^2}
\right] 
.
\label{AmpTree_a}
\end{eqnarray}

\noindent
At this order, the relation between the two sets
of parameters is simple,
\begin{eqnarray}
a_{00}\,=\,\frac{\alpha_{00} M_{\pi^0}^2}{16\pi F_{\pi}^2}
\, ,
\quad a_{x}\,=\,\frac{\beta_{x}}{24\pi F_{\pi}^2}
( M_{\pi^0}^2 - 5 M_{\pi}^2 )\,-\,
\frac{\alpha_{x} M_{\pi^0}^2}{48\pi F_{\pi}^2} 
\, ,
\quad b_x\,=\,- \frac{\beta_{x}}{16\pi}
\nonumber
\end{eqnarray}
\begin{eqnarray}
{a}_{{\mbox{\tiny{$+$}}}0}\,=\,-\frac{\beta_{x}}{24\pi F_{\pi}^2}
( M_{\pi^0}^2 + M_{\pi}^2 )\,+\,
\frac{\alpha_{x} M_{\pi^0}^2}{48\pi F_{\pi}^2}
\ ,\quad b_{{\mbox{\tiny{$+$}}}0}\,=\,- c_{{\mbox{\tiny{$+$}}}0}\,=\,- \frac{\beta_{x}}{32\pi}
\nonumber
\end{eqnarray}
\begin{eqnarray}
a_{\mbox{\tiny{$+-$}}}\,=\,\frac{\beta_{\mbox{\tiny{$+-$}}}}{12\pi F_{\pi}^2}
\, M_{\pi}^2 \,+\,
\frac{\alpha_{\mbox{\tiny{$+-$}}} M_{\pi^0}^2}{24\pi F_{\pi}^2} 
\, ,
\quad b_{\mbox{\tiny{$+-$}}}\,=\,c_{\mbox{\tiny{$+-$}}}\,=\, \frac{\beta_{\mbox{\tiny{$+-$}}}}{32\pi}
\nonumber
\end{eqnarray}
\begin{eqnarray}
{a}_{\mbox{\tiny{$++$}}}\,=\,-\frac{\beta_{\mbox{\tiny{$+-$}}}}{6\pi F_{\pi}^2}
\, M_{\pi}^2 \,+\,
\frac{ \alpha_{\mbox{\tiny{$+-$}}} M_{\pi^0}^2}{24\pi F_{\pi}^2}
\, ,
\quad b_{\mbox{\tiny{$++$}}}\,=\, - \frac{\beta_{\mbox{\tiny{$+-$}}}}{16\pi}
.
\end{eqnarray}

\noindent
The quantities $a = a_x, a_{\mbox{\tiny{$+-$}}}, a_{00}$ etc., are
scattering lengths to the extent that the tree-level amplitudes
(\ref{AmpTree_a}) satisfy
\begin{equation}\label{eq:threshold}
{\mbox{Re}}\,A(s,t,u)\big\vert_{\mbox{\scriptsize{thr}}}\,=\,16\pi a
.
\end{equation}
The parameters $a$ will keep their meaning up to next-to-next-to-leading
order if the above relation still holds for the two-loop amplitudes.
This can be achieved upon adjusting the subtraction polynomials accordingly.
In practice, this is done through the following choice:
\begin{eqnarray}
P^{00}(s,t,u) &=& 16\pi a_{00} \,-\,w_{00} 
 +\,
\frac{3\lambda_{00}^{(1)}}{F_{\pi}^4}\left[
s(s-4M_{\pi^0}^2) + t(t-4M_{\pi^0}^2) + u(u-4M_{\pi^0}^2)
\right]
\nonumber\\
&& \!\!\!\!\!\!\!
+\,
\frac{3\lambda_{00}^{(2)}}{F_{\pi}^6}\left[
s(s-4M_{\pi^0}^2)(s-2M_{\pi^0}^2) + t(t-4M_{\pi^0}^2)(t-2M_{\pi^0}^2) + u(u-4M_{\pi^0}^2)(u-2M_{\pi^0}^2)
\right]
\nonumber
\end{eqnarray}
\begin{eqnarray}
P^{x}(s,t,u) &=& 16\pi a_x \,+\,w_x
 \,+\,16\pi b_x\, \frac{s - 4 M_{\pi}^2}{F_{\pi}^2}
 -\,
\frac{\lambda_{x}^{(1)}}{F_{\pi}^4}\,s(s-4M_{\pi}^2)
\nonumber\\
&& \!\!\!\!\!\!\!\!\!\!\!\!\!
 -\,
\frac{\lambda_{x}^{(2)}}{F_{\pi}^4}\left[
(t + M_{\pi}^2 - M_{\pi^0}^2)(t - 3 M_{\pi}^2 - M_{\pi^0}^2) + 
(u + M_{\pi}^2 - M_{\pi^0}^2)(u - 3 M_{\pi}^2 - M_{\pi^0}^2)
\right]   
\nonumber\\
&& \!\!\!\!\!\!\!\!\!\!\!\!\!
 -\,
\frac{\lambda_{x}^{(3)}}{F_{\pi}^6}\,2s(s-4 M_{\pi}^2)(s - M_{\pi}^2 -  M_{\pi^0}^2)  
\nonumber\\
&& \!\!\!\!\!\!\!\!\!\!\!\!\!
 -\,
\frac{\lambda_{x}^{(4)}}{F_{\pi}^6}\left[
(t + M_{\pi}^2 - M_{\pi^0}^2) \lambda (t) 
\,+\,
(u + M_{\pi}^2 - M_{\pi^0}^2) \lambda (u)  
\right]   
\nonumber
\end{eqnarray}

\begin{eqnarray}
P^{{\mbox{\tiny{$+$}}}0}(s,t,u) &=& 16\pi a_{{\mbox{\tiny{$+$}}}0} \,-\, w_{{\mbox{\tiny{$+$}}}0} 
\,+\, 16\pi b_{{\mbox{\tiny{$+$}}}0}\,\frac{s-(M_{\pi} + M_{\pi^0})^2}{F_{\pi}^2}\,
+\,16\pi c_{{\mbox{\tiny{$+$}}}0}\,\frac{t-u+(M_{\pi} - M_{\pi^0})^2}{F_{\pi}^2}
\nonumber\\
&& \!\!\!\!\!\!\!\!\!\!\!\!\!
+\,
\frac{\lambda_{x}^{(1)}}{F_{\pi}^4}\,t (t - 4M_{\pi}^2)  
 \,+\,
\frac{\lambda_{x}^{(2)}}{F_{\pi}^4}\left[
\lambda (s)\,+\,\lambda (u) 
\right]   
\nonumber\\
&& \!\!\!\!\!\!\!\!\!\!\!\!\!
 +\,
\frac{\lambda_{x}^{(3)}}{F_{\pi}^6}\,2t (t - 4M_{\pi}^2)(t - M_{\pi}^2 - M_{\pi^0}^2) 
 +\,
\frac{\lambda_{x}^{(4)}}{F_{\pi}^6}\left[
 (s + M_{\pi}^2 - M_{\pi^0}^2) \lambda (s)\,+\,
 (u + M_{\pi}^2 - M_{\pi^0}^2) \lambda (u) 
\right]   
\nonumber
\end{eqnarray}

\begin{eqnarray}
P^{\mbox{\tiny{$+-$}}}(s,t,u) &=& 16\pi a_{\mbox{\tiny{$+-$}}} \,-\, w_{\mbox{\tiny{$+-$}}}  
\,+\,16\pi b_{\mbox{\tiny{$+-$}}} \frac{s-4M_{\pi}^2}{F_{\pi}^2}\,+\, 16\pi c_{\mbox{\tiny{$+-$}}} \frac{t-u}{F_{\pi}^2}
\nonumber\\
&& \!\!\!\!\!\!\!\!\!\!\!\!\!
+\,\frac{\lambda_{\mbox{\tiny{$+-$}}}^{(1)} + \lambda_{\mbox{\tiny{$+-$}}}^{(2)}}{F_{\pi}^4}
\left[s ( s - 4 M_{\pi}^2 )\,+\, t( t - 4 M_{\pi}^2 )\right] \,+\,
\frac{2 \lambda_{\mbox{\tiny{$+-$}}}^{(2)}}{F_{\pi}^4}\,u( u - 4 M_{\pi}^2 )
\nonumber\\
&& \!\!\!\!\!\!\!\!\!\!\!\!\!
+\,\frac{\lambda_{\mbox{\tiny{$+-$}}}^{(3)} + \lambda_{\mbox{\tiny{$+-$}}}^{(4)}}{F_{\pi}^6}
\left[s ( s - 4 M_{\pi}^2 ) (s - 2 M_{\pi}^2) \,+\, 
t ( t - 4 M_{\pi}^2 ) (t - 2 M_{\pi}^2)
\right]
+\,\frac{2 \lambda_{\mbox{\tiny{$+-$}}}^{(4)}}{F_{\pi}^6}\,u ( u - 4 M_{\pi}^2 ) (u - 2 M_{\pi}^2)
\nonumber
\end{eqnarray}

\begin{eqnarray}
P^{\mbox{\tiny{$++$}}}(s,t,u) &=& 16\pi a_{\mbox{\tiny{$++$}}} \,-\, w_{\mbox{\tiny{$++$}}}  
\,+\,16\pi b_{\mbox{\tiny{$++$}}} \frac{s - 4 M_{\pi}^2}{F_{\pi}^2}
\nonumber\\
&& \!\!\!\!\!\!\!\!\!\!\!\!\!
+\,\frac{\lambda_{\mbox{\tiny{$+-$}}}^{(1)} + \lambda_{\mbox{\tiny{$+-$}}}^{(2)}}{F_{\pi}^4}
\left[t ( t - 4 M_{\pi}^2 )\,+\, u( u - 4 M_{\pi}^2 )\right] \,+\,
\frac{2 \lambda_{\mbox{\tiny{$+-$}}}^{(2)}}{F_{\pi}^4}\,s( s - 4 M_{\pi}^2 )
\nonumber\\
&& \!\!\!\!\!\!\!\!\!\!\!\!\!
+\,\frac{\lambda_{\mbox{\tiny{$+-$}}}^{(3)} + \lambda_{\mbox{\tiny{$+-$}}}^{(4)}}{F_{\pi}^6}
\left[ t ( t - 4 M_{\pi}^2 ) (t - 2 M_{\pi}^2)\,+\,
u ( u - 4 M_{\pi}^2 ) (u - 2 M_{\pi}^2) 
\right]
+\,\frac{2 \lambda_{\mbox{\tiny{$+-$}}}^{(4)}}{F_{\pi}^6}\,s ( s - 4 M_{\pi}^2 ) (s - 2 M_{\pi}^2) 
\end{eqnarray}
where we have subtracted the values of the one-loop integrals at the appropriate kinematical points
to ensure eq.~(\ref{eq:threshold})
\begin{eqnarray}
w_{00} &=& {\mbox{Re}}\,
\left[W^{00}_0(4M_{\pi^0}^2)\,+\, W^{00}_0(0)\,+\, W^{00}_0(0)\right]
\nonumber\\
w_x &=& {\mbox{Re}}\,\left[
W^x_{0}(4M_{\pi}^2) + 2 W^{{\mbox{\tiny{$+$}}}0}_{0}(M_{\pi^0}^2 - M_{\pi}^2)
+ 6 (5 M_{\pi}^2 - M_{\pi^0}^2) W^{{\mbox{\tiny{$+$}}}0}_{1}(M_{\pi^0}^2 - M_{\pi}^2)
\right]
\nonumber\\
w_{{\mbox{\tiny{$+$}}}0} &=& 
{\mbox{Re}}\,\left[ W^{{\mbox{\tiny{$+$}}}0}_{0}((M_{\pi^0} + M_{\pi})^2)
- 3 (M_{\pi^0} - M_{\pi})^2 W^{{\mbox{\tiny{$+$}}}0}_{1}((M_{\pi^0} + M_{\pi})^2)
\right]
\nonumber\\
&& 
\!\!\!\!\!\!
+\,{\mbox{Re}}\,\left[ W^{{\mbox{\tiny{$+$}}}0}_{0}((M_{\pi^0} - M_{\pi})^2)
- 3 (M_{\pi^0} + M_{\pi})^2 W^{{\mbox{\tiny{$+$}}}0}_{1}((M_{\pi^0} - M_{\pi})^2)
\right] 
\,+\, {\mbox{Re}}\,W^x_0(0) 
\nonumber\\
w_{\mbox{\tiny{$+-$}}} &=& {\mbox{Re}}\left[W^{\mbox{\tiny{$++$}}}_0(0) 
\,+\,
W^{\mbox{\tiny{$+-$}}}_{0}(4 M_{\pi}^2) 
\,+\, W^{\mbox{\tiny{$+-$}}}_{0}(0) \,+\, 12 M_{\pi}^2 W^{\mbox{\tiny{$+-$}}}_{1}(0)\right]
\nonumber\\
w_{\mbox{\tiny{$++$}}} &=& {\mbox{Re}}\,\left[ 2 W^{\mbox{\tiny{$+-$}}}_{0}(0) 
\,+\, W^{\mbox{\tiny{$++$}}}_{0}(4 M_{\pi}^2)
\,-\,24 M_{\pi}^2 W^{\mbox{\tiny{$+-$}}}_{1}(0)
\right]
.
\end{eqnarray}

\noindent
These expressions should involve the same number (fifteen) of independent subtraction constants (among them now
the scattering lengths) as the ones given in eqs. (\ref{polynomials_P}).
This means that there exist six relations between the twenty-one parameters occurring in the above polynomials,
which stem from crossing symmetry [by construction, the unitarity parts 
satisfy separately the crossing relations], $P^{x}(t,s,u)+P^{{\mbox{\tiny{$+$}}}0}(s,t,u)=0$
and $P^{\mbox{\tiny{$+-$}}}(u,t,s)-P^{\mbox{\tiny{$++$}}}(s,t,u)=0$. This yields
\begin{eqnarray}
b_{{\mbox{\tiny{$+$}}}0} \,+\, c_{{\mbox{\tiny{$+$}}}0}\,=\,0\ ,\quad b_x\,-\,2 b_{{\mbox{\tiny{$+$}}}0}\,=\, 0
\ ,\quad b_{\mbox{\tiny{$+-$}}} \,-\, c_{\mbox{\tiny{$+-$}}}\,=\,0\ ,
\quad b_{\mbox{\tiny{$++$}}}\,+\,2 b_{\mbox{\tiny{$+-$}}}\,=\,0 
,
\end{eqnarray}
and 
\begin{eqnarray}
16\pi \left( a_{x} + a_{{\mbox{\tiny{$+$}}}0} - 4\,\frac{M_{\pi}^2}{F_\pi^2}\,b_{x} \right) &=& 
w_{{\mbox{\tiny{$+$}}}0}\,-\,w_{x}
\,-\,8\,\frac{M_{\pi}^2 (M_{\pi}^2 - M_{\pi^0}^2)}{F_\pi^4}\,\lambda_{{\mbox{\tiny{$+$}}}0}^{(2)}
\nonumber\\
16\pi \left( a_{\mbox{\tiny{$+-$}}} - a_{\mbox{\tiny{$++$}}} + 4\,\frac{M_{\pi}^2}{F_\pi^2}\,b_{\mbox{\tiny{$++$}}} \right) &=& 
w_{\mbox{\tiny{$+-$}}}\,-\,w_{\mbox{\tiny{$++$}}}
.
\end{eqnarray}
For the time being, it is convenient not to make use of the two last relations, and to treat all the $S$-wave
scattering lengths as independent. Notice also that in the chiral counting the scattering 
lengths are of order ${\cal O}(E^2)$, whereas $b_{x}$,  $b_{\mbox{\tiny{$+-$}}}$ and $b_{\mbox{\tiny{$++$}}}$ are of order 
${\cal O}(E^0)$. One may now repeat the computation of the relevant partial-wave projections
starting with the expressions of the lowest-order amplitudes in terms of the scattering lengths 
$a_i$ and effective range parameters $b_i$ $(i= 00 , \pm 0 , x , +- , ++)$, considered as independent
quantities, following the procedure outline in sec.~\ref{general} and fig.~\ref{iterconst}. The results can still be brought 
into the representations (\ref{psi_00}), (\ref{psi_+-_0}),
(\ref{psi_+-_1}), or (\ref{psi_x_0}), but the expressions of the polynomials involved are different from
the ones given in appendix \ref{app:polynomials}. For the scattering of neutral pions, the polynomials for $\psi^{00}_0(s)$
 now read
\begin{eqnarray}
\xi_{00}^{(0)}(s) &=& 
\frac{\lambda_{00}^{(1)}}{2 M_\pi^4}\,
({5}s\,+\,{4}M_{\pi^0}^2) (s\,-\,4M_{\pi^0}^2)\,-\,128 \pi^2 \frac{F_\pi^4}{M_\pi^4}
\left( a_x - 4 b_x\frac{\Delta_\pi}{F_\pi^2}\right)^2 {\mbox{Re}}\,{\bar J}(4 M_{\pi^0}^2) 
\nonumber\\
&+&
\frac{8 b_x^2}{9 M_\pi^4}\left( 32 s^2 - 112 s M_{\pi^0}^2 + 39 s M_{\pi}^2 
+ 224 M_{\pi^0}^4 -732 M_{\pi^0}^2 M_{\pi}^2 + 1260 M_{\pi}^4\right)
\nonumber\\
&&
\!\!\!\!\!
-\,
\frac{8 a_x b_x F_\pi^2}{M_\pi^4}\left( s - 20 M_{\pi^0}^2 + 68 M_{\pi}^2 \right) \,+\,
8\,\frac{F_\pi^4}{M_\pi^4} \left( 3 a_{00}^2 + 8 a_x^2 \right)
\nonumber\\
\xi^{(1;0)}_{00}(s) &=& 4 a_{00}^2 \,\frac{F_\pi^4}{M_\pi^4}
\nonumber\\
\xi^{(1;{\mbox{\tiny$\nabla$}})}_{00}(s) &=&
\frac{8 b_x^2}{3 M_\pi^4}\left( s^2  - 8 s M_{\pi^0}^2 + 13 s M_{\pi}^2 
+ 16 M_{\pi^0}^4 - 52 M_{\pi^0}^2 M_{\pi}^2 + 66 M_{\pi}^4\right)
\nonumber\\
&&
\!\!\!\!\!
-\,
\frac{8 a_x b_x F_\pi^2}{M_{\pi}^4}\left( s - 4 M_{\pi^0}^2 + 10 M_{\pi}^2 \right) \,+\, 8 a_x^2 \,\frac{F_\pi^4}{M_\pi^4}
\nonumber\\
\xi^{(2;0)}_{00}(s) &=& 8 a_{00}^2 \,\frac{F_\pi^4}{M_\pi^4}
\nonumber\\
\xi^{(2;{\mbox{\tiny{$\pm$}}})}_{00}(s) &=& 4 \,\frac{F_\pi^4}{M_\pi^4}
\left[
b_x \,\frac{ s - 4 M_{\pi}^2}{F_\pi^2} \,+\, a_x
\right]^2
\nonumber\\ 
\xi^{(3;0)}_{00}(s) &=& -\,\frac{8}{3}\, a_{00}^2 \,\frac{F_\pi^4}{M_\pi^4}
\nonumber\\
\xi^{(3;{\mbox{\tiny$\nabla$}})}_{00}(s) &=&  -\,\frac{160}{3}\, b_x^2
\,+\,32 a_x b_x\,\frac{F_{\pi}^2}{M_\pi^2}\,-\,\frac{16}{3}\, a_x^2 \,\frac{F_\pi^4}{M_\pi^4}
.
\end{eqnarray}

For the scattering of charged pions, the polynomials involved in the expression for $\psi^{\mbox{\tiny{$+-$}}}_0 (s)$ ($S$-wave) read
\begin{eqnarray}
\xi_{{\mbox{\tiny{$+-$}}};S}^{(0)}(s) &=& \frac{\lambda_{\mbox{\tiny{$+-$}}}^{(1)}}{3 M_{\pi}^4}\,
(s - 4 M_{\pi}^2)(2 s + M_{\pi}^2)
+
\frac{\lambda_{\mbox{\tiny{$+-$}}}^{(2)} }{M_{\pi}^4}\,(s - 4 M_{\pi}^2)(s + M_{\pi}^2) 
- 64 \pi^2 \frac{F_\pi^4}{M_\pi^4}\, a_x^2 \, {\mbox{Re}}\,{\bar J}_{0}(4 M_{\pi}^2) 
\nonumber\\
&&
+\,\frac{4s^2}{9 M_\pi^4} \left(25 b_x^2 + 7 b_{\mbox{\tiny{$++$}}}^2 + \frac{119}{3}b_{\mbox{\tiny{$+-$}}}^2\right)
\,+\,\frac{2 s M_{\pi^0}^2}{3 M_\pi^4}\,b_x^2 +
\frac{2 s }{9 M_\pi^2} \left(71 b_{\mbox{\tiny{$++$}}}^2 - 220 b_x^2 - \frac{1736}{3} b_{\mbox{\tiny{$+-$}}}^2 \right)
\nonumber\\
&&
+\,\frac{2 s F_\pi^2}{M_\pi^4} \left(3 a_x b_x - 5 b_{\mbox{\tiny{$++$}}} a_{\mbox{\tiny{$++$}}} + 
6 b_{\mbox{\tiny{$+-$}}} a_{\mbox{\tiny{$+-$}}} \right)
 +\frac{32}{9} \left(29 b_{\mbox{\tiny{$++$}}}^2 + 59 b_x^2 + \frac{448}{3} b_{\mbox{\tiny{$+-$}}}^2 \right)
 -\,\frac{8 M_{\pi^0}^4}{ M_\pi^4}\,b_x^2 + \,\frac{88 M_{\pi^0}^2}{3 M_\pi^2}\,b_x^2
\nonumber\\
&&
-\,\frac{8 F_{\pi}^2}{M_\pi^2} \left(8 a_{\mbox{\tiny{$++$}}} b_{\mbox{\tiny{$++$}}} + 15 a_x b_x + 
32 a_{\mbox{\tiny{$+-$}}} b_{\mbox{\tiny{$+-$}}} \right)
-\,\frac{8 M_{\pi^0}^2 F_\pi^2}{M_\pi^4}\,a_x b_x \,+\, 4 \left( 3 a_{\mbox{\tiny{$++$}}}^2 + 5 a_x^2 + 
6 a_{\mbox{\tiny{$+-$}}}^2 \right)\frac{F_\pi^4}{M_\pi^4}
\nonumber\\
\xi^{(1;{\mbox{\tiny{$\pm$}}})}_{{\mbox{\tiny{$+-$}}};S}(s)  &=& \frac{2s^2}{9 M_\pi^4} 
\left(3 b_{\mbox{\tiny{$++$}}}^2 + 2 b_{\mbox{\tiny{$+-$}}}^2\right) +
\,\frac{2 s }{9 M_\pi^2} \left(15 b_{\mbox{\tiny{$++$}}}^2 + 4 b_{\mbox{\tiny{$+-$}}}^2 \right)
-\,\frac{2 s F_\pi^2}{M_\pi^4} \left( b_{\mbox{\tiny{$++$}}} a_{\mbox{\tiny{$++$}}} 
+ 2 b_{\mbox{\tiny{$+-$}}} a_{\mbox{\tiny{$+-$}}} \right)
\nonumber\\
&&
 +\,\frac{20 }{3 }
\left(3 b_{\mbox{\tiny{$++$}}}^2 + 8 b_{\mbox{\tiny{$+-$}}}^2 \right)
-\, \frac{12 F_{\pi}^2}{M_\pi^2} \left(a_{\mbox{\tiny{$++$}}} b_{\mbox{\tiny{$++$}}} 
+  2 a_{\mbox{\tiny{$+-$}}} b_{\mbox{\tiny{$+-$}}} \right)
\,+\, 2 \left( a_{\mbox{\tiny{$++$}}}^2 + 2 a_{\mbox{\tiny{$+-$}}}^2 \right) \frac{F_\pi^4}{M_\pi^4}
\nonumber\\
\xi^{(1;{\mbox{\tiny$\Delta$}})}_{{\mbox{\tiny{$+-$}}}; S}(s) &=& \frac{4s^2}{3 M_\pi^4} \, b_x^2 +\,
\frac{2 s (M_{\pi^0}^2 + 4 M_{\pi}^2)}{3 M_\pi^4}\,b_x^2 -\,\frac{2 s F_\pi^2}{M_\pi^4}\,a_x b_x
+ \,\frac{4 (8 M_{\pi}^4 + 10 M_{\pi}^2 M_{\pi^0}^2 - 3 M_{\pi^0}^4)}{3 M_\pi^4} \, b_x^2
\nonumber\\
&&
-\, \frac{4 F_\pi^2 (2 M_{\pi}^2 + M_{\pi^0}^2)}{M_\pi^4}\,a_x b_x \,+\, 2 a_x^2\,\frac{F_\pi^4}{M_\pi^4}
\nonumber\\
\xi^{(2;0)}_{{\mbox{\tiny{$+-$}}};S}(s) &=& 
2\,\frac{F_\pi^4}{M_\pi^4} \left[b_{x}\,\frac{s - 4 M_{\pi}^2}{F_\pi^2} 
\,+\ a_{x}
\right]^2
\nonumber\\
\xi^{(2;{\mbox{\tiny{$\pm$}}})}_{{\mbox{\tiny{$+-$}}};S}(s) &=& 
4\,\frac{F_\pi^4}{M_\pi^4} \left[b_{\mbox{\tiny{$+-$}}}\,\frac{s - 4 M_{\pi}^2}{F_\pi^2} 
\,+\ a_{\mbox{\tiny{$+-$}}}
\right]^2
\nonumber\\
\xi^{(3;{\mbox{\tiny{$\pm$}}})}_{{\mbox{\tiny{$+-$}}};S}(s) &=& \frac{16 s}{3 M_\pi^2} \, b_{\mbox{\tiny{$+-$}}}^2 
\,-\, \frac{40 }{9 }
\left(3 b_{\mbox{\tiny{$++$}}}^2 + 8 b_{\mbox{\tiny{$+-$}}}^2 \right)
\,+\,\frac{8 F_{\pi}^2}{M_\pi^2} \left(a_{\mbox{\tiny{$++$}}} b_{\mbox{\tiny{$++$}}} +  
2 a_{\mbox{\tiny{$+-$}}} b_{\mbox{\tiny{$+-$}}} \right)
-\, \frac{4}{3} \left( a_{\mbox{\tiny{$++$}}}^2 + 2 a_{\mbox{\tiny{$+-$}}}^2 \right) \frac{F_\pi^4}{M_\pi^4}
\nonumber\\
\xi^{(3;{\mbox{\tiny$\Delta$}})}_{{\mbox{\tiny{$+-$}}};S}(s) &=&  
-\,\frac{4}{3} \left[
2 \,\frac{M_{\pi^0}^4 - 4 M_{\pi^0}^2 M_{\pi}^2 + 8 M_{\pi}^4}{ M_\pi^4}\,b_x^2
\,+\, 2 \,\frac{(M_{\pi^0}^2 - 4 M_{\pi}^2)F_\pi^2 }{ M_\pi^4}\, a_x b_x \,+\, a_x^2\,\frac{F_\pi^4}{M_\pi^4}
\right]
,
\end{eqnarray}
while for the $P$-wave contribution, $\psi^{\mbox{\tiny{$+-$}}}_1 (s)$, we obtain
\begin{eqnarray}
\xi_{{\mbox{\tiny{$+-$}}};P}^{(0)}(s) &=& 
- \frac{\lambda_{\mbox{\tiny{$+-$}}}^{(1)} - \lambda_{\mbox{\tiny{$+-$}}}^{(2)}}{12 M_{\pi}^4}\,s\left(s-4 M_{\pi}^2\right) 
+\,\frac{s^2}{18 M_\pi^4} \left(25 b_{\mbox{\tiny{$++$}}}^2 - 25 b_x^2 - \frac{16}{3}b_{\mbox{\tiny{$+-$}}}^2\right)
-\,
\frac{4 s }{3 M_\pi^2} \left(\frac{7}{6} b_{\mbox{\tiny{$++$}}}^2 -  b_x^2 + \frac{208}{9} b_{\mbox{\tiny{$+-$}}}^2 \right)
\nonumber\\
&&
+\,\frac{2 s M_{\pi^0}^2}{9 M_\pi^4}\,b_x^2
+\,\frac{22 s F_\pi^2}{9 M_\pi^4} \left( a_x b_x - b_{\mbox{\tiny{$++$}}} a_{\mbox{\tiny{$++$}}} 
+ 2 b_{\mbox{\tiny{$+-$}}} a_{\mbox{\tiny{$+-$}}} \right)
 +\,\frac{8}{9}
\left( 55 b_x^2 + 250 b_{\mbox{\tiny{$+-$}}}^2 - \frac{291}{4} b_{\mbox{\tiny{$++$}}}^2 \right) 
\nonumber\\
&&-\,\frac{14 M_{\pi^0}^4}{3 M_\pi^4}\,b_x^2
+\, \frac{184 M_{\pi^0}^2 }{9 M_\pi^2}\,b_x^2
+\,\frac{8 F_{\pi}^2}{9 M_\pi^2} \left(35 a_{\mbox{\tiny{$++$}}} b_{\mbox{\tiny{$++$}}} 
- 29 a_x b_x - 70 a_{\mbox{\tiny{$+-$}}} b_{\mbox{\tiny{$+-$}}} \right)
-\,\frac{16 M_{\pi^0}^2 F_\pi^2}{3 M_\pi^4}\,a_x b_x 
\nonumber\\
&&
-\, 2 \left( a_{\mbox{\tiny{$++$}}}^2 - a_x^2 - 2 a_{\mbox{\tiny{$+-$}}}^2 \right) \frac{F_\pi^4}{M_\pi^4} 
\nonumber\\
\xi^{(1;{\mbox{\tiny{$\pm$}}})}_{{\mbox{\tiny{$+-$}}};P}(s)  &=& \frac{ s^2}{3 M_\pi^4}
\left( b_{\mbox{\tiny{$++$}}}^2
-\,\frac{4}{3}\,b_{\mbox{\tiny{$+-$}}}^2\right)
  -\,\frac{8 s }{3 M_\pi^2}\,b_{\mbox{\tiny{$+-$}}}^2
+\,\frac{2 s F_\pi^2}{3 M_\pi^4}\,(2a_{\mbox{\tiny{$+-$}}} b_{\mbox{\tiny{$+-$}}} 
- a_{\mbox{\tiny{$++$}}} b_{\mbox{\tiny{$++$}}})
- 10 b_{\mbox{\tiny{$++$}}}^2
+\,
\frac{112}{3}\,b_{\mbox{\tiny{$+-$}}}^2
\nonumber\\
&&
-\,\frac{4 F_{\pi}^2 }{M_\pi^2}\,(2a_{\mbox{\tiny{$+-$}}} b_{\mbox{\tiny{$+-$}}} 
- a_{\mbox{\tiny{$++$}}} b_{\mbox{\tiny{$++$}}})
\nonumber\\
\xi^{(1;{\mbox{\tiny$\Delta$}})}_{{\mbox{\tiny{$+-$}}}; P}(s) &=& -\frac{ s^2}{3 M_\pi^4}\,b_{x}^2
+\,\frac{2 s F_\pi^2}{3 M_\pi^4}\,a_x b_x +\, 
2 b_x^2\,\frac{8 M_{\pi}^4 + 8 M_{\pi}^2 M_{\pi^0}^2 - M_{\pi^0}^4}{3 M_\pi^4}
\,-\, 4 a_x b_x\,\frac{(2 M_{\pi}^2 + M_{\pi^0}^2) F_\pi^2}{3 M_\pi^4}
\nonumber\\
\xi^{(2;{\mbox{\tiny{$\pm$}}})}_{{\mbox{\tiny{$+-$}}};P}(s) &=& \frac{4}{9}\,
\left(\frac{s}{M_\pi^2} \,-\, 4 \right)^2 b_{\mbox{\tiny{$+-$}}}^2 
\nonumber\\
\xi^{(3;{\mbox{\tiny{$\pm$}}})}_{{\mbox{\tiny{$+-$}}};P}(s) &=& \frac{16 s}{3 M_\pi^2}\,b_{\mbox{\tiny{$+-$}}}^2 
+\,\frac{40}{3}\,b_{\mbox{\tiny{$++$}}}^2 \,-\,\frac{256 }{9}\,b_{\mbox{\tiny{$+-$}}}^2 
+ \,8\,\frac{F_{\pi}^2}{ M_\pi^2}\,(2 a_{\mbox{\tiny{$+-$}}} b_{\mbox{\tiny{$+-$}}} 
- a_{\mbox{\tiny{$++$}}} b_{\mbox{\tiny{$++$}}})
+\,\frac{4}{3}\left( a_{\mbox{\tiny{$++$}}}^2  - 2 a_{\mbox{\tiny{$+-$}}}^2 \right) \frac{F_\pi^4}{M_\pi^4}
\nonumber\\
\xi^{(3;{\mbox{\tiny$\Delta$}})}_{{\mbox{\tiny{$+-$}}}; P}(s) &=& - 8 \,
\frac{8 M_{\pi}^4 - 4 M_{\pi}^2 M_{\pi^0}^2 + M_{\pi^0}^4}{3 M_\pi^4}\, b_x^2
+\, 8  \,\frac{(4 M_{\pi}^2 - M_{\pi^0}^2) F_\pi^2}{3 M_\pi^4} a_x b_x
-\, \frac{4}{3}\, a_x^2  \frac{F_\pi^4}{M_\pi^4}
\nonumber\\
\xi^{(4;{\mbox{\tiny{$\pm$}}})}_{{\mbox{\tiny{$+-$}}};P}(s) &=& 40 b_{\mbox{\tiny{$++$}}}^2 
\,-\,\frac{64 }{3 }\,b_{\mbox{\tiny{$+-$}}}^2
\,+\,32\,\frac{F_{\pi}^2}{ M_\pi^2}\,(2 a_{\mbox{\tiny{$+-$}}} b_{\mbox{\tiny{$+-$}}} 
- a_{\mbox{\tiny{$++$}}} b_{\mbox{\tiny{$++$}}})
\,+\,8 \left( a_{\mbox{\tiny{$++$}}}^2 - 2 a_{\mbox{\tiny{$+-$}}}^2 \right) \frac{F_\pi^4}{M_\pi^4}
\nonumber\\
\xi^{(4;{\mbox{\tiny$\Delta$}})}_{{\mbox{\tiny{$+-$}}}; P}(s) &=& -8 \,
\frac{16 M_{\pi}^4 - 16 M_{\pi}^2 M_{\pi^0}^2 + 5 M_{\pi^0}^4}{M_\pi^4} \,b_x^2
+\, 32 \,\frac{(2 M_{\pi}^2 - M_{\pi^0}^2) F_\pi^2}{M_\pi^4} \,a_x b_x 
-\, 8 a_x^2 \,\frac{F_\pi^4}{M_\pi^4}
.
\end{eqnarray}

Finally, in the case of a scattering involving two neutral pions and two charged pions, we obtain the expression
for the polynomials describing $\psi^x_0$
\begin{eqnarray}
\xi_{x}^{(0)}(s) &=& 
-\,\frac{\lambda_{x}^{(1)}}{2 M_{\pi}^4}\,s(s - 4 M_{\pi}^2) 
\,-\,
\frac{\lambda_{x}^{(2)} }{3 M_{\pi}^4}\,(s - 4 M_{\pi}^2)
(s + 3 M_{\pi^\pm}^2 -  M_{\pi^0}^2) 
-\,64\pi^2 \frac{F_\pi^4}{M_\pi^4}\,a_x a_{00}\,  {\mbox{Re}}\,{\bar J}_{0}(4 M_{\pi}^2)
\nonumber\\
&&
+\,\,(16\pi)^2 \left[
4\,\frac{7 M_{\pi}^4 + 2 M_{\pi}^2 M_{\pi^0}^2 -  M_{\pi^0}^4}{3 M_\pi^4}\,b_{{\mbox{\tiny{$+$}}} 0}^2
-\,8 a_{{\mbox{\tiny{$+$}}} 0} b_{{\mbox{\tiny{$+$}}} 0}\,\frac{F_{\pi}^2}{M_\pi^2}
\,+\, a_{{\mbox{\tiny{$+$}}} 0}^2\,\frac{F_\pi^4}{M_\pi^4}
\right]
{\mbox{Re}}\,
{\bar J}_{{\mbox{\tiny{$+$}}} 0}(- \Delta_\pi)
\nonumber\\
&&
+\, \frac{ \Delta_\pi^2}{3 M_{\pi}^4}\,(32\pi)^2 \frac{F_\pi^4}{M_\pi^4}\, b_{{\mbox{\tiny{$+$}}} 0}^2
\,  {\mbox{Re}}\ \,   {\bar{\!\!{\bar J}}}_{{\mbox{\tiny{$+$}}} 0}(- \Delta_\pi)
+\,
\frac{8 s^2}{27 M_\pi^4} \left(54 b_x b_{{\mbox{\tiny{$+-$}}} } - 11 b_{{\mbox{\tiny{$+$}}} 0}^2 \right)
\nonumber\\
&&
+\,\frac{16 s M_{\pi^0}^2 }{27 M_\pi^4}\,b_{{\mbox{\tiny{$+$}}} 0}^2 \,+\,
\frac{16 s}{27 M_\pi^2} \left( b_{{\mbox{\tiny{$+$}}} 0}^2 - 216 b_x b_{{\mbox{\tiny{$+-$}}}} \right)
+\,\frac{8 s F_\pi^2}{M_\pi^4} \left( b_x a_{00} + 2 b_x a_{{\mbox{\tiny{$+-$}}}} + 2 a_{x} b_{{\mbox{\tiny{$+-$}}}} + 
5 a_{{\mbox{\tiny{$+$}}} 0} b_{{\mbox{\tiny{$+$}}} 0} \right)
\nonumber\\
&&
+\,32 \left(8 b_{{\mbox{\tiny{$+-$}}}} b_x - 7 b_{{\mbox{\tiny{$+$}}} 0}^2 \right) 
-\,\frac{224 M_{\pi^0}^4}{ M_\pi^4}\,b_{{\mbox{\tiny{$+$}}} 0}^2
-\,\frac{2752}{27}\,\frac{M_{\pi^0}^2 }{M_\pi^2}\,b_{{\mbox{\tiny{$+$}}} 0}^2
+\,128\,\frac{M_{\pi^0}^2 F_\pi^2}{M_\pi^4}\,a_{{\mbox{\tiny{$+$}}} 0} b_{{\mbox{\tiny{$+$}}} 0}
\nonumber\\
&&
+\,\frac{32 F_{\pi}^2}{M_\pi^2} \left(4 a_{{\mbox{\tiny{$+$}}} 0} b_{{\mbox{\tiny{$+$}}} 0} 
- 2 a_x b_{{\mbox{\tiny{$+-$}}}} 
- 2 b_{x} a_{{\mbox{\tiny{$+-$}}}} - b_x a_{00}\right)
+\, 8 \left(  a_x a_{00} - 6 a_{\pm 0}^2 \right) \frac{F_\pi^4}{M_\pi^4}
\nonumber\\
\xi_{x}^{(1)}(s) &=& -\, \frac{8 s (s + 2 M_{\pi^0}^2 )}{9 M_\pi^4}\,b_{{\mbox{\tiny{$+$}}} 0}^2  
+\,\frac{8 s F_\pi^2}{M_\pi^4} \, a_{\pm 0} b_{\pm 0}  
-\,\frac{16}{3}\, \frac{ 5 M_{\pi}^4 + 4 M_{\pi}^2 M_{\pi^0}^2 + 11 M_{\pi^0}^4}{M_\pi^4}\,b_{{\mbox{\tiny{$+$}}} 0}^2 
\nonumber\\
&&
+\,16\,\frac{(M_{\pi}^2 + 2 M_{\pi^0}^2) F_\pi^2}{M_\pi^4}\, a_{{\mbox{\tiny{$+$}}} 0} b_{{\mbox{\tiny{$+$}}} 0} 
-\, 8 \, a_{{\mbox{\tiny{$+$}}} 0}^2 \,\frac{F_\pi^4}{M_\pi^4}  
\nonumber\\
\xi_{x}^{(2;0)}(s) &=& 2 a_{00}\,\frac{F_\pi^4}{M_\pi^4}
\left[ \frac{s - 4 M_\pi^2}{F_\pi^2}\,b_x \,+\, a_x \right]
\nonumber\\
\xi_{x}^{(2;\pm)}(s) &=& 4\,\frac{F_\pi^4}{M_\pi^4}
\left[ \frac{s - 4 M_\pi^2}{F_\pi^2}\,b_x \,+\, a_x \right]
\left[ \frac{s - 4 M_\pi^2}{F_\pi^2}\,b_{\mbox{\tiny{$+-$}}} \,+\, a_{\mbox{\tiny{$+-$}}} \right]
\nonumber\\
\xi_{x}^{(3)}(s) &=& 
-\,\frac{32 s}{9 M_\pi^2} \, b_{{\mbox{\tiny{$+$}}} 0}^2 
\left(1 + \frac{M_{\pi^0}^2}{M_{\pi}^2} + \frac{M_{\pi^0}^4}{M_{\pi}^4} \right)
 +\,\frac{176 }{9 }\,b_{{\mbox{\tiny{$+$}}} 0}^2 \left(1 + \frac{M_{\pi^0}^6}{M_{\pi}^6}\right)
 \,+\,\frac{16 M_{\pi^0}^2(M_{\pi}^2 + M_{\pi^0}^2)}{ M_\pi^4}\,b_{{\mbox{\tiny{$+$}}} 0}^2  
\nonumber\\
&&
 -\,32 a_{{\mbox{\tiny{$+$}}} 0} b_{{\mbox{\tiny{$+$}}} 0}\,\frac{F_{\pi}^2}{3 M_\pi^2} 
 \left(1 + \frac{M_{\pi^0}^2}{M_{\pi}^2} + \frac{M_{\pi^0}^4}{M_{\pi}^4} \right)
 \,+\, 8 a_{{\mbox{\tiny{$+$}}} 0}^2 \left(1 + \frac{M_{\pi^0}^2}{M_{\pi}^2} \right)\frac{F_\pi^4}{M_\pi^4} 
\end{eqnarray}
In addition, eq. (\ref{psi_x_0}) involves two other contributions, one of order $O(\Delta_\pi)$ which reads
\begin{eqnarray}
16\pi \Delta_1 \psi^{x}_0(s) &=&  
 \frac{16}{F_\pi^2}
\left\{
\frac{ s}{9 F_\pi^2}\, b_x^2  \,-\, b_x a_x 
\,+\, 2 b_x^2 \, \frac{M_{\pi}^2 + M_{\pi^0}^2}{F_\pi^2}\right\}
\nonumber\\
&&
\times
\left[
\left(\sqrt{\frac{s - 4 M_{\pi}^2}{s- 4 M_{\pi^0}^2}}\,-\,1\right)
\lambda^{1/2}(t_{\mbox{\tiny $-$}}(s)) {\cal L}_{\mbox{\tiny $-$}} (s)
\,-\,
\left(\sqrt{\frac{s - 4 M_{\pi}^2}{s- 4 M_{\pi^0}^2}}\,+\,1\right)
\lambda^{1/2}(t_{\mbox{\tiny $+$}}(s)) {\cal L}_{\mbox{\tiny $+$}} (s)
\right]
.\qquad{ }
\end{eqnarray}
We do not give the explicit expression of $\Delta_2 \psi^{x}_0$, since it represents a tiny contribution of order $O(\Delta_\pi^2)$ 
which can be neglected for practical purposes, as indicated in section \ref{numerics}~\cite{wip}.

At this stage, one can follow the discussion of section \ref{IB_in_phases} and determine the isospin-breaking differences 
$\Delta\delta_0^{\pi}$, $\Delta\delta_0^{\pi^0}$, $\Delta\delta_1^{\pi}$
in terms of the different scattering lengths and effective range parameters. Even though one might hope to determine 
all these parameters from high-precision data on the different channels involved, it seems more realistic to express them in terms of the 
subthreshold parameters $\alpha_i, \beta_i,\lambda^{(n)}_i$ with $i= 00 , \pm 0 , x , +- , ++$
\begin{eqnarray}
a_{00} &=& \frac{\alpha_{00} M_{\pi^0}^2}{16\pi F_{\pi}^2} \,+\,
\frac{9}{4\pi}\,\lambda_{00}^{(1)}\,\frac{M_{\pi^0}^4}{F_\pi^4}
\,+\,\frac{1}{32 \pi}\left(\frac{\alpha_{00} M_{\pi^0}^2}{F_{\pi}^2}\right)^2 {\bar J}_0(4M_{\pi^0}^2)
\nonumber\\
&&\!\!\!\!\!
+\,\frac{1}{144 \pi F_\pi^4}
\left[2 \beta_{x}(5 M_{\pi^0}^2 - M_{\pi}^2)\,+\,\alpha_{x} M_{\pi^0}^2\right]^2
{\bar J} (4 M_{\pi^0}^2)
\nonumber\\
a_x &=& \frac{\beta_{x}}{24\pi F_{\pi}^2}
( M_{\pi^0}^2 - 5 M_{\pi}^2 )\,-\,
\frac{\alpha_{x} M_{\pi^0}^2}{48\pi F_{\pi}^2} 
\,
-\,\frac{\lambda_{x}^{(1)}}{4 \pi}\,\frac{M_{\pi}^2 (2 M_{\pi}^2 - M_{\pi^0}^2)}{F_\pi^4}
\,-\,\frac{\lambda_{x}^{(2)}}{2 \pi}\,\frac{M_{\pi}^4 }{F_\pi^4}
\nonumber\\
&&\!\!\!\!\!
-\,\frac{1}{72 \pi F_\pi^4} ( 2 \beta_{\mbox{\tiny{$+-$}}} M_{\pi}^2 + \alpha_{\mbox{\tiny{$+-$}}} M_{\pi^0}^2 )
\left[2 \beta_{x} ( 5M_{\pi}^2 - M_{\pi^0}^2) + \alpha_{x} M_{\pi^0}^2 \right]
{\bar J}(4 M_{\pi}^2)
\nonumber\\
&&\!\!\!\!\!
-\,\frac{1}{96 \pi F_\pi^4} \,\alpha_{00} M_{\pi^0}^2
\left[2 \beta_{x} ( 5 M_{\pi}^2 - M_{\pi^0}^2) + \alpha_{x} M_{\pi^0}^2 \right]
{\mbox{Re}} \,{\bar J}_{0}(4 M_{\pi}^2)
\nonumber\\
&&\!\!\!\!\!
-\,\frac{1}{72 \pi F_\pi^4} \left[\beta_{x}^2\,(M_{\pi}^4 + M_{\pi^0}^4 - 10 M_{\pi}^2 M_{\pi^0}^2)
\,+\,4 \beta_{x} \alpha_{x} M_{\pi^0}^2 ( 2 M_{\pi}^2 - M_{\pi^0}^2) \,+\, 
\alpha_{x}^2 M_{\pi^0}^4 \right] {\bar J}_{{\mbox{\tiny{$+$}}} 0} (M_{\pi^0}^2 - M_{\pi}^2)
\nonumber\\
&&\!\!\!\!\!
-\,\frac{1}{24 \pi F_\pi^4} \,\beta_{x}^2\,(M_{\pi}^2 - M_{\pi^0}^2)^2
\ {\bar{\!\!{\bar J}}}_{{\mbox{\tiny{$+$}}} 0} (M_{\pi^0}^2 - M_{\pi}^2)
\nonumber\\
{a}_{{\mbox{\tiny{$+$}}}0} &=& -\frac{\beta_{x}}{24\pi F_{\pi}^2}
( M_{\pi^0}^2 + M_{\pi}^2 )\,+\,
\frac{\alpha_{x} M_{\pi^0}^2}{48\pi F_{\pi}^2}\,+\,
\frac{1}{4 \pi F_\pi^4}\,( \lambda_{x}^{(1)} + 2 \lambda_{x}^{(2)}) M_{\pi}^2 M_{\pi^0}^2
\nonumber\\
&&\!\!\!\!\!
+\,\frac{1}{144 \pi F_\pi^4} \left[\beta_{x}^2\,(M_{\pi}^4 + M_{\pi^0}^4 - 10 M_{\pi}^2 M_{\pi^0}^2
+ 12 M_{\pi}^3 M_{\pi^0} + 12 M_{\pi} M_{\pi^0}^3)
\right.
\nonumber\\
&&\qquad
\left.
\,-\,4 \beta_{x} \alpha_{x} M_{\pi^0}^2 ( M_{\pi}^2 + M_{\pi^0}^2) \,+\, 
\alpha_{x}^2 M_{\pi^0}^4 \right] {\bar J}_{{\mbox{\tiny{$+$}}}0} \left( (M_{\pi} + M_{\pi^0})^2\right)
\nonumber\\
&&\!\!\!\!\!
+\,\frac{1}{144 \pi F_\pi^4} \left[\beta_{x}^2\,(M_{\pi}^4 + M_{\pi^0}^4 - 10 M_{\pi}^2 M_{\pi^0}^2
- 12 M_{\pi}^3 M_{\pi^0} - 12 M_{\pi} M_{\pi^0}^3)
\right.
\nonumber\\
&&\qquad
\left.
\,-\,4 \beta_{x} \alpha_{x} M_{\pi^0}^2 ( M_{\pi}^2 + M_{\pi^0}^2) \,+\, 
\alpha_{x}^2 M_{\pi^0}^4 \right] {\bar J}_{{\mbox{\tiny{$+$}}}0} \left( (M_{\pi} - M_{\pi^0})^2\right)
\nonumber\\
&&\!\!\!\!\!
+\,\frac{\beta_{x}^2}{48\pi F_{\pi}^4}\,( M_{\pi} - M_{\pi^0})^4
\, {\bar{\!\!{\bar J}}}_{{\mbox{\tiny{$+$}}}0} \!\left( (M_{\pi} + M_{\pi^0})^2\right) \,+\,
\frac{\beta_{x}^2}{48\pi F_{\pi}^4}\,( M_{\pi} + M_{\pi^0})^4
\, {\bar{\!\!{\bar J}}}_{{\mbox{\tiny{$+$}}}0} \!\left( (M_{\pi} - M_{\pi^0})^2\right)
\nonumber\\
{a}_{\mbox{\tiny{$+-$}}} &=& \frac{\beta_{\mbox{\tiny{$+-$}}}}{12\pi F_{\pi}^2}
\, M_{\pi}^2 \,+\,
\frac{\alpha_{\mbox{\tiny{$+-$}}} M_{\pi^0}^2}{24\pi F_{\pi}^2}\,+\,
\frac{1}{2 \pi F_\pi^4}\,( \lambda_{\mbox{\tiny{$+-$}}}^{(1)} + 2 \lambda_{\mbox{\tiny{$+-$}}}^{(2)}) M_{\pi}^4
\nonumber\\
&&\!\!\!\!\!
+\,\frac{1}{36\pi F_\pi^4}\,(2\beta_{\mbox{\tiny{$+-$}}} M_{\pi}^2 + 
\alpha_{\mbox{\tiny{$+-$}}} M_{\pi^0}^2 )^2 {\bar J} (4 M_{\pi}^2)
\,+\,\frac{1}{288\pi F_\pi^4}\,(8\beta_{x} M_{\pi}^2 + \alpha_{x} M_{\pi^0}^2 )^2 
{\mbox{Re}} \,{\bar J}_{0}(4 M_{\pi}^2)
\nonumber\\
{a}_{\mbox{\tiny{$++$}}} &=& -\frac{\beta_{\mbox{\tiny{$+-$}}}}{6\pi F_{\pi}^2}
\, M_{\pi}^2 \,+\,
\frac{ \alpha_{\mbox{\tiny{$+-$}}} M_{\pi^0}^2}{24\pi F_{\pi}^2}\,+\,
\frac{1}{2 \pi F_\pi^4}\,( \lambda_{\mbox{\tiny{$+-$}}}^{(1)} + 2 \lambda_{\mbox{\tiny{$+-$}}}^{(2)}) M_{\pi}^4
\,+\,\frac{1}{72\pi F_\pi^4}\,(4\beta_{\mbox{\tiny{$+-$}}} M_{\pi}^2 - \alpha_{\mbox{\tiny{$+-$}}} M_{\pi^0}^2 )^2 {\bar J}(4 M_{\pi}^2)
\,.\qquad{ }
\end{eqnarray}
These expressions can be exploited, by relying on appendix~\ref{app:subtraction} and expressing the subthreshold parameters 
$\alpha_i, \beta_i,\lambda^{(n)}_i$ with $i= 00 , \pm 0 , x , +- , ++$ in terms of the isospin-limit parameters $\alpha,\beta,\lambda^{(n)}$. 
The latter could be taken as the fundamental parameters of the analysis, but they can also be traded for the two $\pi\pi$ scattering lengths 
$a_0^0$ and $a_0^2$ (up to higher-order corrections that can be estimated using Chiral Perturbation Theory). This series of matching will be 
indeed the point of view adopted for the analysis of $K_{\ell 4}$ decays, allowing us to reexpress the isospin-breaking correction to be applied 
to the phase-shift difference in terms of the two scattering lengths $a_0^0$ and $a_0^2$~\cite{wip}.


\end{document}